# Machine Learning and Variational Algorithms for Lattice Field Theory

by

## Gurtej Kanwar

B.S.E., Massachusetts Institute of Technology (2015)
M.Eng., Massachusetts Institute of Technology (2016)

Submitted to the Department of Physics
in partial fulfillment of the requirements for the degree of

Doctor of Philosophy

at the

Massachusetts Institute of Technology

June 2021

Author . . . . . . . . . . . . . . . . . . . . . . . . . . . . . . . . . . . . . . . . . . . . . . . . . . . . . . . . . . . . . . . . . . . . . . . . .
Department of Physics
May 21, 2021

Certified by . . . . . . . . . . . . . . . . . . . . . . . . . . . . . . . . . . . . . . . . . . . . . . . . . . . . . . . . . . . . . . . . . . .
William Detmold
Associate Professor of Physics
Thesis Supervisor

Accepted by . . . . . . . . . . . . . . . . . . . . . . . . . . . . . . . . . . . . . . . . . . . . . . . . . . . . . . . . . . . . . . . . . . .
Deepto Chakrabarty
Professor of Physics
Associate Department Head

## Machine Learning and Variational Algorithms for Lattice Field Theory
### by Gurtej Kanwar

## Abstract

Discretizing fields on a spacetime lattice is the only known general and non-perturbative regulator for quantum field theory. The lattice formulation has, for example, played an important role in predicting properties of QCD in the strongly coupled regime, where perturbative methods break down. To recover information about continuum physics, parameters defining the lattice theory must be tuned toward criticality. However, Markov chain Monte Carlo (MCMC) methods commonly used to evaluate the lattice-regularized path integral suffer from critical slowing down in this limit, restricting the precision of continuum extrapolations. Further difficulties arise when computing the energies and interactions of physical states by measuring correlation functions of operators widely separated in spacetime: for most correlation functions, an exponentially severe signal-to-noise problem is encountered as the operators are taken to be widely separated, limiting the precision of calculations.

This dissertation introduces two new techniques to address these issues. First, we define a novel MCMC algorithm based on generative flow-based models. Such models utilize machine learning methods to describe efficient approximate samplers for distributions of interest. Independently drawn flow-based samples are then used as proposals in an asymptotically exact Metropolis-Hastings Markov chain. We also construct models that flexibly parameterize families of distributions while capturing symmetries of interest, including translational and gauge symmetries. By variationally optimizing the distribution selected from these families, one can maximize the efficiency of flow-based MCMC. We secondly introduce an approach to 'deform' Monte Carlo estimators based on contour deformations applied to the domain of the path integral. The deformed estimators associated with an observable give equivalent unbiased measurements of that observable, but generically have different variances. We define families of deformed manifolds for lattice gauge theories and introduce methods to efficiently optimize the choice of manifold (the 'observifold') so that the variance of the associated deformed observable is minimized. Finally, we demonstrate that flow-based MCMC can mitigate critical slowing down and observifolds can exponentially reduce variance in proof-of-principle applications to scalar $\phi^4$ theory and U(1) and SU($N$) lattice gauge theories.

Thesis Supervisor: William Detmold
Title: Associate Professor of Physics

## Acknowledgments

The trajectory I have taken through grad school, culminating in the work presented here, has been influenced and encouraged by several people, without whom I would certainly have fallen far short of where I am today. I owe a great debt of gratitude to everyone who supported me along the way.

First and foremost, the guidance and inspiration provided by my supervisor Will Detmold and co-supervisor (in all but name) Phiala Shanahan have been invaluable in my progress over these five years. Phiala and Will managed to strike that delicate balance between leaving me free to explore those topics that seemed the most interesting each day (most of them dead ends), while also guiding my focus when it was needed. The energy, humor, and wisdom that they both bring to group meetings and our joint projects has never failed to keep myself or the lattice group moving forward.

I was also fortunate to have no shortage of role models throughout grad school. My early experiences with the lattice group were shaped by many conversations with Andrew Pochinsky, who holds a good fraction of the responsibility in bringing me over to physics in the first place. I was inspired early on by the trailblazing work of the postdocs that I had the opportunity to interact with at the time (in particular Zohreh Davoudi, Mike Endres, and Phiala Shanahan), who helped make the MIT lattice group such a welcoming environment from the start. These interactions played no small role in getting me excited about physics, lattice QCD, and MIT. More recently, I have had the pleasure of working with Mike Wagman on a number of projects. From our collaborations, I have learned by example that it is always possible to bring more excitement and dedication to your work and to juggle a seemingly endless number of research endeavors without losing focus. The physics conversations we have had along the way have kept my interest in the field going strong.

While research kept me engaged, my time as a grad student would have been nothing without endless coffee breaks and late nights full of idle (and sometimes more serious) physics speculation. I owe (what is left of) my sanity at the end of this endeavor to my peers who made these moments — as well as the casual hangouts and crazy adventures outside the office — an entertaining staple of my life. In particular, I have to thank Adam Bene Watts for constantly being willing to chat about completely tangential research directions and for the many random ventures over the years.

Finally, I am grateful to my brother, Satjiv, for being a friend, an honest critic, and a reliable supporter through it all.

# Contents

# Publications Covered

## Journal Publications

## Conference Proceedings and Other Works

# Chapter 1

# Introduction

The search for a fundamental description of the underlying laws describing the physical phenomena observed in Nature has a dynamic history full of wrong turns and dead ends. Nevertheless, this description has steadily been refined over the centuries through careful experimental measurements and theoretical developments. Repeatedly, theories which once appeared fundamental were determined to in fact be effective descriptions generated by theories at a deeper level: the atoms of the atomic theory developed in the early 19$^{\text{th}}$ century were soon discovered to be divisible into electrons and a compact nucleus, the nucleus itself then turned out to be divisible into protons and neutrons, and in the mid 20$^{\text{th}}$ century it was discovered that protons and neutrons are themselves constructed of more fundamental constituents — quarks and gluons — though the strong nuclear force binding these constituents together prevents their individual extraction. In our modern understanding, the quarks and gluons are excitations of continuous *quantum fields*. The Standard Model of particle physics is a quantum field theory that represents our cumulative understanding of Nature at the smallest scales, giving a precise mathematical description not only of the quark and gluon fields but also of a number of other fields whose excitations describe all other observed particles, including leptons and the other force carriers.

The Standard Model is a spectacularly successful theory. Beyond providing explanations of measurements already known at the time of its development, the Standard Model has repeatedly provided predictions of quantities that have subsequently been measured and found to agree. The discovery of the Higgs boson in 2012 is perhaps the most striking instance of Standard Model predictions being borne out in observations of Nature. This discovery provided a confirmation of the existence of the Higgs field predicted by the Standard Model (the Higgs boson being an excitation of this field), with further evidence also supporting its predicted relation to the electromagnetic force and the weak nuclear force. It is clear that the Standard Model provides an accurate effective description — up to the highest energies and shortest distance scales accessible so far — of whatever fundamental laws ultimately govern Nature.

Nevertheless, the Standard Model is known to be incomplete. For one, astronomical evidence of dark matter is not easily described by known particles. The Standard Model also fails to account for the magnitude of the observed matter-antimatter asymmetry. In addition, measurements of neutrino oscillations imply that neutrinos have non-zero mass, whereas the Standard Model describes neutrinos as massless. These observations point to gaps in the description of Nature provided by the Standard Model. Further tensions with Standard Model predictions have also been observed in the past two decades. For example, measurements of decay rates of $B$ mesons (quark-antiquark bound states containing a bottom quark) to particular final states have long been in tension with predicted rates of decay. Similarly, experimental determinations of the anomalous magnetic moment of the muon (and the electron, to a lesser extent) are in tension with the values predicted by the Standard Model. Though such tensions do not yet rise to the level of conclusive departures from the Standard Model, both the known discrepancies and measured tensions underscore the importance of fully constraining the theoretical predictions from the Standard Model as experimental determinations crystallize.

The coupling parameters associated with the electroweak sector of the Standard Model are small at the energy scales relevant for current experiments, allowing systematically improvable theoretical predictions of electroweak phenomena to be made by perturbative expansion in the value of these couplings. On the other hand, the coupling parameter of the strong nuclear force, described by the *quantum chromodynamics (QCD)* sector of the Standard Model, is large at low energies. This results in the confinement of quarks and gluons discussed above. It also prevents the application of perturbative methods when seeking to describe confined QCD states such as the proton and neutron, the $B$ mesons, or nuclear targets used in neutrino oscillation and dark matter experiments. Instead, one must non-perturbatively calculate strong force effects by first discretizing the quark and gluon fields over a four-dimensional spacetime lattice and then utilizing Monte Carlo methods to compute the average values of physical observables under sampled fluctuations of these fields. The discretization must also preserve the gauge symmetry of QCD which is crucial to connecting results on the lattice to the physics of continuous quark and gluon fields. This non-perturbative approach, known as 'lattice QCD', allows a systematically improvable study of QCD effects. In contrast to systematic improvement in the order of expansion, as in the case of perturbation theory, these calculations are systematically improved by increasing the number of samples, reducing the lattice spacing to extrapolate to the continuum limit, increasing the lattice volume to describe physics in macroscopic spaces, and tuning the bare lattice parameters to reproduce physical couplings and masses.

Lattice QCD has recently entered an era of precision calculations with full control of all systematic effects. State-of-the-art calculations have made significant progress in describing how QCD states interact with each other (described by scattering lengths and effective ranges), how QCD states interact with electroweak probes (described by electromagnetic/weak matrix elements and form factors), and how QCD states are structured (described by decay constants, parton distribution functions, and more general structure functions). However, performing such calculations requires an incredible investment of computational resources. At the fine lattice spacings and large volumes that anchor extrapolations to the continuum and infinite-volume limits, typical calculations work with field configurations that involve on the order of $10^9$ degrees of freedom. Ensembles of at least $10^2$–$10^3$ such field configurations must be generated using Monte Carlo sampling, and further costly calculations of observables must be performed on each sampled field configuration. The demanding nature of these calculations remains an obstacle to achieving desired levels of precision in many lattice QCD calculations. For example, to make *ab initio* Standard Model predictions that could definitively identify new physics effects in the $B$ meson tensions or the muon anomalous magnetic moment, relevant lattice QCD calculations would require an increase in statistics by a factor of a few to an order of magnitude relative to the current state of the art.

Reaching these tantalizingly near physics goals either requires a significant increase in the computational resources employed or the development of more efficient methods. In this dissertation, we focus on the latter and detail the development of new algorithms addressing obstacles that presently limit the efficiency of lattice calculations. Two computational bottlenecks motivate the particular algorithms developed here. First, generating ensembles of field configurations is hindered by *critical slowing down* which results in a diverging computational cost per Monte Carlo sample as the lattice parameters are tuned towards smaller lattice spacings (and, ultimately, the continuum limit). Second, precise measurements of many observables are further hindered by a *signal-to-noise problem*: for a fixed ensemble size, the Monte Carlo noise grows exponentially larger than the signal to be measured as particular limits of observables are taken. An example is provided by measurements of two-, three-, and four-point correlation functions of operators that are used to determine physical information about the energies and interactions of states accessed by the operators under study. To compute the most pertinent information from correlation functions, the spacetime separations between the operators must be taken large, but the signal-to-noise ratio falls off exponentially exactly in this limit.

Lattice methods have also been applied in many contexts beyond QCD, and both critical slowing down and the signal-to-noise problem hinder calculations in these other contexts as well. For example, lattice methods are used to determine properties of strongly coupled quantum field theories that are candidate models for describing dark matter or more fundamental physics beyond the Standard Model. In these calculations, critical slowing down is also encountered when taking the continuum limit and similar signal-to-noise problems arise in measurements of correlation functions. Lattice methods also have a long history of application to models of statistical mechanics and condensed matter. In this context, the discretization is over a spatial lattice, rather than a spacetime lattice, and the sampling is performed over microscopic states of the system. Critical slowing down for such models is encountered as the temperature and other parameters are tuned towards criticality, which is necessary to extract universal properties of such systems. For some of these theories, critical slowing down or the signal-to-noise problem have been mitigated or entirely solved by careful construction of Monte Carlo sampling or measurement procedures. Unfortunately, previously known approaches are based on the specific nature of the theory under study and do not extend straightforwardly to lattice QCD and similar lattice quantum field theories.

In this dissertation, we develop efficient approaches to Monte Carlo sampling and observable measurements that tackle critical slowing down and the signal-to-noise problem in a manner that can be extended to generic lattice field theory settings. First, we introduce an approach based on 'flow-based' models for Monte Carlo sampling of lattice field theory configurations, with a particular focus on its application to lattice gauge theories like lattice QCD. These flow-based models describe transformations of field configurations that in general reshape the structure of the distribution and allow it to be more easily sampled. The Jacobian of the transformation can be exactly computed, and as a result, the transformed probability density is known and can be related exactly to the distribution of interest regardless of the particular flow-based transformation applied. Second, we introduce methods that utilize complex contour deformations of the lattice-regularized path integral to mitigate signal-to-noise problems in observables. Contour deformations also describe transformations of field configurations, albeit in the larger space of complexified field configurations. Here, exactness under a deformed contour of integration is guaranteed for any holomorphic path integrand by higher-dimensional analogues of Cauchy's theorem.

Whereas solutions to critical slowing down and the signal-to-noise problem in other theories have relied largely on *zero- or few-parameter* algorithms — algorithms that perform sampling or observable measurement in nearly a fixed way, perhaps with a few tunable knobs — the approaches discussed here are exclusively *parametric* approaches described by a large set of parameters and architectural choices that can be searched over to find optimal solutions. Such an approach is made possible by the exactness guarantees discussed above, which apply independently of the parameter values and architectural choices defining the flow-based transformation or contour deformation. As a result, the classes of algorithms defined in this dissertation extend to a wide range of theories, including lattice gauge theories like lattice QCD. Flexibility in the architectural choices defining these algorithms also allows the incorporation of physical principles to maximize the efficiency of the resulting algorithms. Throughout this work we detail architectures that account for symmetries, locality, and the structure of field fluctuations.

Since the choice of architecture and parameterization can only affect the efficiency of the method, the remaining task is to identify choices that result in highly efficient Monte Carlo and observable measurements algorithms. In this dissertation, we construct expressive classes of field transformations — classes encoding a wide variety of possible transformations — using machine learning architectures described by large numbers of parameters. These parameters can then be effectively optimized using gradient-based methods applied to a 'loss function' measuring the quality of any particular choice of parameters. Appropriate loss functions and approaches to efficiently optimize parameters are discussed for both flow-based sampling and path integral contour deformations.

These methods are demonstrated to significantly improve the efficiency of sampling and observable measurements in the lattice gauge theories and other lattice field theories explored in our proof-of-principle studies. In these studies, we restrict to theories for which analytical or existing Monte Carlo methods already allow accurate estimates of observables and thus confirm the validity our approaches. The ensemble generation and observable measurement techniques laid out in this dissertation are, however, fully general. This work thus lays the stepping stones for the application of highly expressive parametric algorithms to ensemble generation and observable measurements in lattice field theories of interest, including lattice QCD. If the results in this work extend with similar impact to calculations at the scale of the state of the art, these techniques have the potential to extend the statistical power afforded by finite computational resources by a factor of a few to several orders of magnitude, bringing the physics goals outlined above well within reach. The remainder of this dissertation is structured as follows.

After an in-depth review of the formulation of lattice field theory and the associated computational difficulties in Chapter 2, we detail our approach to sampling field configurations from the distributions associated with these lattice field theories in Chapters 3 and 4. The former chapter discusses the application of flow-based models to lattice field theory. We define several approaches to producing asymptotically exact ensembles of field configurations with the same guarantees as traditional methods of Monte Carlo sampling applied in this context. A proof-of-principle application to a scalar lattice field theory is detailed, and we demonstrate that critical slowing down can be mitigated by this approach. The latter chapter explores the exact inclusion of gauge symmetries in these models to facilitate efficient sampling from lattice gauge theories. Using this approach, proof-of-principle applications to $U(1)$ and $SU(N)$ lattice gauge theories are presented, demonstrating that near-elimination of topological freezing and efficient sampling are possible.

Chapter 5 then explains our approach to signal-to-noise problems in observable measurements, covering the relation of signal-to-noise problems to fluctuations in the complex phase of the observable and the use of contour deformations of the path integral to reduce these phase fluctuations. We contrast our approach of defining 'observifolds' — deformations of the path integral integration manifold that are specifically designed to reduce phase fluctuations in observables — to contour deformations that have previously been applied to reduce 'extensive' sign problems associated with theories involving real-time evolution or a non-zero chemical potential. Our observable-focused deformations assume an ensemble has been generated with respect to a fixed distribution and that deformations describe a rewriting of the observable estimator rather than a change in the path integral weights. We detail the form of the resulting 'deformed observable', demonstrate that the expectation value is unchanged, and discuss minimizing the variance of the deformed observable, which is generically modified. Significant variance reduction is demonstrated for observables in proof-of-principle applications to $U(1)$ and $SU(N)$ lattice gauge theories. By a small modification of the approach applied in our study of $U(1)$ lattice gauge theory, these results are also extended to the lattice discretization of a quantum mechanical anharmonic oscillator.

Chapter 6 discusses the common threads in these applications. There we further present the outlook for future developments based on this work.

Finally, Appendix A presents an array of exact analytical results for integration over group-valued variables and solvable lattice gauge theory path integrals. These results are used for comparison to numerical results throughout this dissertation and are referenced where relevant in the main text.

# Chapter 2

# Lattice field theory

We use the term *lattice field theory* to describe the general procedure of discretizing continuous fields on geometric lattices, applying Monte Carlo sampling over a distribution associated with such fields, and using these sampled field configurations to measure observables of interest. This section reviews the formulation, implementation, and practical difficulties of lattice field theory calculations from the perspective of applications to quantum field theory; see Ref. [10] for one of many introductory treatments of the subject of lattice field theory and Ref. [11] for a detailed coverage of lattice QCD in particular. Because of the quantum field theory (QFT) motivations for this work, we work with notation specific to lattice QFT throughout. A connection can be made to statistical mechanics by interpreting expectation values described by lattice-regularized path integrals as thermodynamic expectation values under the Boltzmann distribution, where the spacetime volume is instead interpreted as a spatial volume and the action is instead interpreted as the Hamiltonian of the system. Practical obstacles highlighted in this section limit the physical impact of state-of-the-art calculations and motivate the work presented in the remainder of this dissertation.

## 2.1 Formulation

Quantum field theories arose originally from the unification of quantum mechanics with the Poincaré symmetry of Minkowski spacetime [12].[1] This unification was initially achieved in the Hamiltonian picture by applying 'second quantization' to arrive at operator-valued fields satisfying canonical commutation relations and acting on a Fock space [13–15]. In this picture the Poincaré symmetry is superficially obscured by the selection of a time direction in which the Hamiltonian evolves states, though the Poincaré symmetry of the theory is still appropriately encoded in transformation properties of the fields. The Poincaré symmetry of the theory can be made manifest by instead formulating QFT from the perspective of Lagrangian mechanics. Classical Lagrangian mechanics describes the dynamics of a theory by the principle of least action, where the action $S[\phi(x)]$ to be minimized is a functional defined by integrating the Lagrangian of the c-number *field configuration* $\phi(x)$ over a region of spacetime. We use the notation $\phi(x)$ throughout this section to denote field configurations of a generic field and we apply the term 'c-number' to distinguish non-operator fields $\phi(x)$ from the operator-valued fields $\hat{\phi}(x)$ involved in second quantization. Lagrangian mechanics can be extended to the quantum domain in terms of the path integral [16]. The path integral defines the vacuum expectation value of an operator $\mathcal{O}$ by an integral over field configurations,

$$\langle \mathcal{O} \rangle \equiv \langle \Omega | \mathcal{O} | \Omega \rangle = \frac{\int \mathcal{D}\phi \ \mathcal{O}[\phi(x)] \, e^{iS[\phi(x)]/\hbar}}{\int \mathcal{D}\phi \ e^{iS[\phi(x)]/\hbar}}, \tag{2.1}$$

where $|\Omega\rangle$ is the vacuum state and $\mathcal{O}$ is assumed to be diagonal in the basis of functionals over $\phi(x)$ so that it can be written as a c-number-valued functional $\mathcal{O}[\phi(x)]$. For bosonic fields, integration over the c-number-valued fields $\phi(x)$ reduces to functional integration over real or complex values; for fermionic fields, integrals must instead be taken over anti-commuting Grassmann numbers [17, 18] (we discuss how these fermionic integrals must be treated after lattice regularization below). The classical limit emerges when $\hbar$ is much smaller than typical fluctuations of the action, in which case the lowest-order stationary phase approximation simply corresponds to the principle of least action. Quantum effects are relevant outside this limit, and the aim of quantum field theory in either the operator or path integral formulation is to quantitatively describe these effects. For the remainder of this dissertation, we work in units in which $\hbar = 1$.

---

[1]Though non-relativistic QFTs satisfying reduced spacetime symmetries can also be defined, we focus here on relativistic QFTs applicable to describing fundamental physics.

Unregularized descriptions of most QFTs are ill-defined due to ultraviolet divergences, and the path integral definition in Eq. (2.1) is no exception. Without regularization, the path integral has an ill-defined measure $\mathcal{D}\phi$. To provide a concrete definition of path integral expectation values, this measure must be addressed. In contrast to approaches like dimensional regularization [19–21] which are specific to perturbation theory, the path integral formalism can be *non-perturbatively* regularized by restricting field configurations to live on a discrete hypercubic spacetime lattice $\mathscr{L}$ [22]. The lattice spacing $a$ between nearest-neighbor sites in $\mathscr{L}$ cuts off fluctuations of the field with frequencies higher than $\sim \pi/a$, and the ill-defined continuum measure is then reduced to a discrete set of integrals $\int_{\mathscr{L}} \mathcal{D}\phi = \int \prod_{x \in \mathscr{L}} d\phi(x)$.

The action and observables must also be defined on the discretized spacetime to fully define the regularized path integral. Observables $\mathcal{O}[\phi(x)]$ can simply be restricted to functions $\mathcal{O}(\phi)$ applied to the discrete field of values $\phi$ consisting of $\phi(x)$ evaluated at lattice sites $x \in \mathscr{L}$. The potential energy terms in the action can similarly be restricted from functionals of the field $\phi(x)$ to functions over the discrete field $\phi$. The kinetic energy terms in the action involve spacetime derivatives $\partial_x \phi(x)$ that must be transformed to discretized derivatives given by functions of field values in the neighborhood of each point $x$. In the following, we use the notation $S(\phi)$ to denote the discretized action evaluated over the discretized field. The choices of discretization can have significant effects on the approach to the continuum [23], but we assume a suitable discretization has already been determined for the purpose of the work presented in this dissertation.

To connect the lattice-regularized path integral to the continuum theory, it is necessary to take the continuum limit $a \to 0$. To do so, we must first transform from Minkowski spacetime to Euclidean spacetime by a Wick rotation of the time coordinate. In Euclidean spacetime, the regularized path integral is instead defined by integration with respect to weights $e^{-S(\phi)}$ as

$$\langle \mathcal{O} \rangle_{\text{Eucl. lattice}} = \frac{\int_{\mathscr{L}} \mathcal{D}\phi \, \mathcal{O}(\phi) \, e^{-S(\phi)}}{\int_{\mathscr{L}} \mathcal{D}\phi \, e^{-S(\phi)}}, \tag{2.2}$$

in terms of the discretized Euclidean action $S$ and discretized field $\phi$ evaluated over sites in a Euclidean spacetime lattice $\mathscr{L}$. Observables can be related to equal-time observables suitable for describing equilibrium properties of the Minkowski theory under mild assumptions [24, 25]. In Euclidean spacetime, the continuum limit corresponds to tuning parameters towards a second order critical point at which all correlation functions diverge in lattice units and $a$ is therefore taken to zero relative to fixed physical correlation lengths. Information about the continuum Minkowski theory can thus be recovered by computing renormalized Euclidean quantities, taking the limit as lattice parameters are tuned towards a second order critical point to remove the regulator, and relating the resulting continuum Euclidean quantities to Minkowski quantities [26].

The Euclidean spacetime lattice $\mathscr{L}$ renders all ultraviolet divergences finite, but does not immediately make the calculation of the path integral tractable: there are an infinite number of sites $x$, even after discretization, and thus an infinite number of integrals to be evaluated. This can be resolved by further restricting the lattice $\mathscr{L}$ to a finite subset of the spacetime volume. Results can then either be matched to finite-volume continuum physics or extrapolated to the infinite volume limit. When working in a finite volume, one also must choose the boundary conditions utilized. A common choice is to apply periodic boundary conditions (anti-periodic in time for fermions) identifying opposite ends of the spacetime volume, which has the advantage of preserving an exact lattice translational symmetry. Open boundary conditions may also be used and may be preferable in some cases [27–34]. We take Eq. (2.2) defined over a Euclidean, discretized, finite-volume lattice $\mathscr{L}$ with appropriate boundary conditions as the starting point for all lattice field theory approaches discussed in this dissertation. For conciseness, we leave the 'Euclidean lattice' and $\mathscr{L}$ annotations implicit in future uses of Eq. (2.2).

Equation (2.2) gives regularized Euclidean observable expectation values as a ratio of integrals that can in principle be evaluated. In practice, one must address the Grassmann-valued integrals over fermion fields and the high dimensionality of the remaining integration over boson fields. Fermionic fields can be 'integrated out' by hand, resulting in a more complicated path integral over the remaining bosonic degrees of freedom (see Ref. [11] for an introductory presentation). Monte Carlo integration provides an efficient approach to evaluating the remaining real- or complex-valued integrals. In particular, Monte Carlo methods utilize importance sampling to draw an *ensemble* of $n$ field configurations $\{\phi_i\}_{i=1}^n$ distributed proportionally to $e^{-S(\phi)}$. Observable vacuum expectation values can then be estimated by

$$\langle \mathcal{O} \rangle \approx \frac{1}{n} \sum_{i=1}^{n} \mathcal{O}(\phi_i), \tag{2.3}$$

which provides an estimate with precision scaling as $1/\sqrt{n}$.

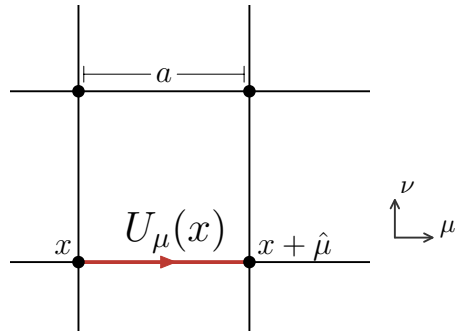

**Figure 2.1:** Discretization of the gauge field. The degrees of freedom $U_\mu(x)$ in the discretized gauge field are each associated with a link $(x, x + \hat{\mu})$ of the spacetime lattice. Non-gauge fields are typically discretized on the sites $x \in \mathscr{L}$.

## 2.2 Lattice gauge theory

The work in this dissertation is motivated primarily by the application to QCD, which is a gauge theory. Gauge theories are defined by the presence of a 'gauge symmetry' that acts by a simultaneous local transformation of matter fields (e.g. quarks) and gauge fields (e.g. gluons) according to a gauge group $G$. For example, QCD is described by the gauge group $G = \mathrm{SU}(3)$ which acts on quarks in the fundamental representation of $\mathrm{SU}(3)$ by unitary rotations of three 'color' components and on gluons by transformations in the eight-dimensional adjoint representation of $\mathrm{SU}(3)$. (See Appendix A for a review of this and other group theory topics encountered in this dissertation.) In general, gauge bosons physically correspond to force carriers of fundamental forces associated with each gauge group. For example, the Standard Model is at its heart a gauge theory that in addition to the $\mathrm{SU}(3)$ gauge symmetry of QCD encompasses the $\mathrm{U}(1) \times \mathrm{SU}(2)$ gauge symmetry associated with the electromagnetic and weak forces, where the gauge bosons associated with each symmetry group are the force carriers.

In a lattice formulation, gauge bosons are commonly discretized in terms of a field of variables $U_\mu(x)$ in the fundamental matrix representation of $G$, where each $U_\mu(x)$ is associated with a link $(x, x + \hat{\mu})$ of the lattice and $\hat{\mu}$ indicates a vector of length $a$ oriented in one of the $D$ spacetime directions, $\mu \in \{0, 1, \dots, D-1\}$. Figure 2.1 depicts this discretization of the gauge field with respect to the lattice geometry. Non-gauge fields are discretized on the sites $x \in \mathscr{L}$ as described for the generic field $\phi(x)$ above.

The continuum action includes pure-gauge terms that describe the propagation and self-interactions of gauge bosons, terms that couple the gauge and matter fields, and terms that describe the propagation and self-interactions of matter fields. For non-Abelian gauge theories, it is possible to have non-trivial self-interactions between gauge fields alone, yielding physically interesting pure-gauge dynamics [35]. The proof-of-principle applications presented in this dissertation are restricted to either non-gauge theories or pure-gauge theories, and as such we focus here on the discretization of the gauge field alone. The inclusion of fermionic matter fields is a very relevant subject for future applications of these techniques; this direction is discussed further in the outlook presented in Chapter 6.

The Wilson gauge action [22] is the simplest discretized action describing SU($N$) gauge fields. In $D$ spacetime dimensions the action is given by

$$S(U) = -\frac{\beta}{N} \sum_x \sum_{\mu=1}^{D} \sum_{\nu=\mu+1}^{D} \operatorname{Re} \operatorname{tr} \left[ P_{\mu\nu}(x) \right], \tag{2.4}$$

where $\beta$ controls the strength of the gauge coupling and the *plaquette* $P_{\mu\nu}(x)$ is given by

$$P_{\mu\nu}(x) = U_\mu(x) \, U_\nu(x + \hat{\mu}) \, U_\mu^\dagger(x + \hat{\nu}) \, U_\nu^\dagger(x). \tag{2.5}$$

Lattice QCD and other phenomenologically relevant lattice gauge theories can be formulated using the Wilson action with $D = 4$. However, it is also interesting to explore gauge theories in dimensions $D \neq 4$. For example, some theories with dimensionality $D < 4$ retain many of the properties of their four-dimensional cousins [36–39] and are thus particularly useful to gain intuition about gauge theories of interest at lower computational cost.

An important feature of the Wilson action is that it is invariant under the gauge symmetry restricted to the sites of $\mathscr{L}$. In the continuum, a gauge transformation is described by the group-valued field $\Omega(x) \in G$ giving the local transformations at each site $x$. The continuum gauge field can be written in matrix form as

$$A_\mu(x) = A_\mu^a(x) \, \tau_a, \tag{2.6}$$

where $A_\mu^a(x)$ are the adjoint-representation components of the gauge field and $\tau_a$ are the generators of the gauge group. In this form, the field transforms as

$$A_\mu(x) \to \Omega(x) \, A_\mu(x) \, \Omega^\dagger(x) + i(\partial_\mu \Omega(x))\Omega^\dagger(x) \tag{2.7}$$

under gauge transformations. The gauge links $U_\mu(x)$ in the discretized theory are parallel transporters

$$U_\mu(x) = \operatorname{P} \exp\left( i \int_x^{x+\hat{\mu}} dx' \, A_\mu(x') \right), \tag{2.8}$$

where $\operatorname{P} \exp(\cdot)$ denotes the path-ordered exponential, and each gauge link thus transforms according to the field $\Omega$ at either endpoint under gauge transformations,

$$U_\mu(x) \to \Omega(x) \, U_\mu(x) \, \Omega^\dagger(x + \hat{\mu}). \tag{2.9}$$

One can explicitly transform the gauge fields in Eq. (2.4) to confirm that the Wilson action is indeed invariant under gauge transformations. This implies that only gauge-invariant components of observables will contribute to their vacuum expectation value when using the Wilson lattice regularization, and this symmetry plays a role in ensuring that a gauge-invariant continuum theory is arrived at when taking the continuum limit.

Even in pure gauge theory, interesting gauge-invariant observables can be measured. By the transformation properties of the gauge links, the trace of any product of gauge links forming a loop starting and ending at the same point will be gauge invariant. These (traced) *Wilson loops* can be shown to form a basis that is dense in the space of all gauge-invariant observables [40]. We write a general Wilson loop as

$$W_\ell(x) \equiv \prod_{dx, \mu \in \ell} U_\mu(x + dx), \tag{2.10}$$

where $\ell$ defines a sequence of offsets $dx$ and directions $\mu$ that form a continuous loop on the lattice starting from an arbitrary base site $x$, and links oriented in the reverse direction are defined by $U_{-\mu}(x) = U_\mu^\dagger(x - \hat{\mu})$. In the proof-of-principle studies presented in this dissertation, we measure two particular classes of Wilson loops: planar loops $W_{\mathcal{A}}$ surrounding a two-dimensional region $\mathcal{A}$ and loops $X_\mu$ winding around the periodic $\mu$ direction of the lattice. Several choices of these types of loops are depicted for example in Fig. 2.2.

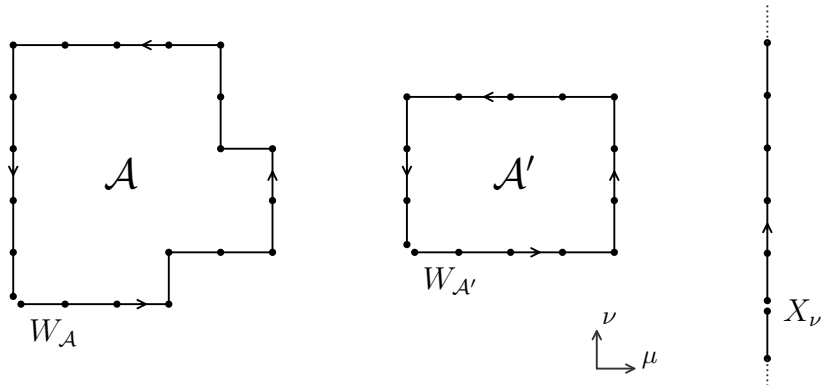

**Figure 2.2:** Several choices of Wilson loops that can be used to compute gauge-invariant quantities. Each Wilson loop is defined by an ordered product over the links shown, where the small gap in each diagram indicates the start and end of the loop. The left two loops, $W_{\mathcal{A}}$ and $W_{\mathcal{A}'}$, are planar loops surrounding the two-dimensional regions $\mathcal{A}$ and $\mathcal{A}'$, respectively. The rightmost loop $X_\nu$ is a Polyakov loop winding around the periodic $\nu$ dimension. The planar loops $W_{\mathcal{A}}$ and $W_{\mathcal{A}'}$ are contractible and do not transform under center symmetry, while $X_\nu$ transforms under center symmetry when $\nu = 0$. Taking the trace of each quantity shown produces a gauge-invariant observable that can be measured to derive physical information from a lattice gauge theory.

An individual Wilson loop $W_{\mathcal{A}}$ surrounding a rectangular region $\mathcal{A}$ has a physical interpretation as a correlation function of a pair of static charges (see Sec. 2.5 for the physics of correlation functions in Euclidean spacetime more generally). For confining theories, the expectation value of the gauge-invariant traced quantity $\frac{1}{N} \operatorname{tr} W_{\mathcal{A}}$ is expected to scale exponentially with the area $A$ of the enclosed region.

Loops $X_\mu$ that wind around periodic boundaries of the lattice, if such boundary conditions are used, are termed *Polyakov loops*. We denote Polyakov loops winding around the periodic boundary in the $\hat{\mu}$ direction by

$$X_\mu(x_\perp) = \prod_{x_\mu} U_\mu(\{x_\perp, x_\mu\}), \tag{2.11}$$

where $x_\perp$ indicates the components of the lattice coordinate orthogonal to $\hat{\mu}$, and the product is taken over all values of the coordinate $x_\mu$ in the $\mu$ direction. In pure-gauge theories and theories with matter in certain representations of the gauge group, there is an exact *center symmetry* given by transformations $U_\mu(\{x_\perp, x_\mu\}) \to z \, U_\mu(\{x_\perp, x_\mu\})$ applied uniformly across all transverse coordinates $x_\perp$ for a fixed direction $\mu$, fixed coordinate $x_\mu$, and any $z$ in the center of the gauge group $G$. Elements in the center of $G$ commute with all elements of $G$ by definition, and center transformations thus cancel for any loops that traverse the coordinate $x_\mu$ an equal number of times in the forwards and backwards $\mu$ direction. However, Polyakov loops have a non-trivial winding number and transform non-trivially under this symmetry as $X_\mu(x_\perp) \to z \, X_\mu(x_\perp)$. The traced Polyakov loop $\eta_\mu(x_\perp) \equiv \operatorname{tr} X_\mu(x_\perp)$ is a gauge-invariant observable that also transforms under center symmetry. The deconfinement transition in theories with a center symmetry is associated with spontaneous breaking of this symmetry, and traced Polyakov loops can therefore be measured as order parameters for this transition [41–44].

## 2.3 Symmetries

The discretized action of a lattice field theory typically encodes discretized versions of the symmetries of the continuum action. These symmetries play a crucial role in tuning towards a continuum field theory described by the desired set of relevant operators, as discussed for the case of gauge symmetry above. Symmetries in the discretized or continuum action manifest as transformations of field configurations $\phi$ that leave the action $S(\phi)$ invariant. We generically write the action of a transformation $t$ on a field configuration $\phi$ as $t \cdot \phi$; a group of transformations $H$ is then a symmetry of the action if

$$S(\phi) = S(t \cdot \phi) \quad \forall t \in H, \, \forall \phi. \tag{2.12}$$

Lattice actions are commonly constructed to preserve geometric symmetries of the discretized Euclidean spacetime in addition to internal symmetries like gauge symmetries. In particular, actions are typically invariant under two geometric symmetry groups described as follows:

1. The set of **discrete translational symmetries** $T = \{T_{\delta x}\}$ consists of translations by all possible offsets $\delta x$ that map the spacetime lattice to itself. This is only an exact symmetry of the finite volume lattice if periodic (or anti-periodic) boundary conditions are employed.

2. The set of **hypercubic symmetries** $R = \{R_i\}$ consists of the $2^D(D!)$ transformations generated by rotations and reflections of the $D$-dimensional hypercube, where the index $i$ enumerates all possible symmetry group elements. These oper-ations are typically defined about a distinguished point on the lattice, implying a non-trivial interaction with the translational symmetry group.

These geometric symmetries together allow a controlled extrapolation to continuum physics with the desired Poincaré and internal symmetries. As the lattice spacing is reduced, the discrete translational symmetry group encompasses translations at decreasing length scales, converging towards a dense subset of the continuum translational symmetry group as $a \to 0$. The hypercubic symmetry group by itself does not necessarily converge towards the $O(D)$ rotational symmetry group of the Euclidean measure, but typical lattice actions are constructed such that the only relevant operators in the continuum limit are invariant under the full rotational symmetry group which can be determined both analytically [23, 45, 46] and numerically [47–50]. The $O(D)$ symmetry of the Euclidean continuum limit ensures that results that are Wick rotated to the Minkowski signature are consistent with the Lorentz subgroup of Poincaré symmetries.

Beyond geometric symmetries, particular actions may also satisfy internal symmetries associated with transformations of components of the fields. These may either be global symmetries given by transformations applied uniformly over all sites of the lattice or local gauge symmetries given by independent transformations at each site. Gauge symmetries were discussed in the previous section. Though the global symmetries depend on the theory under consideration, a prototypical example for a single-component scalar field is complex phase rotation $\phi(x) \to e^{i\theta}\phi(x)$ for complex fields or negation $\phi(x) \to -\phi(x)$ for real fields. For gauge theories, center symmetry may also be considered a global symmetry, as it is uniform across the transverse coordinate $x_\perp$.

## 2.4 Ensemble generation

A Monte Carlo importance sampling approach to evaluating the path integral in Eq. (2.2) requires drawing an ensemble of samples $\phi$ according to a distribution proportional to $e^{-S(\phi)}$, where $\phi$ stands in for a generic set of lattice fields and may, for example, be the gauge field $U$ when considering lattice gauge theory. The desired normalized distribution is defined for an arbitrary action by the probability density

$$p(\phi) = e^{-S(\phi)}/Z, \quad \text{where} \quad Z = \int \mathcal{D}\phi \, e^{-S(\phi)}. \tag{2.13}$$

Though the action $S(\phi)$ is known analytically, algorithms to draw independent samples directly from $p(\phi)$ are typically not known except in the simplest of cases. This issue can be circumvented by applying *Markov chain Monte Carlo (MCMC)* sampling. In the (all-important) continuum limit, and in general when tuning lattice theories to criticality, most MCMC methods encounter *critical slowing down* — the computational cost to draw a fixed number of nearly independent samples diverges in the critical limit. We review both Markov chain Monte Carlo and critical slowing down in the following subsections.

### 2.4.1 Markov chain Monte Carlo

A *Markov chain* is a sequence of random variables

$$\phi^{(1)} \rightarrow \phi^{(2)} \rightarrow \cdots \rightarrow \phi^{(i)} \rightarrow \ldots \tag{2.14}$$

defined by a conditional transition probability $T(\phi^{(i)} \rightarrow \phi^{(i+1)})$ that must, by definition, be independent of all prior states in the chain [51]. We restrict our focus to cases in which the transition probability is also independent of the Markov chain index $i$. In practice, a Markov chain is constructed by producing an initial state $\phi^{(1)}$ by arbitrary means (e.g. by choosing a particular fixed state) then sequentially applying a stochastic transition function to produce each subsequent state in the chain. Under the conditions discussed below, a Markov chain converges to an equilibrium distribution $p(\phi)$ over the states $\phi^{(i)}$ as $i \rightarrow \infty$. The Markov chain thus provides a Monte Carlo sampling approach for the distribution $p(\phi)$ that is asymptotically exact. Taking the random variables $\phi^{(i)}$ to be lattice field configurations and the equilibrium distribution to be $p(\phi) = e^{-S(\phi)}/Z$, Markov chain Monte Carlo provides a sampling procedure for lattice field theory.

**Balance, equilibrium distribution, and convergence.** The distribution $p$ is an equilibrium distribution of the Markov chain if the transition function satisfies the global balance condition

$$\int \mathcal{D}\phi \, p(\phi) \, T(\phi \rightarrow \phi') = p(\phi'). \tag{2.15}$$

Explicitly evaluating the integral in Eq. (2.15) to verify global balance is typically difficult, and in practice the stronger *detailed balance* condition

$$p(\phi) \, T(\phi \rightarrow \phi') = p(\phi') \, T(\phi' \rightarrow \phi) \tag{2.16}$$

is often used to guarantee the correct equilibrium distribution. Detailed balance is synonymous with 'reversibility' of the Markov chain, i.e. the property that the joint equilibrium distribution over $(\phi^{(i)}, \phi^{(i+1)})$ is symmetric under exchange of $\phi^{(i)}$ and $\phi^{(i+1)}$, and this alternate term often appears in the literature.

A Markov chain is *ergodic*, and will asymptotically converge to the equilibrium distribution identified by the balance condition, if it is both irreducible and aperiodic [52, 53].[2] Irreducibility requires a non-zero asymptotic transition probability between any pair of states. Given a particular transition function, irreducibility can be checked by verifying that after some number of steps $n \geq 1$ the transition probability from any state $\phi$ to any subset of the state space has non-zero probability [54]. Aperiodicity is the requirement that the Markov chain does not cyclically move probability mass around the state space, in which case the asymptotic distribution would fail to 'relax' to the equilibrium distribution. Though this property is more difficult to check, it can be verified for particular transition functions considered below. See Ref. [55] for a complete mathematical treatment of convergence in Markov chains and Ref. [54] for further details on Markov chains with continuous state spaces.

---

[2] The term 'ergodicity' is sometimes used to refer only to irreducibility. Here, we follow the convention of Ref. [53] and use the term to indicate both irreducibility and aperiodicity.

**Metropolis-Hastings.** There are many approaches to constructing transition functions with a desired equilibrium distribution $p$. Arbitrary transition proposals described by the conditional probability $P(\phi \to \phi')$ can be made 'exact', i.e. incorporated in a Markov chain transition function that satisfies detailed balance for a distribution $p$, using the Metropolis-Hastings algorithm [56, 57]. In a Metropolis-Hastings step, a proposed transition $\phi \to \phi'$ is accepted with probability

$$p_{\text{acc}}(\phi \to \phi') = \min\left(1, \frac{p(\phi') \, P(\phi' \to \phi)}{p(\phi) \, P(\phi \to \phi')}\right), \qquad (2.17)$$

in which case the subsequent state in the Markov chain is assigned to $\phi'$. Otherwise, the proposal is rejected and the subsequent state in the Markov chain is set to the previous state $\phi$.

Simple transition functions can be constructed for lattice field theory distributions using a Metropolis-Hastings Markov chain based on transition proposals described by simple perturbative updates to field configurations. We use the term 'diffusive' to describe Metropolis-Hastings Markov chains based on such proposals because these Markov chains typically exhibit statistics consistent with random-walk behavior [58, 59]. For example, a transition proposal for a scalar field $\phi(x)$ consisting of a single real number per site $x$ could be given by sampling $\phi'(x) \in [\phi(x) - \epsilon, \phi(x) + \epsilon]$ uniformly at random for any site $x$. Often transition proposals are symmetric in the forward and reverse directions, as is the case in this example, and in these cases the acceptance probability takes the more familiar form

$$p_{\text{acc}}(\phi \to \phi') = \min(1, p(\phi')/p(\phi)) = \min\left(1, e^{-S(\phi') + S(\phi)}\right). \qquad (2.18)$$

Symmetric proposals analogous to the example for scalar field theory are easily sampled and the acceptance probability in Eq. (2.18) can be computed in terms of the known action $S(\phi)$. Under mild assumptions,[3] this results in an ergodic Markov chain suitable for producing ensembles for lattice field theory calculations.

Non-symmetric proposal distributions are also possible, as long as the conditional transition probabilities $P(\phi \to \phi')$ and $P(\phi' \to \phi)$ can be calculated. The independence Metropolis algorithm [52] utilized in Chapters 3 and 4 is an example of a Metropolis-Hastings Markov chain based on non-symmetric proposals. In particular, proposals in an independence Metropolis sampler are generated independently of the previous configuration in the chain according to a distribution $q(\phi)$. In this case the proposal and acceptance probabilities respectively reduce to

$$P_{\text{ind}}(\phi \to \phi') = q(\phi') \qquad (2.19)$$

---

[3]The state space must be connected, for example.

and

$$p_{\text{acc}}^{\text{ind}}(\phi \to \phi') = \min\left(1, \frac{p(\phi')q(\phi)}{p(\phi)q(\phi')}\right). \tag{2.20}$$

**Hybrid Monte Carlo.** Diffusive Metropolis-Hastings Markov chains tend to exhibit a slow convergence to the equilibrium distribution and a high cost of drawing independent samples (as discussed further in the following section). This diffusive scaling can be partially avoided by the use of *Hybrid Monte Carlo (HMC)*, in which transitions are proposed by evolving states according to a fictitious molecular dynamics system [59, 60]. In HMC sampling, the state space of the Markov chain is expanded to include fictitious 'conjugate momentum' variables $\pi(x)$ associated with each primary variable $\phi(x)$. Rather than sampling the target distribution $p(\phi) = e^{-S(\phi)}/Z$ over variables $\phi$, one instead samples from the joint distribution

$$p(\phi, \pi) = \frac{1}{Z}e^{-S(\phi)} \times \frac{1}{\sqrt{2\pi}^{\mathscr{N}}}e^{-\pi^2/2}, \tag{2.21}$$

where $\mathscr{N}$ indicates the number of components of $\pi(x)$. The distribution $p(\phi)$ is recovered from Eq. (2.21) by marginalizing over $\pi$,

$$p(\phi) = \int \mathcal{D}\pi \; p(\phi, \pi), \tag{2.22}$$

thus samples from $p(\phi)$ can be acquired by drawing from this joint distribution and discarding the field $\pi$.

A Markov chain for Hybrid Monte Carlo sampling is given by two transition steps alternatingly applied to the joint state $(\phi, \pi)$:

1. A 'momentum refresh' step resamples $\pi$ conditioned on $\phi$. This conditional distribution has trivial dependence on $\phi$, and the momentum variables are thus sampled independently from the Gaussian distribution proportional to $e^{-\pi^2/2}$. This step satisfies detailed balance for the joint distribution.

2. A molecular dynamics trajectory evolves the pair $(\phi, \pi)$ over a duration $\tau$ of molecular dynamics time according to the (fictitious) Hamiltonian $\pi^2/2 + S(\phi)$. If this trajectory were integrated perfectly, the value of the Hamiltonian would be unchanged and Liouville's theorem would ensure that the joint probability density is not modified [59]. In practice, a volume-preserving integrator with discrete time step — such as the leapfrog integrator [61] — is applied to numerically integrate the equations of motion. This introduces a small error $\Delta H$ in the value of the Hamiltonian over the course of the trajectory while preserving the volume element $\mathcal{D}\phi\mathcal{D}\pi$. Applying a Metropolis-Hastings accept/reject step with probability

$$p_{\text{acc}}^{\text{HMC}}(\Delta H) = \min(1, e^{-\Delta H}) \tag{2.23}$$

is then sufficient to ensure that this step satisfies balance.

The equations of motion required to integrate the HMC trajectory can be straightfor­wardly derived from the Hamiltonian, giving

$$\frac{d\phi(x)}{d\tau} = \pi(x) \quad \text{and} \quad \frac{d\pi(x)}{d\tau} = -\frac{\partial S(\phi)}{\partial \phi(x)}. \tag{2.24}$$

Integrating these equations of motion in a volume-preserving scheme is straightforward for simple domains of integration, such as those associated with scalar lattice field theory; see Ref. [11] for further details on applying these methods to the compact spaces associated with lattice gauge theory.

The resulting Markov chain satisfies ergodicity when these two transition steps are composed, assuming there is a path between any pair of configurations $\phi$ and $\phi'$ that traverses only regions of finite action. Since one is interested only in Monte Carlo evaluation of integrals with respect to the marginal distribution $p(\phi)$, a typical HMC calculation proceeds by (1) sampling new momenta $\pi$ according to step 1 above, (2) evolving $(\phi, \pi)$ for a time $\tau$ to produce $(\phi', \pi')$, (3) applying a Metropolis-Hastings accept/reject step based on the error $\Delta H$, and (4) storing $\phi'$ while discarding $\pi'$.

### 2.4.2 Critical slowing down

The dependence of each Markov chain step on the previous state in the chain intro­duces correlations between successive states. While these correlations do not compro­mise asymptotic correctness, they determine the efficiency of using the Markov chain for Monte Carlo estimation of observables. As the bare parameters defining the lattice field theory are tuned towards criticality, Markov chain steps based on stochastic trans­formations of previous configurations result in increasingly strong correlations between successive samples and a vanishing efficiency of sampling [62]. This *critical slowing down (CSD)* is a major obstacle to lattice QCD calculations [27, 63–66], among other theories. The aim of the first half of this dissertation is to address this limit to the statistics achievable by Monte Carlo sampling near the continuum limit, which de­termines the precision with which continuum physics may be extracted from lattice calculations of such theories.

**Autocorrelations and CSD.** The correlations between successive configurations in a Markov chain at equilibrium can quantitatively be described by the normalized *autocorrelation* function $\rho_{\mathcal{O}}(\tau)$ of any observable $\mathcal{O}$ [67, 68], defined to be the normalized covariance between measurements of $\mathcal{O}$ on Markov chain configurations separated by $\tau$ Markov chain steps. Labeling the value of $\mathcal{O}$ on the $i$th configuration in an equilibrated Markov chain by $\mathcal{O}_i \equiv \mathcal{O}(\phi^{(i)})$, the autocorrelation function can be estimated in practice by

$$\widehat{\rho_{\mathcal{O}}}(\tau) = \frac{\frac{1}{n-\tau}\sum_{i=1}^{n-\tau}(\mathcal{O}_i - \bar{\mathcal{O}})(\mathcal{O}_{i+\tau} - \bar{\mathcal{O}})}{\frac{1}{n}\sum_{i=1}^{n}(\mathcal{O}_i - \bar{\mathcal{O}})^2}, \tag{2.25}$$

where $n$ is the total number of configurations in the Markov chain and $\bar{\mathcal{O}} = \frac{1}{n}\sum_{i=1}^{n}\mathcal{O}_i$. This autocorrelation function describes Markov chain correlations as a function of sep­aration $\tau$; at large $\tau$ the autocorrelation function typically decays exponentially, and the rate of decay of the slowest mode determines the theoretical mixing time of the Markov chain [67]. This theoretical mixing time is typically difficult to compute, and the *integrated autocorrelation time*

$$\tau_{\mathcal{O}}^{\text{int}} = \frac{1}{2} + \lim_{\tau_{\max} \to \infty} \sum_{\tau=1}^{\tau_{\max}} \rho_{\mathcal{O}}(\tau) \tag{2.26}$$

is typically computed instead as a practical measure of the scale of autocorrelations [68, 69]. In the presence of autocorrelations, statistical uncertainties on estimates of $\mathcal{O}$ scale as $\sqrt{2\tau_{\mathcal{O}}^{\text{int}}/n}$, and $n/(2\tau_{\mathcal{O}}^{\text{int}})$ serves as an estimate of the number of independent samples contributing to the measurement of $\mathcal{O}$ [68]. The number of Markov chain steps required to achieve a fixed statistical precision using MCMC thus scales with the integrated autocorrelation time of measurements of observables of interest.

Accurately estimating Eq. (2.26) from noisy MCMC data can be a challenge. In particular, there is a variance-bias tradeoff in the choice of where to truncate $\tau_{\max}$ in the summation — the autocorrelation function is estimated less well at large $\tau$ but also decays quickly to contribute little to the value of $\tau_{\mathcal{O}}^{\text{int}}$. There are several possible 'windowing' procedures that may be used to determine the truncation point. For example, Ref. [67, App. C] and Ref. [68] outline two different approaches; calculations of autocorrelation times presented in this dissertation follow the automatic windowing procedure for the former. The choice of $\tau_{\max}$ in this approach is given by the smallest value of $W$ at which

$$W \geq c\,\tau_{\mathcal{O}}^{\text{int}}\big|_{\tau_{\max}=W}, \tag{2.27}$$

where typically $c = 4$ is chosen to give a practical bias-variance tradeoff.

As critical parameters are approached, MCMC procedures such as HMC and diffusive Metropolis-Hastings encounter critical slowing down [62]. For lattice field theory calculations in particular, CSD is encountered as a second order critical point is approached to take the continuum limit $a \to 0$. In this limit, the integrated autocorrelation time for observables typically scales as a power law in $1/a$, and a *dynamical critical exponent* $z$ can be fit to the scaling of the integrated autocorrelation time for most observables in the small $a$ regime,

$$\tau_{\mathcal{O}}^{\text{int}} \sim a^{-z}. \tag{2.28}$$

Large dynamic critical exponents correspond to severe slowing down as the continuum limit is approached. Diffusive Metropolis-Hastings updates tend to give rise to a universal dynamical critical exponent $z \approx 2$ [70–74]. This can qualitatively be explained by random-walk behavior, for which the distance accessed after $\tau$ updates scales as $\sqrt{\tau}$, and mixing information over a fixed physical scale on the lattice therefore requires $O(a^{-2})$ Markov chain steps. Topological observables have been demonstrated to be consistent which a much more severe exponential scaling $\tau_{\mathcal{O}}^{\text{int}} \sim e^{z/a}$, resulting from 'topological freezing' suppressing transitions between topological sectors in MCMC simulations of lattice field theories with well-defined topological properties [63–65, 75, 76]. Power-law critical slowing down and topological freezing are significant obstacles to making the crucial extrapolation to the continuum ($a = 0$) in lattice field theory calculations.

**Existing approaches.** The problem of critical slowing down has motivated a wide range of solutions. Better tuning of existing Markov chain algorithms [77–79] and faster practical implementations [80–87] have reduced the computational cost, allowing progress to made in some cases despite CSD. However, the diverging computational cost cannot be avoided unless the MCMC algorithm itself is modified. Improved MCMC algorithms that reduce or eliminate critical slowing down have been developed for particular theories by applying cluster updates [72, 88–95], worm algorithms [96–98], overrelaxation [99–101], multigrid methods [102, 103], event-chains [104–108], dual variables [109–111], and other model-specific methods [30, 67].

Hybrid Monte Carlo itself can be considered a significant improvement upon local updating methods because it avoids diffusive behavior if tuned well [59], but HMC also suffers from critical slowing down in the continuum limit because the fictitious molecular dynamics system still only mixes information locally. Much effort has been devoted to finding solutions to CSD within the framework of HMC; if successful, mitigating CSD would have significant impact on simulating state-of-the-art lattice gauge theories, as HMC is the primary tool in this setting. Approaches have been proposed to reduce CSD by altering HMC [112–116], using multi-scale extensions to HMC-based Markov chains [117–119], applying open boundary conditions or non-orientable manifolds [28, 31, 120], or applying meta-dynamics [121].

Recently, machine learning approaches have been applied to find improved Markov chain updates [122–137] and to directly generate samples for a variety of physical systems [124, 138–151]. This promising direction motivates the work on ensemble generation detailed in this dissertation. Chapters 3 and 4 introduce a method for efficient ensemble generation in lattice field theory using flow-based models [152–154], a class of generative machine-learning models based on invertible transformations of field space. This method uses field configurations generated from a flow-based model as proposals within a Metropolis-Hastings Markov chain to guarantee asymptotically correct statistics. We demonstrate the reduction of CSD and topological freezing in proof-of-principle applications presented in these chapters.

## 2.5 Measuring observables

The path integral evaluated over a Euclidean time extent $L_t$ corresponds in the quantum operator picture to evolution by $e^{-HL_t}$, where $H$ is the Hamiltonian of the system. This operator picture allows measured observables to be interpreted physically. For example, for a time-ordered product of $n$ temporally localized operators $\mathcal{O}_1 \mathcal{O}_2 \ldots \mathcal{O}_n$, where $\mathcal{O}_i$ is restricted to be a function of the field values at a single temporal coordinate $t_i$, the path integral over a periodic Euclidean time extent is given by

$$
\begin{aligned}
\langle \mathcal{O}_n \ldots \mathcal{O}_2 \mathcal{O}_1 \rangle &= \frac{1}{Z} \int \mathcal{D}\phi \, \mathcal{O}_1(\phi) \, \mathcal{O}_2(\phi) \ldots \mathcal{O}_n(\phi) \, e^{-S(\phi)} \\
&= \frac{1}{\text{tr}[e^{-HL_t}]} \, \text{tr}[\mathcal{O}_1 \, e^{-H(t_2-t_1)} \, \mathcal{O}_2 e^{-H(t_3-t_2)} \ldots].
\end{aligned}
\tag{2.29}
$$

Such expectation values are termed *n-point correlation functions.* Equation (2.29) takes the form of an expectation value with respect to the density matrix $e^{-HL_t}$. We can decompose the spectral representation of this operator in terms of eigenstates $|n_i\rangle$, for $i \in \{0, 1, \dots\}$, associated with eigenvalues $E_i$ of the Hamiltonian, giving

$$e^{-HL_t} = \sum_i e^{-E_i L_t} |n_i\rangle \langle n_i|. \tag{2.30}$$

This spectrum exactly corresponds to the expectations from thermal field theory if the Euclidean time $L_t$ is interpreted as the inverse temperature of the system. The path integral in Eq. (2.29) thus describes the thermal expectation value of the time-ordered product of operators $\mathcal{O}_1 \mathcal{O}_2 \dots$, and lattice evaluations of the Euclidean path integral can be used to extract physical properties of QFTs either in the thermodynamic regime (using an appropriate finite Euclidean time extent) or in the zero-temperature vacuum (by extrapolating the Euclidean time extent to infinity).

### 2.5.1 Correlation functions

The simplest correlation function is a two-point function, generally given by a pair of operators $\mathcal{O}_1(0)$ and $\mathcal{O}_2(t)$ localized respectively to the Euclidean times 0 and $t$, where we have exploited time translation invariance to shift $\mathcal{O}_1$ to time 0 for simplicity. Evaluating the expectation value of such a two-point correlation function using Eq. (2.29) gives

$$\langle \mathcal{O}_2(t)\mathcal{O}_1(0)\rangle = \frac{\text{tr}[\mathcal{O}_2(t)\, e^{-Ht}\, \mathcal{O}_1(0)\, e^{-H(L_t-t)}]}{\text{tr}[e^{-Ht} e^{-H(L_t-t)}]}. \tag{2.31}$$

For large $L_t$, the factors of $e^{-H(L_t-t)}$ are dominated by $e^{-E_0(L_t-t)} |n_0\rangle \langle n_0|$, giving

$$\langle \mathcal{O}_2(t)\mathcal{O}_1(0)\rangle \approx \frac{1}{e^{-E_0 t}} \langle n_0|\mathcal{O}_2(t)e^{-Ht}\mathcal{O}_1(0)|n_0\rangle. \tag{2.32}$$

The spectral decomposition of $e^{-Ht}$ gives further insight into the structure of two-point correlator,

$$\langle n_0|\mathcal{O}_2(t)e^{-Ht}\mathcal{O}_1(0)|n_0\rangle = \sum_i e^{-E_i t} \langle n_0|\mathcal{O}_2(t)|n_i\rangle \langle n_i|\mathcal{O}_1(0)|n_0\rangle. \tag{2.33}$$

The lowest energy states $|n_i\rangle$ that are respectively created and annihilated by the operators $\mathcal{O}_1$ and $\mathcal{O}_2$ will dominate the sum, and considering the (multi-)exponential scaling of $\langle \mathcal{O}_2(t)\mathcal{O}_1(0)\rangle$ with $t$ in the large-$t$ regime allows extraction of the energy differences $E_i - E_0$ relative to the vacuum of these low-lying states.

At times $t' \in [0, t]$ in the intervening region of imaginary time, the single lowest state can be expected to dominate if there is significant separation between all temporal coordinates, $0 \ll t' \ll t$. In this case, constructing correlation functions by the insertion of additional operators at intermediate times $t'$ can be used to investigate the structure of states of interest. For example three-point functions are regularly employed in lattice QCD studies of hadronic parton distributions functions [155, 156] and electroweak form factors and decay constants [157–159].

### 2.5.2 Noise and sign problems

To extract controlled estimates of physical properties of particular states, it is important to ensure that the separation $t$ between operators is large enough to suppress 'contamination' from higher excited states accessed by $\mathcal{O}_1$ and $\mathcal{O}_2$. For structure studies involving 3-point correlation functions, one must also ensure that $t'$ and $t - t'$ are similarly large. Unfortunately, statistical noise increases exponentially in exactly these limits resulting in 'signal-to-noise problems', as we discuss next. In practice, one must choose the operators carefully and generate enough measurements to find a window of separations in which excited state contamination is controlled and statistical noise is suppressed. Fits to extract physical information can then be performed in this window.

Signal-to-noise problems appear in many observables of interest in phenomenologically relevant lattice field theories, including correlation functions in lattice QCD [160–174], nuclear many-body theories [175–181], condensed matter theories [182–188], and cold atom physics [189, 190]. The aim of the latter half of this dissertation is to address the growing statistical noise that determines the precision with which physical information can be extracted from such theories in state-of-the-art lattice calculations.

**Statistical noise.** Monte Carlo estimates of observables are given as functions of random samples of field configurations. Fluctuations over possible realizations of these random samples result in uncertainties in any estimate computed from a finite set of samples. In practice these uncertainties can be estimated either by error propagation [69] or bootstrap resampling [191]. In this dissertation, reported uncertainties are given by bootstrap resampling unless otherwise indicated. For quantities that are determined by taking the sample mean of random variates, the variance of the random variates controls the scale of statistical uncertainties. Understanding the variance of Monte Carlo estimators thus informs one of the practical difficulty of deriving precise estimates.

The structure of the growing statistical noise in two-point correlators can be understood in terms of the variance by a concise physical argument put forward by Parisi and Lepage in Refs. [192, 193]. A standard Monte Carlo estimator of the expectation value $\langle \mathcal{O}_2(t)\mathcal{O}_1(0)\rangle$ consists of the sample mean of the components $\mathrm{Re}[\mathcal{O}_2(t)\mathcal{O}_1(0)]$ and $\mathrm{Im}[\mathcal{O}_2(t)\mathcal{O}_1(0)]$ used to respectively extract the real and imaginary components. The variance of these two Monte Carlo estimators can be expressed as expectation values with respect to the same path integral weights:

$$
\begin{aligned}
\mathrm{Var}[\mathrm{Re}[\mathcal{O}_2(t)\mathcal{O}_1(0)]] &= \frac{1}{2}\left\langle \mathcal{O}_2(t)\mathcal{O}_2^*(t)\mathcal{O}_1^*(0)\mathcal{O}_1(0)\right\rangle \\
&\quad + \frac{1}{2}\mathrm{Re}\left\langle \mathcal{O}_2^2(t)\mathcal{O}_1^2(0)\right\rangle - \left\langle \mathcal{O}_2(t)\mathcal{O}_1(0)\right\rangle^2 \\
\mathrm{Var}[\mathrm{Im}[\mathcal{O}_2(t)\mathcal{O}_1(0)]] &= \frac{1}{2}\left\langle \mathcal{O}_2(t)\mathcal{O}_2^*(t)\mathcal{O}_1^*(0)\mathcal{O}_1(0)\right\rangle \\
&\quad - \frac{1}{2}\mathrm{Re}\left\langle \mathcal{O}_2^2(t)\mathcal{O}_1^2(0)\right\rangle - \left\langle \mathcal{O}_2(t)\mathcal{O}_1(0)\right\rangle^2 .
\end{aligned}
\tag{2.34}
$$

The term $\langle \mathcal{O}_2(t)\mathcal{O}_2^*(t)\mathcal{O}_1^*(0)\mathcal{O}_1(0)\rangle$ in the variances of both the real and imaginary components is itself a two-point correlation function. This correlation function also has a spectral representation given by the form in Eqs. (2.32) and (2.33) in terms of the composite operator $\mathcal{O}_1^*(0)\mathcal{O}_1(0)$ at time 0 and $\mathcal{O}_2^*(t)\mathcal{O}_2(t)$ at time $t$. The composite operators both involve the produce of an operator and its adjoint and therefore have vacuum quantum numbers. As a result, the spectral decomposition of the variance is dominated by the vacuum state, and the variance is asymptotically constant as $t \to \infty$. In contrast, the two-point correlator itself scales exponentially as $e^{-Et}$ in terms of the energy gap $E$ (with respect to the vacuum) of the state under study. For large values of $t$, the variance of Monte Carlo estimates of two-point correlation functions will then be exponentially larger than the expectation value, making these expectation values exponentially difficult to estimate.

This issue can be made precise by considering the *signal-to-noise (StN)* ratios of each Monte Carlo estimator,

$$\mathrm{StN}[\mathrm{Re}[\mathcal{O}_2(t)\mathcal{O}_1(0)]] = \frac{|\mathrm{Re}\,\langle \mathcal{O}_2(t)\mathcal{O}_1(0)\rangle|}{\sqrt{\frac{1}{n}\mathrm{Var}[\mathrm{Re}[\mathcal{O}_2(t)\mathcal{O}_1(0)]]}} \sim \sqrt{n}e^{-Et}$$

$$\mathrm{StN}[\mathrm{Im}[\mathcal{O}_2(t)\mathcal{O}_1(0)]] = \frac{|\mathrm{Im}\,\langle \mathcal{O}_2(t)\mathcal{O}_1(0)\rangle|}{\sqrt{\frac{1}{n}\mathrm{Var}[\mathrm{Im}[\mathcal{O}_2(t)\mathcal{O}_1(0)]]}} \sim \sqrt{n}e^{-Et}, \tag{2.35}$$

where $n$ is the number of Monte Carlo samples. The StN ratio determines the statistical precision of a Monte Carlo estimator, and values above unity are needed to resolve the expectation value from zero. As one can see from Eq. (2.35), two-point correlation functions are generically affected by a signal-to-noise problem, in that the StN ratios of these Monte Carlo estimators decrease exponentially quickly with the separation $t$. For these estimators, this can be counteracted only by increasing statistics. Unfortunately, resolving a signal up to a maximum separation $t$ requires exponentially large numbers of statistics for larger values of $t$. A similar argument extends to higher-order correlation functions, and for example implies that signal-to-noise ratios for three-point correlation functions are exponentially small in terms of the separations $t - t'$ and $t'$.

What can be done to remedy this situation? If one works with Monte Carlo estimators of two-point correlation functions taking the form $\mathcal{O}_2(t)\mathcal{O}_1(0)$, the 'Parisi-Lepage' scaling is unavoidable. The only escape from the bleak scaling of Eq. (2.35) is then to use Monte Carlo estimators that do *not* factorize into functions of the field localized to the Euclidean times 0 and $t$. However, it is simultaneously important to preserve the physically meaningful spectral representation in Eqs. (2.32) and (2.33) that is afforded by this structure. Any approach to reducing the exponential signal-to-noise problem of correlation functions must therefore carefully modify the Monte Carlo estimators utilized while maintaining this precise correspondence to the spectral representation.

It is interesting to note that this exponential scaling is partially avoided in lattice field theory calculations involving fermionic fields. For example, Monte Carlo estimates of the zero-momentum two-point correlation function of the pion in lattice QCD do not encounter an exponential signal-to-noise problem, though other states such as the nucleon are still affected by severe residual exponential signal-to-noise problems that restrict the precision of state-of-the-art calculations. This partial avoidance of an exponentially severe signal-to-noise problem is possible because fermions are analytically integrated out in the path integral, meaning observables involving fermions are resolved to concrete functions of bosonic degrees of freedom throughout the entire lattice. Monte Carlo estimators for fermionic two-point correlation functions therefore take the form of complicated functions of the sampled degrees of freedom, and the Parisi-Lepage scaling is avoided. The key to this (partial) improvement is starting with an observable with a well-defined spectral representation — the observable written using fermionic operators — then analytically manipulating the path integral to give a Monte Carlo estimator depending on field values throughout spacetime. A similar idea underpins the approach taken in Chapter 5.

**Sign and complex phase problems.** The signal-to-noise problem can also be related to the structure of sign/complex phase fluctuations of the Monte Carlo estimators under study [6, 9, 194, 195]. From Eq. (2.34), it is clear that the term dominating the variance is a function only of the magnitude $|\mathcal{O}_2(t)\mathcal{O}_1(0)|$ of the observable. The constant scaling of this term with $t$ implies that the magnitudes of each Monte Carlo sample are on average $O(1)$ quantities, while the true average value is exponentially small. This structure results in a *sign problem*, in which exponentially large statistics are required to precisely measure the delicate cancellation between $O(1)$ values with strongly varying signs/complex phases required to produce the much smaller expectation value.

Sign problems have long been recognized as obstacles preventing Monte Carlo simulations of lattice field theory at non-zero chemical potential [196–199] and in real time [200–202] and have inspired several related approaches in these settings. In these cases, the sign problem is extensive in the volume because the total complex phase of the integrand acquires fluctuating contributions from the field fluctuations throughout the spacetime volume. The distribution over complex phases that must be averaged in these approaches is then the result of 'wrapping' modulo $2\pi$ the very broad distribution given by the sum of these contributions. The resulting wrapped distribution is a nearly uniform distribution over complex phases which gives rise to strong cancellations in expectation values [194, 203–205]. Cumulant expansion methods based on the 'unwrapped' or 'extensive' phase given by adding the complex phase contributions without projecting into the $[0, 2\pi]$ domain have been applied to resolve sign problems in some theories with non-zero chemical potential [203–208], though the method introduces a source of systematic error and is not easily extended to complicated theories like lattice QCD.

An analogy can be made to the sign problem in correlation functions. The phase distributions of correlation functions between Euclidean times 0 and $t$ can likewise be attributed to an extensive sum over contributions, in this case from values in the intervening times $t' \in [0, t]$. As $t$ is taken large, these phase distributions also converge to nearly uniform wrapped distributions [194]. Pubs. [6, 9] developed an analogous phase unwrapping approach for the measurement of two-point correlation functions. In this approach, unwrapped phases are defined as functions of the phase fluctuations of the two-point correlation function at intervening times $t' \in [0, t]$. Low-order cumulants of the unwrapped phase distribution can be precisely estimated and truncated cumulant estimates of the two-point correlation function were found to significantly improve precision in proof-of-principle demonstrations. Truncating the cumulant expansion, however, introduces uncontrolled systematic uncertainties as in the case of phase unwrapping approaches for extensive sign problems. We can nevertheless extract some insight from these investigations:

1. Phase fluctuations causing the signal-to-noise problem in two-point correlation functions can be attributed to fluctuating contributions throughout the interval $[0, t]$ of Euclidean time, each of which fluctuates much less severely than the phase of the overall correlation function.

2. Introducing a dependence on intervening values of field configurations is necessary to avoid Parisi-Lepage scaling, but in particular depending on the complex phases of these values is sufficient to produce Monte Carlo estimators that are not afflicted by severe sign and signal-to-noise problems.

**Contour deformations.** In addition to the unwrapping method discussed above, extensive sign problems have also been addressed in several lattice theories using complex contour deformations of the path integral (see Ref. [209] for a review and Chap. 5 for a detailed discussion). Deformed path integrals integrate over manifolds in complexified field space which generally result in very different typical values of the integrand, and as a result Monte Carlo integration over these manifolds can mitigate or remove sign problems. Path integral contour deformations can be shown to leave the total integral value unchanged by higher-dimensional analogues of Cauchy's theorem for holomorphic integrands, so no systematic bias is introduced as a result of these methods.

Inspired by these approaches to the extensive problem and insights from phase unwrapping, Chapter 5 introduces a contour deformation approach to defining Monte Carlo estimators for two-point correlation functions. More generally, we detail how this approach can be applied to any observables in theories with a real action and, in contrast to methods for extensive problems, does not require additional Monte Carlo sampling. We apply insights based on the structure of the observable signal-to-noise problem and phase fluctuations to introduce families of contour deformations that result in significantly improved Monte Carlo estimators. These new Monte Carlo estimators generally acquire a dependence on field values throughout the spacetime volume, modifying the scaling of the variance while having an identical expectation value to the original correlators with well-defined spectral representations. As a result, the physical interpretation of measured values is preserved, but this approach mitigates the Parisi-Lepage scaling and can exponentially improve the signal-to-noise ratio for observables.

**Related approaches to the noise problem.** Besides the methods based on complexification of field variables discussed in Chapter 5, several techniques have been considered to mitigate noise problems (and associated sign problems) in lattice field theory. In dual-variable approaches [111, 196, 210–225] noisy fluctuations are analytically integrated out using specific mathematical features of particular theories. Multi-level methods [226–234] exploit factorization of observables and the action to exponentially increase Monte Carlo averaging over fluctuations. Density of states / histogram methods [204–207, 235, 236] factor path integrals into sign-problem-free estimates of density/histogram weights and numerical evaluations of the remaining integration over these weights. Tensor networks [237–241] can be applied in low-dimensional systems to explicitly represent ground and excited states and evaluate expectation values without a sign or noise problem. An auxiliary field method was found to remove the sign problem in the Nambu-Jona-Lasinio model [242]. Finally, the choice of operators used to access physical states can be optimized to reduce noise [167].

Promising recent results have been observed for several of these approaches. However, these methods have not yet been found to make significant progress in the phenomenologically relevant setting of four-dimensional lattice gauge theories, such as lattice QCD, in most cases due to intractable computational costs in higher spacetime dimensions or a reliance on particular mathematical properties that do not extend to this setting. This gap motivates our investigation of path integral contour deformations in Chapter 5. Our proof-of-principle applications include $SU(N)$ gauge theory, and though we work with low-dimensional theories, the method can be straightforwardly extended to higher spacetime dimensions.

# Chapter 3

# Flow-based ensemble generation

*Content in this chapter is partially adapted with permission from:*

- Pub. [7]: M. S. Albergo, D. Boyda, D. C. Hackett, G. Kanwar, K. Cranmer, S. Racanière, D. J. Rezende, and P. E. Shanahan, "Introduction to Normalizing Flows for Lattice Field Theory", (2021), arXiv:2101.08176 [hep-lat]

- Pub. [2]: D. Boyda, G. Kanwar, S. Racanière, D. J. Rezende, M. S. Albergo, K. Cranmer, D. C. Hackett, and P. E. Shanahan, "Sampling using SU($N$) gauge equivariant flows", Phys. Rev. D **103**, 074504 (2021), arXiv:2008.05456 [hep-lat]

- Pub. [5]: M. S. Albergo, G. Kanwar, and P. E. Shanahan, "Flow-based generative models for Markov chain Monte Carlo in lattice field theory", Phys. Rev. D **100**, 034515 (2019), arXiv:1904.12072 [hep-lat]

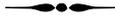

Critical slowing down of sampling is tied to local/diffusive updates within the framework of MCMC (see Sec. 2.4.2). This chapter introduces a new approach to generating ensembles for lattice field theory which circumvents such diffusive updates by directly sampling from approximations to target distributions. These approximations are provided by *flow-based models* [152–154], a class of machine learning models that give flexible parameterizations of high-dimensional distributions that can be variationally optimized and efficiently sampled. These models have been successfully applied to sampling in a variety of applications outside of physics [243–250], and they have also recently been used to generate samples from molecular/many-body systems [124, 142–146, 251] and from some lattice systems [138–140, 252]. In contrast to other generative machine learning approaches applied to physical systems [123, 125, 128–136, 147–151, 253, 254], the probability density associated with a flow-based model distribution can be measured, allowing the method to be made exact by methods described in this chapter. In particular, we describe a *flow-based Markov chain* that utilizes samples from a flow-based model as proposals for Metropolis-Hastings Markov chain steps, guaranteeing exactness in the limit of an asymptotically long Markov chain. These guarantees are analogous to traditional methods and hold regardless of the quality of the flow-based model.

Machine learning methods have also recently been applied to lattice calculations to extract information about phase transitions and the renormalization group flow [255–308], predict values of correlation functions [309, 310], apply inverse Fourier and Laplace transformations [311–313], and measure the topological charge [314, 315]. These promising methods to improve observable estimates and extract useful human-interpretable information from sampled field configurations are complementary to the work presented in this chapter. We further explore complementary machine learning methods for observable estimation in Chapter 5.

If successfully applied to state-of-the-art lattice field theory calculations, flow-based models promise several advantages over traditional approaches to ensemble generation:

1. Despite reintroducing the use of a sequential Markov chain, proposals from such a flow-based model are completely uncorrelated with the previous state in the chain. Thus when a high acceptance rate can be consistently achieved, these provide global updates that have the potential to completely resolve critical slowing down.

2. The exactness of this method is decoupled from the quality of approximation of the flow-based model: optimizing the probability distribution encoded in the flow-based model improves the efficiency of the Markov chain by reducing the rejection rate, but does not affect correctness. By factorizing correctness from efficiency from the start, one can flexibly decide on a trade-off between initial investment in optimization of the model versus the additional cost of sampling using a model that poorly approximates the target distribution.

3. Models optimized for a specific set of physical parameters are also expected to serve as useful starting points for targeting distributions associated with other similar physical parameters. In many lattice field theory calculations, ensembles must be generated over a range of physical parameters to perform controlled extrapolation/interpolation, for example to the continuum limit, infinite volume limit, or towards physical values of the quark masses (in the case of lattice QCD). For these studies, it may therefore be advantageous to invest significant time optimizing a model in the neighborhood of several choices of target parameters or volumes, because this cost of training can be amortized over the generation of multiple ensembles.

4. The generation of proposals is an 'embarrassingly parallel' process, and is well suited to execution at scale on large-scale supercomputing resources.

5. Flow-based sampling is structured terms of training and evaluation of machine learning models which can be performed using increasingly optimized libraries and specialized hardware architectures [316–318].

6. Ensemble storage costs can be significantly reduced or eliminated: in cases where the model optimization step dominates the cost of ensemble generation, it may be practical to only store the model for long-term usage.

The remainder of this chapter reviews flow-based models (Secs. 3.1 and 3.2), details our methods to construct and optimize such models for lattice field theory distributions (Secs. 3.3 and 3.4), introduces a method to draw unbiased samples from such models (Sec. 3.5), presents a proof-of-principle study performed for lattice scalar field theory (Sec. 3.6), and discusses future scaling prospects (Sec. 3.7).

## 3.1 Flow-based models

In this section we review the use of flow-based models (or 'normalizing flow models') for sampling, with a focus on approaches well-suited for sampling lattice field theory distributions. Refs. [319, 320] thoroughly cover flow-based models from a general standpoint and give a review of the established and recent work on the topic. As a relatively new field, some of the terminology surrounding flow-based models is not well-established, and alternative terms used in the literature are mentioned where appropriate.

Flow-based models are inspired by the idea that the density of points in a continuous space can be squished/stretched by a change of variables mapping the space to itself. One can use this map to transform from a *prior distribution* (or 'base distribution') to a new output distribution over that space. The change-of-variables map is termed a *flow*, typically labeled $f$, and should be invertible and differentiable in both directions (i.e. a diffeomorphism).[1] In constructing a flow-based model, the aim is to choose a flow and prior distribution such that the model output distribution approximates a target distribution of interest. In the following, the functions $r(\cdot)$, $q(\cdot)$, and $p(\cdot)$ will be used to specify the probability densities associated with the prior, model, and target distributions, respectively.

With flow-based models one can efficiently:

1. **Measure the model probability density** $q(x)$ associated with a sample $x$ in the output space. This is determined by the prior probability density and the Jacobian determinant of the flow as

$$q(x) = r(z)[J(z)]^{-1} = r(z) \left| \det_{kl} \frac{\partial f_k(z)}{\partial z_l} \right|^{-1},\qquad(3.1)$$

where $z \equiv f^{-1}(x)$ is the preimage of $x$, $k$ and $l$ label real degrees of freedom defining these samples, and $J(z)$ is the Jacobian determinant of the transformation. The factor $J(z)$ gives the change in an infinitesimal volume element as it is mapped through the function, and therefore it is the inverse of this quantity that appears in the formula for the change in density (which is inversely related to the volume).

2. **Draw samples** from the model distribution. This can be achieved by sampling $z$ from $r(z)$, which we denote by $z \sim r(z)$, and applying the flow to produce an output sample $x = f(z)$. The probability density associated with the produced sample can be computed simultaneously using the prior sample $z$ and Eq. (3.1). Note that the inverse function $f^{-1}$ does not need to be implemented or applied if this is the only operation required in practice.

These two operations are typically associated with different settings and uses of the flow-based model. Measuring the probability density on an existing set of samples is useful when attempting to reproduce a distribution implicitly defined by a 'training set'. Drawing samples with associated densities directly from the model is useful when attempting to reproduce a distribution explicitly given by a computable output probability density. In the lattice field theory setting, large training sets of existing samples are often not available, while the explicit probability density function (encoded in the action) is easily computed. Efficiently drawing samples is thus the relevant operation for the application of flow-based models to lattice field theory, and simultaneously generating the density associated with sampled points is necessary to ultimately produce unbiased samples from the target density given the flow-based approximation.

---

[1] The term 'flow' originates from an initial approach based on applying a differential equation to continuously transform the density in the space [152]. In more recent literature this term has come to be used in the sense described here. Continuous versus discrete flows are discussed in more detail below.

### 3.1.1 A simple example: the Box-Muller transform

Many codes for sampling from the normal distribution use a surprisingly common instance of flow-based sampling — the Box-Muller transform [321]. We cover this simple example to shed some light on how flows can be used to efficiently sample from particular distributions.

The Box-Muller algorithm simultaneously samples two independent and identically distributed (IID) Gaussian samples with unit-variance, following Algorithm 3.1:

---

**Algorithm 3.1:** Generate two IID, unit-variance Gaussian samples using the Box-Muller transform (a simple flow-based sampler)

---

1. Draw $U_1$ and $U_2$ from the uniform distribution over $[0, 1]$.

2. Compute and return

$$Z_1 = \sqrt{-2 \log U_1} \cos(2\pi U_2), \quad Z_2 = \sqrt{-2 \log U_1} \sin(2\pi U_2). \qquad (3.2)$$

---

The uniform distribution over $[0, 1]^2$ can be considered the prior distribution of the Box-Muller procedure, and the associated probability density is simply $r(U_1, U_2) = 1$. The invertible transformation defined by Eq. (3.2) can be considered the flow. Figure 3.1 depicts how this transformation results in samples from the normal distribution. In particular, the transformation maps $U_1$ to a radial coordinate and maps $U_2$ to an angular coordinate, with the precise mapping constructed to increase the density of points near the peak of the normal distribution and decrease the density far from the peak. We can confirm that the unit-variance normal distribution is produced by explicitly computing the probability density $q(Z_1, Z_2)$ at the output point $(Z_1, Z_2)$ using the change-of-variables formula given in Eq. (3.1):

$$
\begin{aligned}
q(Z_1, Z_2) &= r(U_1, U_2) \left| \det_{kl} \frac{\partial Z_k(U_1, U_2)}{\partial U_l} \right|^{-1} \\
&= 1 \times \left| \det \begin{pmatrix} \frac{-1}{U_1 \sqrt{-2 \ln U_1}} \cos(2\pi U_2) & -2\pi \sqrt{-2 \ln U_1} \sin(2\pi U_2) \\ \frac{-1}{U_1 \sqrt{-2 \ln U_1}} \sin(2\pi U_2) & 2\pi \sqrt{-2 \ln U_1} \cos(2\pi U_2) \end{pmatrix} \right|^{-1} \\
&= \left| \frac{2\pi}{U_1} \right|^{-1}.
\end{aligned}
\tag{3.3}
$$

Rearranging the change of variables, one can identify $U_1 = \exp(-(Z_1^2 + Z_2^2)/2)$, finally arriving at the output density

$$
q(Z_1, Z_2) = \frac{1}{2\pi} e^{-(Z_1^2 + Z_2^2)/2}.
\tag{3.4}
$$

This is exactly the probability density associated with two IID, unit-variance Gaussian variables. For the Box-Muller transform, one can see in Eq. (3.3) that the Jacobian determinant of the flow is entirely responsible for the structure of the model distribution; in general flow-based models, a combination of the structure of the prior distribution and Jacobian determinant of the flow may play a role in producing the model distribution. In this low-dimensional example, it was also possible to explicitly form the Jacobian and compute the determinant. To extend the method to high-dimensional spaces, flow functions must be constructed that allow efficient calculation of the density without a 'brute-force' evaluation of the Jacobian determinant which generically requires $O(\mathcal{N}^3)$ operations for $\mathcal{N}$-dimensional samples.

### 3.1.2 The general approach

The Box-Muller transform uses a fixed transformation that exactly produces the desired model distribution. In contrast, an exact transformation producing the target distribution is typically not known for general applications to lattice field theories. A key component of many flow-based models is to instead work with a variational family of flow functions $f_\omega$, parameterized by a collection of parameters $\omega$. The output distribution can then be considered a variational ansatz with a parameterized probability density $q_\omega(\cdot)$ that can be optimized to closely approximate the target probability density $p(\cdot)$.

We expect the optimal choice to be a good approximation to the target distribution if the family of functions is *expressive*, meaning that a large variety of possible functions are included in the family. All choices of flow functions $f_\omega$ in the family must also (1) be invertible, (2) be differentiable, and (3) have a tractable Jacobian determinant, as outlined above. Composition of functions is a powerful approach to achieving expressive functions satisfying all three properties: we can define the flow function by

$$
f_\omega = g_{\ell; \omega_\ell} \circ \cdots \circ g_{1; \omega_1},
\tag{3.5}
$$

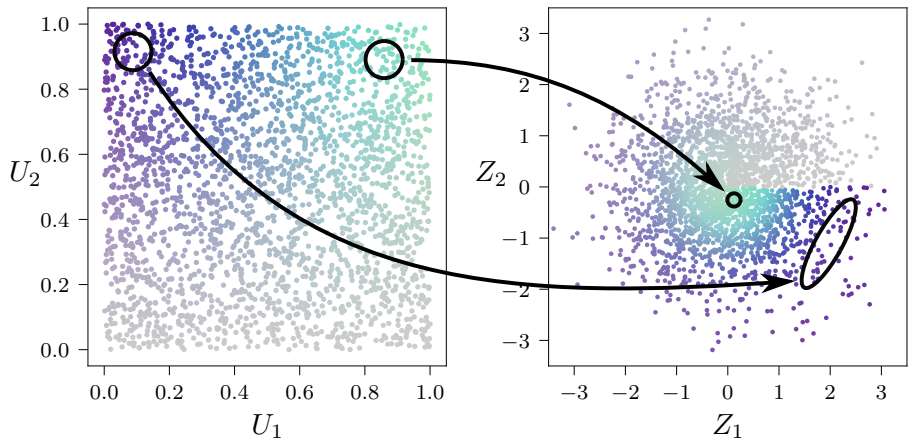

**Figure 3.1:** Box-Muller transformation from the uniform prior distribution over $[0, 1]^2$ to the uncorrelated, unit-variance normal distribution in $\mathbb{R}^2$. The mapping defined in the main text is illustrated using hue and saturation to respectively distinguish the $U_1$ and $U_2$ coordinates. The output distribution for the Box-Muller transformation is entirely a result of the Jacobian determinant of the map itself, and the overlaid arrows depict the effect of this change of measure on two groups of points. The measure is concentrated for regions mapped near the origin, while it is stretched for regions mapped far from the origin.

with $\omega = \bigcup_{i=1}^{\ell} \omega_i$ consisting of the union of parameters in each 'layer' $g_i$. In the following, the explicit parameterization of functions will often be suppressed to keep notation concise, but the (sometimes implicit) variational parameterization is a key feature of all flow-based models discussed throughout the remainder of this dissertation. The composed flow will satisfy all three desired properties when each individual layer satisfies these properties. This is immediate for invertibility and differentiability by the rules of function composition, and the chain rule further gives a prescription for efficiently computing the composed Jacobian determinant as a product of the Jacobian determinants of each layer:

$$J_f(z) = J_{g_\ell}(z_\ell) \dots J_{g_2}(z_2) J_{g_1}(z_1), \tag{3.6}$$

where

$$z_1 \equiv z \quad \text{and} \quad z_i \equiv [g_i \circ \dots \circ g_1](z). \tag{3.7}$$

Composition enables expressive functions to be constructed from relatively simple components that each can be much more easily confirmed to satisfy the desired properties.

A number of approaches have been developed to construct expressive families of flows based on this principle. The variety of options and their relevance for sampling distributions associated with lattice field theory are reviewed below.

**Autoregressive flows** — Distributions over a multi-dimensional variable can generally be decomposed into a product over conditional distributions for each compo-nent,

$$p(x) \equiv p(x_1, \ldots, x_{\mathcal{N}}) = p(x_1) \, p(x_2|x_1) \, \ldots \, p(x_{\mathcal{N}}|x_{\mathcal{N}-1}, \ldots, x_1), \tag{3.8}$$

where $\mathcal{N}$ indicates the number of components in the multi-dimensional variable $x$. Autoregressive methods [322–332] aim to sample approximately according to $p(x)$ by conditionally sampling components $x_i$ from approximations to the conditional distributions given in Eq. (3.8). This can be accomplished using flows by the composition of layers acting on individual components,

$$g_i^{\mathrm{AR}}(z_1, \ldots, z_{\mathcal{N}}) = (z_1, \ldots, z_i', \ldots, z_{\mathcal{N}}), \tag{3.9}$$

where the transformation of the $i$th component $z_i \to z_i'$ is conditioned only on $z_1, \ldots, z_{i-1}$ (i.e. $z_{i+1}, \ldots, z_{\mathcal{N}}$ are simply ignored by the layer). Typically a simple invertible operation is applied to $z_i$ to produce $z_i'$; for example, an affine transformation could be applied as

$$z_i' = \sigma_i(z_1, \ldots, z_{i-1}) \, z_i + \mu_i(z_1, \ldots, z_{i-1}), \tag{3.10}$$

with the parameters $\sigma_i$ and $\mu_i$ of the transformation given by (potentially complicated) functions of the conditioned variables. Note that the functions $\sigma_i$ and $\mu_i$ need not be invertible, as they depend on components of $z$ which are unmodified by the action of $g_i^{\mathrm{AR}}$ and thus the output of $g_i^{\mathrm{AR}}$ contains enough information to reproduce the values of $\sigma_i(z_1, \ldots, z_{i-1})$ and $\mu_i(z_1, \ldots, z_{i-1})$ that are required for the inverse transformation. The expressivity of the flow lies in the choice of invertible operation and the parameterization of the functions applied to the conditioned variables.

In an autoregressive flow, each transformation has a triangular Jacobian. For the transformation formulated in Eq. (3.9), exactly one row of the Jacobian will differ from the identity matrix and will only contain non-zero entries to the left of the diagonal. The Jacobian determinant in each layer can thus be calculated quite efficiently from the single non-unit diagonal entry in each layer. In the affine transformation in Eq. (3.10), for example, the Jacobian determinant would be given by $J_{g_i^{\mathrm{AR}}}(z) = |\sigma_i(z_1, \ldots, z_{i-1})|$.

There are many variations on this idea. One example of such a variation is to update each component $z_i$ in parallel, conditioned on the original $z_1, \ldots, z_{i-1}$, rather than updating sequentially, conditioned on the final $z_1', \ldots, z_{i-1}'$. In principle both have the same expressive power in the limit of fully general one-component transformations, but in practice these approaches can have differing efficiency and expressivity and they are each suited to distinct settings [333].

While autoregressive flows are effective in situations in which the components of the variables can be sensibly ordered, in the context of lattice field theory these components are associated with specific sites of a geometric lattice, and the locality and translational invariance of the lattice is lost by an autoregressive approach. This does not necessarily preclude an autoregressive flow being used to draw samples for lattice field theory, but this approach may require complicated functions that only recover locality and translation invariance of the target distribution after optimization. The importance of symmetries in efficiently learning lattice field theory distributions is discussed in detail in Sec. 3.3 and motivates alternate approaches to flow-based models for lattice field theory.

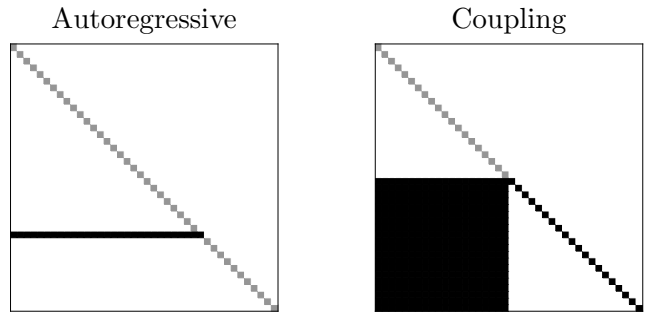

**Figure 3.2:** Structure of non-zeros in the Jacobian for an autoregressive layer (left) versus a coupling layer (right), where gray cells represent the value 1 and black cells represent a generic non-zero value. Whereas the transformation in an autoregressive layer depends on all prior components in the canonical ordering, each transformed variable in a coupling layer depends only on components within the frozen subset of components, resulting in a non-trivial lower left block. In both cases, the determinant can be efficiently computed from the diagonal entries.

**Coupling layers** —   Rather than conditionally update components in a canonical order, one can choose to update more components simultaneously by conditioning on fewer variables in the definition of the transformation. In doing so, geometric symmetries may be enforced and the parallelism of the transformation is improved. Coupling layers [243, 244] strike this compromise. In each coupling layer, components of the input are split into two subsets — a *frozen* subset that is unmodified by the layer and an *updated* subset that is invertibly transformed elementwise, conditioned on the frozen subset. As with autoregressive layers, this ensures the Jacobian is triangular and the Jacobian determinant can be efficiently computed by accumulating the contributions from the diagonal elements. Figure 3.2 compares the structure of non-zeros in the Jacobian for an autoregressive layer versus a coupling layer.

To incorporate translational symmetry in a coupling layer, the subsets can be chosen to be symmetric under (a subset of) the lattice translational group. For example, a checkerboard decomposition of the lattice preserves half of the translational invariance when fields are defined on sites of the lattice. The elementwise transformations of the coupling layer can then be constructed to preserve translational invariance and include a degree of locality. These desirable features of coupling layers lead us to choose this approach for our studies of lattice ensemble generation below.

**Residual flows** —   Residual flows are based on the idea of enforcing invertibility using Lipschitz bounds rather than architectural constraints. In general, a residual flow layer is written as a 'residual network' [334] which takes the form of a perturbation on the identity function, $g^{\mathrm{Res}}(z) = z + \hat{g}(z)$. To guarantee invertibility, it is sufficient to further enforce that $\hat{g}$ is a contractive function [335], i.e. that it is Lipschitz continuous with Lipschitz constant 1. This can be enforced by controlling the spectral norm of every operation [336, 337]. In this form, tractably computing the Jacobian determi-nant for training and unbiased sampling is not necessarily straightforward. Ref. [338] introduces an unbiased and efficient estimator of the log Jacobian determinant, which enables training of the model in cases where previous biased estimators failed. However, exactly evaluating the Jacobian determinant is still expensive, making residual flows ill suited for the lattice field theory setting where exact measurements are necessary to produce asymptotically exact observable estimates. Other methods can also be used to guarantee invertibility of the residual flow layer [153, 154, 339], though these methods have limited expressivity.

**Continuous flows** — Taking the limit of composing a large number of near-identity transformations, one arrives at a continuous flow. Continuous flows [340] define an invertible transformation by the integration of an ordinary differential equation (ODE) over a period of fictitious 'flow time':

$$\frac{dz_i}{dt} = [g(z(t); t)]_i \quad \rightarrow \quad z(t) = \int_0^t dt' \frac{dz}{dt'} = \int_0^t dt' \, g(z(t'); t'). \tag{3.11}$$

The resulting variable $z(T)$ at some final time $T$ determines the output of the continuous flow. The reversibility of ODEs ensures that the transformation is invertible. A similar approach based on a linearized form of the Monge-Ampère equation can be applied to give an ODE in terms of the gradient of a scalar function $g(z(t); t) = \nabla \varphi(z(t); t)$, rather than constructing and optimizing the vector-valued function $g(z(t); t)$ directly [139].

In this continuous formulation, the total change in density can be specified using a second differential equation,

$$\frac{\partial \log q(z(t))}{\partial t} = -\text{tr}_{kl} \frac{\partial g_k(z(t))}{\partial z_l(t)}, \tag{3.12}$$

under the assumption that $g$ is uniformly Lipschitz continuous in $z$ and is continuous in $t$ [340]. This can in principle be integrated over the time interval $[0, T]$ to arrive at the total change in probability density analogous to the Jacobian factor in Eq. (3.1). Whereas the determinant factor could, for a dense Jacobian, require $O(\mathcal{N}^3)$ operations to compute for an $\mathcal{N}$-dimensional sample, the trace in Eq. (3.12) can be evaluated in $O(\mathcal{N})$ operations.

A further advantage particularly associated with Monge-Ampère flows for the lattice field theory setting is that the scalar function receives all components of the sample as input and can easily be designed to be invariant to all symmetries of interest [139]. By Eq. (3.12) this results in an invariant probability density $q(z(T); T)$ if the boundary condition $q(z(0); 0)$ is suitably chosen.

On the other hand, the accuracy of the density estimates produced from continuous flows depends on the accuracy of the integration scheme applied. Many small steps may be required to achieve the desired accuracy in estimation of the density to ensure unbiased sampling for lattice field theory distributions using exact flow-based sampling approaches discussed in Sec. 3.5. In contrast, in an approach applying composition of discrete invertible steps it is possible to exactly compute the model probability density associated with samples. This difficulty motivates the use of discrete coupling layers in the remainder of this work.

### 3.1.3   Relation to the trivializing map

The term 'normalizing flow' originates from the idea that estimation of the probability density of an unknown distribution is possible if it is transformed by an (inverse) flow to a tractable distribution, such as a normal distribution [152]. This original method is based on evolving samples — and implicitly their corresponding distribution — according to a differential equation that has the normal distribution as a stationary solution. A similar line of thought was followed in the introduction and study of the 'trivializing map' that takes a lattice gauge theory distribution to the distribution of a trivial theory [341]. A trivial lattice gauge theory is defined by a uniform distribution over the (compact) sample space, thus this can be considered the limit of an infinite-width Gaussian distribution, and the method achieves similar goals to that of Ref. [152]. The trivializing map is in essence a very particular continuous normalizing flow defined by integrating a differential equation over a large interval of fictitious flow time, with the asymptotic behavior guaranteed to converge towards the trivial theory.

The trivializing map has been tested in the context of accelerating HMC simulations of the $CP^{N-1}$ model [342], which suffers from severe critical slowing down. In this approach, field configurations were sampled according to the nearly trivial distribution given by applying the trivializing map with a finite integration time. Though the autocorrelation times were reduced by simulating according to this nearly trivial distribution, the cost of simulation was also increased because HMC forces had to be computed by repeatedly integrating the differential equation defining the trivializing map. The tradeoff between the improvements from simulating distributions closer to trivial and the cost of integrating the trivializing map for longer intervals was found to quickly be unfavorable, such that only small improvements in overall simulation complexity could be made, and the scaling towards the continuum was not improved.

Despite the modest results achieved by the trivializing map and the similarity to the original formulation of normalizing flows, modern flow-based architectures allow progress beyond these early results. The discrete and continuous flows discussed in the previous section have departed from the approach used in Refs. [152, 341]. Figure 3.3 contrasts these approaches with the trivializing map; rather than following a fixed continuous flow that only asymptotically converges to the desired map, these flow-based architectures explore general maps between the prior and target distributions. Complex and high-dimensional distributions have been successfully captured by these modern approaches [139, 146, 243–245]. This dissertation demonstrates in proof-of-principle applications that these generalized flow-based approaches can also sample efficiently from the distributions associated with lattice field theories, in contrast to the trivializing map investigated in previous studies.

### 3.1.4   Prior distributions

The only constraints on the choice of prior distribution in a flow-based model are that it must have an efficiently computed probability density and must be tractable to sample from. As discussed below in Sec. 3.3, it is also useful, though not necessary, to choose a prior distribution satisfying the symmetries of the lattice field theory under study. For lattice fields with real components, a simple choice is an IID Gaussian distribution with zero mean and unit variance, given by the probability density

$$r(\phi) = \prod_{x,i} \frac{1}{\sqrt{2\pi}} e^{-(\phi_x^i)^2/2}, \qquad (3.13)$$

where we use the generic notation $\phi_x^i \in \mathbb{R}$ for the $i$th field component associated with lattice site $x$. Using an IID distribution ensures that the prior distribution is invariant under geometric symmetries of the lattice, including translations, rotations, and reflections. This distribution is also invariant under a $\mathbb{Z}_2$ symmetry corresponding to negation of the fields and under unitary rotations between field components $i$.

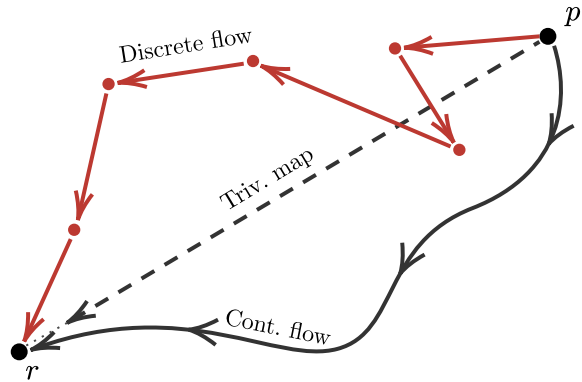

**Figure 3.3:** Schematic comparison between flow-based models consisting of discrete layers (red points/arrows), continuous flow-based models (curved gray path), and the trivializing map (dashed gray path). Arrows are oriented to indicate mapping the target distribution $p$ to a trivially sampled prior distribution $r$, consistent with the definition of the trivializing map. However, either direction can be computed (the maps are invertible) and for example the opposite direction of the map would be used when sampling. The trivializing map defines a specific continuous path through the space of distributions that asymptotically approaches the prior distribution when starting from the target distribution. The layers of a discrete flow-based model or the ODE of a continuous flow-based model are optimized to find *some* path approximately from the target distribution to the prior distribution. We specifically construct and apply discrete flow-based models for ensemble generation in this chapter.

For lattice gauge theories, the prior distribution must be defined over the compact space associated with gauge field configurations. A simple choice is given by an IID uniform distribution (with respect to the Haar measure) over all gauge links. The definition of this prior distribution and the associated sampling procedure are deferred to the construction of flow-based models for lattice gauge theory in Chapter 4.

One can also write a variety of possible probability densities that satisfy the symmetries of the lattice field theory while relaxing the IID nature of the prior distribution and introducing correlations. For example, the Boltzmann distribution associated with any choice of physical parameters is in principle a suitable prior distribution. For many of these choices, however, it far from clear how to efficiently and exactly sample from the distributions (this is after all the goal of much of this dissertation!) but this inspires a few possible alternatives to the IID prior distributions applied in this work. For one, gradient flow [343, 344] cuts off high-frequency fluctuations and increases correlations over a longer length scale by a diffusive process; one could therefore choose to work with the distribution resulting from applying a number of steps of gradient flow to samples from an IID probability measure. Gradient flow can be performed using an invertible integrator and the resulting probability density can then be exactly computed by exactly accumulating the Jacobian of the gradient flow operations [341]. A second option is to use a flow-based model that has been trained for another set of physical parameters as the prior distribution for the training of a new flow. This can either be interpreted as a fixed prior distribution, or as a prefix to the flow-based model that is just initialized using previous training. In the first case the weights in the model defining this prior distribution would be held constant, while in the second case the weights would be optimized along with the weights of the remaining flow. These approaches can be thought of 'preconditioners' for the flow, in that they provide a constant prefix to the flow that preemptively transforms the fields in a way that aims to partially construct the desired correlation structure.

## 3.2 Coupling layers using neural networks

Of the approaches to defining flow-based models discussed in the previous section, coupling layers are the only choice for which translational invariance can be exactly captured, the probability density can be exactly evaluated, and the flow can be efficiently applied using parallelized updates of field components. Coupling layers are defined in detail in this section.

### 3.2.1 Masking patterns and context functions

Coupling layers in general divide the components of their input into an updated subset and a frozen subset, applying an invertible elementwise transformation to the updated subset conditioned on the frozen subset. For a lattice field theory, the inputs and outputs of coupling layers are fields defined over a discretized spacetime lattice. It is convenient to define the frozen and updated subsets of a coupling layer by a *masking pattern* [244]

$$
m_x^i = \begin{cases} 0 & \text{indicating } \phi_x^i \text{ will be updated} \\ 1 & \text{indicating } \phi_x^i \text{ will be frozen and conditioned upon.} \end{cases} \tag{3.14}
$$

In terms of the masking pattern, the elementwise product $[m\phi]_x^i \equiv m_x^i \phi_x^i$ contains only information from frozen degrees of freedom. This construct then allows the action of the coupling layer to be structured to accept the masked lattice field $m\phi$ (many components of which may be zero), regardless of the choice of masking pattern and the number of frozen degrees of freedom.

We write the elementwise transformation defining the coupling layer as a manifestly invertible one-dimensional function parameterized by the output of *context functions* applied to the frozen subset of the field, $m\phi$. For example, a coupling layer $g^{\text{scale}}$ could be defined to apply a scaling operation to updated components as

$$g^{\text{scale}}(\phi) \equiv m\phi + (1 - m)[e^{s(m\phi)}\phi], \tag{3.15}$$

where the scaling factor is defined by the context function $s(\cdot)$. Note that this context function is defined to produce an output with the same number of degrees of freedom as $\phi$ and elementwise operations are assumed throughout Eq. (3.15). The use of the masking pattern and its complement in Eq. (3.15) ensures that the frozen components of the field are not updated, $mg^{\text{scale}}(\phi) = m\phi$. As a result, the coupling layer is manifestly invertible, with the inverse function explicitly given by

$$[g^{\text{scale}}]^{-1}(\phi') \equiv m\phi' + (1 - m)[e^{-s(m\phi')}\phi']. \tag{3.16}$$

While we require the elementwise operation on the updated subset to be an invertible one-dimensional function, context functions defining the transformation parameters need not be invertible and in general have few restrictions placed upon them; here the context function $s(\cdot)$ is used as a scaling factor, so any function producing real-valued output is acceptable. The context functions are generally variationally defined by a collection of parameters $\omega$ (see Sec. 3.1.2).

By fixing the form of the elementwise function and placing the flexibility and variational parameterization entirely within the context functions, the coupling layer is guaranteed to be invertible, be differentiable, and have a tractable Jacobian determinant for all choices of variational parameters. Explicitly writing the dependence on the variational parameters $\omega$, we can write an arbitrary coupling layer and its inverse as

$$\begin{aligned}
g_\omega(\phi) &= m\phi + (1 - m)\left[\hat{g}(\phi;\ s_{1;\omega}(m\phi), s_{2;\omega}(m\phi), \dots)\right] \\
g_\omega^{-1}(\phi') &= m\phi' + (1 - m)\left[\hat{g}^{-1}(\phi';\ s_{1;\omega}(m\phi'), s_{2;\omega}(m\phi'), \dots)\right],
\end{aligned} \tag{3.17}$$

where the variational parameters appear only in the context functions $s_1, s_2, \dots$ that parameterize the invertible elementwise function $\hat{g}$. The invertibility of the coupling layer rests upon the invertibility of $\hat{g}$ and the division into a frozen subset $m\phi$ and updated subset $(1 - m)\phi$. The Jacobian determinant of this arbitrary coupling layer can be efficiently computed in terms of the diagonal elements of the Jacobian as

$$J_g(\phi) = \prod_{\substack{x,i \\ m_x^i = 0}} \left| \frac{\partial [\hat{g}(\phi)]_x^i}{\partial \phi_x^i} \right|. \tag{3.18}$$

Coupling layers satisfy the desired properties of invertibility and a tractable Jacobian determinant independently of the context functions used, and thus the expressivity of the coupling layer can be systematically improved by utilizing more general variational ansätze for the context functions.

### 3.2.2 Neural networks

To define expressive and systematically improvable context functions, feed-forward neural networks are commonly used in their implementation. We briefly review the relevant properties of neural networks in this section; see Ref. [345] for an in-depth introduction and review of recent methods. Despite the deep connections to the field of artificial intelligence, feed-forward neural networks in isolation can simply be thought of as highly expressive function approximators consisting of a composition of linear and non-linear functions parameterized by a collection of continuous parameters. The term 'feed-forward' here distinguishes an architecture based on sequential composition of functions that can be evaluated by 'feeding' an input forward through the functions; other forms of neural networks may involve passing outputs as inputs to earlier functions and for example generating a sequence of outputs by iterative evaluation [346].

The simplest example of a feed-forward neural network is a 'single-layer perceptron'. The action of a single-layer perceptron on an input vector $\boldsymbol{x}$ can be written as $\boldsymbol{y} = f(W\boldsymbol{x} + b)$, where the matrix of 'weights' $W$ and vector of 'biases' $b$ linearly transform the input, and the non-linear 'activation function'[2] $f$ acts elementwise on the components of $W\boldsymbol{x} + b$. The weights $W^{ij}$ and biases $b^i$ are free parameters that can be optimized to cause the action of the single-layer perceptron to approximate a desired function from $\boldsymbol{x}$ to $\boldsymbol{y}$. By composing linear transformations and non-linear activation functions, expressivity of the neural network can be increased. The simplest 'multi-layer perceptron' consists of a layer producing the 'hidden' variable $\boldsymbol{h} = f_1(W_1\boldsymbol{x} + b_1)$, followed by a layer producing the final outputs $\boldsymbol{y} = f_2(W_2\boldsymbol{h} + b_2)$. Remarkably, this network with a single hidden layer is a universal function approximator in the limit of making $\boldsymbol{h}$ arbitrarily high-dimensional [348]. In practice, multiple layers with finite-dimensional hidden vectors $\boldsymbol{h}_1, \boldsymbol{h}_2, \ldots$ are often used to achieve sufficient expressivity. Figure 3.4 depicts the relation between inputs, hidden variables, and outputs in a single-layer perceptron and a multi-layer perceptron.

Feed-forward networks with no restrictions placed on the weights matrices $W$ are labeled 'fully-connected' because of the potentially non-trivial coupling between each input and output component of every linear operation. While this allows a high degree of expressivity, the computational cost scales quadratically with the dimension of the input, hidden variables, and output. If used as context functions in flow-based models for lattice field theory, for which the dimension of the input and output scale as the lattice volume, fully-connected neural networks would incur significant cost for lattices with large volumes.

*Convolutional neural networks (CNNs)* are an alternative to fully-connected networks that have been applied in contexts where there is a natural sense of locality to the problem, such as in computer vision problems in which nearby pixels in an image jointly determine features of the image. CNNs can be considered a restriction of the more general fully-connected architecture: rather than allowing output components of a linear operation to depend on arbitrary components of the input, linear operations are restricted to only build linear combinations of components within a fixed distance of the position of each output component. Figure 3.5 depicts an example of a CNN acting on an input with a two-dimensional grid geometry. In the figure, output components $h_{k;(i,j)}$ of $\boldsymbol{h}$ are determined from a linear operation on the components $x_{(i',j')}$ of $\boldsymbol{x}$ with $i - 1 \leq i' \leq i + 1$ and $j - 1 \leq j' \leq j + 1$. A similar locality restriction applies to the determination of $\boldsymbol{y}$ from $\boldsymbol{h}$. The $k$ index enumerates 'channels' of the hidden variable $\boldsymbol{h}$; this terminology originates from application to computer vision problems in which these may be used to represent the color channels of images. In the example shown in the figure, the input and output variables are shown with only a single channel for simplicity, though variables in each layer may have multiple channels in general. Note that despite the additional structure to the components of $\boldsymbol{x}$, $\boldsymbol{h}$, and $\boldsymbol{y}$, they can still be considered vectors where the components are enumerated by a particularly complicated index $(i, j)$ for $\boldsymbol{x}$ and $\boldsymbol{y}$ or $(k; (i, j))$ for $\boldsymbol{h}$.

---

[2]Many non-linear functions have been considered for activation functions; common examples include the hyperbolic tangent defined by $f(x) = \tanh(x)$, the logistic function defined by $f(x) = 1/(1 + e^{-x})$, or the rectified linear unit [347] defined by $f(x) = \max(0, x)$.

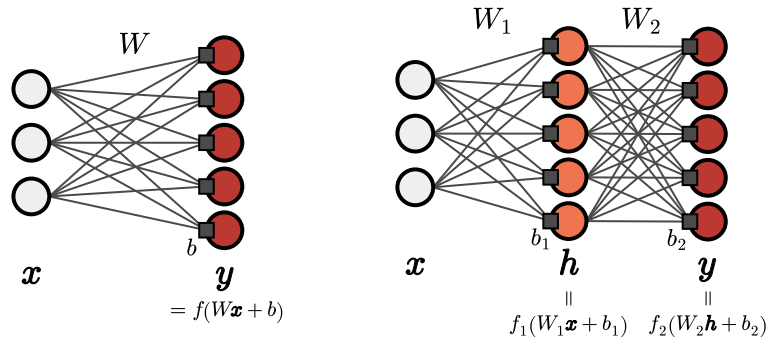

**Figure 3.4:** Single-layer (left) and multi-layer (right) feed-forward neural networks. Evaluating the output $\boldsymbol{y}$ proceeds by sequentially evaluating the linear operations and activation functions from left to right. Linear operations are labeled by weights matrices $W$, $W_1$, $W_2$ and biases $b$, $b_1$, $b_2$. Activation functions $f$, $f_1$, $f_2$ are applied elementwise following each linear operation. Gray edges corresponding to components of weights matrices and gray squares corresponding to biases together constitute the free parameters of the neural networks. These can be optimized to approximate the desired function mapping from $\boldsymbol{x}$ to $\boldsymbol{y}$ as described in the main text.

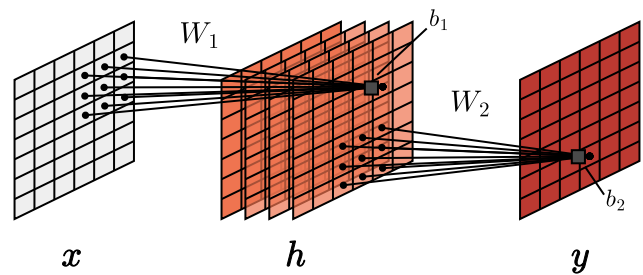

**Figure 3.5:** Geometric structure of a convolutional neural network. Each convolution operation (equivalent to a tensor multiplication) is shown in solid lines. The multiple layers shown for $\boldsymbol{h}$ correspond to distinct channels $h_k$. In general, $\boldsymbol{x}$ and $\boldsymbol{y}$ may also include multiple channels and multiple hidden layers may be used. Non-linear elementwise operations are typically applied in between convolutions (not shown for simplicity), similarly to the case of an arbitrary feed-forward neural network.

CNNs further impose translational invariance on the weights and biases used to define the linear operation in each layer; in the example in Figure 3.5 this corresponds to an identical set of nine weights and one bias determining the linear combinations of components of the input $\boldsymbol{x}$ that produce each component of $\boldsymbol{h}$ across all possible translations of input/output components; likewise a fixed set of nine weights and one bias would produce each component of $\boldsymbol{y}$ in terms of components of $\boldsymbol{h}$. Mathematically, these operations can be described in terms of convolutions as

$$
\begin{aligned}
h_{k;(i,j)} &= f_1\Big( \sum_{\substack{i-1 \leq i' \leq i+1 \\ j-1 \leq j' \leq j+1}} W_1^{k;(i'-i,j'-j)} x_{(i',j')} + b_1 \Big) \\
y_{(i,j)} &= f_2\Big( \sum_{\substack{i-1 \leq i' \leq i+1 \\ j-1 \leq j' \leq j+1}} W_2^{k;(i'-i,j'-j)} h_{k;(i',j')} + b_2 \Big)
\end{aligned}
\tag{3.19}
$$

for the example shown in the figure, where $f_1$ and $f_2$ are non-linearities applied elementwise as in the fully-connected case. The convolutional structure uses weights tensors that only depend on the differences in coordinates $(i' - i, j' - j)$ and therefore do not scale with the size of the input. Instead, the weights in a CNN capture an operation with local structure applied over the entire input. This structure is naturally suited to the setting of lattice field theory, and using CNNs with appropriate choices of masking patterns can be used to construct coupling layers that respect translational symmetry, as discussed in Sec. 3.3 below. These advantages motivate the exclusive use of CNNs in the ensemble generation applications explored in this dissertation.

Figure 3.5 also shows a fixed lattice geometry for all variables in the convolutional neural network. In general, the lattice geometry may vary depending on how boundary conditions are handled, i.e. based on the choice of 'padding' to be applied before the convolution. For applications to lattice field theories with periodic boundary conditions, using periodic padding for convolutions in each CNN allows translational symmetry to be preserved. Anti-periodic padding can be implemented by a simple modification of periodic padding, and this may be useful in future work implementing flow-based models for lattice theories involving fermion fields, which are anti-periodic in time. The lattice geometry may also vary across the layers of a CNN when 'pooling' operations [349–351] are used to coarse-grain the components of $\boldsymbol{x}$ or $\boldsymbol{h}$ as part of the action of the neural network. This coarse-graining can be useful in applications that seek to extract a few scalar quantities capturing global information describing an input lattice field (e.g. for classifying images), but when applied as context functions for coupling layers CNNs must instead return a lattice of parameters defining the elementwise transformations of all updated components. Pooling operations are thus never used in the CNNs defining flow-based models in any of the applications considered here.

### 3.2.3 Optimizing neural networks

A key pillar underwriting the significant success of machine learning methods is the ability to apply gradient-based optimization on the parameters of the model. When a measure of the 'loss' of the neural network — i.e. the failure of the model to capture the desired functionality — can be encoded as a scalar *loss function*, iterative gradient descent can be used to optimize parameters towards a local minimum. For example, when a training set of $n$ desired input/output pairs $\{(\boldsymbol{x}^i, \boldsymbol{y}^i)\}_{i=1}^n$ are given, a mean-squared-error loss function can be constructed as

$$\mathcal{L}^{\text{MSE}} = \frac{1}{n} \sum_{i=1}^n |\text{NN}_\omega(\boldsymbol{x}^i) - \boldsymbol{y}^i|^2 \tag{3.20}$$

where $\text{NN}_\omega(\cdot)$ denotes the application of the neural network with free parameters $\omega$. Gradients $\nabla_\omega \mathcal{L}^{\text{MSE}}$ can be used to iteratively improve $\omega$, minimizing this loss function and thereby maximizing agreement between model outputs and the desired outputs on the given inputs. Several iterative optimization techniques are commonly applied:

- **Mini-batch stochastic gradient descent:** To capture fluctuations in the underlying distribution over inputs of interest, subsets of input/output pairs can be randomly resampled from the training set. A sum over these random choices of $i$ replaces the sum over all $n$ input/output pairs for each evaluation of Eq. (3.20). Subsets chosen for each evaluation are termed 'mini-batches'. On each step of optimization, the gradient of the loss function using the given mini-batch is evaluated and a small step is taken in the direction opposite to the gradient.

- **Gradient descent with momentum:** Rather than taking a step of fixed size along gradients, a measure of the recent direction of motion in the space of parameters can be accumulated using a fictitious momentum vector. This momentum vector is then added to the computed gradient when updating the parameters in each iteration. Momentum allows more rapid convergence along nearly flat directions of the loss function, along which many more steps of simple gradient descent would be required to make progress. The use of momentum is orthogonal to the use of mini-batches for stochastic evaluation of the loss function, and these two approaches may be combined.

- **Adam:** The Adam optimizer [352] (derived from 'adaptive moment estimation') can be applied to handle cases in which different components of the gradient may fluctuate with varying magnitudes and scales, which would otherwise result in very different dynamics in each of the parameters of the model. The Adam procedure maintains estimates of the per-component first and second moments of the gradient and normalizes the update to each component using these moments. Several 'hyperparameters' determine the precise behavior of this optimizer, as detailed in Ref. [352].

The combination of mini-batching with Adam has seen wide use due to its stability and efficiency at converging to the minima of loss functions. As such, we use mini-batched Adam for optimization throughout this dissertation, with the Adam hyperparameters fixed to the defaults specified in the PyTorch [353] or JAX [354] libraries utilized (which define identical defaults). Hyperparameter tuning is a practical consideration that may be useful to study before applications of the methods in this dissertation at larger scales.

For a feed-forward neural network composed of independently differentiable components, *backpropagation* (or 'reverse-mode differentiation') can be used to efficiently compute the gradient of the loss function with respect to network parameters. In essence, backpropagation is an application of the chain rule to rewrite gradients in terms of a product of individual Jacobian factors, allowing pointwise evaluation of gradients for given model parameters by (1) forward evaluation of all intermediate quantities and (2) propagation of a gradient vector in reverse to all model parameters. Ref. [355] gives a detailed introduction to the definition and use of backpropagation (and more generally, automatic differentiation) in machine learning algorithms. In practice, all machine learning frameworks, including those used in this dissertation, automate the process of forward and reverse evaluation to determine gradients. Manual implementation of reverse-mode differentiation is only required for one component of the flow-based models implemented for gauge theory in Chapter 4 (see Sec. 4.5.3).

### 3.2.4 Neural networks as context functions

To use neural networks as context functions in coupling layers of a flow-based model, one simply needs to ensure that the inputs and outputs live in the correct domain. For real-valued lattice fields and real-valued outputs parameterizing elementwise invertible transformations in coupling layers, the feed-forward neural networks described above are immediately applicable using the masked field variables $m\phi$ as inputs.[3] For a single context function output per field component, as in the simple scaling coupling layer described in Eq. (3.15) above, the output consists of one component $[s(m\phi)]_x^i$ per component $\phi_x^i$ of the input. Functions from a lattice field to a lattice field can be achieved using convolutional neural networks with a lattice-shaped input consisting of one channel per component $i$ on each lattice site and an identical output shape. Hidden variables in intermediate layers can have more or fewer channels than the input and output, but to preserve the most translational symmetry they should maintain an identical lattice geometry throughout, as addressed above. Chapter 4 details the necessary modifications to coupling layers and context functions to treat group-valued inputs living in compact domains. Figure 3.6 gives a high-level view of how coupling layers are composed together to define a flow, highlighting the use of neural networks as context functions.

---

[3]On a practical note, some computational cost could be saved by removing the operations applied to elements set to zero by the masking pattern. However, this only affects the first layer of a feed-forward network, and it is thus not expected to provide significant reductions in cost in most cases.

When coupling layers are defined using both tileable masking patterns and context functions implemented using convolutional neural networks, the resulting coupling layers are applicable to lattices with any geometry compatible with the masking pattern tiling. Combined with an IID prior distribution, this architecture can be used to define a fixed flow-based model architecture applicable to arbitrary lattice geometries (subject to the same masking pattern constraint). This property is used to apply identical model architectures to varying lattice sizes in some of the applications presented in this dissertation.

### 3.2.5 Universality and expressivity

Given the simply stated universality results for neural networks in general, a natural question is whether normalizing flows based on coupling layers are similarly suitable to universally describe distributions in some limit. Unlike autoregressive approaches, in which it is clear that there is a description of the joint distribution in terms of products of conditionals that can be learned, it is not necessarily the case that composition of many coupling layers taking a particular form can produce arbitrary joint distributions. Some particular results have, however, been shown for particular classes of coupling layers. For example, restricting to *affine transformations*, in which each updated component is transformed by both scaling and an offset, is a common choice [243, 244]. Ref. [356] argues that flows composed entirely of affine coupling layers are non-universal and Ref. [357] similarly demonstrates non-universality for a few simpler classes of normalizing flow architectures. On the other hand, Ref. [358] proves that interposing invertible linear transformations between affine coupling layers results in a class of normalizing flows that is universal in the limit of an infinite number of layers and arbitrarily expressive context functions. Unfortunately, in practice it is costly to include general linear transformations because computing the Jacobian determinant of an arbitrary linear transformation requires $O(\mathcal{N}^3)$ operations in terms of the dimensionality $\mathcal{N}$ of the input vector. For example, this is intractable for state-of-the-art lattice field theories defined in terms of approximately $10^9$ degrees of freedom [359].

Despite these mixed results, ultimately it is the practical expressivity of flows that determines whether they can provide a useful sampling procedure for distributions of interest. There are two clear ways to systematically improve the expressivity of any coupling layer architecture — improve the expressivity of context functions (e.g. by increasing the size of the neural networks defining these functions) and increase the number of coupling layers composed to produce the flow. In the proof-of-principle application to scalar lattice field theory in Sec. 3.6, we demonstrate that simple neural networks and few coupling layers are already sufficient to learn effective approximations to the distributions relevant for lattices of size up to $14 \times 14$. Scaling results to larger and finer lattices is discussed in Sec. 3.7, and is expected to require increasing expressivity using the approaches mentioned above.

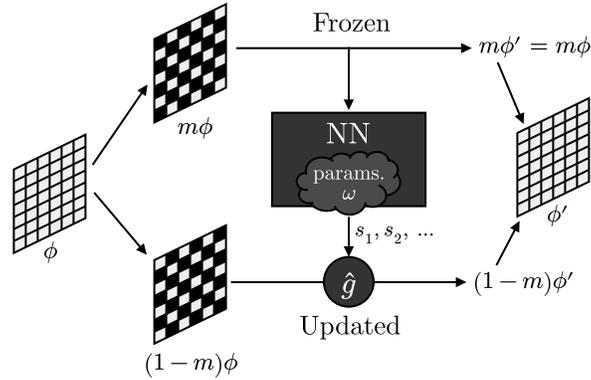

**(a)** A coupling layer.

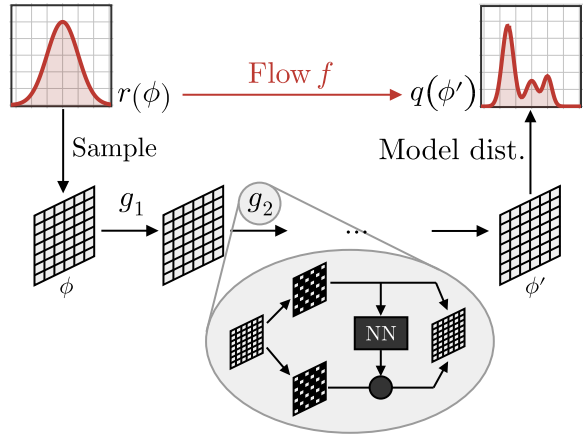

**(b)** A flow composed of coupling layers.

**Figure 3.6:** A high-level view of the definition of flow-based models using coupling layers. Each coupling layer is implemented using a masking pattern dividing the components of the input into a frozen subset and an updated subset, as shown schematically for a checkerboard masking pattern in (a). Neural networks (NNs) are used to parameterize the elementwise invertible transformation $\hat{g}$ applied to the updated subset, and the weights $\omega$ contained in these neural networks collectively parameterize the flow. A flow $f$ composed from coupling layers $g_1, g_2, \ldots$ incrementally transforms each sample. This transformation changes the probability density over the space of samples from the prior distribution $r(\cdot)$ to a model distribution $q(\cdot)$, as shown in (b). The parameters $\omega$ in each coupling layer can be optimized to shape $q(\cdot)$, typically with the aim of closely approximating a target distribution $p(\cdot)$. Figure adapted from Fig. 1 of Pub. [5].

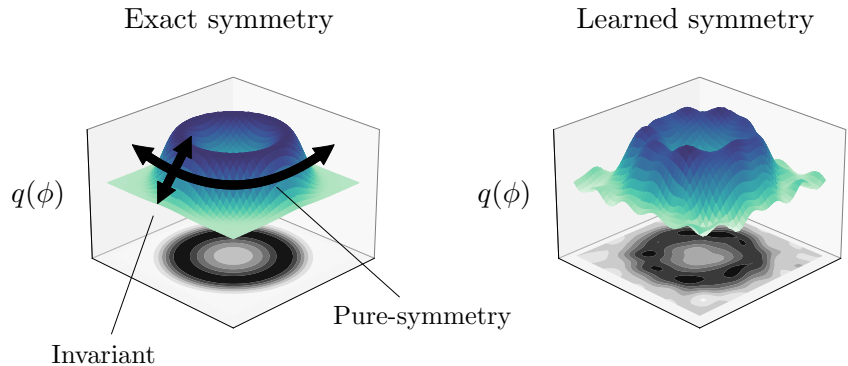

Exact symmetry

Learned symmetry

$q(\phi)$

$q(\phi)$

Invariant

Pure-symmetry

**Figure 3.7:** Continuous symmetries correspond to 'flat' directions that leave the probability density invariant. The left figure shows a schematic view of a symmetric distribution. In this simple depiction the flat directions correspond to pure-symmetry rotational degrees of freedom while the non-trivial structure of the distribution is encoded in the invariant radial degrees of freedom. A flow-based model that uses equivariance to exactly enforce a symmetry will only produce distributions of the form shown on the left by only transforming the invariant degrees of freedom. A general flow-based model trained to reproduce a distribution with such a symmetry may produce distributions such as those shown on the right, where the symmetry is reproduced by manipulating all degrees of freedom to approximately distribute the density in a symmetric fashion. Figure adapted from Fig. 1 of Pub. [2].

## 3.3 Symmetries

Lattice field theories may satisfy a number of discrete and continuous symmetries (see Chapter 2):

- **Geometric symmetries** including discrete translations and often reflections and rotations;

- **Internal symmetries** that modify components on each lattice site, including global negation, phase rotation, or unitary rotation of field components;

- **Gauge symmetries** described by simultaneously transforming gauge and matter fields according to a field of transformation parameters $\Omega(x)$.

A theory satisfies a symmetry when the associated transformations of the field variables leave the action — and therefore the probability distribution — invariant. For continuous symmetries, the generators of the symmetry correspond to directions in field configuration space along which the probability density is invariant. Schematically, these can be thought of as directions in which the probability density is 'flat', as shown in the left panel of Figure 3.7.

Any expressive flow-based model should approximately reproduce the symmetries of the original action after optimization, even if these symmetries are not imposed in the model. Exact symmetries are recovered on average in the sampled distribution after applying the corrective steps described in Sec. 3.5 below. Nevertheless, any breaking of the symmetries in the model reflects differences between the model and target distribution and is thus associated with sampling inefficiencies. Imposing symmetries explicitly in the form of the model effectively reduces the variational parameter space to include only symmetry-respecting maps, i.e. those that factorize the distribution. A schematic comparison of the factorized distributions that can be accessed by factorized models versus those that are available to general models is illustrated in Figure 3.7. This abstract reduction in the variational space of distributions does *not* decrease the useful expressivity of the model because the target distribution satisfies the restricted form of the symmetric distributions being searched over. Instead, removing unnecessary degrees of freedom can reduce the number of free parameters in the flow-based model, and in many machine learning contexts it has been found that explicitly preserving the symmetries of interest in models improves both the optimization costs and ultimate model quality [139, 141, 360–364].

It is typically not possible to simply remove non-symmetric distributions from the space of possible distributions parameterized by an existing model. Instead, one must carefully architect the flow-based model to exactly preserve the symmetry. To do so, it is sufficient to (1) sample from a prior distribution that is exactly invariant under the symmetry and (2) use a flow that is *equivariant* under the symmetry [251, 365–368], meaning that all symmetry transformations $t$ commute with application of the flow $f$,

$$f(t \cdot U) = t \cdot f(U). \tag{3.21}$$

Section 3.1.4 introduced prior distributions for lattice field theory that are symmetric under translations and hypercubic symmetries. The Gaussian prior distribution for real-valued fields discussed in that section also satisfies many internal symmetries, including negation and unitary rotation of fields, and the uniform prior distribution for lattice gauge theory satisfies gauge symmetry. For the study of theories with other symmetries, it may be necessary to choose different prior distributions satisfying the relevant symmetries. In the context of lattice field theory, these symmetries can be divided into spacetime and internal/gauge symmetries, and using IID prior distributions allows a neat factorization of the treatment of these classes of symmetries: an IID distribution will necessarily be invariant under geometric symmetries of the lattice, and the marginal distribution applied to field components on each site of the lattice can be independently chosen to satisfy the desired internal symmetries. Complicated one-dimensional distributions are also straightforward to directly sample from when the cumulative density function is known [369], allowing a wide class of distributions to be chosen.

Equivariance of the flow $f$ composed of coupling layers $g_i$ can be enforced by making each coupling layer equivariant, since $g_i(t \cdot U) = t \cdot g_i(U)$ for all $i$ implies

$$f(t \cdot U) = g_n(g_{n-1}(\ldots g_1(t \cdot U) \ldots)) = t \cdot f(U). \tag{3.22}$$

Coupling layers decompose the components of a field configuration into updated and frozen subsets by spacetime location, thus it is natural to separately and simultaneously address making coupling layers equivariant to spacetime symmetries (translational and hypercubic symmetries) and making coupling layers equivariant to internal symmetries (such as gauge symmetry).

Section 3.2.2 introduced the idea that convolutional neural networks satisfy translational symmetry. We can now precisely state this property: CNNs with periodic boundary conditions are equivariant under lattice translations [360, 370].[4] This is apparent from the structure of each convolutional operation, in which the linear operation only depends on relative differences of coordinates. The elementwise application of activation functions is similarly equivariant to translations. By further restricting the form of convolutions, CNNs can also be made equivariant under rotations and reflections [360, 361].

For geometric symmetries, combining equivariant CNNs in context functions with a symmetric masking pattern ensures that each coupling layer is equivariant under the relevant geometric symmetry group. In particular, using tiled masking patterns in addition to CNNs results in coupling layers equivariant under a large subgroup of the translational symmetry group. In choosing the masking patterns, one must balance the aim of encoding the largest possible subgroup of the geometric symmetries with the need to provide sufficient coupling between frozen and updated degrees of freedom. For example, choosing to freeze all sites on the lattice results in a masking pattern with the maximum translational symmetry but obviously prevents the model from expressing any transformations at all, while choosing to update all sites on the lattice leaves no frozen components to be used as input to context functions and only simple elementwise operations can be described in this case. A checkerboard masking pattern is a practical compromise in the case of non-gauge lattice theories involving fields discretized on lattice sites, as this preserves invariance under half of the translational symmetry group and results in large frozen and updated subsets that can be utilized in an expressive coupling layer transformation.

For global internal symmetries, symmetry transformations alter degrees of freedom uniformly across the lattice. To enforce equivariance under such symmetries in coupling layers, the elementwise transformations of all updated components on the lattice must each be made equivariant under the symmetry (transformations of frozen degrees of freedom trivially commute through application of the layer). A possible general approach to achieving this equivariance is to define all context functions to be invariant to symmetry transformations and carefully structure the elementwise transformations to commute with symmetry operations. For example, equivariance under global sign flips could be achieved by restricting coupling layers to apply an elementwise scaling by $e^s$, as in Eq. (3.15) above, and only providing the absolute value of all field variables as inputs to context functions. While this indeed results in an equivariant coupling layer, in many ways it is overly restrictive: the described context functions are invariant under *local* sign flips as well as global sign flips, losing much of the information in the input field, and the use of a positive scaling operation means that the overall signs of field variables will never be updated by the coupling layer. Other constructions are possible, but this hypothetical example highlights the tradeoff between expressiv-ity and symmetrization that must always be addressed when attempting to enforce symmetries in flow-based models. Given these subtleties, and that the dimension of internal symmetry groups does not scale with the lattice volume, in this dissertation we do not focus on enforcing global symmetries and only do so when such symmetries can be straightforwardly incorporated. Global symmetries have been considered in other approaches to sampling [138, 139] and developing a careful understanding of the expressivity/symmetrization tradeoff would be an interesting subject of future work on flow-based models.

---

[4]More general CNNs than the ones introduced above may include pooling operations or convolutions that coarsen the lattice geometry, reducing the number of sites. These must also be excluded to preserve translational equivariance.

Finally, all lattice gauge theories are symmetric under a high-dimensional gauge symmetry group. We defer discussion of gauge symmetries to the following chapter.

## 3.4 Optimizing flow-based models

For a fixed prior probability density $r(\cdot)$, the weights $\omega$ parameterizing the flow $f_\omega$ can be optimized — i.e. the model can be *trained* — to bring the variational ansatz $q_\omega(\cdot)$ close to the target probability density $p(\cdot)$. This training is undertaken by minimizing a *loss function* measuring the discrepancy between $q_\omega$ and $p$, analogous to loss functions such as the MSE loss in Eq. (3.20) utilized for optimization of neural networks in other contexts. There are many possible ways to measure this discrepancy in general; in the lattice field theory context, our aim is to utilize the known form of the probability density, $p(x) = e^{-S(x)}/Z$, and the ability to efficiently sample from the model. This idea of attempting to 'distill' the information contained in the known action into a trained model is a specific instance of Probability Density Distillation [330, 371].

Given these constraints, we choose an unnormalized form of the Kullback-Leibler (KL) divergence [372] as a measure of the discrepancy. In terms of the probability densities $q_\omega(x)$ and $p(x) = e^{-S(x)}/Z$, the resulting loss function is written

$$
\begin{aligned}
L(q) &= D_{KL}(q||p) - \log Z \\
&= \int \prod_j d\phi_j \, q(\phi) \left( \log q(\phi) - \log p(\phi) - \log Z \right) \\
&= \int \prod_j d\phi_j \, q(\phi) \left( \log q(\phi) + S(\phi) \right),
\end{aligned}
\tag{3.23}
$$

where $D_{KL}(q||p)$ is the KL divergence. Subtracting off the unknown normalizing constant $\log Z$ allows writing the final loss function in terms of explicitly computable quantities without affecting the gradients or locations of the minima. By non-negativity of the KL divergence, the lower bound on the loss is $(-\log Z)$, and this minimum is achieved exactly when $q = p$. In practice, the loss is stochastically estimated by drawing batches of $m$ samples from the model $\left\{ \phi^{(i)} \sim q \right\}_{i=1}^m$ and computing the sample mean:

$$
\widehat{L(q)} = \frac{1}{m} \sum_{i=1}^m \left( \log q(\phi^{(i)}) + S(\phi^{(i)}) \right).
\tag{3.24}
$$

This stochastically estimated loss function has been successfully applied in related generative approaches to statistical physics models [139, 142].

The loss function given in Eq. (3.23) and stochastically estimated in Eq. (3.24) measures the average discrepancy between the lattice action $S$ and an 'effective' action associated with the model, which we define as

$$S_q^{\text{eff}}(\phi) \equiv -\log q(\phi). \qquad (3.25)$$

The term in parentheses in Eq. (3.24) can be rewritten using the effective action as $\left(-S_q^{\text{eff}}(\phi^{(i)}) + S(\phi^{(i)})\right)$. The form of the loss function does not include an absolute value or square around this term, and at first glance it appears that minimizing this loss function would trivially cause the model to produce an effective action that grows without bound. However, it is important to note that the effective action is always derived from the properly normalized probability density sampled by the model. The fixed normalization prevents the effective action from trivially being rescaled to reduce the loss. Given that $(-\log Z)$ is a lower bound on this loss function, optimizing this loss function actually drives the flow-based model towards finding an effective action satisfying $S_q^{\text{eff}}(\phi) = S(\phi) + \log Z$, which corresponds to exactly reproducing the target action up to the normalization. Noting that the term in parentheses in Eq. (3.24) is evaluated using samples drawn from the model, optimizing this loss function amounts to trading off the goal of maximizing the average effective action and minimizing the target action across these samples. The first term encourages a diffusion of the model density while the second term encourages concentration of the model density in regions where the target density is large (i.e. the target action is small).

The stochastically estimated loss can be efficiently computed because it is written in terms of samples drawn from the flow-based model itself. The training process thus does not require existing samples from the desired distribution as training data. This *self-training* is a key feature of the proposed approach to Monte Carlo sampling for field theories, where samples from the desired distribution are often computationally expensive to obtain. If samples do exist, they can be used to 'pre-train' the network, although in the numerical studies undertaken in this dissertation this was not found to be markedly more efficient in model optimization than using only self-training. In contrast to other machine learning applications, we note that drawing new samples for each training iteration implies that there is no 'over-training' regime in which the model becomes specialized for a specific training set rather than the general problem. In principle, the model quality continues to improve for an arbitrary number of training iterations, assuming the training dynamics give stable convergence towards a minimum of the loss function.

The concept of self-training is not new, and has been applied in other learning contexts including applications to lattice theories. In policy learning contexts, such as the construction of artificial intelligence players for a game, reinforcement learning techniques iteratively improve models using information about the current policy of the agent [373]; self-play has also played a role in the significant success of artificial intelligence agents such as AlphaGo and AlphaZero [374, 375]. In the physics context, Self-learning Monte Carlo (SLMC) methods have been applied fairly successfully to one- to three-dimensional Ising, fermionic, scalar bosonic systems [128–136, 376]. These methods construct, by a variety of techniques, an effective Hamiltonian for a theory that can be more easily simulated using traditional methods than the original Hamiltonian. The effective Hamiltonian is learned using supervised learning techniques based on training data drawn from a combination of existing MCMC simulations, randomly-mutated samples, and the accelerated Markov chain itself (hence the term "self-learning").

Self-training is an appealing approach because it can be performed independently of any training set of samples drawn from the target distribution. However, it has associated drawbacks: in the case of flow-based models, if important regions of configuration space are severely under-sampled by a model, self-training methods will not be sensitive to these regions because they are rarely seen during training.[5] If the probability density has relatively smooth support over the space of field configurations, the diffusion term in the loss function described above can be expected to give the necessary coverage of the target probability distribution. If the distribution has isolated modes, i.e. local maxima of the probability density, this term may not be enough to drive the model to move significant density into these regions; however, this is also the regime in which diffusive MCMC approaches similarly encounter difficulties. For lattice field theories with physical parameters in a regime involving multiple vacua, for example due to spontaneous breaking of a symmetry or instanton effects, studying the ability of flow-based models to learn the associated distributions is an interesting subject of future work.

Lastly, a key distinction must be drawn between the process of training these models and the process of sampling using traditional MCMC methods: the training process for flow-based models does not need to satisfy any exactness criteria. There is significant freedom in how flow-based models are trained, and any of the following approaches may be used if it benefits the quality of the variational approximation determined by the flow-based model:

- Other loss functions may be used either in addition to, or in place of, the loss function defined in Eq. (3.24).

- The choice of loss function may be modified over the course of training in arbitrary ways.

- Additional samples constructed by arbitrary means may also be used during training. For example, a small number of samples from existing traditional simulations may be used in addition to self-training samples to 'anchor' important regions of parameter space. Anchoring samples could also be artificially constructed through other techniques, for example by using gradient flow [343, 344] from a fixed distribution to capture the long-range fluctuations of typical configurations.

- The gradient-based optimizer may be modified over the course of training. A standard approach is to apply 'scheduling' to modify the learning rate determining the step size used for each iteration of the optimizing procedure [377].

---

[5]Theoretical ergodicity is nevertheless guaranteed by invertibility of the flow; these issues are purely practical.

- Model parameters may be arbitrarily initialized, for example based on previously trained models or to initially reproduce a trivial flow.

## 3.5 Ensemble generation

Given a flow-based model for sampling the distribution $q$, our aim is to derive unbiased estimates of observables under the target distribution $p$, which for lattice field theory has a probability density proportional to $e^{-S(\phi)}$. The probability densities reported along with each sample drawn from the model are key to correcting from the model distribution to the target distribution, as described below.

### 3.5.1 Asymptotically unbiased measurements

There are several possible approaches to utilizing model samples $\{\phi_i\}_{i=1}^n$, model densities $q(\phi_i)$, and the Boltzmann weights $e^{-S(\phi_i)}$ to estimate observables:

1. **Reweight** observables estimated under samples drawn from $q(\phi)$. Given an ensemble of $n$ configurations $\{\phi_i\}_{i=1}^n$ drawn from the model distribution, the expectation value of an observable $\mathcal{O}$ is estimated by

$$\langle \mathcal{O} \rangle \approx \frac{\frac{1}{n} \sum_{i=1}^n \mathcal{O}(\phi_i) \frac{e^{-S(\phi_i)}}{q(\phi_i)}}{\frac{1}{n} \sum_{i=1}^n \frac{e^{-S(\phi_i)}}{q(\phi_i)}}. \tag{3.26}$$

Note that the denominator estimates the normalizing constant $Z \approx \frac{1}{n} \sum_{i=1}^n \frac{e^{-S(\phi_i)}}{q(\phi_i)}$.

2. **Resample** configurations from an ensemble drawn from the model distribution. The resampled ensemble is produced by drawing configurations $\phi_i$ with replacement from the original ensemble with probability proportional to $e^{-S(\phi_i)}/q(\phi_i)$.

3. **Apply MCMC** using configurations drawn from the model as proposals for independence Metropolis Markov chain updates, as discussed below, to acquire unbiased samples in a sufficiently long Markov chain. These samples can be used to measure observables as usual.

All three methods give unbiased estimates of observables in the asymptotic limit $n \to \infty$.

Each option has advantages and drawbacks. Reweighting allows sampling and observable measurements to be performed using a simple, fully parallelizable procedure. However, when sample weights $e^{-S(\phi)}/q(\phi)$ fluctuate significantly, computational effort may be wasted on measuring observables on configurations that will be severely downweighted. Resampling avoids this drawback by very infrequently including configurations with low weights in the output ensemble; there will on average be fewer unique configurations in the output ensemble than the original ensemble, reducing the cost of observable measurements at the cost of some bookkeeping. The MCMC approach also has this benefit, while additionally allowing flow-based proposals and updates to be mixed with traditional updates that may have complementary properties. MCMC sampling does introduce a sequential step in configuration generation, but this can be mostly circumvented by generating proposals from the model in parallel and only sequentially deciding whether to accept or reject each proposal. In the lattice field theory context, the asymptotic properties of Markov chains have been well explored, useful existing Markov chain updates have been developed, and observable estimates can be significantly expensive; we therefore focus on the flow-based MCMC approach below.

### 3.5.2   Flow-based MCMC

The target probability distribution $p(\phi)$ can be asymptotically exactly sampled using a flow-based Markov chain. We define each step of the Markov chain as an independence Metropolis sampler [52] by independently proposing configurations from the flow-based model and applying a Metropolis-Hastings accept/reject step (see Sec. 2.4.1). The proposal probability for this Metropolis-Hastings step is given by $P(\phi \to \phi') = q(\phi')$ and the acceptance probability was given in Sec. 2.4.1 by

$$p_{\text{acc}}^{\text{ind}}(\phi \to \phi') = \min\left(1, \frac{p(\phi')q(\phi)}{p(\phi)q(\phi')}\right).$$  (3.27)

The Metropolis-Hastings step ensures that this Markov chain step satisfies detailed balance for $p(\phi)$. The Markov chain is ergodic if all states $\phi$ are proposed by the model with non-zero density, and this is sufficient to guarantee that the independence Metropolis Markov chain will converge asymptotically to $p(\phi)$ [52]. Since flows are defined to be diffeomorphisms, a flow-based model will satisfy this requirement if the flow is applied to a prior distribution that has support everywhere in the space. One can easily check whether this property is satisfied for any particular prior distribution; all prior distributions considered for flow-based models in this dissertation have support everywhere in the domain of field configurations and thus result in ergodic Markov chains.

### 3.5.3   Acceptance rate and autocorrelations

Unlike typical uses of the Metropolis-Hastings accept/reject step in lattice field theory, the proposal distribution for flow-based MCMC is not symmetric in the forward and reverse direction,

$$P(\phi \to \phi') = q(\phi') \ \neq \ P(\phi' \to \phi) = q(\phi).$$  (3.28)

This results in the non-canceling factors of $q(\phi)$ and $q(\phi')$ in the acceptance rate in Eq. (3.27). In principle, proposals can be accepted with probability 1 if the model distribution exactly matches the target distribution, in which case the factors $p(\phi')/q(\phi')$ and $q(\phi)/p(\phi)$ are exactly equal to 1. Because the proposals are independent of the previous configuration in the Markov chain, a high acceptance rate means that independent proposals will frequently cause large changes in the sampled field configuration. On the other hand, a non-trivial rejection rate results in autocorrelations in the Markov chain arising from the duplicated configurations.

This intuition can be made more precise under some limiting assumptions. We assume for now that we are working with a Markov chain in equilibrium and that the probability of accepting subsequent configurations is independent of the distribution over values of an observable $\mathcal{O}$ at the current configuration. In practice this assumption does not hold exactly, but it allows the derivation of an estimate of the scale of autocorrelations in an independence Metropolis chain that gives a complementary measure of the performance of the Markov chain independent of autocorrelations of particular observables. Given that proposed configurations are independently generated, accepting a proposed configuration results in a completely independent measurement of the observable $\mathcal{O}$ under the assumption above. On the other hand, if all proposals between two positions in the Markov chain are rejected, the observable values at these two positions are 100% correlated. We define the $\tau$-rejection rate as the probability that $\tau$ proposals will be sequentially rejected,

$$p_{\tau\mathrm{rej}} \equiv \left\langle \prod_{i=1}^{\tau} \mathbb{1}_{\mathrm{rej}}(i) \right\rangle, \tag{3.29}$$

where $\mathbb{1}_{\mathrm{rej}}(i)$ is an indicator function equal to 1 if the $i$th proposal after the current reference point is rejected in the Metropolis step. The expectation denoted by $\langle \cdot \rangle$ is taken with respect to the current state $\phi$ distributed according to the equilibrium distribution $p(\phi)$ and the randomness in the realization of the next $\tau$ steps of the Markov chain. For any observables $\mathcal{O}$ satisfying the assumption of independence from the probability of rejection, the associated autocorrelation function is given exactly by the $\tau$-rejection rate,

$$\rho_{\mathcal{O}}(\tau) = \frac{\left\langle (\mathcal{O}(\phi_i) - \bar{\mathcal{O}})(\mathcal{O}(\phi_{i+\tau}) - \bar{\mathcal{O}}) \right\rangle}{\left\langle (\mathcal{O}(\phi) - \bar{\mathcal{O}})^2 \right\rangle} = p_{\tau\mathrm{rej}}. \tag{3.30}$$

In practice, an estimator can be defined for a Markov chain of length $n$ as

$$\widehat{p_{\tau\mathrm{rej}}} \equiv \frac{1}{n-\tau} \sum_{j=0}^{n-\tau-1} \prod_{i=1}^{\tau} \mathbb{1}_{\mathrm{rej}}(i+j) \tag{3.31}$$

to measure the $\tau$-rejection probability. For a Markov chain near equilibrium, this estimator approaches the true value of $p_{\tau\mathrm{rej}}$ as $n \to \infty$. A corresponding 'universal' integrated autocorrelation time $\tau_{\mathrm{acc}}^{\mathrm{int}}$ may be defined by summing $\widehat{p_{\tau\mathrm{rej}}}$ and applying the usual self-consistent windowing procedures (see Sec. 2.4.2).

The $\tau$-rejection rate is closely related to, but cannot be derived from, the average single-step acceptance rate $a \equiv 1 - p_{1\mathrm{rej}}$. For all values of $\tau \neq 1$ the acceptance rate $a$ does imply a lower bound on the $\tau$-rejection rate,

$$p_{\tau\mathrm{rej}} = \mathbb{E}_{\phi \sim p} \left[ (1 - p_{\mathrm{acc}}(\phi))^\tau \right] \geq \left( \mathbb{E}_{\phi \sim p} \left[ 1 - p_{\mathrm{acc}}(\phi) \right] \right)^\tau = (1-a)^\tau, \tag{3.32}$$

where $\mathbb{E}_{\phi \sim p}$ indicates the expectation value with respect to samples $\phi$ drawn from $p(\phi)$. Depending on the particular distribution of the weight factors $p(\phi)/q(\phi)$, this bound may either be nearly saturated or the actual $\tau$-rejection rates could be significantly higher. This however suggests that the acceptance rate is also a useful metric of performance — since it gives a lower bound, it is necessary, though not sufficient, to improve the acceptance rate to reduce autocorrelations for all observables satisfying the independence assumption above. For observables not satisfying the independence assumption, we cannot generically relate the acceptance rate to autocorrelations. Regardless, when it is possible to increase the acceptance rate without increasing the associated cost of drawing samples from the flow-based model (e.g. by training a fixed model architecture), this tends to be an effective way to improve the performance of the flow-based Markov chain, as we explore in the context of scalar field theory in Sec. 3.6.

### 3.5.4 Other metrics

The acceptance rate together with autocorrelations in particular observables are useful methods of characterizing the performance of a flow-based Markov chain. During optimization of a particular flow-based model, however, it can be impractical to repeatedly generate Markov chains to evaluate performance. We discuss several other metrics that provide useful benchmarks of performance:

**KL divergence.** As described in Sec. 3.4, we use the KL divergence as a loss function which is minimized during optimization. Though it is not normalized, this can provide a useful relative measure of performance when comparing different models or comparing different points during the training of a particular model. The KL divergence is very directly a measure of the discrepancy between distributions (though far from the only way to measure such a discrepancy), and therefore gives relative information about how far model distributions are from the target.

**Variance of log reweighting factors.** To derive a measure of model performance that gives absolute information, one can instead consider the sample variance of the log reweighting factors $\log q(\phi_i) - \log p(\phi_i)$ used in the estimate of the KL divergence. Noting that the expectation values of these factors is just the KL divergence, this is in effect a measure of the variance of the KL divergence estimate. This measure of performance can be useful because it is absolutely normalized: when the model distribution perfectly matches the target distribution, the probability densities must agree sample-by-sample, and the variance must be zero.

**Partition function and normalized KL divergence.** Flow-based models provide normalized values of the model probability density on arbitrary field configurations. This allows the partition function to be computed as an expectation value under the model distribution,

$$Z = \int d\phi \, q(\phi) \left[ \frac{e^{-S(\phi)}}{q(\phi)} \right]. \tag{3.33}$$

The sample mean of $e^{-S(\phi)}/q(\phi)$ measured on samples drawn from the model thus provides an unbiased estimator of $Z$. Though this estimate will have large variance for a poorly-trained model, it can be used to give an estimate of the normalized KL divergence that becomes self-consistently accurate when it is close to zero. Note, however, that for finite-sample estimates of the partition function the distribution of factors $e^{-S(\phi)}/q(\phi)$ can have very long tails and care should be taken to avoid underestimating errors.

**Effective sample size.** In the context of applying reweighting to compute asymptotically exact observable estimates, the effective sample size (ESS) gives a measure of the statistical power achieved on the target distribution given an ensemble drawn from the model distribution [378]. In terms of the reweighting factors $w_i \equiv p(\phi_i)/q(\phi_i)$ of each sample $\phi_i$ contained in the ensemble, the normalized ESS is given by

$$\text{ESS} \equiv \frac{\left(\frac{1}{n}\sum_i w_i\right)^2}{\frac{1}{n}\sum_i w_i^2}. \tag{3.34}$$

When the model distribution equals the target distribution, all reweighting factors are equal to 1 and the ESS achieves a maximum value of 1. When there is a mismatch between the model and target distributions, the value is strictly less than 1 and characterizes the fractional statistical power afforded by each sample drawn from the model.

## 3.6 Scalar field theory in (1+1)D

Scalar field theory with a quartic $\phi^4$ interaction provides a simple testing ground for novel lattice field theory algorithms, including the flow-based MCMC algorithm detailed in this chapter. In this section we apply flow-based MCMC to ensemble generation for a real scalar field theory in $(1 + 1)$D and study the properties of our flow-based sampler in comparison to two traditional sampling methods — HMC and a local Metropolis sampler.

### 3.6.1 Action, observables, and parameters

Despite the apparent simplicity of a scalar theory discretized on a two-dimensional lattice, the theory considered here in fact gives access to both a Gaussian fixed point and a fixed point in the Ising universality class [379, 380]. In our study, we focus on the approach to the Gaussian fixed point as this captures the approach to a continuum scalar field theory analogous to the continuum limit of lattice field theories of phenomenological interest.

**Action.** The discretized action for a massive scalar field with a $\phi^4$ self-interaction is given in $(1 + 1)$ dimensions by

$$S(\phi; m^2, \lambda) = \sum_x \left( \sum_y \phi(x)\Delta(x, y)\phi(y) + m^2\phi(x)^2 + \lambda\phi(x)^4 \right), \tag{3.35}$$

|       | E1      | E2      | E3      | E4      | E5      |
|-------|---------|---------|---------|---------|---------|
| $L$   | 6       | 8       | 10      | 12      | 14      |
| $m^2$ | $-4$    | $-4$    | $-4$    | $-4$    | $-4$    |
| $\lambda$ | 6.975 | 6.008 | 5.550 | 5.276 | 5.113 |
| $m_\phi L$ | 3.96(3) | 3.97(5) | 4.00(4) | 3.96(5) | 4.03(6) |

**Table 3.1:** Choices of physical parameters defining the five target distributions that were simulated. We choose a range of lattice sizes $L$ while tuning $\lambda$ to hold the physical correlation length fixed with respect to the box size using the condition $m_\phi L = 4$. Table adapted from Table I of Pub. [5].

in terms of the bare parameters $m^2$ and $\lambda$ and the lattice discretization of the Laplace operator defined in two dimensions as

$$\sum_y \Delta(x,y)\phi(y) \equiv \sum_{\mu \in \{0,1\}} (2\phi(x) - \phi(x-\hat{\mu}) - \phi(x+\hat{\mu})). \tag{3.36}$$

For brevity, the lattice spacing $a$ has been absorbed into the definitions of the field variables and couplings. The sums over $x$ and $y$ in Eq. (3.35) run over the sites of a two-dimensional lattice which is taken to have a square geometry consisting of $L$ lattice sites in each direction in the cases studied here; we use the variable $V = L^2$ to denote the total number of lattice sites. The action in Eq. (3.35) is symmetric under all geometric symmetries of the lattice and under a global sign flip, $\phi \to -\phi$. Depending on the choice of parameters, the global sign flip symmetry may be spontaneously broken; we choose to focus on the symmetric phase for this proof-of-principle study. To study scaling towards the continuum limit, we work with a range of choices of $L$ with $m^2 = -4$ held fixed and $\lambda$ tuned to fix the ratio of the correlation length to the lattice size $L$ by the condition $m_\phi L = 4$ in terms of the scalar particle pole mass $m_\phi$ defined below. Table 3.1 gives the particular choices of $L$, $m^2$ and $\lambda$ defining the target distributions considered here. Parameter sets E2 and E4 match the parameters used in a study of critical slowing down in Ref. [379], and results measured on these ensembles were confirmed to match the results in the reference.

**Observables.** We measure the following choices of observables on all ensembles.

1. **Two-point function:** The volume-averaged two-point function of the scalar field is defined as

$$G(x) \equiv \frac{1}{V} \sum_y \langle \phi(y)\phi(y+x) \rangle. \tag{3.37}$$

   Symmetry of the theory under a global sign flip implies that the expectation value of a single insertion of the scalar field is exactly zero, $\langle \phi(x) \rangle = 0$. This two-point function is thus equivalent to the connected two-point function defining the propagation of excitations of the scalar field, $G_c(x) = \frac{1}{V} \sum_y [\langle \phi(y)\phi(y+x) \rangle - \langle \phi(y) \rangle \langle \phi(x+y) \rangle] = G(x)$.

2. **Pole mass:** The correlation function giving access to the zero-momentum state is given by

$$C(t) = \sum_{x_1=1}^{L} G(\{t, x_1\}).$$ (3.38)

The pole mass of the associated scalar particle is given in a lattice volume with infinite time extent by

$$m_\phi = \lim_{t \to \infty} \left[-\partial_t \log C(t)\right].$$ (3.39)

For lattices with finite time extents, an 'effective' mass

$$m_\phi^{\text{eff}}(t) = \text{arccosh}\left(\frac{C(t-1) + C(t+1)}{2C(t)}\right)$$ (3.40)

gives an estimate of the true pole mass $m_\phi$ as $t$ is taken large.

3. **Two-point susceptibility:** A measure of the total response of the scalar field to an impulse is given by the two-point susceptibility

$$\chi_2 = \sum_x G(x).$$ (3.41)

4. **Ising energy density:** For $m^2 < 0$ and $\lambda > 0$, the local potential $m^2\phi^2 + \lambda\phi^4$ has the shape of a 'double well' with two distinct minima at $\phi_* = \pm\sqrt{-m^2/2\lambda}$. Taking the limit of an infinitely deep double-well potential by holding $m^2/\lambda$ fixed and taking $\lambda \to \infty$ gives access to the Ising limit of the theory, in which each site is essentially restricted to one of the minima $\phi_*$. Interpreting this model as a statistical model rather than a quantum field theory, the associated energy density is defined in two dimensions by

$$E = \frac{1}{4} \sum_{\mu \in \{0,1\}} \left[G(\hat{\mu}) + G(-\hat{\mu})\right].$$ (3.42)

Though the value of $E$ does not have significance in the field-theoretical continuum limit, we nonetheless measure this quantity to compare against results in Ref. [379] and as a measure of local fluctuations in the scalar field complementary to other observables.

These observables provide a useful cross-check of results produced from the flow-based MCMC procedure. Autocorrelations of these observables are also used in our study of critical slowing down.

### 3.6.2 Benchmark MCMC methods

As a baseline against which flow-based models were compared, we applied HMC and a local Metropolis sampler to produce ensembles associated with each choice of physical parameters. There are many other MCMC methods that specifically address critical slowing down in scalar field theory and may be expected to perform better than HMC and local Metropolis in the continuum limit, including worm algorithms [379], multigrid methods [103], Fourier-accelerated Langevin updated [112], and cluster updates [92]. However, our aim is to consider methods that can be extended to phenomenologically relevant lattice field theories such as QCD, and such specialized methods are not known in these cases.

|              | E1   | E2   | E3   | E4   | E5   |
|--------------|------|------|------|------|------|
| $\tau$       | 1.4  | 1.4  | 1.3  | 1.2  | 1.1  |
| Acc. rate (%)| 79   | 72   | 70   | 70   | 71   |

**Table 3.2:** Choices of total trajectory length and the corresponding measured acceptance rate for each ensemble generated using HMC. All acceptance rates were measured to sub-percent-level precision. Trajectory lengths were tuned to give a 70 % acceptance rate to within 2 % in all cases except ensemble E1. A higher acceptance rate was selected for ensemble E1 as discussed in the main text.

**HMC.** HMC is a commonly used, general-purpose MCMC sampler (see Chap. 2) which is also applicable in gauge theory and thus provides an interesting comparison to the flow-based approach investigated here. For each set of physical parameters, we produced 10 replica ensembles, each consisting of 100 000 configurations drawn from Markov chains defined using HMC trajectories. In each Markov chain, the first 10 000 steps were discarded for thermalization and every 10th subsequent configuration was saved. In each case except ensemble E1, the length $\tau$ of each HMC trajectory was tuned to achieve approximately a 70 % acceptance rate when integrated using 10 leapfrog integration steps, as shown in Table 3.2. For ensemble E1, it was not possible to sufficiently reduce the acceptance rate without encountering instability of the leapfrog integrator, and a higher acceptance rate was chosen for this ensemble. Ensemble E1 does not affect our estimates of critical slowing down in the HMC algorithm and the differing acceptance rate is not an issue for observable comparisons.

**Local Metropolis.** A local perturbative update with Metropolis accept/reject step is a cost-effective MCMC sampler for theories with simple local actions, including puregauge lattice gauge theories; we applied this as a second point of comparison for this two-dimensional scalar field theory. A single Markov chain step was defined by the application of one local update to each site of the lattice. For all choices of parameters, the local update to the field value $\phi(x)$ was defined by uniformly proposing a new value $\phi'(x) \in [\phi(x) - \delta, \phi(x) + \delta]$ and applying a Metropolis accept/reject step. In each case, the parameter $\delta$ was tuned to achieve an average acceptance rate of approximately 70 % as shown in Table 3.3. We produced 10 replica ensembles for each set of physical parameters. Each ensemble consisted of 100 000 configurations drawn from Markov chains defined by local Metropolis steps, where in each Markov chain the first 10 000 steps were discarded for thermalization and every 10th subsequent configuration was saved.

|          | E1   | E2   | E3   | E4   | E5   |
|----------|------|------|------|------|------|
| $\delta$ | 0.51 | 0.50 | 0.50 | 0.50 | 0.50 |
| Acc. rate (%) | 70 | 71 | 71 | 70 | 70 |

**Table 3.3:** Choices of $\delta$ defining Metropolis updates and the corresponding measured acceptance rate for each ensemble generated using a local Metropolis sampler. All acceptance rates were measured to sub-percent-level precision. The values of $\delta$ were tuned to give a 70 % acceptance rate to within 2 % in all cases.

### 3.6.3 Model architecture, training, and MCMC

For each choice of physical parameters, we constructed an independent, identical flow-based model[6] and optimized the parameters of the model to approximate the target distribution $p(\phi) \propto e^{-S(\phi; m^2, \lambda)}$ defined by the scalar action $S$ and parameters $m^2$ and $\lambda$. We detail the construction and training of each model below. All models were implemented using the PyTorch framework [353].

**Prior distribution.** We chose an IID unit-variance Gaussian distribution with probability density

$$r(\phi) = \prod_{x,i} \frac{1}{\sqrt{2\pi}} e^{-(\phi_x^i)^2/2}. \tag{3.43}$$

as the prior distribution for all flow-based models. This choice is easily sampled and is invariant under all geometric symmetries of the lattice (see Sec. 3.1.4). Though this prior distribution is also invariant under the global sign flip symmetry, coupling layers defined in the following section are not explicitly constructed to respect the symmetry, and thus this symmetry is not explicitly enforced in the flow overall. Regardless, flow-based models that are trained to provide a good approximation to the target distributions under study will approximately reproduce this symmetry, and by the exactness of the flow-based MCMC sampler the symmetry will be full restored on average in the generated ensemble.

**Coupling layers.** We implemented the invertible flow within each flow-based model using a composition of coupling layers (see Sec. 3.2). In each coupling layer, a checkerboard mask was used to select the updated and frozen subsets of sites, with an alternating parity in successive layers ensuring that all sites were updated once after the application of each 'stack' of two coupling layers. We composed six stacks of coupling layers, totaling 12 coupling layers overall.

Each coupling layer was constructed as a real non-volume-preserving (Real NVP) transformation [244]. A Real NVP coupling layer applies an elementwise affine transformation to the updated subset of components with the scaling parameter $s$ and offset parameter $t$ given by the output of context functions applied to the frozen subset. In

---

[6]A fixed flow-based model architecture can be applied to arbitrary lattice volumes when using convolutions as discussed in Sec. 3.2.4.

terms of a masking pattern $m$, the action of each Real NVP coupling layer is defined as

$$g^{\mathrm{RNVP}}(\phi) \equiv m\phi + (1-m)[e^{s(m\phi)}\phi + t(m\phi)].  \qquad (3.44)$$

The inverse transformation can be written in closed form as

$$[g^{\mathrm{RNVP}}]^{-1}(\phi') \equiv m\phi' + (1-m)[e^{-s(m\phi')}(\phi' - t(m\phi'))].  \qquad (3.45)$$

A convolutional neural network with two channels of output was used to simultaneously define the context functions $s(\cdot)$ and $t(\cdot)$ in each coupling layer, as shown in Fig. 3.8. The masked field $m\phi$ was given as a single-channel input to each convolutional neural network. Each convolutional layer was defined in terms of kernels of size $3 \times 3$, and convolutions employed periodic padding and a stride of one to ensure that the lattice geometry was unmodified by the convolution. The convolutional network included two hidden layers with eight channels each between the input and output layers. Between each convolution, an elementwise leaky rectified linear unit (ReLU) [347] defined by $f(x) = \max(0, x) + 0.01 \min(0, x)$ was applied as a non-linear activation function. The hyperbolic tangent function was applied as an activation function after the final convolution, restricting the output of each context function to lie in the interval $[-1, 1]$. Though this restricts the expressivity of each coupling layer, it also prevents large scaling factors $e^{s(\cdot)}$ which could potentially result in numerical instability. Using a checkerboard mask and convolutional neural networks exactly enforced translational symmetry under all even-parity translations for these models (see Sec. 3.3). Many other architectural choices are possible in defining these coupling layers; in this proof-of-principle study we did not extensively tune the architecture because the aim was to investigate the general performance of flow-based MCMC. To produce models that are highly tuned for particular theories in future work, it may be fruitful to scan over possible choices to empirically determine whether significant improvements in model expressivity are possible.

**Training.** We trained one flow-based model to approximate each of the target distributions studied. In each case, the KL divergence between the model distribution and target distribution was used as the loss function to be minimized. Batches of 1024 field configurations were sampled from the model and used to stochastically estimate the KL divergence and its gradient with respect to model parameters in each optimization step. We used the Adam optimizer to determine updates to the model parameters as a function of the gradients computed at each step, with the Adam hyperparameters fixed to the defaults assigned by the PyTorch implementation and a learning rate fixed to $10^{-3}$. A total of 100 000 training iterations were applied for each model, which was sufficient to give approximate convergence to a plateau of the model quality under this training scheme. For example, Fig. 3.9 shows convergence of the measured normalized KL divergence to a nearly constant final value over the course of training for all five models. The normalization of these measurements is given by the log partition function $\log Z$, and as shown in Fig. 3.10 this quantity quickly converges to a precise estimate using the unbiased estimator defined in the previous section, indicating that estimates of the normalized KL divergence are reliable.

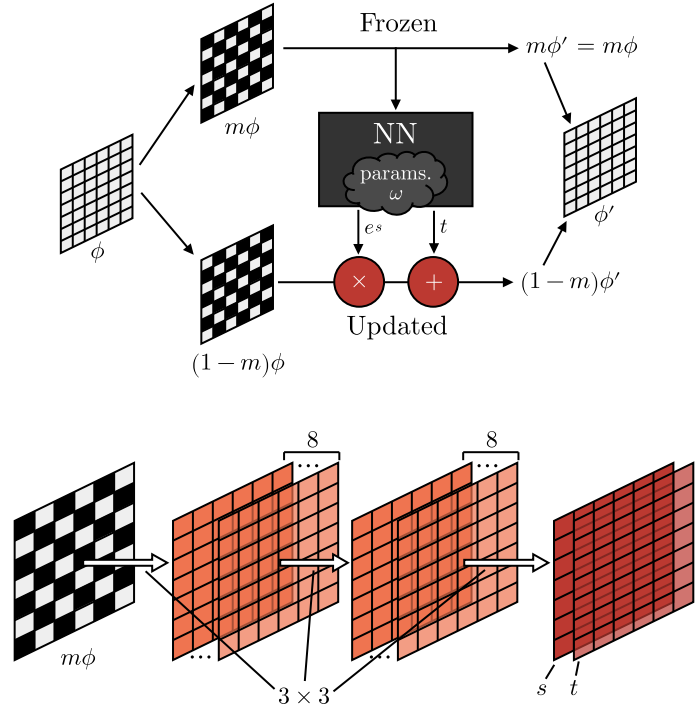

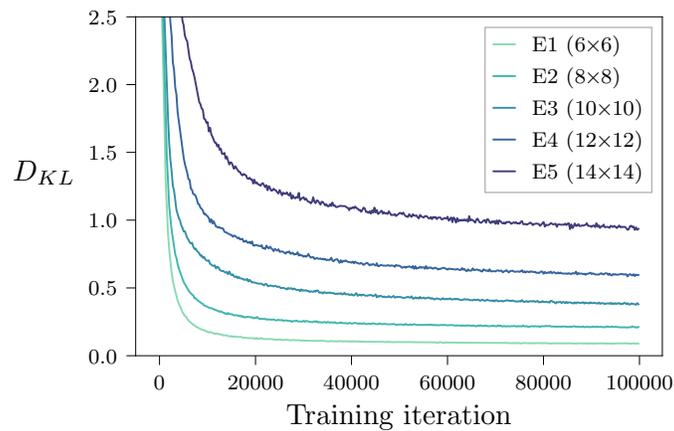

**Figure 3.8:** Architectural details for the Real NVP coupling layers applied in the scalar field theory application discussed in the main text. Upper: checkerboard masking and affine transformation using a Real NVP coupling layer. Lower: architecture of the convolutional neural network acting on the frozen degrees of freedom given by the masked field $m\phi$.

**Figure 3.9:** Convergence of the normalized KL divergence $D_{KL}$ for each flow-based model over the course of training. The normalization $\log Z$ is well estimated after iteration 10 000 for all models and the measured KL divergences thus give reliable estimates of model quality. The varying final KL divergence values are discussed in the main text.

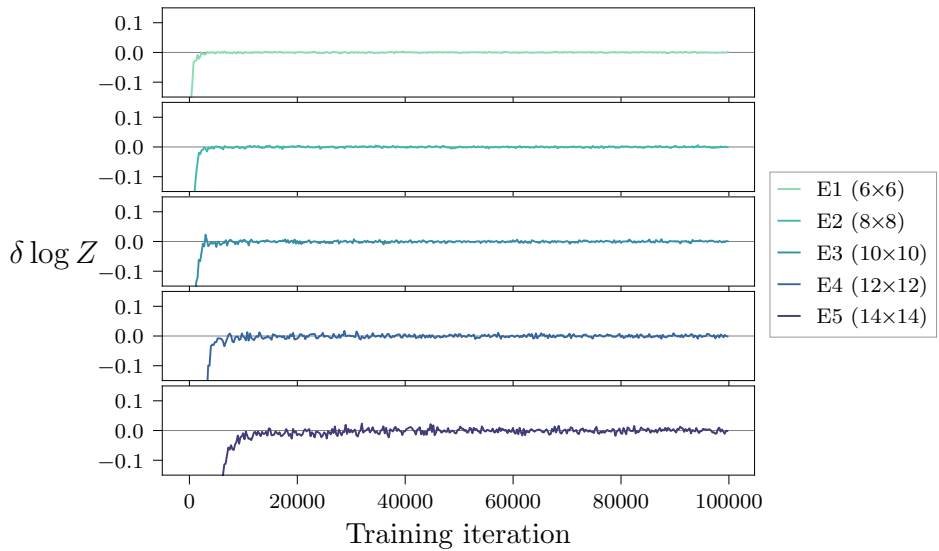

**Figure 3.10:** Convergence of the estimates of $\log Z$ measured using each flow-based model over the course of training. In all cases, a very precise estimate is quickly possible after less than 10 % of the total training time.

As one can see in Fig. 3.9, the fixed architecture studied here gives better convergence to the target distributions associated with coarser lattice spacing and a smaller number of lattice sites. A thorough comparison of the possible ways to scale the model complexity in the continuum limit is beyond the scope of this proof-of-principle application. In this study, however, we specifically considered the alternative option of fixing the flow-based MCMC acceptance rate across all models. To do so, we additionally saved 'early stopping' versions of all models at points in their respective training history at which a fixed acceptance rate of 35 % was achieved to within 2 %, as detailed in Table 3.4. This choice matches the final acceptance rate of the flow-based model targeting parameter set E5. This set of early stopping models is compared to the same models trained for the full 100 000 iterations in the study of critical slowing down below. This by no means represents all possible methods of applying flow-based models to scaling towards the continuum limit (working with a fixed model architecture is severely restrictive). Section 3.7 discusses a variety of considerations related to scaling, and addressing these considerations is a well-motivated subject of future work.

**MCMC.** The trained flow-based models were used to generate ensembles for their respective target distributions using flow-based MCMC. A total of 1 024 000 proposals were generated from each model. We divided these proposals into 10 subsets of 102 400 configurations, each of which were independently composed into a Markov chain by applying the independence Metropolis accept/reject step to each proposal in turn. The first 2400 steps of each resulting Markov chain were discarded for thermalization and the remaining 100 000 configurations were reserved as an ensemble for measurements. The 10 independent replica ensembles were used to reliably estimate errors on autocorrela-tion measurements in the investigation of critical slowing down presented below. Note that in contrast to saving only every 10th Markov chain state, as we did in the cases of the HMC and local Metropolis ensembles, we saved every flow-based Markov chain configuration after thermalization so that the $\tau$-rejection autocorrelation estimator could be measured. When comparing autocorrelations between HMC, local Metropolis, and flow-based MCMC, the autocorrelation time is thus rescaled by a factor of 10 for HMC and local Metropolis to provide a fair comparison. When comparing observables, the flow-based Markov chain is thinned by reducing to every 10th configuration and correspondingly the HMC and local Metropolis measurements are reduced to one of the 10 replicas to provide equivalent statistics in all cases.[7]

|            | E1   | E2   | E3   | E4     | E5      |
|------------|------|------|------|--------|---------|
| Iters.     | 2000 | 3000 | 6000 | 16 000 | 100 000 |
| Acc. rate (%) | 35 | 35   | 35   | 35     | 33      |

**Table 3.4:** Number of training iterations at which models were saved for the study of fixed acceptance rate discussed in the main text. Reported acceptance rates were measured with sub-percent-level precision in all cases. All acceptance rates agree with a target final value of 35 % to within 2 %.

### 3.6.4 Monte Carlo studies

We next discuss the correctness of results generated using flow-based MCMC and investigate critical slowing down using measurements of observables on the HMC, local Metropolis, and flow-based ensembles.

**Tests of physical observables and statistical scaling.** Flow-based MCMC is necessarily correct in the asymptotic limit of an infinitely long Markov chain. For the finite ensembles studied here, we confirm that flow-based MCMC practically converges to sampling from the target distribution by comparing measurements of the four observables listed above. Figure 3.11 compares the two bulk observables, $\chi_2$ and $E$, measured on ensembles generated by all three MCMC samplers. One can see agreement between flow-based MCMC and the benchmark methods, up to statistical fluctuations, across all five choices of physical parameters, which correspond to a wide range of the values of these bulk observables. Figure 3.12 depicts the measured values of the zero-momentum two-point correlation function $C(t)$ over all possible time separations $t$ and the corresponding effective mass on the parameter set E3, corresponding to the middle lattice spacing studied. A similar level of agreement can be seen for all choices of time separation $t$, indicating that fluctuations in the scalar field at all lengths scales are captured by the flow-based model.

We also confirm the statistical scaling properties of the flow-based ensemble by comparing the statistical uncertainties on estimates of $\chi_2$ as the number of measurements included is increased. Figure 3.13 shows the measured bootstrap uncertainty on estimates of $\chi_2$ as a function of the number of measurements $n$ for the flow-based ensemble corresponding to parameter set E3. We compare the measured uncertainties against the theoretical curve given by $\sqrt{\mathrm{Var}(\chi_2)/n}$, finding good agreement across a wide range of the number of measurements $n$. The full ensemble was used to measure the variance of the $\chi_2$ distribution fixing the normalization of the theoretical curve in the figure.

---

[7]A direct comparison could be made for observables, but the resulting measurements do not accurately convey the relative statistical power of ensembles constructed with equivalent statistics across the methods.

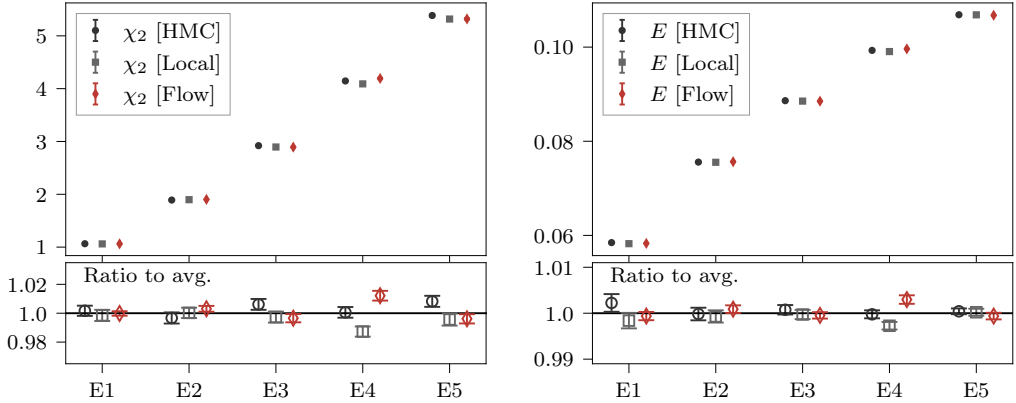

**Figure 3.11:** A comparison of the bulk observables $\chi_2$ and $E$ measured on all ensembles. The ratio of the measured values with with the average central value is shown for each choice of parameters and each observable. Error bars are estimated using bootstrap resampling and are rescaled by $\sqrt{2\tau_{\chi_2}^{\text{int}}}$ and $\sqrt{2\tau_E^{\text{int}}}$ for $\chi_2$ and $E$, respectively, to account for autocorrelations in the measurements. Some fluctuations can be seen for particular ensembles, most prominently in E4, but these fluctuations appear across all three methods and are consistent with statistical fluctuations.

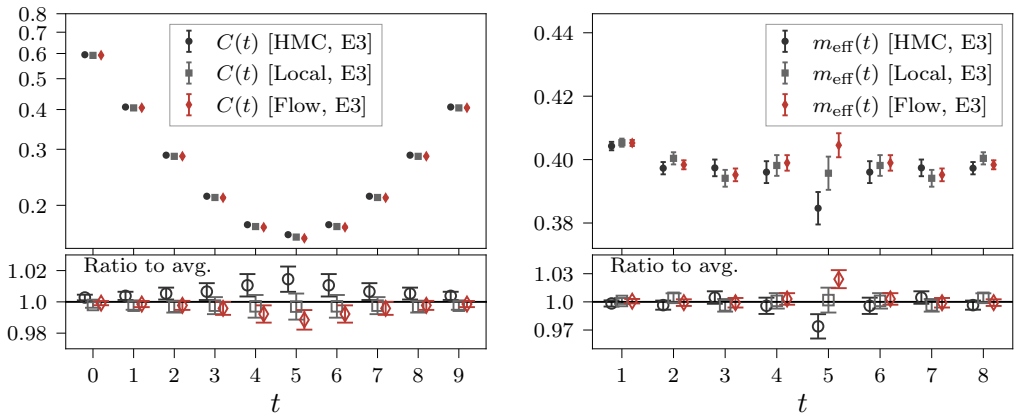

**Figure 3.12:** A comparison of the zero-momentum correlation function $C(t)$ and derived effective mass $m_{\text{eff}}(t)$ for ensembles generated using all three MCMC approaches on parameter set E3. The ratio of the measured values with with the average central value is shown for each choice of parameters and each observable. Error bars are estimated using bootstrap resampling and are rescaled by $\sqrt{2\tau_{C(0)}^{\text{int}}}$ to account for autocorrelations in the measurements.

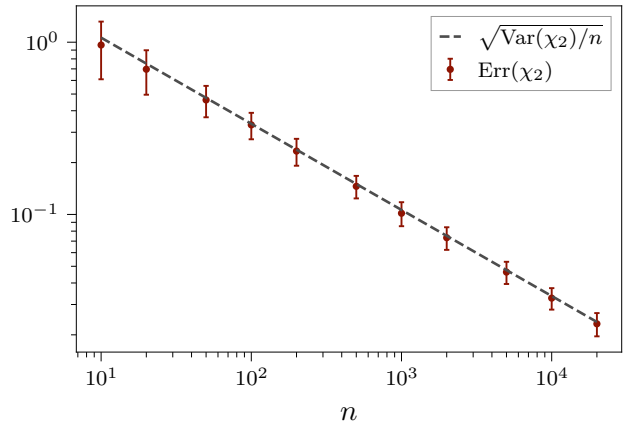

**Figure 3.13:** Comparison of the statistical uncertainties on estimates of $\chi_2$ over a range of total measurements $n$ used in the estimate, where samples are taken from the flow-based ensemble associated with the parameter set E3. Uncertainties on the uncertainties were measured using bootstrap resampling over different subsets of size $n$ in each case. The measured uncertainties are compared to the theoretical $\sqrt{\mathrm{Var}(\chi_2)/n}$ scaling curve, where $\mathrm{Var}(\chi_2)$ is estimated on the full ensemble to compute the normalization.

**Critical slowing down.** Finally, we investigated the integrated autocorrelation times for a number of observables to study the ability of flow-based MCMC to mitigate the effects of critical slowing down. We considered both the flow-based models trained for a fixed number of iterations and the flow-based models trained to a fixed acceptance rate discussed above. Figure 3.14 shows the measured values of $\tau^{\mathrm{int}}_{\mathcal{O}}$ for three of the observables considered, $E$, $G(\{0,0\})$, and $\chi_2$, as well as the value of $\tau^{\mathrm{int}}_{\mathrm{acc}}$ corresponding to estimates of the universal autocorrelation time arising from the rejection statistics of the flow-based Markov chain. The integrated autocorrelation times $\tau^{\mathrm{int}}_{\mathcal{O}}$ were computed using the self-consistent estimator defined in Sec. 2.4.2 with a multiplicative factor $c = 4$. Uncertainties on $\tau^{\mathrm{int}}_{\mathcal{O}}$ and $\tau^{\mathrm{int}}_{\mathrm{acc}}$ were determined by applying bootstrap resampling over the 10 replicas generated for all three approaches. To account for the reduced measurement frequency in both the HMC and local Metropolis calculations, the values of $\tau^{\mathrm{int}}_{\mathcal{O}}$ computed from the saved measurements are given in units of Markov chain steps by multiplying by a factor of 10.

To study critical slowing down, the values of $\tau^{\mathrm{int}}_{\mathcal{O}}$ on the ensembles with the three finest lattice spacings were fit to a power law $\alpha L^z$ to quantitatively determine the rate of critical slowing down with respect to the lattice dimension $L$. We restricted to fitting the flow-based models trained to a fixed acceptance rate, as these represent the more realistic use case of scaling model quality as the continuum limit is approached. One can clearly see in Fig. 3.14 that instead choosing to work with a fixed architecture while scaling towards the continuum results in more rapid scaling of the autocorrelation time, though it is noteworthy that the actual values of the integrated autocorrelation time are less than those of the local Metropolis method for all physical parameters, and are less than those of HMC for the coarser lattices. The best-fit parameters, uncertainties, and values of $\chi^2$/d.o.f. are reported in Table 3.5. The best-fit power law results are also shown as straight lines overlaid on the log-log plots shown in Fig. 3.14.

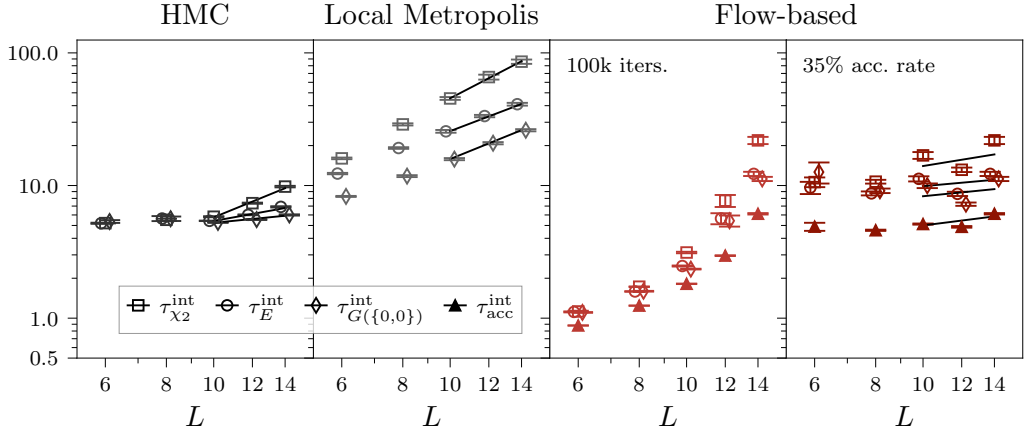

**Figure 3.14:** Integrated autocorrelation times $\tau_{\mathcal{O}}^{\text{int}}$ for ensembles generated using HMC, local Metropolis, and flow-based MCMC. The estimator $\tau_{\text{acc}}^{\text{int}}$ is based on the acceptance statistics in the independence Metropolis scheme used for flow-based MCMC and is only plotted in the two flow-based panels on the right. The rightmost two panels compare results for ensembles generated from flow-based models trained for 100 000 iterations each (third panel) to results for ensembles generated from flow-based models trained to a fixed 35 % acceptance rate (fourth panel). The latter case represents a more realistic approach to scaling models towards the continuum limit, as discussed in the main text. Thin black lines depict the central values of power law fits to the largest three lattice sizes for the HMC, local Metropolis, and fixed-acceptance flow-based ensembles.

| $\mathcal{O}$ | HMC | | | Local Metropolis | | |
|---|---|---|---|---|---|---|
| | $\alpha$ | $z$ | $\frac{\chi^2}{\text{d.o.f.}}$ | $\alpha$ | $z$ | $\frac{\chi^2}{\text{d.o.f.}}$ |
| $E$ | 1.16(8) | 0.67(3) | 2.77 | 1.02(26) | 1.40(10) | 0.05 |
| $\chi_2$ | 0.18(2) | 1.50(5) | 4.72 | 0.56(17) | 1.91(13) | 0.09 |
| $G(\{0,0\})$ | 2.27(22) | 0.36(4) | 1.56 | 0.51(10) | 1.49(8) | 0.01 |

| | Flow-based (35 % acc. rate) | | |
|---|---|---|---|
| $\mathcal{O}$ | $\alpha$ | $z$ | $\frac{\chi^2}{\text{d.o.f.}}$ |
| $E$ | 4.6(2.0) | 0.33(17) | 14.49 |
| $\chi_2$ | 3.5(2.2) | 0.60(25) | 15.04 |
| $G(\{0,0\})$ | 3.6(1.4) | 0.37(16) | 31.46 |
| acc | 1.71(19) | 0.47(5) | 26.81 |

**Table 3.5:** Best fit parameters and corresponding $\chi^2/\text{d.o.f.}$ for power law fits to the integrated autocorrelation time $\tau_{\mathcal{O}}^{\text{int}}$ associated with each observable $\mathcal{O}$. Fits were performed to the form $\alpha L^z$ over the largest three lattice sizes $L \in \{10, 12, 14\}$ in all three cases. The 'Flow-based' columns indicate fits to models at a fixed 35 % acceptance rate; the large $\chi^2$ values indicate significant tension with the fit for the reasons discussed in the main text.

As one can see from the figure and table, both HMC and local Metropolis results are consistent with power law scaling for the observables measured. On the other hand, the integrated autocorrelation times measured for flow-based models with fixed acceptance rates have significant deviations, outside the measured uncertainties, from the fits. This result is expected: there is no reason *a priori* to expect a particular sequence of flow-based models trained to produce independent global proposals at five different choices of parameters to produce flow-based ensembles with autocorrelations following a smooth scaling curve. However, there is a sense in which a smooth scaling curve may be expected from flow-based MCMC — if we consider marginalizing over the flow-based model parameters determined from all possible optimization trajectories, the underlying model architecture and prescription for training give a fixed algorithm that could be expected to produce autocorrelation times that in expectation smoothly vary with the choice of physical parameters. In other words, the fluctuations in training procedure could be considered an additional source of the scatter in the measured integrated autocorrelation times that is not captured by the statistical uncertainties measured in the context of particular flow-based models. In this light, it is clear that the measured autocorrelation times broadly follow the trends of the fit lines and that the poor quality of the fits can be attributed to this additional scatter.[8] Despite the poor quality of the fits, the best-fit exponents $z$ of the plotted fit lines do give a measure of the expected critical slowing down of flow-based MCMC under the training scheme to fixed acceptance rate defined above. These best-fit exponents correspond to a scaling that is comparable to or milder than that of measurements using HMC and local Metropolis ensembles for all observables.

The approach of training flow-based models to a fixed acceptance rate and applying flow-based MCMC for ensemble generation thus holds clear promise. In this study we worked with a fixed model architecture and achieved fixed acceptance rates by modifying the training procedure, but in continuing to scale towards the continuum limit of a theory it is not obvious whether similar fixed model architectures can be further optimized to continue to achieve a target acceptance rate. Instead, it may be necessary to scale the model expressivity as the number of degrees of freedom are increased and the lattice spacing is decreased. The landscape of tradeoffs surrounding

---

[8]Note that these additional fluctuations outside of statistical uncertainties appear only in estimates of autocorrelations. Measurements of observables are unbiased and fully constrained by a Markov chain constructed using proposals from any particular flow-based model.

this scaling to the continuum limit is discussed in the following section.

## 3.7   Scaling

Scaling the methods presented in this chapter to a larger number of degrees of freedom may be of interest for two distinct scaling directions:

1. Scaling up the physical lattice volume by holding bare parameters fixed and increasing the number of lattice sites; or

2. Scaling parameters towards the continuum limit, or criticality in general, while holding the physical volume fixed in terms of some scale-setting measure such as the correlation length of a state of interest.

We discuss considerations for scaling in both of these directions below.

### 3.7.1   Increasing the physical volume

Increasing the physical volume of the lattice while holding the bare parameters fixed corresponds generally to the case of increasing the number of lattice degrees of freedom while maintaining the local correlation structure at any point on the lattice. This story is modified slightly if there are strong finite volume effects. In this case it is difficult to analyze the expected scaling behavior and in many ways this situation is in closer analogy to the scaling towards the continuum limit discussed below — the local correlation structure may change in non-trivial ways as we change the volume. We therefore focus on the regime in which finite volume effects are suppressed in this section. In this regime, performing simulations at multiple physical volumes might be of interest to extrapolate away these residual finite volume effects, but their effect on local correlation structure is mild.

In this chapter, we have emphasized the utility of encoding translational symmetry in flow-based models. By using convolutional neural networks and coupling layers based on masking patterns, a large subgroup of translational symmetries can be exactly encoded in a flow-based model. The result is that a fixed model can be applied to all lattice geometries compatible with the masking pattern, because the model parameterizes the local correlation structure. For model architectures taking this form, the effective action $S_{\text{eff}}$ of a given flow-based model can be decomposed into a sum of identical local terms as

$$S_{\text{eff}}(\phi) = \sum_x \mathcal{L}_{\text{eff}}(\phi(y - x)), \tag{3.46}$$

where the function $\mathcal{L}_{\text{eff}}$ can be thought of as a discretized effective Lagrangian that acts as a function of the $\phi$ field translated by $x$. Translationally invariant target actions can also be decomposed in a similar way as

$$S(\phi) = \sum_x \mathcal{L}(\phi(y - x)), \tag{3.47}$$

and this description can be extended straightforwardly to multiple fields. The quality of a translationally invariant flow-based model is thus determined by how well $\mathcal{L}_{\text{eff}}$ captures $\mathcal{L}$.

We can now consider applying a translationally invariant flow-based model architecture to target distributions defined by the same bare parameters but different lattice volumes. In this case, the target $\mathcal{L}$ remains the same and the optimal flow-based model is the same at all lattice volumes: the aim of optimization is always to produce the best agreement between $\mathcal{L}_{\text{eff}}$ and $\mathcal{L}$. As a small caveat to this statement, the measure of 'best agreement' depends on the samples drawn from the model. If there are significant finite volume effects that give rise to differing local correlation structure for the same $\mathcal{L}$ or $\mathcal{L}_{\text{eff}}$, then the optimal choice of model parameters may vary significantly across volumes; under our assumptions above, this is not a concern in the settings of interest. Thus, if one increases the physical volume under study while holding the model architecture fixed, the optimal choice of $\mathcal{L}_{\text{eff}}$ remains the same, and importantly the discrepancy between the effective action of the model, $S_{\text{eff}}$, and the target action, $S$, will be an extensive quantity. Reweighting factors and acceptance probabilities scale exponentially with the typical difference in action $S_{\text{eff}} - S$, and fixed (translationally invariant) model architectures can therefore be expected to scale exponentially poorly with volume.

This superficially pessimistic outlook is tempered by a few observations. First, when considering increasing volume to capture small residual finite volume effects, one does not need to scale the lattice volume significantly to understand these suppressed finite volume effects and remove them by extrapolation. In practice lattice volumes that are a multiple of several correlation lengths in each dimension are typically sufficiently large to perform such extrapolations. Secondly, the extensive nature of the overlap between $S_{\text{eff}}$ and $S$ also works to our advantage. If the model expressivity is increased such that the average agreement between $\mathcal{L}$ and $\mathcal{L}_{\text{eff}}$ is improved by some factor, then the variation in reweighting factors or the acceptance rate in a flow-based MCMC approach will be improved by a factor that is exponential in the physical volume times that factor. The parameterization of flow-based models that use convolutional neural networks and coupling layers based on masking patterns describes $\mathcal{L}_{\text{eff}}$ directly and one can expect such significant gains as a result of fixed factors of improvement in the model expressivity, so long as the model can be trained to exploit the additional expressivity.

Ultimately, the tradeoff between increased cost of sampling and training from more expressive models and the improved autocorrelations resulting from higher quality models needs to be explored as a practical question in theories of interest.

### 3.7.2  Decreasing the lattice spacing

The second interesting scaling direction is to reduce the lattice spacing while holding the physical volume fixed. In this direction, the number of degrees of freedom is increased to capture field fluctuations at progressively shorter distances. This requires tuning the bare parameters of the model towards criticality, corresponding to a diverging correlation length, while also increasing the number of sites in each dimension of the lattice to hold the total physical volume fixed.

In terms of the discrete Lagrangian description given in Eqs. (3.46) and (3.47), decreasing the lattice spacing or approaching criticality corresponds to modifying the discrete Lagrangian $\mathcal{L}$ to more strongly couple degrees of freedom that are a fixed number of lattice sites apart. For example, for actions involving only nearest-neighbor coupling, this corresponds to increasing the strength of the nearest-neighbor interaction.

Unlike in the case of scaling the physical volume, when modifying the lattice spacing, different sets of model parameters are expected to give the best performance as $\mathcal{L}$ is varied. How well the resulting optimized choices of $\mathcal{L}_{\text{eff}}$ agree with $\mathcal{L}$ depends entirely on the model architecture. The scaling of the performance (e.g. the acceptance rate or autocorrelation time) of a model with a fixed architecture and size thus depends on the particular architecture under study. In the study of scalar field theory presented in the previous section, the autocorrelation time was found to increase significantly for the models with fixed architecture trained to convergence (see Fig. 3.14). This demonstrates that, for some choices of model architecture and training schemes, this scaling under fixed model architecture may not be favorable. It is also obvious that it is not possible to extend a fixed convolutional coupling layer architecture towards arbitrarily fine lattice spacings: for a finite number of coupling layers and a fixed number of convolutional layers, there is simply a limiting radius beyond which correlations cannot be imposed at all.

These arguments make clear that, as in the case of scaling physical volume, it is likely that the most efficient approach to scaling is based on increasing model expressivity as the scaling limit is taken. The maximum 'radius of influence' described above suggests that at least one way in which models should be scaled is by increasing this radius of influence as the continuum limit is taken. This could be achieved in a number of ways:

1. Increasing the number of coupling layers

2. Increasing the size of each convolutional kernel

3. Increasing the number of convolutional layers in each neural network

4. Passing some form of global information to context functions

All of these approaches will increase the cost of training and sampling to some extent. The remaining question is whether the correspondingly increased model expressivity gives sufficient improvements in the efficiency of the resulting flow-based sampler, and in particular whether this scaling is improved relative to existing methods. The nearly constant autocorrelations observed for models with a fixed acceptance rate in the previous section suggest that a 'rule of thumb' for scaling towards the continuum limit may be to fix a target acceptance rate in a flow-based Markov chain and to scale expressivity in models to meet that target.

Finally, we note that scaling the expressivity of flow-based models to fix a similar acceptance rate across all choices of parameters is analogous to scaling the integrator step size and total number of steps in HMC trajectories to maintain a fixed acceptance rate. In contrast to HMC, there is a great deal of freedom in the design and training of flow-based models, all of which is factorized from the question of asymptotic exactness.

There are many opportunities to apply physical principles to exploit this freedom and produce both efficient and expressive models.

# Chapter 4

# Flow-based sampling for gauge theories

*Content in this chapter is partially adapted with permission from:*

- Pub. [2]: D. Boyda, G. Kanwar, S. Racanière, D. J. Rezende, M. S. Albergo, K. Cranmer, D. C. Hackett, and P. E. Shanahan, "Sampling using SU($N$) gauge equivariant flows", Phys. Rev. D **103**, 074504 (2021), arXiv:2008.05456 [hep-lat]

- Pub. [3]: G. Kanwar, M. S. Albergo, D. Boyda, K. Cranmer, D. C. Hackett, S. Racanière, D. J. Rezende, and P. E. Shanahan, "Equivariant flow-based sampling for lattice gauge theory", Phys. Rev. Lett. **125**, 121601 (2020), arXiv:2003.06413 [hep-lat]

- Pub. [8]: D. J. Rezende, G. Papamakarios, S. Racanière, M. S. Albergo, G. Kanwar, P. E. Shanahan, and K. Cranmer, "Normalizing flows on tori and spheres", (2020), arXiv:2002.02428 [stat.ML]

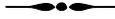

In this chapter, we extend the methods of Chapter 3 to the setting of lattice gauge theory. Lattice gauge theories are defined using path integrals over gauge configurations consisting of a collection of gauge variables $U_\mu(x)$, each associated with the link $(x, x+\hat{\mu})$ of the lattice (see Sec. 2.1). These gauge configurations live in the compact manifold $G^{DV}$, where $G$ is the manifold of the gauge group, $D$ is the spacetime dimension, and $V$ is the number of lattice sites. The action of a lattice gauge theory is invariant under *gauge transformations* defined for arbitrary group-valued fields $\Omega(x) \in G$ by[1]

$$U_\mu(x) \ \rightarrow \ \Omega(x) U_\mu(x) \Omega^\dagger(x + \hat{\mu}). \tag{4.1}$$

Defining invertible transformations suitable for the compact gauge group and exactly encoding this gauge symmetry are the two main aims of this chapter. These goals are inextricably linked: to design equivariant coupling layers (see Sec. 3.3) that exactly encode the gauge symmetry requires understanding how one can transform the compact space, and to design appropriate compact transformations requires understanding how they ultimately will be used to create gauge-equivariant coupling layers.

---

[1]Here and below it is assumed that group-valued fields are given in a unitary representation, thus $\Omega^\dagger = \Omega^{-1}$ may be used here.

In the remainder of this chapter, we detail our approach to encoding gauge symmetry in flow-based models using a gauge-invariant prior distribution and gauge-equivariant coupling layers (Secs. 4.1, 4.2, and 4.3), then we construct invertible operations suitable for manifolds associated with U(1) and SU($N$) variables (Secs. 4.4 and 4.5), and finally we demonstrate a proof-of-principle application of these methods to U(1) and SU($N$) lattice gauge theories (Secs. 4.6 and 4.7).

## 4.1   Encoding gauge symmetry in flow-based models

In principle, a simple way to achieve factorization between pure-gauge and gauge-invariant degrees of freedom would be to employ a gauge fixing procedure. Maximally gauge fixing reduces configurations to gauge-invariant degrees of freedom only, and flows acting only on this reduced subset of degrees of freedom would automatically result in invariant distributions. When considering approaches to encode gauge symmetry into flow-based models, however, it is important to consider interactions between symmetry groups.

We consider the two classes of gauge fixing procedures that can be used for lattice gauge theory:

- **Explicit gauge fixing** approaches like maximal tree gauge fixing [381, 382] explicitly separate pure-gauge and gauge-invariant degrees of freedom. Gauge-invariant degrees of freedom in this approach can be directly manipulated without leaving the subspace of gauge-fixed configurations. Figure 4.1 gives an example of a maximal tree gauge fixing pattern that can be used to factorize the degrees of freedom on the lattice.

- **Implicit gauge fixing** approaches use implicit differential equation constraints instead of an explicit factorization to remove pure-gauge degrees of freedom. Fixing to Landau or Coulomb gauge are typical choices in this class [383–385].

Difficulties with applying flows to gauge-fixed degrees of freedom arise in both cases.

Using explicit gauge fixing, it is straightforward to restrict the flow-based model to act on gauge-invariant degrees of freedom only by defining a prior distribution over this subset of degrees of freedom and restricting flows to act on the same subset. As seen in Figure 4.1, however, the gauge fixing pattern does not satisfy translational invariance. Gauge fixing in this way does not break the translational symmetry of gauge-invariant observables in the theory, but it does complicate translational equivariance for flow-based models acting on the reduced subset of variables. In particular, translations act on this subset by a complicated set of interrelated equations between degrees of freedom, and satisfying equivariance under these transformations, while possible, is not simple to achieve using CNNs and invariant masking patterns as was done in the case of scalar field theory in Chapter 3.

More fundamentally, the gauge-invariant links remaining after gauge fixing represent non-local physical quantities. To understand this problem, note that maximal tree gauge fixing is defined by fixing a maximal set of links that can be included without forming any closed loops. As indicated by the marked link and region in Fig. 4.1, the remaining degrees of freedom in a maximal tree gauge fixing scheme correspond to the loop that 'would be' closed, if the link were added to the set. For most links in the gauge-fixed configuration, this corresponds to a non-local set of the underlying degrees of freedom. The correlation structure and distribution over these remaining links is thus intricate and can be expected to be highly non-local. Learning the distributions over such variables is not a task well-suited for coupling layers with local updating structures (e.g. using convolutional neural networks); other structures of coupling layers or general flow-based architectures are possible, but it is not clear how to design flows to best capture these distributions.

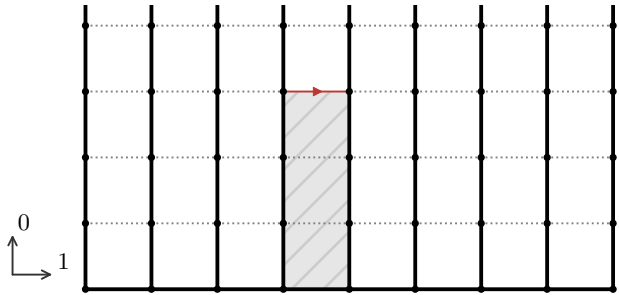

**Figure 4.1:** A maximal-tree gauge fixing scheme. Bold links indicate pure-gauge degrees of freedom that are fixed to the identity, while dotted links correspond to the remaining physical gauge-invariant degrees of freedom. The indicated link (red arrow) is one such gauge-invariant variable which corresponds to the loop surrounding the gray hatched region in the non-gauge-fixed configuration. The particular choice of gauge fixed links shown in the diagram can be thought of as temporal gauge (fixing a maximal set of links oriented in the $\hat{0}$ direction) with additional gauge fixing applied on the first timeslice; an analogous gauge-fixing pattern can be applied in higher spacetime dimensions.

Implicit gauge fixing schemes, on the other hand, can be constructed to preserve translational invariance [386]. Landau or Coulomb gauge fixing, for example, are based on numerically solving a differential equation that is itself translationally symmetric. Translations of a gauge-fixed configuration are thus also valid gauge-fixed configurations, unlike in the case of maximal tree gauge fixing. The implicit nature of these gauge fixing schemes, however, does not give a clear factorization between pure-gauge and gauge-invariant degrees of freedom. It is not obvious how to restrict flows to invertibly transform gauge configurations while remaining within the gauge-fixed subspace.

Similar questions have arisen in recent attempts to implement gauge theories on quantum computers, though in this case one is interested in the Hamiltonian picture. Recent work in this direction has demonstrated that the Hamiltonian-representation degrees of freedom can be projected into a gauge-invariant subspace for gauge theories in two and three spatial dimensions [387–390], in some cases preserving the spatial translational invariance structure of the theory. However, it is not obvious whether these approaches can be extended to apply to the path integral formulation.

These tensions between gauge fixing and maintaining manifest locality and translational invariance are a significant obstacle to implementing flow-based sampling for gauge-fixed configurations. Though in principle translational invariance can be discarded to sample in the gauge-fixed subspace, encoding both symmetries exactly into flow-based models can be expected to improve the data efficiency of training and the final quality of the resulting models. Locality also allows flow-based models to be efficiently defined in terms of fewer parameters through the use of convolutional neural networks. This motivates our construction of flow-based models that maintain gauge symmetry while working with the *non-gauge-fixed* representation of gauge configurations below. As in the previous chapter, we encode the gauge symmetry in flow-based models by implementing a prior distribution that is invariant to the symmetry and coupling layers that are equivariant to the symmetry. Invariant prior distributions in the space of non-gauge-fixed configurations are straightforward to construct. Constructing gauge-equivariant (and expressive) coupling layers is a more significant challenge and is one of the main developments presented in this chapter.

## 4.2 Gauge-invariant prior distributions

Following the strategy of constructing invariant prior distributions for scalar field theory in Chapter 3, we construct prior distributions for gauge theories that are invariant under geometric symmetries of the lattice by using an IID distribution over each gauge link. Lattice translations and hypercubic symmetries permute links of the lattice and therefore leave this distribution invariant, regardless of the marginal distribution per link. Such distributions are also easily sampled by independently drawing each gauge link from the specified marginal distribution, and the total probability density can be computed as a product of the marginal densities of each gauge link.

Invariance under gauge transformations can then be encoded in the prior distribution by the choice of marginal distribution. Gauge transformations correspond to conjugation by arbitrary group-valued elements, as given in Eq. (4.1), and any marginal distribution that is invariant under these transformations will result in a gauge-invariant prior distribution overall. Invariance under this transformation property turns out to be quite restrictive: the Haar measure is the only normalized probability measure that is invariant under left and right multiplication by arbitrary group elements [391], and the choice of marginal distribution is thus completely fixed by gauge invariance to be the Haar measure. Intuitively, the Haar measure can be thought of as the most natural 'uniform' measure over the group space, and the IID product of Haar measures over all links can also be identified with the infinitely strong coupling limit of pure gauge theory [392]. The probability density associated with this prior distribution is simply $r(U) = 1$, because the IID Haar measure is exactly the path integral measure.

One can straightforwardly sample from this distribution for a number of common groups. We concretely describe sampling from the Haar measure for the groups U(1) and SU($N$) required for the majority of phenomenologically relevant lattice studies as well as the applications described below in this chapter:

- **The** U(1) **Haar measure:** For the U(1) gauge group, group elements can be written as $U = e^{i\phi}$ in terms of an angle $\phi \in [0, 2\pi]$. With this parameterization, the normalized Haar measure is written as $dU = \frac{d\phi}{2\pi}$. This can be verified to be

---

**Algorithm 4.1:** Sample an SU($N$) matrix according to the Haar measure

---

1. Sample an $N \times N$ matrix $V$ with complex entries distributed independently according to the unit-variance Gaussian distribution with mean zero.

2. Perform a QR decomposition of $V$, producing a unitary matrix $Q$ and an invertible, upper-triangular matrix $R$ satisfying $V = QR$. Any QR decomposition routine may be used.

3. Multiply the $i$th column of $Q$ by the unit-norm complex number $R_{ii}/|R_{ii}|$, to produce a matrix $U'$:

$$U'_{ij} = Q_{ij} \frac{R_{jj}}{|R_{jj}|}. \tag{4.3}$$

   The matrix $U'$ is distributed according to the Haar measure of U($N$) [393].

4. Divide by $\det(U')^{1/N}$ and multiply by a random center element $z = e^{i2\pi k/N} \mathbb{1}$ of SU($N$) selected by sampling $k$ from the uniform distribution over $\{1, \ldots, N\}$. This produces a matrix $U$ with unit determinant:

$$U = \frac{z \, U'}{\det(U')^{1/N}}. \tag{4.4}$$

   The matrix $U$ is distributed according to the Haar measure of SU($N$).

---

left and right invariant by explicit calculation (these are the same because $U(1)$ is Abelian): multiplication by an arbitrary $V = e^{i\phi'}$ transforms the measure to

$$d(VU) = \frac{d(\phi + \phi')}{2\pi} = \frac{d\phi}{2\pi} = dU. \tag{4.2}$$

The U(1) Haar measure can be sampled by drawing $\phi$ uniformly from the interval $[0, 2\pi]$ and computing $U = e^{i\phi}$.

- **The SU($N$) Haar measure:** For SU($N$) gauge groups, the parameterization and derivation of the Haar measure is more involved. Ref. [393] presents a comprehensive discussion of the Haar measure and sampling from compact groups including U($N$). These results can be applied to SU($N$) by a straightforward extension, and the procedure for sampling an SU($N$) matrix according to the Haar measure is given in Algorithm 4.1.

The Haar measure provides a prior distribution that is invariant under the symmetries of interest for lattice gauge theory, but it is not the only prior distribution that can be used. To construct other prior distributions, the choice of an IID distribution must be relaxed; several options for non-IID prior distributions were discussed in Sec. 3.1.4, and these options are equally applicable to the case of lattice gauge theory.

## 4.3 Gauge-equivariant coupling layers

The heart of our approach is the construction of coupling layers that are equivariant under gauge transformations, i.e. for which gauge transformations commute through application of the coupling layer. A gauge transformation by definition only affects the pure-gauge degrees of freedom. Thus to be gauge equivariant it is sufficient for a coupling layer to only transform gauge-invariant degrees of freedom. As discussed in Sec. 4.1, an explicit factorization of these degrees of freedom does not interact well with translational invariance. Instead, our approach will be to construct a coupling layer that acts on the full set of degrees of freedom (i.e. at the level of non-gauge-fixed configurations) that can however be guaranteed to only result in a transformation of gauge-invariant quantities.

### 4.3.1 Equivariance using transformations of loops

Section 2.2 described the construction of traced Wilson loops — a set of gauge-invariant observables in lattice gauge theory — and noted that functions of traced Wilson loops form a basis for gauge-invariant observables [40]. This suggests that coupling layers restricted to act on only these quantities would be both gauge equivariant and expressive. However, traced Wilson loops with different geometries are highly interrelated and are not in one-to-one correspondence with the set of underlying gauge links. It is far from obvious whether and how one can implement invertible transformations acting directly on these quantities. Instead, our approach to constructing gauge-equivariant coupling layers is based on working with matrix-valued *untraced* Wilson loops, which contain strictly more information than traced Wilson loops, but can be more easily manipulated to enforce gauge equivariance as follows.

We first introduce terminology to more clearly distinguish between these representations of Wilson loops in this chapter:

- We define **open loops** $W_\ell(x) \in G$ to be untraced Wilson loops, given as in Sec. 2.2 by

$$W_\ell(x) \equiv \prod_{dx, \mu \in \ell} U_\mu(x + dx), \qquad (4.5)$$

  where $\ell$ defines a sequence of offsets $dx$ and spacetime directions $\mu$ that form a loop starting and ending at a base site $x$.

- We define **closed loops** $\operatorname{tr} W_\ell(x) \in \mathbb{C}$ to be the trace over an open loop.

The distinction between these loops is depicted in Figure 4.2. By the structure of the gauge transformation given in Eq. (4.1), open loops transform by matrix conjugation as

$$W_\ell(x) \to \Omega(x) W_\ell(x) \Omega^\dagger(x), \qquad (4.6)$$

and closed loops are gauge invariant.[2]

Using this terminology, the advantage of working with matrix-valued open loops, rather than closed loops, is that it allows a change of variables back and forth between subsets of gauge links and subsets of possible open loops. We stress that it is *not* straightforward to fully change variables to open loops once and for all, because there are again non-trivial relations between the possible open loops one can construct. Instead, we will construct coupling coupling layers that specify transformations on *subsets* of open loops that each form an independent set and we will then utilize composition of coupling layers over different subsets to produce expressive flows. Gauge equivariance can be enforced by ensuring that transformations of matrix-valued open loops are equivariant under Eq. (4.6). Transformations of open loops can also be made to act locally by acting on loops with local extent and in a translationally equivariant way by applying equivalent transformations to open loops everywhere on the lattice. This approach therefore circumvents the difficulties associated with gauge fixing discussed above.

---

[2]For Abelian gauge groups there is no distinction between open and closed loops.

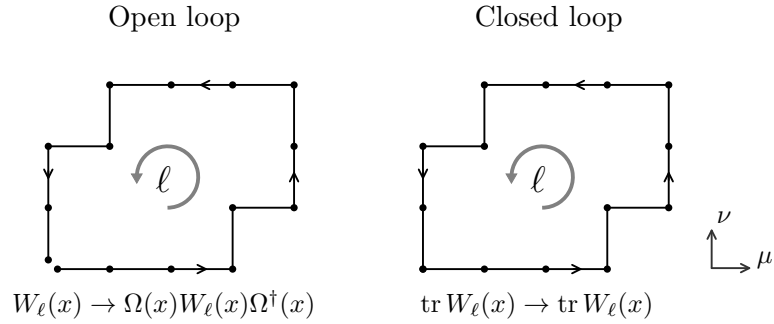

**Figure 4.2:** Open loops (left) versus closed loops (right). Open loops are matrix-valued elements that transform by conjugation under gauge transformations, while closed loops are traced and gauge invariant.

At a high level, we use a nested structure to define the action of coupling layers. Each outer coupling layer is structured to:

1. Compute a set of open loops associated with each site of the lattice;

2. Apply an inner coupling layer to update a subset of the open loops, where the elementwise operation applied by the inner coupling layer is equivariant under Eq. (4.6); and

3. Solve for a corresponding transformation at the level of link variables that enacts the desired transformation of the open loops.

Open loops and links are not in one-to-one correspondence, meaning that care must be taken to propose a transformation of open loops that does not overconstrain the link-level representation. In addition, the transformation at the link level will necessarily affect other loops not included in the updated subset of open loops in step 2. We say the loops that are updated directly by the inner coupling layer are 'actively updated', while loops that transform as a by-product of solving for and updating the corresponding links are 'passively updated'. To ensure invertibility of the coupling layer, the frozen loops used in the definition of step 2 must not be passively updated; i.e. frozen loops must remain unchanged by the coupling layer even after solving for and applying the link-level transformation.

The outer coupling layer acting at the link level and inner coupling layer acting on open loops (and satisfying appropriate matrix-conjugation equivariance) can be constructed as follows:

**Outer coupling layer** — To define the outer coupling layer, links of the gauge configuration are divided into updated and frozen subsets using a masking pattern as usual. The individual components of the matrix-valued links are not distinguished by the masking pattern; instead, the matrix-valued links are each treated as single components of the gauge configuration for the purposes of this division into subsets. For gauge fields, masking patterns applied to links can thus be defined by

$$m_\mu(x) = \begin{cases} 0 & \text{indicating } U_\mu(x) \text{ will be updated} \\ 1 & \text{indicating } U_\mu(x) \text{ will be frozen and possibly conditioned upon.} \end{cases} \quad (4.7)$$

Much like in the case of real-valued field variables, the product $mU$ defined by the components $[mU]_\mu(x) = m_\mu(x)U_\mu(x)$ can be used to derive inputs for context functions in the inner coupling layer.

All links are then used to compute a set of open loops defined by a collection of $k$ geometries $\ell_1, \ldots, \ell_k$, giving the fields $W_{\ell_1}(x), \ldots, W_{\ell_k}(x)$. These loop variables are partitioned into three sets — components that will be actively updated, components that will be passively updated, and components that will be frozen and conditioned upon. We specify these sets by defining a pair of masking patterns

$$\hat{m}_i^{\bar{A}}(x) = \begin{cases} 0 & \text{indicating } W_{\ell_i}(x) \text{ will be actively updated} \\ 1 & \text{otherwise} \end{cases} \quad (4.8)$$

and

$$\hat{m}_i^F(x) = \begin{cases} 0 & \text{indicating } W_{\ell_i}(x) \text{ will be updated (actively or passively)} \\ 1 & \text{indicating } W_{\ell_i}(x) \text{ will be frozen} \end{cases} \quad (4.9)$$

to be applied to the loops. By convention the parities here are selected to closely mirror Eq. (4.7), in that updated quantities are associated with a value of 0. The label $\bar{A}$ in Eq. (4.8) indicates that a non-zero value corresponds to loops that are not actively updated and the label $F$ in Eq. (4.9) similarly indicates that a non-zero value corresponds to loops that are frozen. Together, products of the masking patterns or their complements can be used to select each of the three possible subsets. We use the product notation $\hat{m}W$ defined by $[\hat{m}W]_i(x) \equiv \hat{m}_i(x)W_{\ell_i}(x)$ to concisely specify the masked versions of the collection of open loop variables analogously to the masked links.

These subsets indicate how the inner coupling layer should transform the open loops, and they must be defined in a way that is consistent with the masking pattern applied to links. We enforce two consistency conditions: (1) each open loop in the active subset must have an updated link as the first link in the loop, while the remaining links must be frozen, and (2) each open loop in the frozen subset must contain no updated links. These consistency conditions are sufficient to allow one to easily compute the update to links after open loops have been transformed and to ensure that the coupling layer is manifestly invertible. In particular, condition (1) allows each actively updated open loop $W_{\ell_i}(x)$ to be written as

$$W_{\ell_i}(x) = U_\mu(x) V_{\ell_i}(x) \tag{4.10}$$

in terms of a link $U_\mu(x)$ marked to be updated and the product of frozen links labeled $V_{\ell_i}(x)$ that complete the loop, as shown in Figure 4.3. The inner coupling layer will define a transformation to a new value $W'_{\ell_i}(x)$ for this loop, as described below. This can then be implemented as a transformation of the link marked to be updated to a new value $U'_\mu(x)$ that can be computed in terms of known quantities as

$$W'_{\ell_i}(x) = U'_\mu(x) V_{\ell_i}(x) \quad \implies \quad \begin{aligned} U'_\mu(x) &= W'_{\ell_i}(x) V^\dagger_{\ell_i}(x) \\ U'_\mu(x) &= W'_{\ell_i}(x) W^\dagger_{\ell_i}(x) U_\mu(x). \end{aligned} \tag{4.11}$$

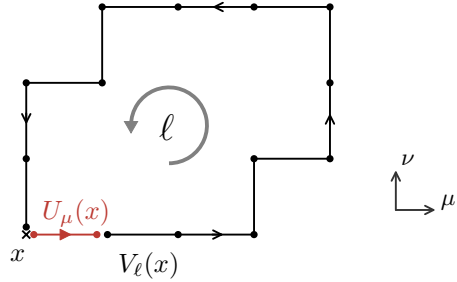

**Figure 4.3:** Decomposition of the open loop $W_\ell(x)$ into the leading link $U_\mu(x)$ and the product $V_\ell(x)$ of remaining frozen links. The inclusion of only one updated link in each actively transformed loop allows the link $U_\mu(x)$ to be uniquely transformed as a function of the transformation $W_\ell(x) \to W'_\ell(x)$ as discussed in the main text.

Each updated link is in one-to-one correspondence with an actively updated loop by condition (1), meaning each updated link can be uniquely updated as in Eq. (4.11). Condition (2) ensures that all loops marked frozen are indeed unaffected by transforming links in this way. Then in the final configuration of links produced, each open loop marked 'active' will be transformed to the proposed value from the inner coupling layer, each open loop marked 'frozen' will be unaffected, and open loops marked 'passive' may be updated as a result of the transformation of links.

The consistency conditions given here are stronger than necessary. For example, one could permit multiple updated links to appear in active plaquettes as long as the transformation of links could still be generically solved in terms of transformations of loops. The transformations of links would then require a more general solution than the one given in Eq. (4.11). We demonstrate in this chapter that the stronger consistency conditions and simpler transformations of links in Eq. (4.11) are sufficient to produce flow-based models that effectively reproduce lattice gauge theory distributions, but this generalization may be a means of increasing expressivity of gauge-equivariant coupling layers in future work.

Finally, the Jacobian determinant of the coupling layer can be computed based on the transformation of each updated link given in Eq. (4.11). The left-invariance of the Haar measure implies that

$$d(U_\mu(x)) = d(U_\mu(x)V_{\ell_i}(x)) = d(W_{\ell_i}(x)), \tag{4.12}$$

thus the change in density of $U_\mu(x)$ is equivalent to the change in density of $W_{\ell_i}(x)$. The Jacobian factors needed are then given by the Jacobian of each actively transformed open loop in the inner coupling layer, and the total Jacobian determinant is given by the product of these factors.

**Inner coupling layer** — The inner coupling layer receives a set of open loops $W_{\ell_1}(x), \ldots, W_{\ell_k}(x)$ as well as the masking patterns $\hat{m}^A$ and $\hat{m}^F$ from the outer coupling layer. The inner coupling layer then functions much like the standard coupling layers described for scalar field theory in Chapter 3: frozen loops are used to derive inputs for context functions that parameterize an invertible transformation of the actively updated loops. In contrast to the coupling layers for scalar field theory, however, we enforce several additional properties to guarantee gauge equivariance and invertibility of the inner coupling layer.

First, the loops marked as passively updated must be ignored entirely by the coupling layer. These loops will be updated as a 'side effect' of the active transformations, so specifying an update for these loops would overconstrain the transformation at the level of link variables. They also cannot be used as inputs to context functions, because these inputs must be unaffected by the coupling layer as a whole to preserve manifest invertibility.[3]

Secondly, the inputs to the context functions must be gauge invariant. This can be achieved by providing only the traces of loops marked as frozen (i.e. only the closed version of frozen loops) to context functions. This ensures that the context functions parameterizing the invertible elementwise transformation of the actively updated open loops are themselves gauge invariant, which simplifies the construction of gauge-equivariant elementwise transformations.

Finally, the invertible elementwise map applied to the actively updated open loops must be equivariant under the transformation of open loops according to Eq. (4.6). This property combined with gauge invariance of the outputs of context functions makes the entire coupling layer gauge invariant, as we prove in Sec. 4.3.2 below. We term this invertible elementwise map the *kernel* of the gauge-equivariant transformation and use the notation $h : G^{N_\ell} \to G^{N_\ell}$ for such kernels, where $N_\ell$ is the number of open loops acted on by the inner coupling layer and is proportional to both the number of lattice sites and number of loop geometries constructed.[4] We further use the notation $h_i(\cdot)$ for the action of the kernel on the $i$th component of the input, where $i \in \{1, \ldots, N_\ell\}$. The kernel $h$ factors into the independent application of these $h_i$ to their respective input components. In terms of the factors $h_i$, a kernel is equivariant to a transformation of its argument according to Eq. (4.6) if and only if each factor is *conjugation equivariant*, as defined by

$$h_i(XWX^\dagger) = Xh_i(W)X^\dagger, \quad \forall X, W \in G. \tag{4.13}$$

---

[3]It is possible that a coupling layer using passively updated loops in context functions may still be invertible in principle, but it is not clear how one would verify/enforce this in practice.

The construction of conjugation-equivariant and invertible kernels for U(1) and SU($N$) variables is described in Sections 4.4 and 4.5.

In terms of a conjugation-equivariant kernel $h$, context functions $s_1, s_2, \ldots$ and masking patterns $\hat{m}^{\bar{A}}$ and $\hat{m}^F$, the action of the inner coupling layer can be generically written as

$$g^{\text{inner}}(W) = \hat{m}^{\bar{A}}W + (1 - \hat{m}^{\bar{A}}) \left[ h(W; \; s_1(\text{tr}[\hat{m}^F W]), s_2(\text{tr}[\hat{m}^F W]), \ldots) \right], \tag{4.14}$$

where the parameterization of the kernel in terms of the outputs of context functions has been made explicit, and $\text{tr}[\hat{m}^F W]$ implies a trace over all loop variables in $\hat{m}^F W$. The Jacobian determinant associated with this transformation of loops reduces to the product of the change of measure associated with each factor $h_i$. We discuss the calculation of these factors for the kernels presented for U(1) and SU($N$) variables below.

### 4.3.2 Proofs of equivariance and invertibility

We next prove that the class of coupling layers constructed indeed satisfies both gauge equivariance and invertibility.

To prove gauge equivariance of this coupling layer architecture, we demonstrate that a generic gauge transformation commutes through application of the coupling layer. Any gauge transformation affects loops by matrix conjugation, as given in Eq. (4.6). We use the notation $W^\Omega$ to specify a field of gauge-transformed open loops, defined for each loop geometry $\ell_i$ given to the inner coupling layer by

$$W_{\ell_i}^\Omega(x) \equiv \Omega(x)W_{\ell_i}(x)\Omega^\dagger(x). \tag{4.15}$$

A similar superscript notation is used for gauge-transformed open loops produced as outputs in the following derivation. Applying the generic form of the inner coupling

---

[4]This notation slightly differs from the notation used in Pub. [3].

layer given in Eq. (4.14) to the gauge transformed open loops gives

$$
\begin{aligned}
g^{\text{inner}}(W^\Omega) &= \hat{m}^{\bar{A}} W^\Omega + (1 - \hat{m}^{\bar{A}}) \left[ h(W^\Omega; \, s_1(\text{tr}[\hat{m}^F W^\Omega]), s_2(\text{tr}[\hat{m}^F W^\Omega]), \dots) \right] \\
&= \hat{m}^{\bar{A}} W^\Omega + (1 - \hat{m}^{\bar{A}}) \left[ h(W^\Omega; \, s_1(\text{tr}[\hat{m}^F W]), s_2(\text{tr}[\hat{m}^F W]), \dots) \right] \\
&= \hat{m}^{\bar{A}} W^\Omega + (1 - \hat{m}^{\bar{A}}) \left[ h(W; \, s_1(\text{tr}[\hat{m}^F W]), s_2(\text{tr}[\hat{m}^F W]), \dots) \right]^\Omega \\
&= \left\{ \hat{m}^{\bar{A}} W + (1 - \hat{m}^{\bar{A}}) \left[ h(W; \, s_1(\text{tr}[\hat{m}^F W]), s_2(\text{tr}[\hat{m}^F W]), \dots) \right] \right\}^\Omega \\
&= [g^{\text{inner}}(W)]^\Omega.
\end{aligned}
\tag{4.16}
$$

To arrive at the second line, we use gauge invariance of the closed loops passed to context functions; the third line uses conjugation equivariance of the kernel. This demonstrates that the inner coupling layer is equivariant under gauge transformations of the open loops given as inputs, and therefore the open loops produced as outputs of the coupling layer transform properly under gauge transformations. We can then demonstrate equivariance of the full coupling layer in two parts. First, gauge transformations of any frozen links necessarily commute through application of the coupling layer. Second, gauge transformations of updated links are given by the transformation of Eq. (4.11):

$$
\begin{aligned}
U'_\mu(x) &= W'_{\ell_i}(x) W^\dagger_{\ell_i}(x) U_\mu(x) \\
&\to [\Omega(x) W'_{\ell_i}(x) \Omega^\dagger(x)][\Omega(x) W^\dagger_{\ell_i}(x) \Omega^\dagger(x)][\Omega(x) U_\mu(x) \Omega^\dagger(x + \hat{\mu})] \\
&= \Omega(x) U'_\mu(x) \Omega^\dagger(x + \hat{\mu}),
\end{aligned}
\tag{4.17}
$$

where the transformation properties of $W'$ produced by the inner flow were used in the first step. Thus both frozen and updated links transform properly when a gauge transformation is applied to the input links, and the coupling layer constructed above is gauge equivariant.

To prove invertibility of the coupling layer, we explicitly construct the inverse coupling layer. First, we note that the consistency conditions applied to the masking patterns ensure that none of the links contained in frozen loops are modified by the coupling layer. This implies that the frozen loops computed from the output of the coupling layer will be identical to the frozen loops computed from the input, $\hat{m}^F W' = \hat{m}^F W$. In the inner coupling layer, the results of evaluating the context functions $s_1, s_2, \dots$ are thus same when given either $W$ or $W'$ as input. This allows us to write the inverse inner coupling layer as

$$
\begin{aligned}
\left[ g^{\text{inner}} \right]^{-1}(W') &= \hat{m}^{\bar{A}} W' + (1 - \hat{m}^{\bar{A}}) \left[ h^{-1}(W'; s_1(\hat{m}^F W'), s_2(\hat{m}^F W'), \dots) \right] \\
&= \hat{m}^{\bar{A}} W' + (1 - \hat{m}^{\bar{A}}) W,
\end{aligned}
\tag{4.18}
$$

where we use the fact that the kernel $h(\,\cdot\,; s_1(\hat{m}^F W), s_2(\hat{m}^F W), \dots)$ is defined to be an invertible function. The inverse inner coupling layer satisfies $[g^{\text{inner}}]^{-1}(g^{\text{inner}}(W)) = W$. The original open loops contained in the actively updated subset can then be recovered

by taking

$$(1 - \hat{m}^{\bar{A}})[g^{\text{inner}}]^{-1}(W') = (1 - \hat{m}^{\bar{A}})W, \tag{4.19}$$

because $1 - \hat{m}^{\bar{A}}$ projects onto the rightmost term in Eq. (4.18). Finally, the original link variables can be computed by

$$U_\mu(x) = W_{\ell_i}(x)W'^{\dagger}_{\ell_i}(x)U'_\mu(x), \tag{4.20}$$

where $W_{\ell_i}(x)$ is in the actively updated subset by the consistency conditions on masking patterns.

### 4.3.3 A specific masking pattern

Many choices of loops and masking patterns give rise to valid gauge-equivariant coupling layers. In general, the requirement that actively updated loops contain as a prefix exactly one link in the updated subset imposes a degree of sparsity to the masking patterns that may be used. As described for scalar field theory in Chapter 3, expressive coupling layers can be achieved by using masking patterns that provide enough frozen components to leverage expressivity of the context functions while updating enough of the remaining components to make non-trivial transformations overall. Aiming to strike this balance, we introduce a 'striped' masking pattern $m_\mu^{\text{stripe}}(x)$ defined by

$$m_\mu^{\text{stripe}}(x) = (1 - \delta_{\mu,\nu})(1 - \delta_{x_\rho \bmod 4, dx}), \tag{4.21}$$

in terms of some direction $\nu$ with a stride of four sites in an orthogonal $\rho$ direction starting from an offset of $dx$. This masking pattern is depicted for a two-dimensional lattice in Figure 4.4. In higher dimensions, the masking pattern defined in Eq. (4.21) has similar structure to what is shown in the figure for two-dimensional slices and repeats in the orthogonal directions. Also shown in the figure are examples of possible closed loops that can be included in the frozen subset for use as inputs to context functions and possible open loops that can be included in the actively updated subset in the inner coupling layer. If the marked open loops are not included in the actively updated subset, they will be passively updated as a 'side effect' of this striped masking pattern.

The choice of a stride of four sites is motivated by the applications to two-dimensional lattice gauge theory in this chapter. When working specifically with two-dimensional lattices, using a smaller stride significantly restricts the possible frozen loops that can be passed to context functions. In higher spacetime dimensions there are several variations of this masking pattern that may be interesting to explore — the stride could be reduced because there are significantly more loops involving orthogonal direction that can be included in the frozen subsets, different strides could be used in the various orthogonal directions, or a combination of links in orthogonal directions could be included in the updated subsets. Exploring which masking patterns and loops result in expressive coupling layers for higher dimensional lattice gauge theory is an interesting subject for future empirical studies.

In analogy with the checkerboard masking pattern used for scalar field theory,

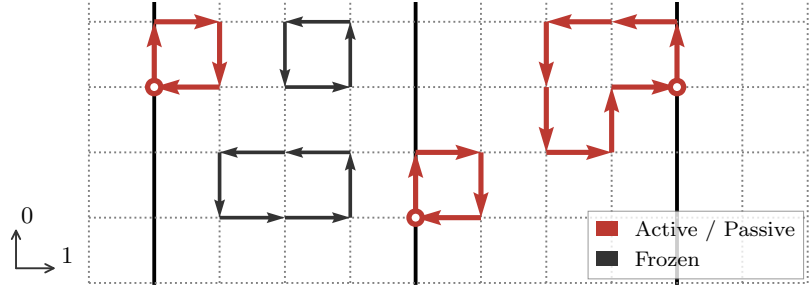

**Figure 4.4:** An example of the striped masking pattern oriented with $\nu = 0$ and $\rho = 1$, using an offset of $dx = 1$ and a stride of four lattice sites. Bold links are marked to be updated by the coupling layer, corresponding to $m_\mu(x) = 0$, while dotted links are marked frozen, corresponding to $m_\mu(x) = 1$. Examples of frozen loops that may be used as inputs for context functions are shown using thin gray arrows, while some examples of loops that will be updated either actively or passively are shown using thick red arrows. Hollow circles indicate the site at which open (i.e. untraced) loops start and end. The loops indicated in red all satisfy the consistency conditions that (1) the initial link of the loop is an updated link and (2) all remaining links are frozen links. Each of these loops are thus in principle suitable to be included in the actively updated subset. A valid masking pattern must also associate exactly one actively updated loop with each updated link, as discussed in the main text. Figure 4.5 shows a concrete choice of loops that satisfy these conditions.

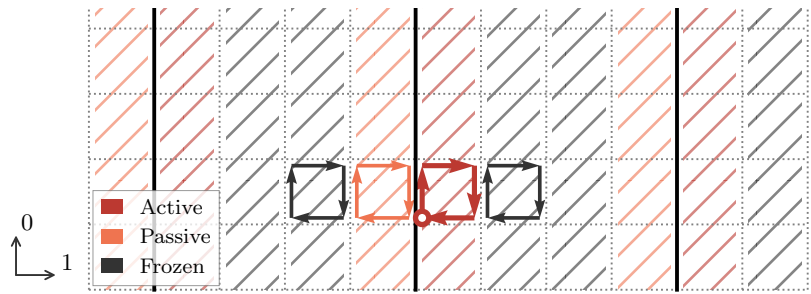

**Figure 4.5:** One choice of loops and loop-level masking patterns that can be associated with the striped link masking pattern shown in Fig. 4.4. This pattern restricts to $1 \times 1$ plaquettes as the actively transformed, passively transformed, and frozen loops. The pattern of active, passive, and frozen plaquettes repeats by tiling as indicated by the hatched colors. This masking pattern is labeled Passive-Active-Frozen-Frozen (PAFF) in the main text, reflecting the structure of the loop-level masking pattern. Figure adapted from Fig. 3 of Pub. [2].

updating local quantities based on nearby physical information is a useful way to incrementally build the correlation structure associated with distributions for lattice field theory. Following this idea, a simple choice of loops that can be used in the inner flow is depicted in Figure 4.5. This set of loops consists exclusively of $1 \times 1$ plaquettes located at all possible sites on the lattice. The pattern can thus be constructed and passed to the inner coupling layer using just a single loop geometry $\ell_1$ describing the $1 \times 1$ plaquette structure. In our applications to two-dimensional lattice gauge theory, for example, the loop geometry is explicitly given by $\ell_1 = [(\hat{0}, (0,0)), (\hat{1}, (1,0)), (-\hat{0}, (1,1)), (-\hat{1}, (0,1))]$. We use the term Passive-Active-Frozen-Frozen (PAFF) to label this structure of link-level masking pattern, choice of loops, and loop-level masking patterns. The PAFF masking structure has free parameters $\rho$, $\nu$, and $dx$ describing the link-level mask which fully constrains the loop-level masks. To update every link on the lattice, PAFF coupling layers using each possible choice of $\rho$, $\nu$, and $dx$ can be composed sequentially. For example, in two dimensions, PAFF coupling layers defined by each of the eight combinations of $\nu \in \{0, 1\}$, $\rho = 1 - \nu$, and $dx \in \{0, 1, 2, 3\}$ are sufficient to update every link once, and multiple 'stacks' of eight such coupling layers can be further composed to increase the expressivity of the model.

### 4.3.4  Context functions

Context functions for gauge-equivariant coupling layers can be implemented using convolutional neural networks analogously to the case of context functions used for scalar field theory, up to one minor caveat: closed loops for most choices of gauge group are complex-valued. A simple approach to handling complex-valued inputs is to use both the real and imaginary components as independent channels given as input to the convolutional neural networks defining the context functions. A notable exception is traced Wilson loops in SU(2) gauge theory, which are purely real by the eigenvalue structure of any SU(2) matrix; in this case, simply restricting to the real component is sufficient.

### 4.3.5  Universality and expressivity

Section 3.2.5 discussed some mixed results for the universality of general flow-based models employing coupling layers. With the further restrictions imposed by requiring gauge equivariance one might expect that universality is even less likely for the coupling layers introduced in this chapter. For one, it is clear that (by construction) distributions that are not gauge invariant cannot be accessed. This loss of expressivity is of course intentional, and it allows flow-based models to practically and efficiently be trained in the proof-of-principle examples studied below. However, a relevant further question is whether these coupling layers give rise to flow-based models that are asymptotically universal within the space of distributions satisfying the symmetries that have been enforced. The question of universality of neural networks satisfying symmetries has just begun to be understood. For example, the criteria necessary to make invariant neural networks universal have been detailed in Ref. [394], but universality for flow-based models constructed using equivariant flows has not yet been investigated.

Interestingly, we can however make a precise statement by relating gauge-equivariant flows to the trivializing map (see Sec. 3.1.3). For lattice gauge theories with a pure-gauge action, the trivializing map reduces to a differential equation describing the update of links in terms of conjugation-equivariant functions applied to loops containing the links [341]. In the limit of arbitrarily many coupling layers, an integrator for this differential equation can be arbitrarily precisely captured by the coupling layers described in this chapter. Thus these coupling layers have the expressivity to asymptotically produce at least one map between the prior distribution and target distributions for pure-gauge lattice gauge theory, in the limit of arbitrarily complex context functions and an arbitrary number of layers.

### 4.3.6 Related approaches

Gauge equivariance has previously been studied in the context of convolutions on non-Euclidean manifolds [360, 361, 395–397]. In this setting, the structure of the manifold is treated as a fixed 'background gauge' on which one aims to apply feed-forward networks that are compatible with the geometry. In contrast, the approaches presented in this chapter are focused on dynamical gauge fields and the setting of invertible flow-based models. The structures used to enforce equivariance in this context are thus quite distinct, though the application of equivariance is inspired by these earlier works.

Non-invertible gauge-equivariant and gauge-invariant feed-forward networks for lattice gauge theory [254, 398–400] have also been explored in parallel to the equivariant architecture presented here. In particular, Ref. [399] extends the methods presented here to the setting of $\mathbb{Z}_2$ lattice gauge theory, Ref. [254] develops an autoregressive description of quantum states based on gauge-invariant conditional probabilities constructed by projecting to gauge-invariant subspaces of the Hilbert space, and Refs. [398, 400] employ an approach based on composition of covariant paths and loops consisting of linear combinations of links between fixed pairs of endpoints. These latter works generalize the notion of working with Wilson loops transforming by matrix conjugation to objects like Wilson lines that transform more generally. The additional complexity in this approach is managed by restricting to structures transforming according to a subset of all possible representations. These works may provide an interesting avenue to generalizing the coupling layers presented here; for example, context functions could conceivably be implemented as functions of gauge-covariant quantities, rather than gauge-invariant quantities, using covariant neural networks that only project to gauge-invariant outputs at the final stages. Gauge-invariant inputs are in principle sufficient to describe arbitrary gauge-invariant functions [40], but in practice the expressivity might be increased by these alternate architectures.

## 4.4 Invertible functions of U(1) variables

We next consider the construction of kernels for specific gauge groups. To construct kernels for U(1) lattice gauge theory, invertible and conjugation-equivariant maps over U(1) are required. Given that the gauge group U(1) is Abelian, conjugation equivariance reduces to a trivial requirement which is satisfied for any function of U(1) variables,

$$h(XWX^\dagger) = Xh(W)X^\dagger$$
$$h(WXX^\dagger) = h(W)XX^\dagger \qquad (4.22)$$
$$h(W) = h(W),$$

for any $W, X \in \mathrm{U}(1)$. One can thus arbitrarily choose parameterized invertible maps over $\mathrm{U}(1)$ as kernels, where any parameters describing the maps will be given in gauge-equivariant coupling layers by the outputs of context functions. In this section we review the results of Pub. [8] giving invertible maps applicable to the $\mathrm{U}(1)$ manifold.

We will use the coordinate $\phi \in [0, 2\pi)$ to describe the $\mathrm{U}(1)$ manifold. This coordinate can be related to the unit-norm group values by $U = e^{i\phi}$, where $\phi = 0$ is therefore identified with $\phi = 2\pi$. The Haar measure for $\mathrm{U}(1)$ is given by $\frac{d\phi}{2\pi}$ and the Jacobian of each invertible map defined below should be computed with respect to this measure to give the relevant factors required to compute the Jacobian of the coupling layer, as described in Sec. 4.3.1 above.

This periodic domain must be respected when defining transformations of $\phi$. For example, the affine transformation, considered for scalar field theory in Chapter 3, would not be suitable for transformations of $\phi$. Affine transformations are defined by scaling and adding an offset to the input, and while adding an offset can be performed modulo $2\pi$, the scaling operation fails to invertibly transform the periodic domain. As a simple example of this issue, we can consider scaling $\phi \to 2\phi \bmod 2\pi$; this results in a double cover of the manifold in which pairs of inputs $\phi$ and $(\phi + \pi) \bmod 2\pi$ are mapped to the same output, violating invertibility. Restricting to only applying an offset modulo $2\pi$ does give an invertible map, but it is strictly volume preserving and cannot squish/stretch the probability density. We thus combine the offset operation with other transformations of the $U(1)$ domain described below to create expressive invertible maps:

**Non-compact projections** — Complications arising from the compact/periodic nature of the $\mathrm{U}(1)$ manifold can be circumvented by using a compactification map. In particular, more complicated and non-volume-preserving maps can be constructed by (1) applying the inverse compactification map, (2) applying a non-compact invertible transformation, and (3) applying the compactification map. This idea is presented for general spherical manifolds in Ref. [401] and is specifically implemented for $S^1 \sim \mathrm{U}(1)$ in Pub. [8] in a method termed *non-compact projection (NCP)*. NCP transformations utilize the map $x : (0, 2\pi) \to \mathbb{R}$ from $\mathrm{U}(1)$ to the non-compact real line given by

$$x(\phi) = \tan\left(\frac{\phi}{2} - \frac{\pi}{2}\right). \qquad (4.23)$$

The inverse map $x^{-1} : \mathbb{R} \to (0, 2\pi)$ is given by

$$x^{-1}(y) = 2\arctan(y) + \pi, \qquad (4.24)$$

where the image of $\arctan(y)$ is defined to live in the domain $[-\pi/2, \pi/2]$.

Given any invertible transformation of $\mathbb{R}$, composing with $x$ and $x^{-1}$ results in an invertible map over all of U(1) except for the identity element. For example, the NCP method with an inner scaling transformation gives the map[5]

$$g(\phi; s) \equiv 2 \arctan\left(e^s \tan\left(\frac{\phi}{2} - \frac{\pi}{2}\right)\right). \qquad (4.25)$$

The endpoint $\phi = 0$ is a singular point of the compactification map and the result of the composition is therefore ill-defined exactly for $U = e^{i \cdot 0} = 1$. For the scaling map, $\lim_{\phi \to 0^+} g(\phi; s) = 0$ and $\lim_{\phi \to 2\pi^-} g(\phi; s) = 2\pi$, allowing the value at the endpoints to be defined by $g(0; s) = 0$ and $g(2\pi; s) = 2\pi$. This is consistent with periodicity and gives rise to a smooth function. Very near these endpoints, the map may be susceptible to numerical instability from the nearly singular behavior, but in the study of U(1) lattice gauge theory undertaken in Sec. 4.6, we find that double precision numbers are sufficiently precise to avoid this issue in practice. Pub. [8] describes a procedure to define the map near this singular point using a linearization of the transformation in case numerical instabilities should arise.

This NCP map has a non-trivial Jacobian and the change in density can be explicitly computed to be

$$[J_g(\phi)]^{-1} = \left|\frac{\partial g}{\partial \phi}\right|^{-1} = e^{-s} \sin^2(\phi/2) + e^s \cos^2(\phi/2). \qquad (4.26)$$

When used as a kernel for a gauge-equivariant coupling layer, the scaling factor $s$ is given as the output of a context function and the parameters of that context function give control over the change in density applied as a result of this transformation.

The single scaling operation in the NCP method gives a very restricted change of density. Modifying the inner invertible map applied within the non-compact space can potentially increase the expressivity of this transformation. However, transformations applied within the non-compact space may be severely modified when mapped back onto the circle; in particular, features such as peaks or valleys of the density are strongly concentrated near the singular point $U = 1$, while they are relatively less modified far from this point, making it difficult to design transformations that construct a change in density with similar fluctuations over the entire domain.

One can, however, somewhat increase the expressivity of NCP transformations by directly mixing the action of the transformations on the circle. This is made possible by monotonicity of the transformation within the finite angular domain. The resulting transformation and Jacobian are then both given by simply averaging the individual NCP transformations in the mixture. Though the inverse transformation is no longer analytically tractable in this case, the function can in practice be inverted using numerical methods, if needed.

**Circular splines** — Circular splines allow a direct transformation on the circular domain, circumventing the issue described above. Piecewise-polynomial transforma-tions have been applied in several related approaches to invertibly transform real-valued variables, including the use of rational quadratic splines [402] and cubic splines [403] for non-compact domains, and linear and quadratic piecewise-polynomial transformations for finite domains [404]. Here we review the application of rational quadratic splines to the circular domain presented in Pub. [8].

---

[5]This map is equivalent to the map defined in Pub. [8] with $\beta$ fixed to zero.

An invertible rational quadratic spline[6] for a non-compact domain is defined by $K + 1$ 'knots' corresponding to the points $(x_i, y_i)$, for $i \in \{0, \ldots, K\}$, and $K - 1$ free 'slopes' $s_i$ associated with the internal knots with indices $i \in \{1, \ldots, K-1\}$. Slopes at the endpoints are fixed to $s_0 = s_K = 1$. Calling the rational quadratic spline $R(\cdot)$, each knot $(x_i, y_i)$ dictates that $R(x_i) = y_i$, and each slope $s_i$ dictates that $dR(x_i)/dx = s_i$. In each interval $[x_i, x_{i+1}]$, the function is defined by a rational quadratic function, which takes the general form

$$R(x) = \frac{a_i x^2 + b_i x + c_i}{a_i' x^2 + b_i' x + c_i'}. \tag{4.27}$$

The coefficients $a_i, b_i, c_i, a_i', b_i', c_i'$ are fixed to ensure $R(\cdot)$ intersects the knots with the appropriate slopes. To guarantee invertibility of the rational quadratic spline, knots are required to be monotonic and the slopes are required to be positive. The Jacobian of the transformation is given by differentiating the piecewise function defined by Eq. (4.27). Further details on invertible rational quadratic splines can be found in Ref. [403].

In a circular spline, this approach is applied to the finite domain $[0, 2\pi]$. To satisfy periodicity, the first and last knot are fixed to $(x_0, y_0) = (0, 0)$ and $(x_K, y_K) = (2\pi, 2\pi)$. The condition on slopes at the endpoints for invertible rational quadratic splines applied to the non-compact domain can be relaxed for circular splines. Instead, the weaker condition that $s_0 = s_K$ is sufficient to enforce smoothness at the identified endpoints of the domain. The circular spline can thus be defined in terms of $3K$ positive parameters:

1. $K$ slopes $s_1, \ldots, s_K$

2. $K$ bin widths, which can be freely specified and normalized to total $2\pi$ through the use of the softmax function

3. $K$ bin heights, which can similarly be freely specified and later normalized

The bin widths and heights respectively define the horizontal and vertical distances between knots, giving a convenient definition of the position of knots without needing to explicitly enforce monotonicity. To apply circular splines as kernels for U(1) gauge-equivariant coupling layers, these $3K$ parameters are specified as the outputs of context functions.

## 4.5   Invertible functions of $\mathrm{SU}(N)$ variables

We next construct invertible transformations that may be used as kernels for gauge-equivariant coupling layers in $\mathrm{SU}(N)$ gauge theory. Unlike in the case of U(1) variables, conjugation equivariance is a non-trivial constraint on kernels for $\mathrm{SU}(N)$ variables. Kernels for $\mathrm{SU}(N)$ must also act as a bijection on the non-trivial manifold of $\mathrm{SU}(N)$ variables. We address both of these issues here.

---

[6]The term 'spline flow' is applied in the literature. In this chapter we avoid using the term 'flow' for invertible elementwise transformations applied within coupling layers to distinguish these transformations from the overall flow produced by composing coupling layers.

The building blocks of a kernel $h$ are the factors $h_i(\cdot)$ acting elementwise on each component of the input to the kernel. Recall that conjugation equivariance for a single factor is defined by

$$h_i(XWX^\dagger) = Xh_i(W)X^\dagger, \quad \forall X, W \in G, \tag{4.13}$$

where in this case $G = \mathrm{SU}(N)$. We construct *spectral maps* satisfying Eq. (4.13) through a sequence of transformations below. To provide some orientation before embarking on the journey, Figure 4.6 presents a high level overview of the steps defining a spectral map. As depicted in the figure, the overall strategy is to steadily simplify the space on which the transformation must act, ultimately arriving at the $(N-1)$-dimensional open box $\Omega = (0,1)^{N-1}$ which can be acted on by the invertible transformations considered previously. All components shown in the figure are motivated and precisely defined in Secs. 4.5.1 and 4.5.2. The construction is summarized and the Jacobian of the overall transformation is defined in Sec. 4.5.3. Finally, we present a numerical investigation of the performance of these spectral maps for single-variable distributions in Sec. 4.5.4.

### 4.5.1 Conjugation equivariance and spectral maps

For any group $G$, conjugation equivariance can be interpreted as a statement about how the kernel transforms group elements within and across *conjugacy classes* of $G$. Conjugacy classes of $G$ are defined as all possible sets $\{XWX^{-1} : X \in G\}$ that can be constructed for $W \in G$. We work with unitary representations of the gauge group throughout this dissertation, and the inverse is thus equivalent to the adjoint, $X^{-1} = X^\dagger$. Conjugation equivariance implies that specifying the action of the kernel on a single element of each conjugacy class is both necessary and sufficient to specify its action on all elements in the group.

For the group $G = \mathrm{SU}(N)$, each conjugacy class can be uniquely defined by the *spectrum* — the unordered set of eigenvalues — shared by all elements in the class. This follows from the fact that all elements of these groups are unitarily diagonalizable: each element is equivalent under conjugation by elements within the group to all diagonal matrices with the same spectrum, and by transitivity each element is equivalent under conjugation to all other group elements with the same spectrum. As a concrete example, the set of eigenvalues $\{e^{i3\pi/12}, e^{i5\pi/12}, e^{-i8\pi/12}\}$ uniquely defines a single conjugacy class within $\mathrm{SU}(3)$ that includes all diagonal matrices constructed from these entries,

$$\begin{pmatrix} e^{i3\pi/12} & & \\ & e^{i5\pi/12} & \\ & & e^{-i8\pi/12} \end{pmatrix}, \quad \begin{pmatrix} e^{i3\pi/12} & & \\ & e^{-i8\pi/12} & \\ & & e^{i5\pi/12} \end{pmatrix}, \quad \dots \tag{4.28}$$

as well as all non-diagonal matrices in $\mathrm{SU}(3)$ with the same spectrum.

The diagonal matrices in each conjugacy class are a natural choice of matrices on which to define the action of a conjugation-equivariant transformation on $\mathrm{SU}(N)$, since all other elements of the conjugacy class can be easily transformed to one of these representative elements by diagonalization. However, since there are multiple diagonal matrices within each conjugacy class, invertible transformations acting on diagonal matrices must give compatible results for all related diagonal matrices. Specifically, a transformation must be equivariant under permutations of the diagonal entries to be consistent with conjugation equivariance. We call a transformation that is insensitive to the order of eigenvalues in this way a *spectral map*.[7]

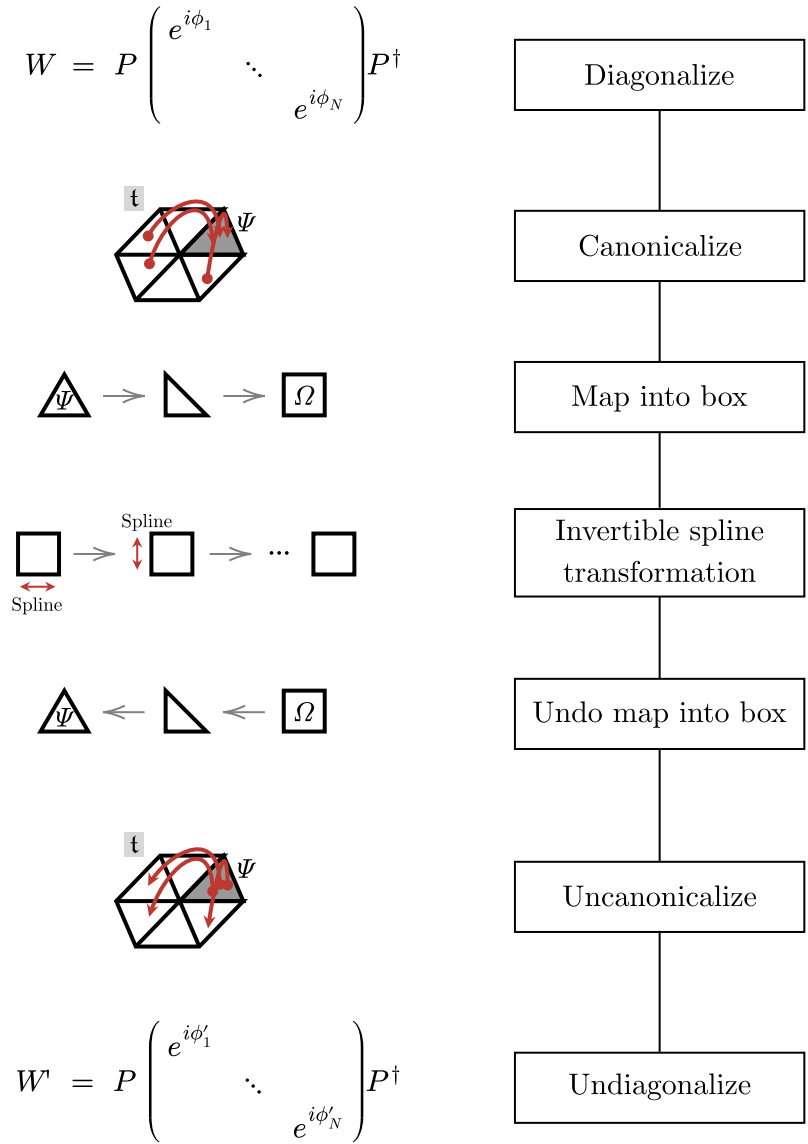

**Figure 4.6:** Overview of the sequence of steps in our construction of a conjugation-equivariant map acting on $W \in \mathrm{SU}(N)$. Section 4.5.1 explains diagonalization, canonicalization, and conjugation equivariance. Section 4.5.2 explains the space $\Psi$, mapping into the box $\Omega$, and applying invertible splines to transform points in $\Omega$. Figure adapted from Fig. 2 of Pub. [2].

The conjugation-equivariant transformation of an arbitrary element $W \in \mathrm{SU}(N)$ can then be given by the procedure:

1. **Diagonalize** $W$, giving matrices $P$ and $D = \mathrm{diag}(\lambda_1, \ldots, \lambda_N)$ satisfying $W = PDP^\dagger$.

2. **Apply a spectral map** to transform $(\lambda_1, \ldots, \lambda_N) \to (\lambda'_1, \ldots, \lambda'_N)$.

3. **'Undiagonalize'** by constructing $W' = PD'P^\dagger$, where $D' = \mathrm{diag}(\lambda'_1, \ldots, \lambda'_N)$.

The inverse of this transformation is given by an analogous procedure in which the inverse spectral map is applied in step 2. Any diagonalization algorithm may be used in step 1 as long a unitary matrix $P$ is returned. The only ambiguities in the diagonalization procedure are the global phases of each column of $P$ and the order of eigenvalues and eigenvectors, and phase rotations of columns of $P$ cancel in the undiagonalization step, while reordering the eigenvalues and columns of $P$ does not modify $W'$ by the assumed permutation equivariance of the spectral map. Though the focus of this section is on $\mathrm{SU}(N)$ variables, it is noteworthy that conjugacy classes in $\mathrm{U}(N)$ are similarly uniquely identified by the spectra of fundamental-representation matrices; the procedure above also gives a conjugation-equivariant invertible transformation of $\mathrm{U}(N)$.

Previous works in the context of point-clouds, collections of particles, and 3D objects have developed a variety of architectures that may be used to define permutation-equivariant invertible maps [251, 367, 368, 405–412]. These approaches could be applied immediately to produce spectral maps for $\mathrm{U}(N)$, for which all $N$ eigenvalues can be considered independent variables. The eigenvalues of $\mathrm{SU}(N)$ matrices, however, are constrained by the unit-determinant condition to multiply to 1. These eigenvalues thus lie in an $(N-1)$-dimensional submanifold within the space all possible unitary eigenvalues; for example, the one- and two-dimensional submanifolds for $\mathrm{SU}(2)$ and $\mathrm{SU}(3)$, respectively, are depicted in Figure 4.7. Spectral maps for $\mathrm{SU}(N)$ must act within the $(N-1)$-dimensional submanifold while being equivariant to permutations of the $N$ eigenvalues defining the overall space. This unique structure motivates the development of a specific architecture suitable for permutation equivariance of $\mathrm{SU}(N)$ eigenvalues in this section.

It is helpful in the following to distinguish between two spaces:

1. The space of all diagonal $\mathrm{SU}(N)$ matrices is defined by

$$T \equiv \{\mathrm{diag}(\lambda_1, \ldots, \lambda_N)\} \subset \mathrm{SU}(N). \tag{4.29}$$

---

[7]In Pub. [2] the term 'spectral flow' was used. Here we avoid the term 'flow' for transformations inside single coupling layers to avoid confusion with the overall flow built by composing coupling layers.

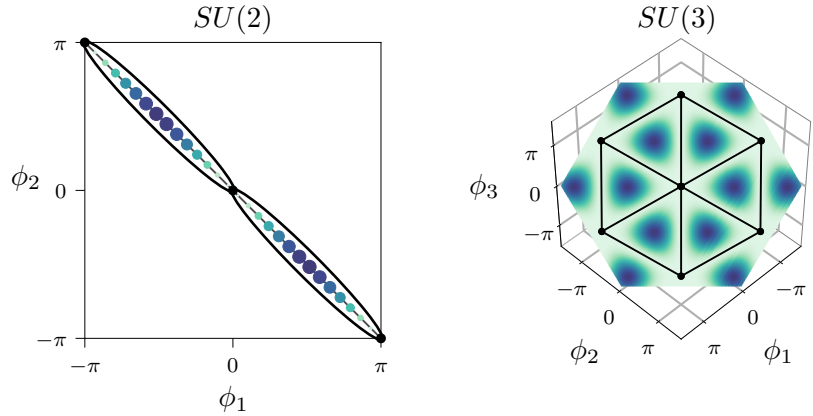

**Figure 4.7:** The submanifolds describing the possible eigenvalues of an SU(2) (left) or SU(3) (right) matrix within the space of arbitrary unitary eigenvalues. The plots are parameterized by the coordinates $\phi_1$, $\phi_2$, and (for SU(3)) $\phi_3$ indicating the complex phase of each eigenvalue $\lambda_i = e^{i\phi_i}$. Both submanifolds are restricted by the unit-determinant constraint to be subspaces of codimension one. For the group SU(3), the range of each angular variable is extended past $[-\pi, \pi]$ to depict the repeating pattern of cells in the hyperplane $\mathbf{t}$ described in the main text. The $N!$ cells depicted with black outlines in both cases are related to each other by permutations of the eigenvalues and form a covering set. Colors indicate the density of the Haar measure with respect to the natural measure on the space of eigenvalues (see Eq. (4.39)), with the dark purple color indicating the highest density; the SU(2) plot shows a representative set of points drawn with size proportional to this density to more clearly indicate its effect in the one-dimensional space. Figure adapted from Fig. 4 of Pub. [2].

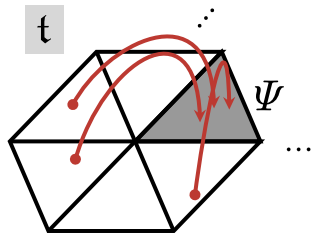

**Figure 4.8:** Schematic depiction of the simplex $\Psi$ within the hyperplane $\mathbf{t}$ for the group SU(3). The diagram shows a minimal set of six cells that cover the space of possible eigenvalues exactly once (a covering set). Ellipses indicate that the hyperplane $\mathbf{t}$ is unbounded and cells resulting in equivalent eigenvalues after exponentiation repeatedly tile the hyperplane. Arrows depict a few examples of the canonicalizing map sending points in the hyperplane $\mathbf{t}$ to the canonical simplex $\Psi$.

2. The space of complex phases of the eigenvalues can be defined by

$$\mathfrak{t} \equiv \left\{ (\phi_1, \ldots, \phi_N) : \sum_i \phi_i = 0 \right\} \subset \mathbb{R}^N, \qquad (4.30)$$

where the sum is not taken modulo $2\pi$.

The function $\mathrm{expi}(\cdot)$ gives a surjective map from $\mathfrak{t}$ onto the set of diagonal matrices $T$.[8] The preimage of any diagonal matrix under this map is an infinite lattice of points related by offsets by integer multiples of $2\pi$ that sum to zero. Rather than working with the space given in Eq. (4.30), one could instead work with angles $(\phi_1, \ldots, \phi_N)$ restricted to satisfy $\sum_i \phi_i \equiv 0 \pmod{2\pi}$ and $0 \leq \phi_i \leq 2\pi$, which can be related to the space of diagonal matrices $T$ by a one-to-one map; we choose to work with $\mathfrak{t}$ because it simplifies the construction of permutation-equivariant maps below.

As can be seen for the examples of SU(2) and SU(3) in Fig. 4.7, permutations act on $\mathfrak{t}$ by relating sets of $N!$ distinct *cells* within the submanifold. For general $N$, each cell is an $(N-1)$-simplex. The surfaces of co-dimension $k$ that bound each simplex are sets of points for which $k$ eigenvalues are degenerate, corresponding to $k$ phases being equivalent modulo $2\pi$. Each set of $N!$ cells that is related by permutations covers the space of diagonal matrices $T$ exactly once. We label such a set of cells a *covering set*. The figure shows a covering set of cells for the group SU(2) consisting of the two circled line segments and a covering set of six triangular cells for the group SU(3). Each cell in the SU(2) diagram is bounded by points corresponding to the group elements $\mathrm{diag}(1,1)$ and $\mathrm{diag}(-1,-1)$. The cells in the SU(3) diagram are bounded by line segments corresponding to points for which a pair of eigenvalues is degenerate and by points at the corners of each cell which correspond to the group elements $\mathrm{diag}(1,1,1)$, $\mathrm{diag}(e^{i2\pi/3}, e^{i2\pi/3}, e^{i2\pi/3})$, and $\mathrm{diag}(e^{i4\pi/3}, e^{i4\pi/3}, e^{i4\pi/3})$.

A spectral map can be constructed to be permutation equivariant by defining its action on a single *canonical cell* and using permutation equivariance to extend uniquely to the other $N!-1$ cells in a covering set. We label the canonical cell $\Psi \subset \mathfrak{t}$. The choice of working with the extended hyperplane $\mathfrak{t}$ ensures that a canonical cell can be chosen to be a contiguous $(N-1)$-simplex without partially 'wrapping' modulo $2\pi$, which would otherwise complicate the implementation of invertible transformations on the cell. Figure 4.8 schematically shows the process of mapping points into a canonical cell $\Psi$ within the space $\mathfrak{t}$. Extending a map on the canonical cell to all diagonal matrices is given by the following procedure:

1. **Extract phases** $\phi_i = \arg(\lambda_i) \in [0, 2\pi]$. These do not necessarily satisfy $\sum_i \phi_i = 0$ to fall within $\mathfrak{t}$ at this step.

2. **Canonicalize** by applying a permutation $P$ and adding integer multiples of $2\pi$

---

[8] A deeper, albeit technical, connection exists between these spaces. The diagonal matrices $T$ form a 'maximal torus' within SU($N$), which is an Abelian subgroup. The space $\mathfrak{t}$ can be identified with the algebra of this subgroup. The surjective map is equivalent to the map $\exp(\cdot)$ from the algebra to the group.

to the list of complex phases,

$$(\phi_1^{\text{canon}}, \ldots, \phi_N^{\text{canon}}) = P \cdot (\phi_1, \ldots, \phi_N) + (2\pi k_1, \ldots, 2\pi k_N),$$

to map to a canonical point $(\phi_1^{\text{canon}}, \ldots, \phi_N^{\text{canon}})$ within $\Psi$.

3. **Apply a map on the simplex** $\Psi$ to transform

$$(\phi_1^{\text{canon}}, \ldots, \phi_N^{\text{canon}}) \rightarrow (\phi_1'^{\text{canon}}, \ldots, \phi_N'^{\text{canon}}).$$

4. **'Uncanonicalize'** by applying the inverse permutation $P^{-1}$ to the transformed phases and map back into $T$ by exponentiating,

$$(\lambda_1', \ldots, \lambda_N') = P^{-1} \cdot (e^{i\phi_1'^{\text{canon}}}, \ldots, e^{\phi_N'^{\text{canon}}}).$$

There are many ways to map to a canonical point in $\mathfrak{t}$ given a set of eigenvalues. A simple possibility is given by numerically sorting the complex phases of the eigenvalues assuming they are defined within a particular interval such as $[0, 2\pi]$. Approaches along these lines do indeed map all points related by permutations to a single canonical point, but they do not in general result in a canonical cell associated with a contiguous simplex $\Psi \subset \mathfrak{t}$. We thus require a *canonicalizing map* that gives a permutation of the eigenvalue phases and a set of offsets by multiples of $2\pi$ that does result in points being mapped into a contiguous simplex $\Psi$ in step 2 of the procedure above. Algorithm 4.2 describes a canonicalizing map satisfying this property:

---

**Algorithm 4.2:** Canonicalizing map into the simplex $\Psi$

---

1. Compute the integer $S = \frac{1}{2\pi} \sum_i \phi_i$.

2. Sort the angles to be numerically ascending, resulting in the list $\phi^{\text{sort}} = \text{sort}([\phi_1, \ldots, \phi_N])$. The angles $\phi_i$ are assumed to live in the domain $[0, 2\pi]$.

3. 'Snap' the angles to satisfy $\sum_i \phi_i^{\text{snap}} = 0$ by applying $2\pi$ offsets giving the list

   $$\phi^{\text{snap}} = [\phi_1^{\text{sort}}, \ldots, \phi_{N-S+1}^{\text{sort}} - 2\pi, \ldots, \phi_N^{\text{sort}} - 2\pi].$$

4. Sort the list of snapped angles to be numerically ascending, giving $\phi^{\text{canon}} = \text{sort}(\phi^{\text{snap}})$, where $\phi^{\text{canon}} \in \mathfrak{t}$.

5. Return $\phi^{\text{canon}}$ and the composition of the permutation used to sort in step 2 with the permutation used to sort in step 4.

---

To check that points are mapped into a contiguous simplex in the hyperplane $\mathfrak{t}$ by Algorithm 4.2, we consider the effect of continuously deforming the list of angles without crossing any points with degenerate eigenvalues associated with the boundaries of a cell. Local interchanges in the order of $\phi^{\text{sort}}$ are not possible without crossing points with degenerate eigenvalues. Continuously deforming a point within the interior of a cell can thus only cause the list $\phi^{\text{sort}}$ to discontinuously change when an angle crosses the boundary between 0 and $2\pi$. This happens in two cases: either the angle at the start of the list cross from 0 to $2\pi$ or the angle at the end of the list crosses from $2\pi$ to 0. In the first case, the components of $\phi^{\text{sort}}$ rotate cyclically to the left by one index and $S$ increments by one. This continuously affects the set of elements $\phi^{\text{snap}}$ because the angle that is now at the end of the list is also now decremented by an additional factor of $2\pi$ and all angles for which a factor of $2\pi$ was subtracted are still appropriately subtracted as a result of $S$ having incremented. Given that the set of angles contained in $\phi^{\text{snap}}$ change continuously without any angles becoming degenerate, the components of $\phi^{\text{canon}}$ also change continuously. Continuity in the second case follows from a similar argument. This confirms that the canonicalizing map is continuous in the inputs. Finally, we note that this indeed gives a *single* canonical simplex, because information on the original ordering of the eigenvalues is removed at step 2.

The image of the canonicalizing map described by Algorithm 4.2 is a simplex $\Psi$ bounded by vertices $y_1, \ldots, y_N \in \mathfrak{t}$ defined by

$$[y_k]_j \equiv 2\pi \left( k/N - \mathbb{1}_{j \leq k} \right), \tag{4.31}$$

where $\mathbb{1}_{j \leq k} = 1$ if $j \leq k$ and $\mathbb{1}_{j \leq k} = 0$ otherwise. These bounding vertices can be derived by noting that each vertex corresponds to a point of maximal eigenvalue degeneracy. These points then correspond to the preimage of the $N$ center elements of $\mathrm{SU}(N)$ under the $\exp i(\cdot)$ map. In the group space, the center elements are given by the diagonal matrices $\mathrm{diag}(e^{i2\pi k/N}, \ldots, e^{i2\pi k/N})$ for $k \in \{1, \ldots, N\}$. The canonicalizing map given in Algorithm 4.2 continuously maps into the interior of the simplex $\Psi$, thus we can identify the bounding vertices by taking the limit as interior points approach the preimage of each of these center elements in $\mathfrak{t}$. For the $k$th center element, consider the family of interior points $(2\pi k/N - \epsilon, 2\pi k/N - \epsilon, \ldots, 2\pi k/N + \epsilon(N-1))$ which approach the preimage of the $k$th center element as $\epsilon \to 0$. We separately evaluate two cases:

1. For $k \neq N$, applying Algorithm 4.2 to these points results in a value of $S = k$ and the list

$$\phi^{\text{snap}}_{k,\epsilon} = \Big[ \underbrace{\frac{2\pi k}{N}, \ldots, \frac{2\pi k}{N}}_{N-k \text{ times}}, \underbrace{\frac{2\pi k}{N} - 2\pi, \ldots, \frac{2\pi k}{N} - 2\pi}_{k \text{ times}} \Big] + O(\epsilon). \tag{4.32}$$

Sorting in step 4 brings the $k$ components from the end to the beginning, giving

$$\phi^{\text{canon}}_{k,\epsilon} = \Big[ \underbrace{\frac{2\pi k}{N} - 2\pi, \ldots, \frac{2\pi k}{N} - 2\pi}_{k \text{ times}}, \underbrace{\frac{2\pi k}{N}, \ldots, \frac{2\pi k}{N}}_{N-k \text{ times}} \Big] + O(\epsilon). \tag{4.33}$$

The limiting point is given by $\lim_{\epsilon \to 0} \phi^{\text{canon}}_{k,\epsilon} = y_k$.

2. In the case of $k = N$, the final component $2\pi + \epsilon(N-1)$ will be greater than $2\pi$, and to satisfy the requirements of step 2 of the algorithm, $2\pi$ must be subtracted to project to $\epsilon(N-1) \in [0, 2\pi]$. This component will be sorted to the front of the list $\phi^{\text{sort}}$. A value of $S = N - 1$ is computed in step 1, and thus $2\pi$ is subtracted from components 2 through $N$ of $\phi^{\text{sort}}$, corresponding to components 1 through $(N-1)$ in the input. The result of step 3 is $\phi_{N,\epsilon}^{\text{snap}} = [0, \ldots, 0] + O(\epsilon)$, giving $\lim_{\epsilon \to 0} \phi_{N,\epsilon}^{\text{canon}} = y_N$.

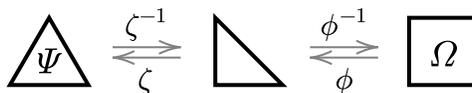

**Figure 4.9:** Schematic depiction of the map from the simplex $\Psi$ into the box $\Omega$ using the composition of the bijections $\phi$ and $\zeta$. Figure adapted from Fig. 8 of Pub. [2].

Having constructed a valid canonicalizing map, any invertible transformation suitable for the simplex $\Psi$ can be extended into a spectral map by canonicalizing and uncanonicalizing, as described above, and can then be extended to a conjugation-equivariant map by diagonalizing and undiagonalizing. The manipulations performed so far have thus significantly simplified the construction of a spectral map: the invertible map on $\Psi$ is not constrained by any symmetry considerations and $\Psi$ is a simply connected domain within $\mathbb{R}^{N-1}$. The map on the simplex $\Psi$ can then be freely designed to be highly expressive.

### 4.5.2 Transformations of the simplex $\Psi$

We now proceed to construct suitable invertible maps for the $(N-1)$-simplex $\Psi$ for arbitrary $N$. An invertible map on the simplex by definition must bijectively map the finite domain $\Psi \subset \mathfrak{t}$ to itself. Invertible transformations on finite domains of single variables are well understood; for example, invertible rational quadratic splines were discussed in Sec. 4.4 for their application to transformations of the finite domain $[0, 2\pi]$ of U(1). To connect single-variable invertible transformations over finite domains to the $(N-1)$-simplex, we apply a final recursive decomposition of the problem. First, the simplex $\Psi$ is bijectively mapped to an open box $\Omega = (0, 1)^{N-1}$; secondly, invertible transformations on $[0, 1]$ are applied to each coordinate of the point within $\Omega$; finally, the transformed point is mapped back to the domain of $\Psi$.

We describe the map sending $\Omega$ to $\Psi$ as the composition of the bijections $\phi$ and $\zeta$, where $\phi$ collapses one end of $\Omega$ along each axis to produce a right-angled $(N-1)$-simplex and $\zeta$ is an affine map that sends the resulting right-angled simplex to $\Psi$. The inverse maps can be used to transform from $\Psi$ to $\Omega$. Figure 4.9 schematically depicts the maps to be constructed and their relationship to these spaces.

We define the map $\phi$ to send a point $\alpha \in \Omega$ to the point $\phi(\alpha)$ with components given by

$$[\phi(\alpha)]_i = \begin{cases} \alpha_1 & i = 1 \\ \alpha_i \prod_{j=1}^{j<i}(1 - \alpha_j) & i > 1. \end{cases} \tag{4.34}$$

The resulting space is the right-angled simplex given by the set of points that have positive coordinates and for which the sum over coordinates is less than 1. This property can be easily verified for the image of an arbitrary point under $\phi$: each coordinate

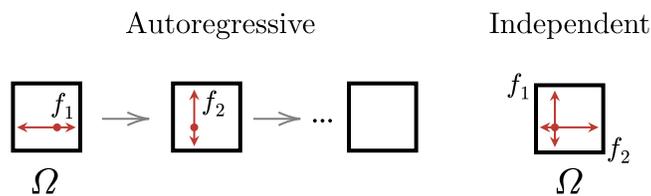

**Figure 4.10:** Schematic depiction of invertible transformations of each coordinate of $\Omega$, either applied autoregressively (left) or independently (right). When autoregressively transforming the coordinates, values of the previous coordinates are passed as additional inputs to context functions defining the parameters of later transformations.

computed in Eq. (4.34) is positive and the sum over coordinates is

$$\sum_i [\phi(\alpha)]_i = \alpha_1 + (1 - \alpha_1)\big[\alpha_2 + (1 - \alpha_2)[\dots]\big],\tag{4.35}$$

which is $\leq 1$ by induction on the term in square brackets. The transformation in Eq. (4.34) is autoregressive and can therefore be inverted by sequentially inverting the transformation of each coordinate in increasing order.

The vertices of the resulting right-angled simplex are given by $\{\kappa_1, \dots, \kappa_N\}$, where $\kappa_1$ is the origin and each other vertex $\kappa_i$ is given by the unit vector along axis $i-1$. The affine transformation $\zeta$ can be constructed by mapping each displacement vector $\kappa_i - \kappa_1 = \kappa_i$ to $y_i - y_1$ then translating by $y_1$. Explicitly, the action of $\zeta$ on a point $\rho$ in the right-angled simplex is given by

$$\zeta(\rho) = y_1 + M\rho\tag{4.36}$$

where $M$ is an $N \times (N-1)$ matrix with the $i$th column given by $y_{i+1} - y_1$. The inverse transformation is given by

$$\zeta^{-1}(x) = (M^T M)^{-1} M^T (x - y_1).\tag{4.37}$$

The matrix $(M^T M)^{-1} M^T$ is the pseudo-inverse of $M$, which gives the inverse of the linear transformation from $\kappa_i$ to $y_i - y_1$.

Changing variables from a point in $\Psi$ to a point in $\Omega$ by the composition of $\phi$ and $\zeta$ reduces the problem to applying an invertible transformation on points in the box $\Omega$. Invertible transformations of points in $\Omega$ can be described by applying either an autoregressive or independent invertible transformation to each of the $(N-1)$ coordinates in the interval $[0, 1]$. By convention, we define an autoregressive transformation of $\Omega$ to transform the $i$th coordinate of a point conditioned upon the values of all coordinates $j < i$ (see Sec. 3.1.2 for a general discussion of autoregressive transformations). Autoregressive transformations serialize the process of computing the coordinate transformations and may slow down evaluation of the coupling layer but in principle provide greater expressivity. Figure 4.10 gives a schematic view of these two options to transform points in $\Omega$.

In the applications to single SU($N$) variables and SU($N$) lattice gauge theory below, we use rational quadratic splines for the transformation of each coordinate of $\Omega$. In contrast to circular splines, these splines act on coordinates with fixed boundaries that are not identified by periodicity. A single invertible rational quadratic spline can be applied to transform a fixed finite interval by working with $K$ knots and $K+1$ arbitrary positive slopes. This transformation is described by $3K+1$ free parameters in contrast to the $3K$ parameters used to describe circular splines for U(1) variables, where the extra parameter gives the additional slope at one endpoint. In total, $(N-1)(3K+1)$ parameters are required to describe the rational quadratic spline transformation of all coordinates of $\Omega$. When used within a gauge-equivariant coupling layer, these parameters are produced from the evaluation of context functions acting on gauge-invariant quantities as described for general gauge-equivariant coupling layers in Sec. 4.3.

### 4.5.3   Summary and the Jacobian determinant of spectral maps

Having thoroughly decomposed the steps required to implement conjugation-equivariant maps on SU($N$) it is worth taking a step back to review the process at a high level. We construct transformations that are conjugation equivariant by incrementally simplifying the problem, as shown in Figure 4.6. At each stage, part of the symmetry, or restrictions on how the transformation must act, are removed from the problem and a simpler transformation with fewer symmetry constraints or restrictions is recursively addressed. This occurs in three layers:

1. An arbitrary conjugation-equivariant map on an SU($N$) variable is described by diagonalizing, applying a permutation-equivariant spectral map on the resulting eigenvalues, and returning a matrix composed from the transformed eigenvalues and original eigenvectors.

2. An arbitrary spectral map on the eigenvalues of SU($N$) is described by applying a canonicalizing map that describes a canonical permutation[9] of eigenvalue phases that projects into an $(N-1)$-simplex $\Psi$, using any invertible map to transform points within $\Psi$, and applying the inverse of the canonical permutation.

3. An arbitrary invertible map on the simplex $\Psi$ is described by changing variables to a box $\Omega$ for which each coordinate can be independently transformed in the interval $[0, 1]$.

As described in Secs. 4.5.1 and 4.5.2, the components used to construct spectral maps are independently invertible. The map is therefore invertible as a whole.

The map is also differentiable, and it is necessary to compute the Jacobian determinant of the transformation to determine the change in density given by using this as a kernel within a gauge-equivariant flow-based model. When considering group-valued variables, we are specifically interested in the change in density with respect to the Haar measure. The spectral map acts in the space of eigenvalues and for SU($N$) the

---

[9]As well as appropriate offsets by $2\pi$ to project into the hyperplane $\mathfrak{t}$.

marginalized Haar measure over eigenvalues is given by (see Appendix A)

$$dU = \left[ \sum_{k=-\infty}^{\infty} \delta \left( \sum_i \phi_i - 2\pi k \right) \right] \text{Haar}(e^{i\phi_1}, \ldots, e^{i\phi_N}) \frac{d\phi_1}{2\pi} \cdots \frac{d\phi_N}{2\pi}, \qquad (4.38)$$

where

$$\text{Haar}(\lambda_1, \ldots, \lambda_N) \equiv \frac{1}{N!} \prod_{i<j} |\lambda_i - \lambda_j|^2. \qquad (4.39)$$

Composing the inverses of the Jacobian factors of (1) the change of variables from $\Psi$ to $\Omega$, (2) the inner transformation on $\Omega$, and (3) the change of variables from $\Omega$ back to $\Psi$ gives the change in density with respect to the 'natural' measure on the space of eigenvalue phases, $\left[ \sum_{k=-\infty}^{\infty} \delta \left( \sum_i \phi_i - 2\pi k \right) \right] d\phi_1, \ldots, d\phi_N$. The affine transformation $\zeta$ has a fixed Jacobian determinant independent of the input, so the Jacobian factor arising from $\zeta$ will cancel with the factor arising from $\zeta^{-1}$. However, the transformation $\phi$ between the right-angled simplex and $\Omega$ has a non-trivial Jacobian factor. Because of the autoregressive structure, only the diagonal components of the Jacobian are needed to calculate the determinant of $\phi$, and the Jacobian determinant is given by

$$J_\phi(\alpha) = \prod_i \prod_{j<i} (1 - \alpha_j) = \prod_j (1 - \alpha_j)^{(N-1)-j}. \qquad (4.40)$$

The map $\phi^{-1}$ is applied in the forward transformation $\Psi \to \Omega$, and $J_{\phi^{-1}}^{-1} = J_\phi$ gives the change of density associated with this step. The map $\phi$ is applied in the reverse transformation $\Omega \to \Psi$, and $J_\phi^{-1}$ gives the corresponding change of density. Labeling the Jacobian of the inner transformation on $\Omega$ by $J_\chi$, the change in density with respect to the natural measure on the eigenvalues is then given by the inverse Jacobian factor

$$J_{\text{eigs}}^{-1} = J_\phi J_\chi^{-1} J_\phi^{-1}. \qquad (4.41)$$

We combine this with the factor $\text{Haar}(\cdot)$ to translate between the Haar measure and the natural measure on the eigenvalues. The total change in the probability density with respect to the Haar measure of the spectral map is then given by the inverse Jacobian factor

$$J_{\text{spec}}^{-1} \equiv \text{Haar}(\lambda_1, \ldots, \lambda_N) \, J_{\text{eigs}}^{-1} \, \text{Haar}(\lambda'_1, \ldots, \lambda'_N)^{-1}. \qquad (4.42)$$

Taking the inverse of Eq. (4.42) would give the Jacobian determinant $J_{\text{spec}}$ of the spectral map with respect to the Haar measure on the group, though one never needs to compute this explicitly.

When used in the context of a flow-based model, the $\text{SU}(N)$ spectral map produces a transformed variable $W'$ and associated change in density $J_{\text{spec}}^{-1}$ which are then composed with other transformations — either other spectral maps in the case of one-variable flows or transformations of other loops and link variables in the case of gauge-equivariant flows. The procedure for training such models involves backpropagation of gradients through the produced sample and the change in density, ultimately giving the gradients of loss functions with respect to all free parameters of the model.

The parameters describing a spectral map are entirely encoded in the choice of transformations applied to the coordinates of the box $\Omega$. For example, invertible splines used either autoregressively or independently are defined in terms of a number of parameters giving the bin widths, heights, and slopes at the knots. The free parameters describing the transformation of $\Omega$ can either be specified directly, if considering applications to single $\mathrm{SU}(N)$ variables, or from the outputs of context functions when the spectral map is used within a gauge-equivariant coupling layer. In either case, gradients must be propagated from $W'$ and $J_{\mathrm{spec}}^{-1}$ to these free parameters and the input variable $W$. Almost all operations involved in the calculation of $W'$ and $J_{\mathrm{spec}}^{-1}$ in a spectral map can be implemented in a framework supporting automatic differentiation. Matrix diagonalization over complex matrices is the only operation not commonly supported in such frameworks. To circumvent this issue, a procedure for defining custom gradient propagation over matrix diagonalization in the context of spectral maps was presented in Appendix C of Pub. [2]. Spectral flows implemented for all $\mathrm{SU}(N)$ studies in this dissertation use this custom step combined with standard autodifferentiation to handle backpropagation of gradients.

### 4.5.4  Application of spectral maps to $\mathrm{SU}(N)$ distributions

Combining conjugation-equivariant maps with a conjugation-invariant prior distribution allows the construction of flow-based models for which the model distribution is exactly invariant under conjugation by group elements. These flow-based models can be used to approximate distributions given by class functions over an $\mathrm{SU}(N)$ variable. In this section, we investigate the ability of these models to capture conjugation-invariant distributions over a single $\mathrm{SU}(N)$ variable.

In particular, we consider a class of target distributions given by the probability densities

$$p_{1\mathrm{var}}^{(i)}(U) = e^{-S_i(U)}/Z_i, \tag{4.43}$$

where $Z_i = \int dU e^{-S_i(U)}$, and the probability densities are defined with respect to the Haar measure. The one-variable 'actions' $S_i(U)$ are generically defined in terms of traces of powers of the $\mathrm{SU}(N)$ variable $U$ by

$$S_i(U) \equiv -\frac{\beta}{N} \operatorname{Re} \operatorname{tr}\left[ \sum_n c_n^{(i)} U^n \right], \tag{4.44}$$

which ensures that the distribution is invariant under conjugation. We worked with three target distributions, defined by the three choices of coefficients $c_n^{(i)}$ given in Table 4.1. The first set of coefficients $c^{(0)}$ gives a distribution that matches the marginal distribution (up to exponentially small finite-volume corrections) over each plaquette variable in the $\mathrm{SU}(N)$ lattice gauge theory applications presented in Sec. 4.7. The remaining two sets of coefficients were produced by randomly generating coefficients and restricting to cases in which the distribution consisted predominantly of a single peak in the space of eigenvalues. The restriction to a single peak is motivated by the structure of typical distributions over Wilson loop observables in non-Abelian gauge theories in the confined phase [413, 414]. It is an interesting subject of future investi-gation to study the ability of these spectral flows to capture multi-modal distributions which could, for example, allow these methods to be extended studies of theories in which symmetries are spontaneously broken.

| Target $i$ | $c_1^{(i)}$ | $c_2^{(i)}$ | $c_3^{(i)}$ |
|------------|-------------|-------------|-------------|
| 0 | 1.00 | 0.00 | 0.00 |
| 1 | 0.17 | $-0.65$ | 1.22 |
| 2 | 0.98 | $-0.63$ | $-0.21$ |

**Table 4.1:** The coefficients used to define three target distributions in terms of the action given in Eq. (4.44). These parameters are used to study performance on the groups SU(2) and SU(3). Table adapted from Table I of Pub. [2].

Flow-based models were trained to approximate all three families of target distributions for three distinct values of the coupling parameter, $\beta \in \{1, 5, 9\}$, for the groups SU(2) and SU(3). Flow-based models were also trained to approximate all three families of target distributions for the groups SU($N$) with $N \in \{4, \ldots, 9\}$ with the coupling parameter fixed to $\beta = 9$, corresponding to distributions with the most significant fluctuations.

Flow-based models were constructed using a uniform prior distribution (with respect to the Haar measure) combined with a spectral map defined following the methods of the previous sections. Invertible rational quadratic splines were applied as the inner transformation used for each coordinate of the box $\Omega = [0, 1]^{N-1}$ for each group SU($N$). In all cases, splines were defined using $K = 3$ bins, corresponding to four knots. For this proof-of-principle study, the inner spline transformation was defined to act independently, rather than autoregressively, on each of the coordinates. The $3K + 1$ parameters per spline transformation together constitute the free parameters of each flow-based model. For each target distribution, a distinct flow-based model was optimized to minimize the KL divergence as a loss function (see Sec. 3.4). Optimization was performed using the Adam optimizer with a learning rate fixed to $10^{-1}$. Batches of 1024 samples were used per optimization step to stochastically evaluate the loss function, and a total of 400 iterations were used to optimize each model.

For each of the three families of target distributions for both SU(2) and SU(3) variables, the effective sample size (ESS; see Sec. 3.5.4) achieved by the model is given in Table 4.2. All ESS values are measured to sub-percent-level precision. For SU(2) variables, the ESS is greater than 97% indicating precise agreement between the learned model density and the target density in each case. For SU(3) variables, the ESS is somewhat lower for the models trained to learn distributions associated with the sets of parameters $c^{(1)}$ and $c^{(2)}$. Despite the relative decrease in ESS, the absolute ESS is above 70% in all cases, indicating a good agreement between the learned distributions. The ESS for the parameter set $c^{(0)}$ associated with the marginal distributions of plaquettes in the gauge theory study below was measured to be significantly higher, with values above 98% for all choices of $\beta$. For other choices of SU($N$), an ESS of greater than 5% was achieved for all target distributions with a significantly higher ESS of greater than 90% observed for the parameter set $c^{(0)}$. The relatively lower ESS for the sets of parameters $c^{(1)}$ and $c^{(2)}$ may be a result of the low expressivity of the models using independent coordinate transformations of $\Omega$ applied in this proof-of-principle study; increasing the number of bins in each spline transformation and using autoregressive conditioning for the transformation of the two coordinates within $\Omega$ provides a systematic way to improve the expressivity of the spectral map and thus the quality of the model. We make these extensions for the main applications to SU($N$) lattice gauge theory in Sec. 4.7.

|  | $c^{(0)}$ | | | $c^{(1)}$ | | | $c^{(2)}$ | | |
| --- | --- | --- | --- | --- | --- | --- | --- | --- | --- |
| $\beta^{\mathrm{SU(2)}}$ | 1 | 5 | 9 | 1 | 5 | 9 | 1 | 5 | 9 |
| ESS(%) | 100 | 100 | 100 | 98 | 98 | 97 | 100 | 99 | 100 |
| $\beta^{\mathrm{SU(3)}}$ | 1 | 5 | 9 | 1 | 5 | 9 | 1 | 5 | 9 |
| ESS(%) | 99 | 98 | 99 | 97 | 80 | 82 | 99 | 91 | 73 |

**Table 4.2:** Measured ESS for flow-based models implementing the spectral map for SU(2) and SU(3). Entries correspond to models optimized to reproduce each of the three target parameters sets for the choices $\beta \in \{1, 5, 9\}$, as indicated. The top two rows show results for SU(2) variables while the bottom two rows show results for SU(3) variables. Table adapted from Tables II & III of Pub. [2].

Figure 4.11 compares the target probability densities with the learned model densities for all distributions studied over SU(2) variables. Figures 4.12 and 4.13 show a similar comparison for SU(3) and SU(9) variables across all target distributions with the coupling fixed to the largest choice explored, $\beta = 9$. Model probability densities were measured as a function of the eigenvalues by constructing diagonal matrices and evaluating the associated probability density; by construction, the probability distributions are invariant under matrix conjugation and thus diagonal matrices were sufficient to fully explore the structure of the distributions. In all cases, the probability density is plotted with respect to the Lebesgue measure of the depicted coordinates, which corresponds to the density that would be seen by plotting a histogram over the shown coordinates. To measure this density, the probability density reported by the model (given with respect to the Haar measure) is multiplied by the factor Haar($\cdot$) given in Eq. (4.39). For the groups SU(2) and SU(3), the density is shown over the entire one- and two-dimensional manifold of eigenvalues, respectively. The eigenvalues of SU(9) matrices range over an eight-dimensional manifold, and the figure depicts the densities on a slice through this manifold corresponding to holding the eigenvalue phases $\phi_3, \ldots, \phi_8$ fixed to random values and varying $\phi_1$ and $\phi_2$. The value of the final eigenvalue is fixed by the first eight, and can for example be computed by assigning $\phi_9 = -\sum_{i=1}^{8} \phi_i$.

In Figs. 4.11 and 4.12, two-fold and six-fold permutation symmetry can respectively be seen in the plotted densities over SU(2) and SU(3) eigenvalues. These symmetries demonstrate the exact permutation invariance of the target and model densities, which is a subset of the full conjugation invariance of the target distributions (by definition) and the model distributions (by construction). Exact conjugation invariance of the model distribution was also checked for all groups studied.

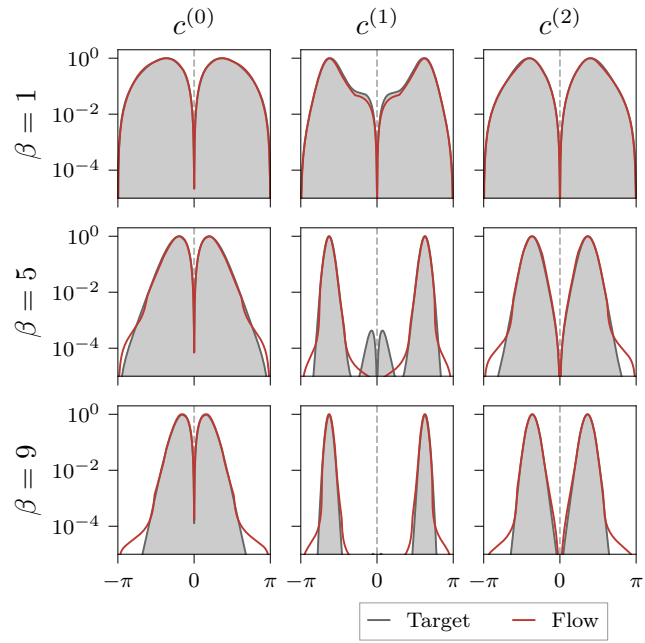

**Figure 4.11:** Comparison of the flow-based model and target densities for all SU(2) distributions. Densities are plotted as a function of the complex phase $\phi_1$ associated with the first SU(2) eigenvalue. The second eigenvalue of each SU(2) matrix can be specified by $\phi_2 = -\phi_1$, and the coordinate $\phi_1$ thus parameterizes the entire manifold of SU(2). The exact invariance under permutations of the eigenvalues can be seen for both target and model distributions as a symmetry under $\phi_1 \leftrightarrow -\phi_1$. Broad agreement between the probability densities can be seen over many orders of magnitude, with deviations restricted to regions of low probability density. Figure adapted from Fig. 5 of Pub. [2].

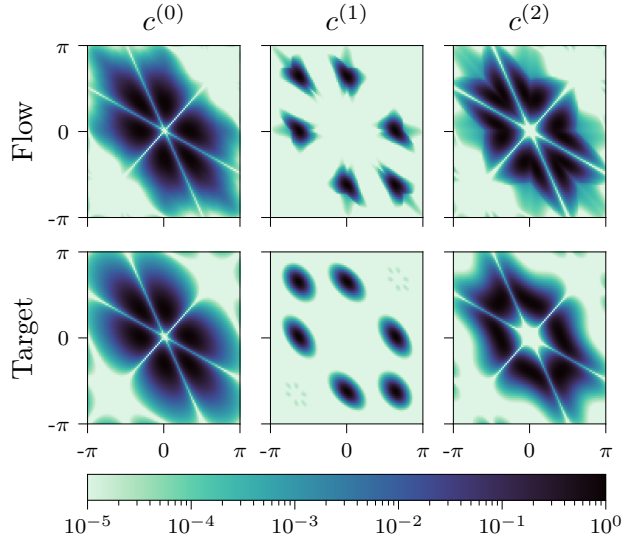

**Figure 4.12:** Comparison of the flow-based model and target densities for the SU(3) distributions with $\beta = 9$. The plotted region is a projection of the manifold of SU(3) eigenvalues into the plane of the first two eigenvalue phases $(\phi_1, \phi_2)$. Six-fold symmetry associated with permutation invariance can be clearly seen for both the target distributions and the model distributions. The flow and target distributions broadly agree in each case, though small deviations are seen in regions of low probability density. Figure adapted from Fig. 7 of Pub. [2].

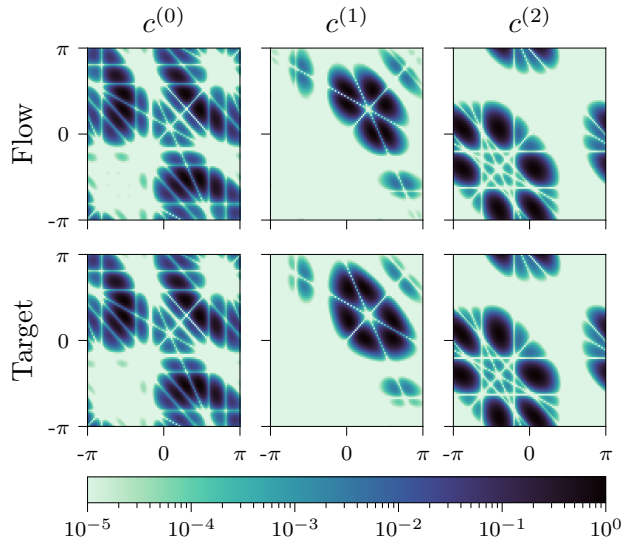

**Figure 4.13:** Comparison of the flow-based model and target densities for the SU(9) distributions with $\beta = 9$. The plotted region is a slice through the manifold of SU(9) eigenvalues given by fixing $\phi_3, \ldots, \phi_8$ to random values, varying $\phi_1$ and $\phi_2$, and assigning $\phi_9 = -\sum_{k=1}^{8} \phi_k$. The figure depicts the resulting densities in the $(\phi_1, \phi_2)$ plane. Lines of zero density correspond to points on the boundaries of cells intersected by this slice. Figure adapted from Fig. 9 of Pub. [2].

## 4.6 U(1) gauge theory in (1+1)D

In this and the following section, we put the pieces together, implementing flow-based models using gauge-equivariant coupling layers and studying their performance for sampling lattice gauge theory distributions. This section describes the application to U(1) lattice gauge theory using the kernels suitable for U(1) variables defined in Sec. 4.4.

### 4.6.1 Action, observables, and parameters

We consider U(1) lattice gauge theory in (1 + 1) spacetime dimensions for this proof-of-principle study of the applicability of gauge-equivariant flow-based models to ensemble generation. This theory can be interpreted as the quenched limit of a lattice discretization of the Schwinger model [36]. With the inclusion of dynamical fermions, i.e. for the full Schwinger model, many interesting features of phenomenologically relevant theories in (3 + 1)D are reproduced, including confinement, topological features, and a form of chiral symmetry breaking. After quenching, the theory retains the properties of confinement and topology. The latter property is particularly interesting, as traditional MCMC methods encounter topological freezing as the lattice spacing is taken to zero. Directly proposing samples has the potential to circumvent this problem, and we study the ability of flow-based models to effectively sample topological sectors below.

**Action.** The distribution of the theory under study is defined by the Wilson gauge action for the U(1) gauge group (see Chap. 2), given in (1 + 1)D by

$$S(U; \beta) = -\beta \sum_x \operatorname{Re} P_{01}(x), \tag{4.45}$$

where $P_{01}(x) \in \mathrm{U}(1)$ are the plaquettes rooted at site $x$,

$$P_{01}(x) = U_0(x) \, U_1(x + \hat{0}) \, U_0^\dagger(x + \hat{1}) \, U_1^\dagger(x). \tag{4.46}$$

The sum over $x$ enumerates the sites of a two-dimensional lattice and periodic boundary conditions are assumed when computing translated sites in Eq. (4.46). We restrict to square lattices with dimensions $L \times L = 16 \times 16$ for this study; we use the variable $V = L^2$ to denote the total number of lattice sites. This quenched two-dimensional theory does not have a meaningful continuum limit, as the distribution can be shown to factorize over each plaquette independently, up to corrections exponentially small in the volume. However, the expected value of each plaquette approaches 1 as the limit $\beta \to \infty$ is taken, which can be interpreted as an increasing correlation length for static quark-antiquark pairs associated with Wilson loops. This limit also corresponds to the limit in which topological freezing sets in. In this study, the bare coupling parameter $\beta$ is thus varied over the values $\{1, 2, 3, 4, 5, 6, 7\}$ to compare the performance of flow-based sampling to traditional MCMC methods as topological freezing sets in.

**Observables.** Functions of closed loops of links define a dense subset of all possible gauge-invariant quantities that can be measured in this (and any) gauge-invariant theory. In two-dimensional gauge theories, many of these observables can be analytically calculated. We fix a set of observables that can be both analytically computed and reliably measured using flow-based MCMC and traditional MCMC methods.

In particular, we select two classes of observables to measure that give information about the local structure of the lattice gauge theory distribution:

1. Plaquettes $P_{01}(x)$ and powers of plaquettes $P_{01}^n(x)$

2. Square Wilson loops $W_{l \times l}(x) = \prod_{\mu, y \in \partial A_{l \times l}(x)} U_\mu(y)$, where $\partial A_{l \times l}(x)$ indicates the ordered set of links bounding an $l \times l$ rectangle rooted at site $x$

For both families of observables, translational invariance of the lattice gauge theory distribution means that the root site $x$ does not affect the expectation value. To increase the precision of Monte Carlo estimates, we average over all sites $x$ to compute expectation values of spatially averaged versions of each of these observables, denoted respectively by $P_{01}^n \equiv \frac{1}{V} \sum_x P_{01}^n(x)$ and $W_{l \times l} \equiv \frac{1}{V} \sum_x W_{l \times l}(x)$. The topological properties of the theory also give rise to observables that depend on the global structure of the gauge fields. We additionally select two topological observables to measure:

3. The integer-valued topological charge $Q = \frac{1}{2\pi} \sum_x \arg (P_{01}(x))$ describing the winding number of the gauge field topology, where $\arg(\cdot)$ is defined to return values in the interval $[-\pi, \pi]$

4. The topological susceptibility $\chi_Q = \langle Q^2/V \rangle$

The expectation values of these observables are analytically derived in Appendix A. These analytical results serve as a baseline to validate the correctness of the flow-based MCMC method.

### 4.6.2 Benchmark MCMC methods

We applied two standard MCMC methods — HMC and heatbath (HB) — as baselines against which flow-based models were compared. HMC is a commonly used, general-purpose MCMC sampler (see Chap. 2) that nonetheless suffers from topological freezing in this theory and thus provides an interesting benchmark for flow-based sampling methods. Heatbath [415–417] is a more specialized MCMC algorithm applicable specifically to some pure-gauge theories. The heatbath algorithm is a local updating algorithm in which each Markov chain step consists of directly resampling a subset of links from the exact conditional distribution over the subset. This direct resampling can result in more efficient updates per unit of computational cost in the context of pure-gauge theories; however, the local nature of the updates still makes this method susceptible to topological freezing and critical slowing down.

In this study, we implemented both methods using the NumPy library [418]. A single HMC molecular dynamics trajectory was defined by leapfrog integration using 5 discrete steps and a total trajectory length tuned to achieve a consistent acceptance rate across all choices of $\beta$. Table 4.3 gives the choices of trajectory length $\tau$ and measured acceptance rate in each case. A single heatbath Markov chain step was defined by the application of heatbath resampling to each link once. For the Wilson action, applying a heatbath update to a given link requires conditioning on the values of all adjacent links.[10] Many links can be updated at once by restricting to subsets of links that are not common to any plaquettes. In our implementation, the heatbath sweep was defined in terms of a sequence of collective updates to four disjoint subsets of the lattice using NumPy operations acting on the whole lattice at once. The underlying implementations were still given by sequential operations on the lattice, but this definition leveraged the efficient library implementation of these operations, placing heatbath on the same footing our implementation of HMC.

| $\beta$ | 1 | 2 | 3 | 4 | 5 | 6 | 7 |
|---|---|---|---|---|---|---|---|
| $\tau$ | 1.50 | 1.00 | 0.75 | 0.60 | 0.55 | 0.50 | 0.45 |
| Acc. rate (%) | 78 | 79 | 80 | 81 | 81 | 81 | 81 |

**Table 4.3:** The choices of total trajectory length $\tau$ and resulting Metropolis acceptance rate for the HMC simulations described in the main text. Acceptance rates were tuned to match a fiducial value of 80 % to within 5 % and are reported with sub-percent-level precision. The number of leapfrog integrator steps was fixed to 5 in all cases.

To understand the scale of practical improvements provided by the flow-based MCMC scheme, we investigated the time taken per Markov chain step in a single-threaded implementation of each MCMC approach. Each heatbath Markov chain step was measured to require half the time to evaluate as each HMC step, to within 10 %, and the runtime of each HMC step was measured to be equivalent to sampling each configuration from the flow-based model (defined below) to within 10 %. These relative runtimes are used to compare practical improvements provided by smaller autocorrelation times in the topological freezing study below.

We used both HMC and heatbath to produce 10 replicas each of ensembles consisting of 100 000 samples for each choice of $\beta$. For HMC, the samples were drawn sequentially from a Markov chain after 5000 thermalization steps were discarded; for heatbath, samples were drawn sequentially from a Markov chain with 1000 thermalization steps discarded. A single ensemble was used for observable measurements at each value of $\beta$ while the collection of replicas was used for our autocorrelation analysis.

### 4.6.3 Model architecture, training, and MCMC

For each choice of $\beta$ we constructed an independent, identical gauge-equivariant flow-based model and trained each model on the distribution $p(U) \propto e^{-S(U;\beta)}$ defined by the Wilson gauge action. We detail the construction and training of each model below. All models were implemented using the PyTorch framework [353].

---

[10]A pair of links is defined to be adjacent if they are both included in any common plaquette.

**Prior distribution.** We chose the uniform distribution with respect to the Haar measure as the prior distribution for each flow-based model. This distribution is invariant under gauge symmetry and all geometric symmetries of the lattice (see Sec. 4.2). The prior probability density factor $r(U) = 1$ was used for all calculations of the model density, which is consistent with our normalized U(1) Haar measure,

$$\int \mathcal{D}U \; r(U) = \int \mathcal{D}U \; = \prod_{x,\mu} \int_0^{2\pi} \frac{d\phi_{x,\mu}}{2\pi} = 1. \qquad (4.47)$$

**Coupling layers.** The invertible flow within each flow-based model was defined by a composition of gauge-equivariant coupling layers. The flow itself was thus gauge-equivariant, and, combined with the gauge-invariant prior distribution, each flow-based model therefore exactly respected the gauge symmetry of the theory by construction. We employed a striped masking pattern with a stride of four lattice sites (see Sec. 4.3.3) in the definition of each gauge-equivariant coupling layer. Coupling layers were composed in sequential 'stacks' of eight coupling layers iterating over all possible orientations and offsets of the masking pattern. In total, we constructed each flow by a composition of three such stacks, corresponding to 24 total coupling layers. These flows were thus defined to transform every link three times. We defined the inner coupling layer of each gauge-equivariant coupling layer to act on plaquettes, using the PAFF masking pattern to satisfy consistency with the striped link mask.

In each inner coupling layer, the kernel was defined in terms of a mixture of six NCP operations and a single offset (see Sec. 4.4). These kernels therefore required seven outputs from context functions to parameterize the scale parameters $s$ of all NCP transformations and the offset $t$. In each coupling layer, we defined these context functions using a convolutional neural network with two input channels — corresponding to the real/imaginary components of the gauge-invariant frozen plaquettes given as input — and seven output channels. Each convolutional layer was defined in terms of kernels of size $3 \times 3$. Convolutions employed periodic padding and a stride of one to preserve the lattice shape through the network. The convolutional network included two hidden layers between the input and output layers, each consisting of eight channels. Leaky ReLU [347] activation functions were applied between each convolution operation.

**Training using parameter bootstrapping.** We used the KL divergence as a loss function to train each model on the respective target distributions. In each training iteration, batches of size ranging from 16 384 to 131 072 samples were used to stochastically estimate the loss function and gradients. The Adam optimizer was used to iteratively update parameters based on these gradient estimates, with the hyperparameters fixed to the defaults defined in the PyTorch implementation. To focus our study on the effects of model expressivity rather than training dynamics, we trained each model until a plateau was reached in the loss function,[11] corresponding to near-convergence to a (local) minimum of the loss function. Total training costs were reduced by us-ing *parameter bootstrapping*, i.e. using optimized model parameters for one choice of $\beta$ to initialize models to be trained for other choices of $\beta$, with the aim of reducing the number of training iterations required to converge to a plateau. In particular, the $\beta = 3$ model was trained first, then the optimized $\beta = 3$ model parameters were used to initialize the models targeting $\beta \in \{1, 2, 4, 5, 6\}$. After 4000 additional iterations of training, the optimized $\beta = 6$ model parameters were used to initialize a model targeting the $\beta = 7$ distribution. In total, the $\beta = 3$ model was trained for 111 000 iterations. The $\beta = 1$ and $\beta = 2$ models were initialized using the parameters of the $\beta = 3$ model partway through training and were additionally trained for 57 000 and 64 000 iterations, respectively. The models for $\beta \in \{4, 5, 6, 7\}$ were trained for between 1000 and 5000 additional iterations after initializing from the final state of the $\beta = 3$ model.

---

[11] No precise stopping criterion was fixed, but training was halted in all cases at a point when no more than a 5% increase in measured ESS was seen over the prior 1500 training iterations.

**MCMC.** The trained flow-based models associated with each target value of $\beta$ were then used to generate ensembles using the flow-based MCMC method. We generated each ensemble using a total of 102 400 proposals from the respective flow-based model, with a Markov chain constructed from these proposals in each case using the asymptotically exact Metropolis accept/reject step described in Sec. 3.5. The first 2400 Markov chain steps were discarded to allow thermalization, and the remaining 100 000 Markov chain samples were used for measurements. A total of 10 replica flow-based ensembles were generated for each choice of $\beta$. Observable measurements were restricted to the first ensemble, while the collection of replicas was used for our autocorrelation analysis.

### 4.6.4 Monte Carlo studies

We next discuss the correctness of results generated using flow-based MCMC and investigate topological freezing across all approaches using measurements of observables on the HMC, heatbath, and flow-based ensembles. The observables described above were measured on all ensembles, along with their respective integrated autocorrelation times (see Chap. 2), and errors were estimated based on bootstrap uncertainties rescaled by the measured autocorrelation time.

**Tests of physical observables.** The Metropolis-Hasting accept/reject step in a flow-based MCMC sampler guarantees asymptotically exact sampling in much the same way that other MCMC methods give exactness guarantees. However, the rate of convergence to asymptotically distributed samples depends on the quality of the model approximation to the target distribution, and for a finite ensemble it may be the case that the Markov chain has not yet converged. By comparing measured observables against analytical results and measurements using HMC and heatbath, we confirm that these observable estimates are consistent with near-convergence of the flow-based Markov chain. Figure 4.14 shows comparisons of the largest powers of plaquettes and largest Wilson loops measured on the $\beta = 7$ ensembles. Figure 4.15 also compares measurements of the topological susceptibility across the ensembles with couplings $\beta \in \{5, 6, 7\}$. The flow-based MCMC estimates are consistent with the analytical results for all choices of observables shown in the figures, as well as all observables not shown. HMC and heatbath are also consistent with analytical results, but for larger Wilson loops and the topological susceptibility these methods are affected by autocorrelation times comparable to the finite Markov chain length and error bars for some measurements appear underestimated. Measurements of the topological susceptibility are most severely affected by this issue, qualitatively demonstrating that flow-based MCMC successfully avoids topological freezing while HMC and heatbath encounter issues.

**Autocorrelations.** The underestimation of error bars for HMC and heatbath estimates of topological observables is suggestive of the greater autocorrelation time for some observables using these sampling approaches in contrast to flow-based MCMC. A striking demonstration of the stark difference in autocorrelation times can be seen in the Markov chain histories of the topological charge $Q$ for the three methods depicted for the coupling $\beta = 7$ in Figure 4.16.

We precise quantify the effect of topological freezing for these three MCMC approaches by measuring the integrated autocorrelation times of the topological charge observable. To acquire reliable estimates of the integrated autocorrelation time even when the finite extent of the Markov chains rival the autocorrelation length, we applied a replica analysis over the ten replica ensembles. Figure 4.17 depicts the measured integrated autocorrelation time with errors computed by bootstrapping over replicas. Both HMC and heatbath suffer from a rapidly growing autocorrelation time for the topological charge as $\beta$ is taken large and topological freezing sets in. The results for HMC and heatbath are both consistent with exponential scaling of the integrated autocorrelation time $\tau_Q^{\mathrm{int}}$. On the other hand, the measured autocorrelation time for the flow-based ensembles scales much more mildly with the value of $\beta$, though it is also consistent with an exponential scaling. At the largest value of $\beta$ studied, the value of $\tau_Q^{\mathrm{int}}$ for HMC is approximately 1500 times as large as that of the flow-based model and the value of $\tau_Q^{\mathrm{int}}$ for heatbath is approximately 400 times as large as that of the flow-based model. Given that the cost of each Markov chain step is equivalent to within a factor of two for all of our implementations of these approaches, this represents an improvement in actual computational cost associated with estimating the distribution of topological charge by using the flow-based sampling approach instead of HMC or heatbath.

## 4.7 SU($N$) gauge theory in (1+1)D

We next follow a similar line of investigation to study gauge-equivariant flow-based models applied to SU($N$) lattice gauge theory. We focus on the case of lattice gauge theory in $(1 + 1)$D, as in the previous section. Unlike for the case of U(1) gauge theory in $(1 + 1)$D, there is no well-defined topology when the gauge group is SU($N$). A comparison of topological freezing effects against standard MCMC approaches is thus not possible. Instead, in this section we characterize the ability of SU($N$) gauge-equivariant flow-based models to capture the lattice gauge theory distributions using absolute metrics such as the ESS associated with reweighting factors of proposed sam-ples. The results produced by the flow-based models are also checked against analytic results to confirm that asymptotic behavior is observed for the finite ensemble sizes studied here.

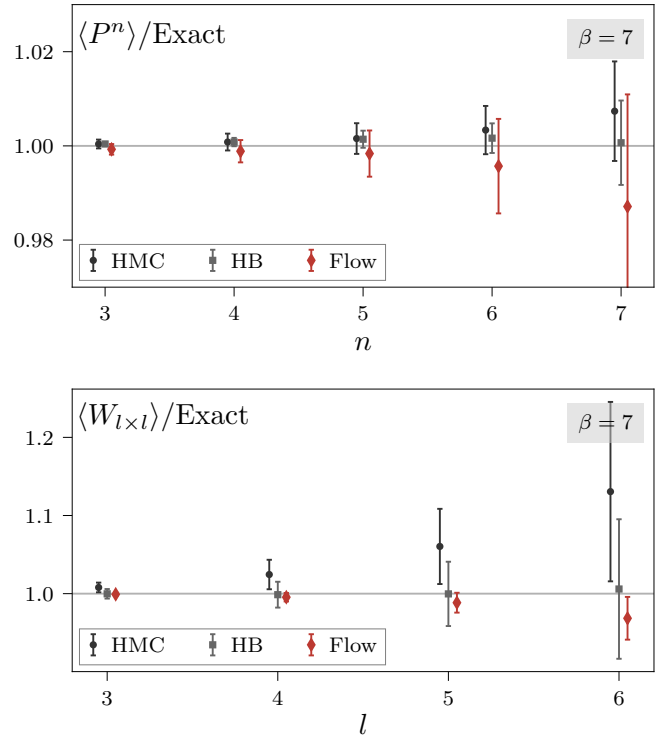

**Figure 4.14:** Monte Carlo estimates of observables relative to the exact analytical values using the $\beta = 7$ HMC, heatbath, and flow-based ensembles. In all cases measurements are consistent with exact analytical results, confirming that the finite Markov chains studied here have converged to asymptotic values of these observables. Powers of the plaquette $\langle P^n \rangle$ are shown above for a range of exponents $n \in \{3, \ldots, 7\}$ selected to compare observables sensitive to higher-order features of the plaquette distribution. Wilson loops $\langle W_{l \times l} \rangle$ are shown below for a range of loop side lengths $l \in \{3, \ldots, 6\}$ selected to compare observables sensitive to fluctuations at longer length scales. Powers of the plaquette are found to be more precisely estimated using HMC or heatbath ensembles in comparison to the flow-based estimates. On the other hand, Wilson loops are found to be more precisely estimated using flow-based ensembles in comparison to HMC and heatbath estimates. Figure adapted from Fig. 3 of Pub. [3].

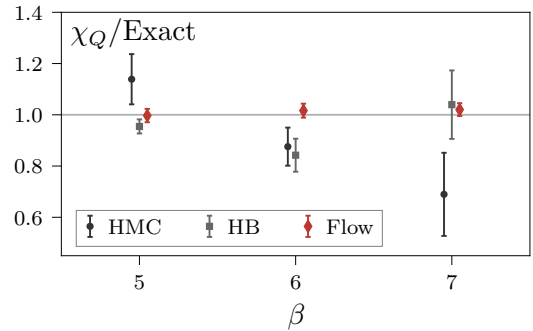

**Figure 4.15:** Monte Carlo estimates of the topological susceptibility $\chi_Q$ relative to the exact analytical values using HMC, heatbath, and flow-based ensembles. Flow-based ensembles very precisely estimate this topological quantity. HMC and heatbath ensembles have much longer autocorrelation times for this topological observable, resulting in much less precise estimates. Error bars appear underestimated in some of the HMC and heatbath measurements, likely due to autocorrelation times longer than the finite lengths of the Markov chains.

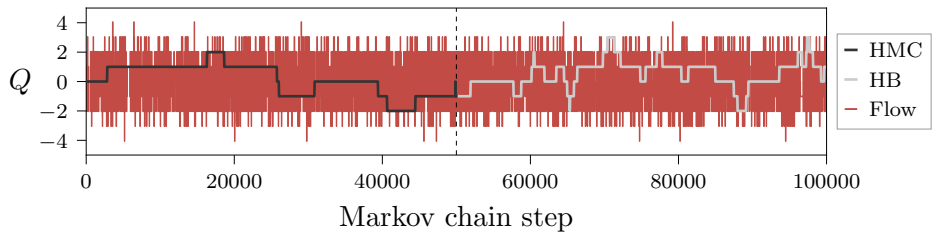

**Figure 4.16:** A comparison of the Markov chain history of the topological charge $Q$ for the Markov chains constructed to sample from the $\beta = 7$ distributions using HMC, heatbath, and flow-based MCMC. The HMC and heatbath Markov chains are severely affected by topological freezing, causing highly autocorrelated (slowly mixing) fluctuations in the topological charge. On the other hand, the flow-based Markov chain very rapidly samples from different topological sectors.

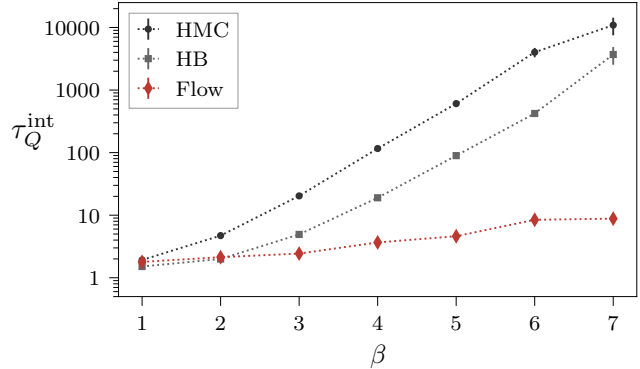

**Figure 4.17:** Integrated autocorrelation time $\tau_Q^{\text{int}}$ for measurements of the topological charge on HMC, heatbath, and flow-based ensembles. Uncertainties are determined by bootstrap resampling applied to the ten replicas of each ensemble. Dotted lines connecting the plot markers are included to guide the eye. Figure adapted from Fig. 4 of Pub. [3].

### 4.7.1 Action, observables, and parameters

Non-Abelian $SU(N)$ gauge theory in $(1+1)$ dimensions can be considered the quenched limit of $QCD_2$, and has previously been studied as a solvable theory that gives insights into the large-$N$ limit [37–39, 419–421]. For our purposes, the analytically computable results for observables make this theory a useful testing ground for gauge-equivariant flow-based models applied to $SU(N)$ variables. Two-dimensional $SU(N)$ gauge theory discretized on a lattice can be defined in much the same way as two-dimensional $U(1)$ gauge theory, as detailed below.

**Action.** A simple choice of discretized gauge action for $SU(N)$ variables is the Wilson action, which in $(1+1)$D is given by

$$S(U; \beta) = -\frac{\beta}{N} \sum_x \text{Re tr}\left[P_{01}(x)\right]. \qquad (4.48)$$

The plaquette $P_{01}(x)$ is computed analogously to $U(1)$ gauge theory as given in Eq. (4.46). The sum over $x$ ranges over sites in a two-dimensional lattice, which for this study is taken to have a square $L \times L = 16 \times 16$ geometry. The 't Hooft coupling $\lambda = 2N^2/\beta$ is the natural coupling to hold fixed as $N$ is varied. In our study we consider the gauge groups $SU(2)$ and $SU(3)$ and match the 't Hooft coupling of the distributions defined in each case. Three target distributions are given for both $SU(2)$ and $SU(3)$ by the three values of the 't Hooft coupling and associated values of $\beta$ listed in Table 4.4. These values are derived from the values $\beta \in \{4, 5, 6\}$ for the $SU(3)$ gauge group, which ranges over bare couplings used in recent studies of QCD; the physical behavior of the theory in $(1 + 1)$ dimensions is expected to be significantly different from $(3 + 1)$ dimensions, but this range of couplings provides a useful starting point at which the marginal distributions over each plaquette are far from uniform.

| SU($N$) | $L$ | $\beta$ | $\lambda = 2N^2/\beta$ | $n_{\mathrm{dof}}$ |
|---------|-----|---------|------------------------|--------------------|
| SU(2)   | 16  | $\{1.8, 2.2, 2.7\}$ | $\{4.4, 3.6, 3.0\}$ | 1536 |
| SU(3)   | 16  | $\{4.0, 5.0, 6.0\}$ | $\{4.5, 3.6, 3.0\}$ | 4096 |

**Table 4.4:** Choices of the coupling parameter defining three distributions for each of SU(2) and SU(3) lattice gauge theory. The choices of couplings were determined by fixing SU(3) couplings to integer values in a typical range and matching 't Hooft couplings to determine the SU(2) couplings. A count of the number of real degrees of freedom in the gauge configuration manifold is provided for reference for both SU(2) and SU(3) lattice gauge theory. Table adapted from Table IV of Pub. [2].

**Observables.** To confirm correctness, we measure the following observables on each of the ensembles studied:

1. Rectangular Wilson loops $\operatorname{tr} W_{l_0 \times l_1}(x)$ defined in terms of an $l_0 \times l_1$ rectangular region rooted at site $x$

2. Polyakov loops $\eta(x) = \operatorname{tr}\left[\prod_t U_0(\{t, x\})\right]$ consisting of a traced product of links wrapping around the temporal boundary of the lattice

3. The squared magnitude of Polyakov loops $|\eta(x)|^2 = \eta^*(x)\eta(x)$

Polyakov loops are included as extensive quantities (in one dimension) in place of the topological quantities measured for the U(1) theory. An exact center symmetry guarantees that $\langle \eta(x) \rangle = 0$ for all choices of parameters, but the squared magnitude $\langle |\eta(x)|^2 \rangle$ has non-zero expectation value. In pure-gauge theories, the distribution of the Polyakov loop observable can be used to derive order parameters for the deconfinement transition (see Sec. 2.2). This theory remains confined at all values of $\beta$, but these observables nonetheless serve as interesting benchmarks related to the center symmetry properties of our flow-based models.

Due to confinement, the expectation values of Wilson loops fall off exponentially with area. For the ensemble sizes used in this study, we were able to resolve Wilson loops with areas $l_0 l_1 \leq 4$ in lattice units. More general two-point functions of Polyakov loops given by $\eta^*(x)\eta(y)$ also fall off sharply with the separation $|x - y|$ and could not be resolved reliably for $x \neq y$, motivating the restriction to the squared magnitude $\eta^*(x)\eta(x)$ only.

Analytical derivations of the expectation values of the selected observables are given in Appendix A.

### 4.7.2   Model architecture, training, and MCMC

For each choice of $\beta$, we constructed an independent, identical gauge-equivariant flow-based model and trained each model on the distribution $p(U) \propto e^{-S(U;\beta)}$ defined by the Wilson gauge action. The construction and training of each model is described below. All models were implemented using the PyTorch framework [353].

**Prior distribution.**   As in the U(1) case, each flow-based model was constructed using the uniform distribution with respect to the Haar measure as the prior distribution. The prior probability density factor $r(U) = 1$ was used for all calculations of the model density, i.e. our definition of the SU($N$) Haar measure is taken to be normalized as

$$\int \mathcal{D}U \; r(U) = \int \mathcal{D}U \; = 1, \tag{4.49}$$

though we never explicitly need to construct it.

**Coupling layers.**   To implement the invertible flow in each flow-based model, we used a composition of coupling layers defined using the gauge-equivariant construction presented in this chapter. The architecture of each coupling layer was defined by the same striped masking pattern on the links, choice of loops, and PAFF loop-level masking pattern as in the case of U(1) lattice gauge theory. For this study, we used twice the number of layers per flow as in the $U(1)$ case, giving a total of six stacks of eight coupling layers corresponding to a total of six updates to every gauge link.

Within each inner coupling layer, we applied conjugation-equivariant SU($N$) kernels implemented using spectral maps (see Sec. 4.5). Each spectral map requires the definition of a transformation applied to the inner $(N-1)$-dimensional box $\Omega$. In the single-variable study above, independent spline transformations were applied to each dimension of $\Omega$ and we trained models to reproduce specific conjugation-invariant distributions as a proof of concept. Here, our aim is to construct expressive flows that can reproduce SU($N$) lattice gauge theory distributions having many more degrees of freedom. As such, we applied spline transformations to points mapped into $\Omega$ using an autoregressive construction for greater expressivity. For SU(2) gauge theory, this is trivially equivalent to independent application of splines, because $\Omega$ is one-dimensional. For the application to SU(3) gauge theory, however, this modifies how the context functions are defined. In our gauge-equivariant coupling layers, the parameters defining the spectral map are given by the outputs of context functions applied to gauge-invariant quantities (in the case of the PAFF masking pattern, these gauge-invariant quantities are traced plaquettes). To support autoregressivity, the parameters defining the spline transformation of each coordinate of $\Omega$ are given by the output of distinct context functions which receive the values of previous coordinates as inputs in addition to the gauge-invariant quantities given to all context functions (see Fig. 4.10).

For the transformation of each SU(2) variable, we defined the spline transformation acting on the single coordinate of $\Omega = (0, 1)$ to use $K = 3$ bins, corresponding to four knots. The parameters of each spline were given by context functions implemented as convolutional neural networks. In the PAFF architecture, the inputs to any context functions are the traced frozen plaquettes specified by the masking pattern; the real and imaginary components of these plaquettes define two input channels to the network. We defined each convolutional neural network to act on these inputs using two hidden layers consisting of 32 channels each and applied convolutions with kernel size $3 \times 3$ with periodic padding to maintain the lattice shape throughout. A leaky ReLU activation function [422] was applied between each convolutional operation.

For SU(3) variables, the two spline transformations used for the coordinates of $\Omega$ were defined by $K = 15$ bins, corresponding to sixteen knots. Each set of parameters for the two spline transformations in each coupling layer was given by the output of a separate convolutional neural network. The network defining the transformation of the horizontal coordinate in $\Omega$ was constructed using the same structure as in the case of SU(2) variables; the network defining the vertical coordinate transformation used the same structure with an additional input channel giving the value of the horizontal coordinate.

The SU($N$) gauge action also satisfies a symmetry under elementwise conjugation $U \to U^*$. For $N = 2$, this is equivalent to a permutation of the eigenvalues and the SU(2) spectral map is equivariant under this symmetry. For $N = 3$, this is a non-trivial symmetry that corresponds to exchanging the two corners of the triangular canonical cell $\Psi$ that are identified with the two non-identity center elements. We chose to additionally enforce this symmetry by a modification of the map from $\Psi$ into $\Omega$. In particular, the transformation of the space $\Psi$ was defined in this case by (1) mapping eigenvalues into a canonical half of the cell by complex conjugating if needed, (2) transforming the half-cell as before by mapping the half-cell to the open box $\Omega$, applying a spline transformation, and mapping back to the half-cell, and then (3) applying a complex conjugation if we did so in the first step.

**Training using volume bootstrapping.** We used the KL divergence as a loss function to train each model on the respective target distributions. In each training iteration, batches of size 3072 were used to stochastically estimate the loss function and gradients. We used the Adam optimizer with hyperparameters set to the defaults defined in the PyTorch library to iteratively update model parameters using the computed gradients. The learning rate determining the step size of the optimizer was initialized to $10^{-3}$ and reduced to $10^{-4}$ over the course of training. A total of between 14 000 and 29 000 optimization steps were used to train each model.

In this training process, we applied *volume bootstrapping* to reduce the total training costs. In each case the model was initially optimized to target the distribution associated with an $8 \times 8$ volume, then further optimization was applied targeting the final $16 \times 16$ volume. This was made possible by the use of convolutional neural networks throughout the flow-based model, which allow any model to be applied to arbitrary lattice volumes. When the local dynamics are not strongly affected by volume — which is the case for these two-dimensional lattice gauge theories at all volumes and for four-dimensional lattice gauge theories in the confined phase — a model trained to reproduce the dynamics at one choice of volume can be expected to produce a good approximation of the distribution at other choices of volumes. In practice, the majority of the total optimization steps used for each model were applied to train at the smaller $8 \times 8$ volume, and the model quality (as measured by the ESS) quickly converged after transferring to the larger volume. For example, Fig. 4.18 compares the model ESS over the course of training on the final $16 \times 16$ volume for a model that was previously trained on an $8 \times 8$ volume versus one that was initialized randomly. As shown in the figure, the transferred model converges to an asymptotic value of the ESS almost immediately whereas a randomly initialized model does not reach a similar value even after more than 10 000 training iterations.

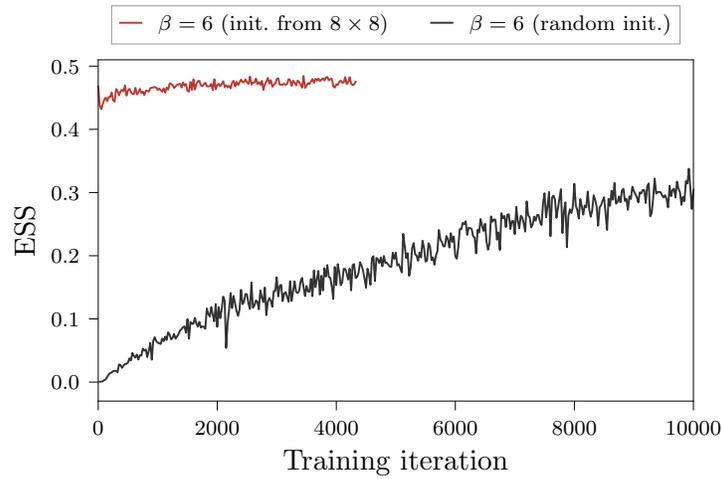

**Figure 4.18:** Comparison of the model quality as a function of training iterations for a model targeting SU(3) gauge theory on a $16 \times 16$ lattice that is respectively initialized randomly and from a model previously trained on an $8 \times 8$ lattice with the same coupling parameter. Values of the ESS are reported after averaging over bins of 25 steps each for clarity. Applying volume bootstrapping from the $8 \times 8$ lattice results in a model that converges to a plateau in model quality (as measured by the ESS) after very few training iterations at the $16 \times 16$ target volume, whereas the randomly initialized model does not completely converge to a similar model quality within the 10 000 training iterations executed for comparison. Figure adapted from Fig. 11 of Pub. [2].

Given the aim of demonstrating the efficacy of gauge-equivariant flow-based models, rather than exploring the practicalities of optimizing for $(1 + 1)$D lattice gauge theories, we did not significantly vary the hyperparameters defining the training of these models. In applications to problems at larger scales and in higher dimensions, carefully tuning the training procedure may be necessary to efficiently produce high-quality models. Automatic hyperparameter tuning frameworks have been developed to reduce the difficulties of tuning in the high-dimensional space associated with all possible hyperparameters, and applying these techniques to training models at larger scales may be quite fruitful [423–426].

**Final model quality.** As a measure of the absolute model quality, we computed the ESS across all flow-based models. The measured values are reported in Table 4.5. A fixed model architecture was used across all values of $\beta$ studied here, and unsurprisingly this results in a decreasing ESS as $\beta$ is increased. These larger values of $\beta$ correspond to distributions with more strongly peaked distributions of the plaquette and thus a stronger correlation structure between the gauge links. In general, as correlations are increased one expects more coupling layers to be required to produce a model of similar quality, because the coupling layers in the architectures used in this study transform each variable based on local information. See Sec. 3.7 for a discussion of the considerations associated with scaling these methods to a greater number of degrees of freedom.

|          | SU(2) |     |     | SU(3) |     |     |
| -------- | ----- | --- | --- | ----- | --- | --- |
| $\beta$  | 1.8   | 2.2 | 2.7 | 4.0   | 5.0 | 6.0 |
| ESS(%)   | 91    | 80  | 56  | 88    | 75  | 48  |

**Table 4.5:** Final ESS values after optimization of each of the six flow-based models constructed in this study. The final ESS decreases for this fixed model architecture as the coupling $\beta$ is made larger, but in all cases the ESS is significantly higher than 0, indicating that models can be used to efficiently draw samples from the target distributions. Table adapted from Table V of Pub. [2].

**MCMC.** The trained flow-based models were used to generate ensembles for their respective target values of $\beta$ using the flow-based MCMC method. A total of $102\,400$ proposals were generated from each model for use in Metropolis accept/reject steps, with 2400 steps of the Markov chain discarded for thermalization and the remaining $100\,000$ configurations used for observable measurements.

### 4.7.3 Monte Carlo tests

We next confirm agreement between measurements utilizing the ensembles of finite size considered here and the analytically determined observable values, check statistical scaling with the number of samples, and finally discuss the impact of symmetries on model quality.

**Tests of physical observables and statistical scaling.** The expectation values of a selection of measured observables are shown in Figure 4.19 for all flow-based ensembles. The observables are shown relative to the analytical values derived in Appendix A, and all measurements were found to be consistent with these analytical values within the statistical uncertainties. To estimate uncertainties on the measured quantities while accounting for autocorrelations, measurements were made on a subset of configurations thinned by the integrated autocorrelation time of each observable. Autocorrelation times were all measured to be smaller than approximately 4. The number of total measurements used to estimate each observable was tuned to give percent-level errors across all choices for a clearer comparison, with between 20 measurements (for the most precise observables) and $15\,000$ measurements (for the least precise observables) used for the data shown in the figure for each observable. Measurements on the full ensemble were confirmed to give agreement with the analytical results for all observables.

To confirm that statistical fluctuations are consistent with asymptotic $1/\sqrt{n}$ scaling with ensemble size $n$, we further measured the statistical errors on estimates of the average plaquette $\langle W_{1\times1}\rangle$ for subsets of the SU(3) $\beta = 6$ ensemble with sizes $n$ ranging from 10 to 20 000. Figure 4.20 depicts the resulting measured errors compared against the theoretical scaling in the asymptotic regime. Good agreement is found even for small subsets of the total ensemble.

**Symmetries.** As argued in Chapter 3, encoding symmetries exactly in flow-based models can reduce training complexity and improve the overall model quality. This chapter has focused on incorporating gauge symmetries into flow-based models without spoiling translational symmetries. The flow-based models in this study use gauge-equivariant coupling layers that accomplish this goal: by construction, the flow-based models used here are exactly symmetric under gauge symmetries, while the combination of a PAFF masking pattern and the use of convolutional neural networks also ensures that each flow-based model is symmetric under translations up to a $\mathbb{Z}_4 \times \mathbb{Z}_4$ symmetry breaking corresponding to translations that are not multiples of the stride of the masking pattern. We explore these and other symmetries qualitatively and quantitatively by comparing the effective action $S_{\text{eff}}(U) \equiv -\log q(U)$ of the model to the target $S(U)$ for samples $U$ related by symmetry transformations.

Figures 4.21 and 4.22 compare the measured effective action $S_{\text{eff}}(\cdot)$ and target action $S(\cdot)$ for sets of 32 samples drawn from the SU(3) model trained to target the coupling $\beta = 6$. The first and second figure respectively compare the measured action on each of the samples to the action for samples related by gauge transformations and translations. These operations leave the target action invariant by definition. In the case of gauge transformations, the effective action is completely invariant as seen by the exactly unchanged values of the effective action over a range of transformed configurations. In the case of translations, the effective action is invariant within each equivalence class of translations modulo shifts by 4 sites, but fluctuates across the choices of translations in different equivalence classes. This can be seen as fluctuations between different 'level sets' for each transformed configuration depicted in the second figure. These fluctuations are significantly smaller than the overall fluctuations in the action across different samples, indicating that a good approximate translational symmetry was learned by the flow-based model. A close agreement between the effective action of each orbit of a configuration can be seen between the effective and target actions in the left and right panels of both figures.

We further investigate the spatial structure of the residual fluctuations in the effective action with respect to the $\mathbb{Z}_4 \times \mathbb{Z}_4$ translational symmetry breaking in Figure 4.23. The values of the fluctuations in the effective action,

$$S_{\text{eff}}(T_{\delta x,\delta y} \cdot U) - \bar{S}_{\text{eff}}(U), \quad \bar{S}_{\text{eff}}(U) \equiv \frac{1}{V} \sum_{\delta x,\delta y} S_{\text{eff}}(T_{\delta x,\delta y} \cdot U), \qquad (4.50)$$

are depicted across all possible lattice translations for three samples drawn from the model. Each of the three plots depicts the deviations of the effective action from its mean value over all translations of the configuration. The scale is normalized to the standard deviation of the target action measured on model samples, as given by

$$\sigma_S \equiv \sqrt{\langle (S - \langle S \rangle)^2 \rangle}. \tag{4.51}$$

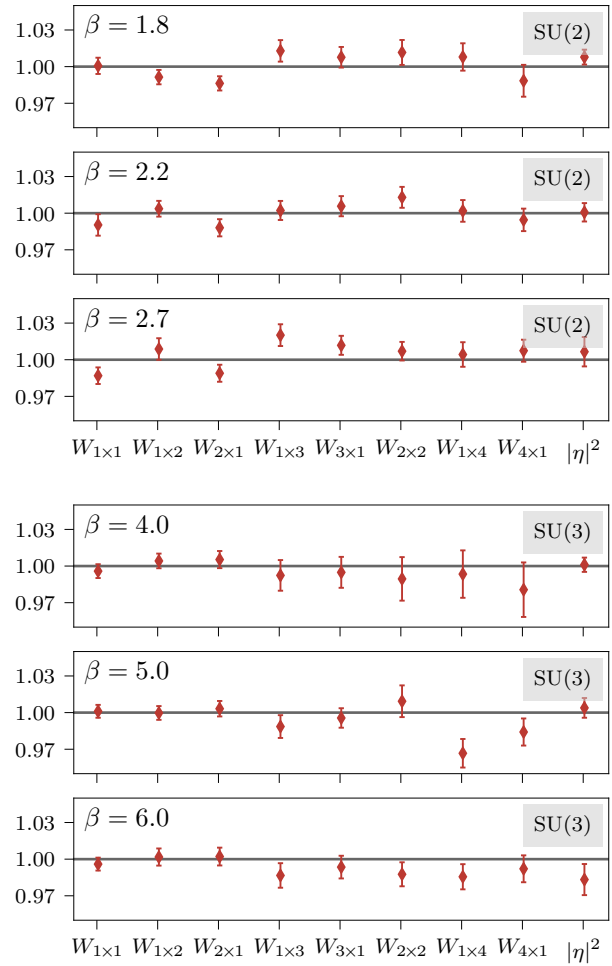

**Figure 4.19:** Observables computed for SU(2) (above) and SU(3) (below) lattice gauge theory on all flow-based ensembles. A subset of configurations was used for each measurement to give approximately equivalent relative uncertainties as described in the main text. Observables are shown relative to the analytically computed values and are consistent with these values to within statistical fluctuations. Figure adapted from Fig. 12 of Pub. [2].

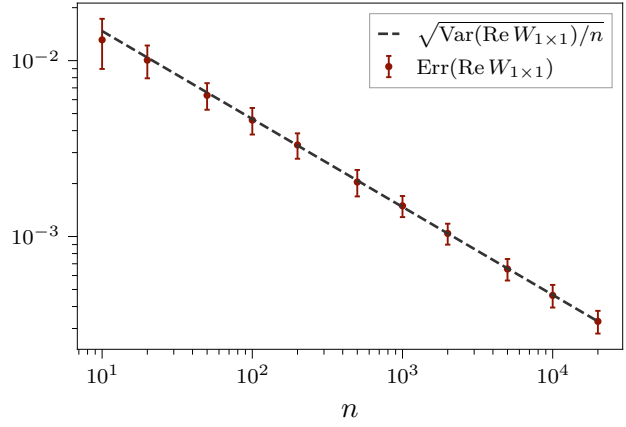

**Figure 4.20:** Measured statistical uncertainties on the observable $\mathrm{Re}\,W_{1\times1}$ for a range of sub-ensemble sizes $n$ drawn from the SU(3) $\beta = 6$ ensemble. The errors are consistent with the expected $1/\sqrt{n}$ scaling plotted with a dashed line. The line is normalized to the rightmost measured point. Figure adapted from Fig. 13 of Pub. [2].

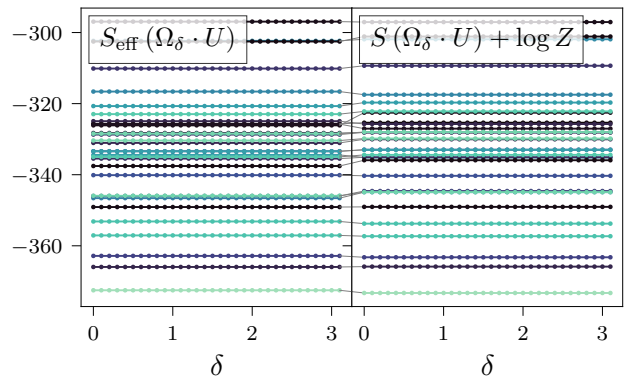

**Figure 4.21:** A comparison of the model effective action $S_{\mathrm{eff}}(U^{\Omega_\delta})$ to the target action $S(U^{\Omega_\delta})$ on a set of 32 gauge configurations sampled from the SU(3) $\beta = 6$ model distribution and transformed by a range of gauge transformations $\Omega_\delta$ parameterized by a continuous parameter $\delta$. The effective and target actions are both exactly invariant under gauge symmetry. Colors are randomly assigned to the sampled configuration and lines connect equivalent configurations in the left and right panels for easier visual comparison. Figure adapted from Fig. 14 of Pub. [2].

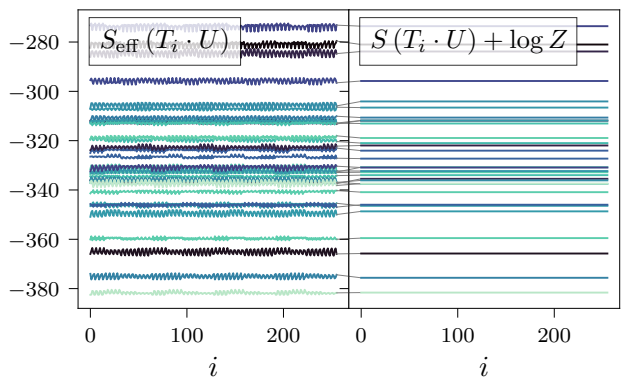

**Figure 4.22:** A comparison of the model effective action $S_{\text{eff}}(T_i \cdot U)$ to the target action $S(T_i \cdot U)$ on a set of 32 gauge configurations sampled from the SU(3) $\beta = 6$ model distribution and transformed by all 256 possible lattice translations $T_i$. The index $i$ corresponds to an offset $(\delta x, \delta y)$ by $i = \delta y + 16\,\delta x$. Colors are randomly assigned to the sampled configuration and lines connect equivalent configurations in the left and right panels for easier visual comparison. Figure adapted from Fig. 15 of Pub. [2].

The residual fluctuations can be seen to be much smaller than the scale of the overall fluctuations in the action, confirming the trend seen in Fig. 4.22 and indicating that the model has learned to produce similar probability density for different translations of the same physical configurations. Configurations related by shifts that are a multiple of the stride 4 of the masking pattern are have exactly equivalent effective action.

The target gauge theory actions are also invariant under both rotations and reflections of the lattice. In two dimensions this is an eight-dimensional symmetry group, but in four dimensions the corresponding hypercubic symmetry group is much larger, consisting of 48 elements. In this study, this symmetry group was not exactly enforced in the flow-based model. Doing so is possible but requires symmetrizing the structure of the masking pattern and the convolutional kernels used in context functions. Convolutional kernels may be symmetrized by weight sharing and masking patterns may be constructed to be rotation and reflection invariant. Neither of these symmetrization steps increase the cost of evaluating the flow, but do require non-trivial modifications of the implementation. Without explicit symmetrization, invariance under this distribution should nevertheless be approximately reproduced for a well-trained flow-based model. Figure 4.24 depicts a comparison of the effective and target actions associated with 32 configurations randomly drawn from the model distribution and transformed by all eight possible symmetry operations. The effective action $S_{\text{eff}}$ is not exactly invariant under reflections and rotations, but near-invariance can be seen by the nearly equivalent values of $S_{\text{eff}}$ across all configurations in each orbit of the symmetry group. Good agreement between $S_{\text{eff}}$ and $S$ across physically inequivalent configurations can also be seen by comparing the left and right panels of the figure.

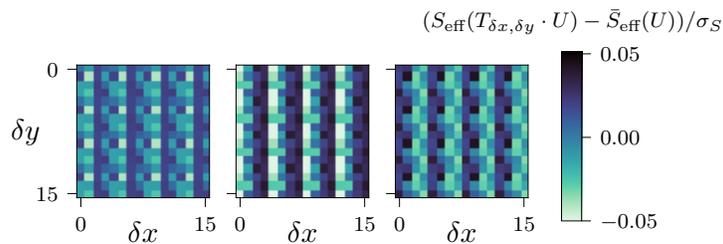

**Figure 4.23:** The measured fluctuations in the model effective action $S_{\text{eff}}$ as a function of the translation $(\delta x, \delta y)$ applied to three configurations randomly drawn from the SU(3) $\beta = 6$ model. Colors are normalized to one standard deviation of the total fluctuations in the action, as defined in the main text. Fluctuations of $S_{\text{eff}}$ are small relative to the scale of typical fluctuations in the action. Offsets that are a multiple of the stride of the masking pattern result in exactly equivalent values of $S_{\text{eff}}$. Figure adapted from Fig. 16 of Pub. [2].

In scaling the approaches of this work to higher dimensions and larger lattices, enforcing larger subsets of the translational symmetry group or a subset of the hypercubic symmetry group may be helpful in improving the training efficiency of flow-based models. The comparisons presented here provide a useful indicator of whether these sorts of improvements will be necessary: if similar measurements of the effective action under orbits of the symmetry group show significant fluctuations, even on models trained to nearly asymptotic performance, then exactly enforcing these symmetries may reduce fluctuations in $S_{\text{eff}} - S$ and improve the acceptance rate in flow-based MCMC.

The pure-gauge theories studied here additionally satisfy center symmetry (see Sec. 2.2) and symmetry under elementwise conjugation, $U \rightarrow U^*$. The flow-based models used for this application employed gauge-equivariant PAFF coupling layers that do not include non-contractible loops like Polyakov loops. These are the only loops that would transform under the center symmetry, and coupling layers that do not include non-contractible loops will thus be equivariant. The model distributions defined in this study were therefore fully invariant under center symmetry. For theories for which center symmetry is explicitly broken — such as theories with fermions in the fundamental representation of the gauge group — this implies that one *must* include non-contractible loops in the coupling layers to allow the flow-based model to produce distributions without center symmetry. The coupling layers were also constructed to be exactly equivariant under conjugation symmetry, resulting in model distributions symmetric under complex conjugation. Though this is a small symmetry group, it is straightforward to incorporate in models without modifying their structure significantly. These two symmetries were confirmed to be exactly enforced in all models produced in this study. In principle, either symmetry could be relaxed if it might result in higher expressivity overall. However, our constructions naturally allowed the enforcement of these symmetries without constraining the expressivity and for these architectures there is no reason to explicitly relax them.

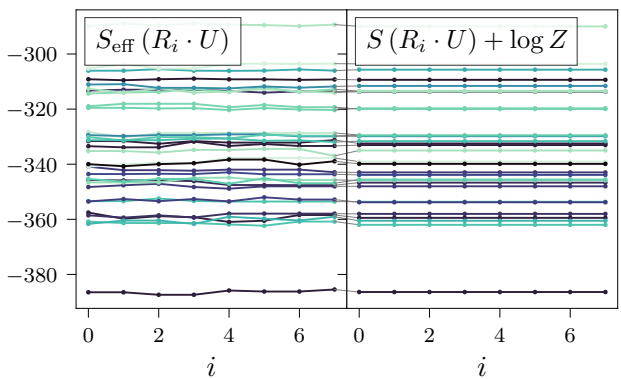

**Figure 4.24:** A comparison of the model effective action $S_{\text{eff}}(R_i \cdot U)$ to the target action $S(R_i \cdot U)$ on a set of 32 gauge configurations sampled from the SU(3) $\beta = 6$ model distribution and transformed by all 8 possible lattice rotations/reflections $R_i$. Colors are randomly assigned to the sampled configuration and lines connect equivalent configurations in the left and right panels for easier visual comparison. Figure adapted from Fig. 17 of Pub. [2].

### 4.7.4 Summary

In contrast to the U(1) lattice gauge theory studied in the previous section, $(1 + 1)$D SU($N$) lattice gauge theories do not possess a well-defined topological charge. There is thus no corresponding topological freezing to consider in this case, so this study instead investigated the absolute quality of models through tests of observables, symmetries, and the ESS. Though flow-based Markov chains must converge in the infinite statistics limit, tests of correctness provided a non-trivial confirmation that flow-based MCMC using the trained models had good asymptotic convergence for the finite Markov chains of size 102 400 employed in this study. In addition, the ESSs of flow-based models were measured to be higher than approximately 50 % across all choices of $\beta$; a 50 % ESS corresponds to roughly one effectively independent sample per two proposed samples if reweighting were applied, indicating that these models give high-quality approximations to the target distributions in an absolute sense. Though we observed a decrease in the final model ESS as $\beta$ was increased, this is not surprising given the fixed model architectures used here. These results are analogous to the scaling for fixed architectures discussed in the previous chapter. As discussed in Sec. 3.7, the model architecture can be scaled to increase expressivity if desired. That said, the highest values of $\beta$ used in this study correspond to typical values of $\beta$ used for the Wilson action in state-of-the-art studies, and if similar results were achieved in higher spacetime dimensions and for larger numbers of lattice sites with similar practical computational cost this would already provide a very effective sampling scheme at practically useful choices of the coupling. Future directions for the flow-based sampling program based on the studies presented in this dissertation are considered in Chapter 6.

# Chapter 5

# Observifolds

*Content in this chapter is partially adapted with permission from:*

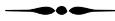

Section 2.5.2 explored the connection between signal-to-noise problems in observables and fluctuations of their complex phases over gauge configurations sampled in the evaluation of their path integral definition. Deforming the contour of integration of the path integral has previously been applied to significant effect in theories suffering from 'extensive' sign problems; these sorts of sign problems (and associated StN problems) appear for example when working with theories at a non-zero chemical potential or in real time. In this chapter, we detail the extension of these methods to reducing noise in *observables* in a lattice field theory that does not itself have an extensive sign problem. In this context, the Monte Carlo weights for ensemble generation are positive definite and the ensemble generation procedure is assumed to be fixed. Our aim is then to use the mathematical formalism of complex contour deformations to transform observables to new 'deformed observables' that have significantly reduced variance while having the same expectation value as their undeformed counterpart within the path integral.

The formalism underlying this approach is the same as the formalism applied to reduce extensive sign problems, but the application to observables results in characteristically different methods. We similarly construct families of deformed manifolds and optimize to identify manifolds with better properties, but the choice of deformed manifold is here motivated by reducing variance in particular observables, and may be different across distinct choices of observables. We use the term *observifold* to identify a manifold optimized for particular observables. A key contribution of the work in this chapter is the identification of useful objective functions and optimization methods to find observifolds that significantly improve the measurements of a range of observables.

Using these manifolds to define deformed observables without modifying the Monte Carlo weights also has the significant advantage of enabling efficient optimization and evaluation of observables defined by the deformed manifolds under study. This is expected to work well for the observables that are not extensive in the volume of the lattice, for which deformations that are localized to the region of influence of these observables should overlap well onto the vacuum distribution. We also find that stochastic optimization methods applied on $O(100)$ configurations are sufficient to determine observifolds resulting in significant variance reductions for observables measured in the applications presented here. In contrast to previous methods, the reuse of existing ensembles and small requirements on training data make it practical to apply these approaches to state-of-the-art calculations in which ensemble generation represents a large fraction of the computational costs, and for which repeatedly producing new Monte Carlo samples to iteratively optimize a deformation manifold would be impractical. Applying the methods described here to state-of-the-art calculations would thus result in significant gains in statistical precision at far lower cost than the costs of generating the exponentially larger ensembles needed to achieve equivalent gains in precision.

The remainder of this chapter is structured as follows. Section 5.1 reviews the application of contour deformations to lattice field theory path integrals. Section 5.2 presents our approach to defining deformed observables based on path integrals over observifolds and optimizing the choice of observifolds for observables of interest. Section 5.3 introduces the families of deformations that we study, including a novel family of deformed manifolds for SU($N$) variables. Finally, Sections 5.4 and 5.5 present applications of these methods to U(1) and SU($N$) lattice gauge theories.

## 5.1 Path integral contour deformations

Lattice-regularized path integrals are simply high-dimensional integrals. As such, standard complex analysis can be applied to these integrals. Recently, the technique of deforming the contour of integration into the domain of complexified field variables has been applied in a number of lattice field theory contexts. We review below why and how such techniques may be applied; see for example Ref. [209] for a comprehensive review of approaches based on complexification of field variables in path integrals and their use in solving sign problems.

### 5.1.1 A motivating example: Gaussian with a sign problem

We motivate the idea of complex contour deformations to solve sign and signal-to-noise problems by first looking at a simple one-variable example. Consider a single real variable $x$ with a quadratic 'action'

$$S_{\text{toy}}(x) = x^2/2. \tag{5.1}$$

We can define a 'path integral' representation of any 'observable' $\mathcal{O}(x)$ according to this action by

$$\langle \mathcal{O} \rangle = \frac{1}{Z} \int dx \, \mathcal{O}(x) e^{-S_{\text{toy}}(x)} = \frac{1}{Z} \int dx \, \mathcal{O}(x) e^{-\frac{x^2}{2}}, \tag{5.2}$$

where $Z = \int dx \, e^{-S_{\text{toy}}} = \sqrt{2\pi}$. This action defines a Gaussian probability measure $e^{-\frac{x^2}{2}}/Z$ that can be sampled to give Monte Carlo estimates of observables.

Many observables can be effectively computed using this Monte Carlo procedure, but observables that have strong phase fluctuations as a function of $x$ will encounter a sign problem preventing precise estimates. A simple representative observable with such a sign problem is the pure-phase quantity $e^{ikx}$, for which the sign problem becomes severe as $k$ is taken large. We demonstrate this by explicitly computing the expectation value and variance of the Monte Carlo estimator. Since the path integral in Eq. (5.2) is a Gaussian integral, the expectation value of the observable can be straightforwardly computed to be

$$\left\langle e^{ikx} \right\rangle = \frac{1}{Z} \int dx \, e^{ikx} e^{-\frac{x^2}{2}} = e^{-k^2/2}, \tag{5.3}$$

which falls to zero faster than exponentially in $k$. The estimator given by taking the sample mean of $\text{Re}[e^{ikx}] = \cos(kx)$ is the natural way to compute this real expectation value using Monte Carlo. (The sample mean of $\text{Im}[e^{ikx}]$ could also be used to compute the imaginary part of the expectation value if it was not known to be exactly zero.) For the estimator of the real component, we can explicitly compute the variance to be

$$\begin{aligned}
\text{Var}[\text{Re}[e^{ikx}]] &= \frac{1}{2} \left\langle |e^{ikx}|^2 \right\rangle + \frac{1}{2} \text{Re} \left\langle (e^{ikx})^2 \right\rangle - \left\langle \text{Re}[e^{ikx}] \right\rangle^2 \\
&= \frac{1}{2} + \frac{1}{2} \text{Re} \left\langle e^{2ikx} \right\rangle - e^{-k^2} \sim \frac{1}{2},
\end{aligned} \tag{5.4}$$

where we isolate the large-$k$ behavior of the variance in the final line. This $O(1)$ variance is exponentially larger than the signal $e^{-k^2/2}$, and the signal-to-noise ratio for a Monte Carlo estimate consisting of $n$ samples is therefore exponentially small in $k$,

$$\text{StN}[\text{Re}[e^{ikx}]] = \frac{\sqrt{n} e^{-k^2/2}}{\sqrt{\text{Var}[\text{Re}[e^{ikx}]]}} \sim \sqrt{n} e^{-k^2/2}. \tag{5.5}$$

A signal-to-noise ratio much smaller than 1 indicates that the quantity is essentially unconstrained by the Monte Carlo estimate because the statistical uncertainties are much larger than the magnitude of the signal. The only handle available to counteract the exponential degradation of the signal-to-noise in this approach is to correspondingly increase $n$; however, $O(e^{k^2})$ samples are required to achieve an $O(1)$ signal-to-noise ratio.

Figure 5.1 demonstrates this signal-to-noise problem in action. In the left panel, estimates of $\text{Re} \left\langle e^{ikx} \right\rangle$ are shown using $10\,000$ Monte Carlo samples. As $k$ is taken large, the signal-to-noise ratio becomes too small to allow any reliable estimates of the expectation value. In the right panel, the value of the real part of the path integrand defining the observable expectation value is shown for the choice of $k = 5$. Though the oscillations are not extremely rapid, the positive and negative values of the integrand provide nearly equal contributions and the expectation value and signal-to-noise ratio are thus exponentially small. This signal-to-noise and sign problem are representative of the issues encountered for noisy observables measured for theories with real actions. We introduce contour deformations and solve this toy sign problem in the following section.

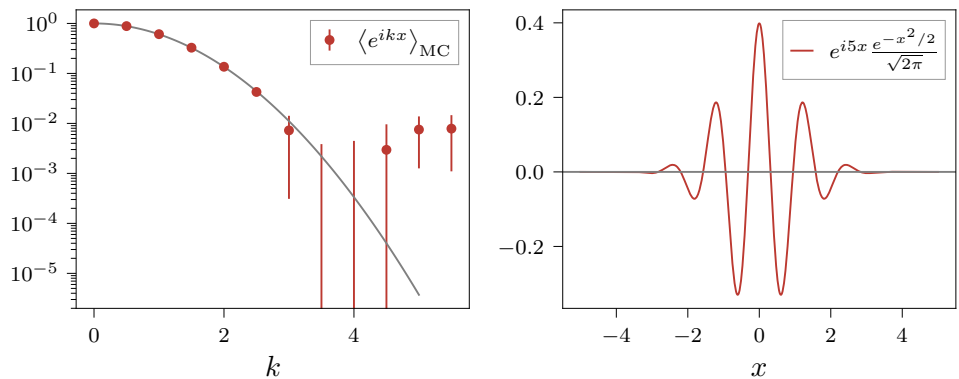

**Figure 5.1:** Left: Monte Carlo estimates of $\langle e^{ikx} \rangle$ on $10\,000$ samples from the toy 'path integral' considered in the main text. As $k$ is taken large, the signal-to-noise ratio becomes exponentially small and statistical uncertainties become too large to allow any reliable estimates. Right: the real part of the path integrand defining the observable for $k = 5$. The near-cancellation of positive and negative terms results in the exponentially small expectation value and signal-to-noise ratio for this value of $k$.

### 5.1.2 Deforming one-dimensional integrals

For a complex function of one variable that is holomorphic in some region $D$, the complex line integral around a closed, piecewise-differentiable curve $\gamma$ contained in that region is equal to zero [427]. This essentially follows from the Cauchy-Riemann equations and Stokes' theorem, and can be considered a special case of Cauchy's integral formula. This also forms the basis of *deformations* of the contour of a complex line integral. The integral of a function $f$ over a curve $\gamma_1$ is equal to the integral over another curve $\gamma_2$,

$$\int_{\gamma_1} f(z)\,dz = \int_{\gamma_2} f(z)\,dz, \tag{5.6}$$

if these curves together from a closed contour contained in a holomorphic region of $f$ for which

$$\int_{\gamma_2 \cdot \gamma_1^{-1}} f(z)\,dz = 0. \tag{5.7}$$

This is schematically depicted in Fig. 5.2.

The sign problem in the one-dimensional example given in Sec. 5.1.1 can in fact be completely solved by a contour deformation. First, we note that the analytic contin-uation of the integrand in the path integral definition of the expectation value $\langle e^{ikx} \rangle$ gives the function $e^{ikz}e^{-z^2/2}$ which is holomorphic for all $z \in \mathbb{C}$. The original integral can be considered a line integral along the real line within the complex domain of $z$. By holomorphy of the integrand, we are free to deform this line integral into the complex plane.

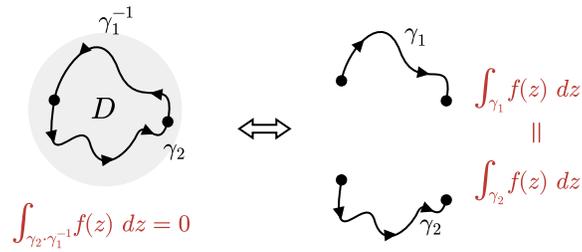

**Figure 5.2:** Schematic illustration of contour deformation within a region $D$ in which the integrand $f$ is holomorphic. Holomorphy implies that integration over a closed contour within this region gives zero, as shown on the left. As a result, integrating on either of the two curves forming the close region gives the same value, as shown on the right.

An effective choice of contour deformation is inspired by considering the form of the observable $e^{ikx}$. Fundamentally, the sign problem and associated signal-to-noise problem both arise because this quantity has magnitude 1 for all $x$. Monte Carlo samples of this unit-norm complex value must then have delicately canceling complex phases to produce the exponentially small expectation value on average. If instead the observable measured on samples of $x$ had magnitude approximately equivalent to the expectation value, the measurements would instead be *required* to coherently average together because any additional cancellation between the samples would spoil the expectation value. This can be directly achieved by deforming $x \to x + ik$.

The points $\mathbb{R} + ik \equiv \{x + ik : x \in \mathbb{R}\}$ consist of a valid contour of integration, and holomorphy of the integrand allows us to rewrite the path integral over this domain without affecting expectation values. Written as complex line integrals,

$$\langle e^{ikx} \rangle = \frac{1}{Z} \int_{\mathbb{R}} dz \, e^{ikz} e^{-\frac{z^2}{2}} = \frac{1}{Z} \int_{\mathbb{R}+ik} dz \, e^{ikz} e^{-\frac{z^2}{2}}. \tag{5.8}$$

Note that this is quite distinct from a change of variables, in that we integrate over an entirely distinct domain $\mathbb{R}+ik \subset \mathbb{C}$ rather than reparameterizing the integration over $\mathbb{R}$. To concretely perform the integration over $\mathbb{R}+ik$, however, coordinates must be chosen for the contour of integration. In all contour deformations discussed in this chapter, it will be convenient to use the original integration variable as the coordinate description of the complex contour of integration. In this simple example, this corresponds to writing $\int_{\mathbb{R}+ik} dz f(z) = \int dx \, f(z(x))$, where we associate each point in the complex contour of integration with a coordinate $x$ by $z(x) = x + ik$. The integral over the deformed contour of integration then simply takes the form of a standard integral over the real line, albeit with a different integrand, defined by

$$\left\langle e^{ikx} \right\rangle = \frac{1}{Z} \int dx \, e^{ikz(x)} e^{-\frac{z(x)^2}{2}} = \frac{1}{Z} \int dx \, e^{ik(x+ik)} e^{-\frac{(x+ik)^2}{2}}. \tag{5.9}$$

Finally, this new integrand can be manipulated to be written as the expectation value of a distinct observable with respect to the *original* Monte Carlo weights:

$$
\begin{aligned}
\left\langle e^{ikx} \right\rangle &= \frac{1}{Z} \int dx \, e^{ikx} e^{-k^2} e^{-\frac{x^2}{2} - ikx + \frac{k^2}{2}} \\
&= \frac{1}{Z} \int dx \, e^{-\frac{k^2}{2}} e^{-\frac{x^2}{2}} = \left\langle e^{-\frac{k^2}{2}} \right\rangle = e^{-\frac{k^2}{2}} \left\langle 1 \right\rangle.
\end{aligned}
\tag{5.10}
$$

As a result of this contour deformation, we have rather spectacularly solved the sign problem in this toy theory! From Eq. (5.10) one can see that the original observable of interest is equivalent to the observable $e^{-\frac{k^2}{2}}$ which has no sign or signal-to-noise problem, as there are no remaining phase fluctuations. In fact, there is no remaining Monte Carlo noise at all — the factor of $e^{-\frac{k^2}{2}}$ was moved outside the integral in Eq. (5.10) and this contour deformation was thus sufficient to completely solve for the value of this observable.

This result is atypical, and in general we do not expect contour deformations to remove all Monte Carlo noise. Contour deformations that shift by values different from $ik$ better demonstrate the effect of general contour deformations in practice. For a shift by $ik'$, following the same manipulations relates the original expectation value to a non-trivial new family of observables, parameterized by $k'$,

$$\left\langle e^{ikx} \right\rangle = \left\langle e^{i(k-k')x} e^{-k'(k-\frac{k'}{2})} \right\rangle. \tag{5.11}$$

For any choice of $k'$, the resulting observable is guaranteed to have identical expectation value. In general, this observable still depends on $x$, but will have milder phase fluctuations when $k' \approx k$. Thus for shifts close to this optimal value, the Monte Carlo noise on these observables would be significantly reduced.

Curvature of the deformed manifold also plays a role in the resulting form of the integrand. Though we have already described an optimal contour deformation above, it is illuminating to work through a second family of contour deformations in this toy theory. We consider in particular a 'Gaussian shift' contour of integration given by the coordinate map

$$z(x) = x + i\alpha e^{-\frac{x^2}{2}}. \tag{5.12}$$

Integrating on this Gaussian-shifted contour will produce identical expectation values as well,

$$\left\langle e^{ikx} \right\rangle = \frac{1}{Z} \int_{\text{G. shift}} dz \, e^{ikz} e^{-\frac{z^2}{2}}. \tag{5.13}$$

However, when using the coordinates given in Eq. (5.12) it is important to account for the Jacobian determinant of the map. The full deformed integral written in terms of

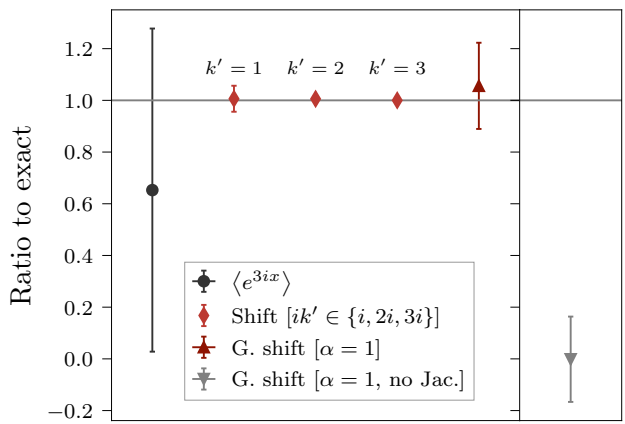

**Figure 5.3:** Comparison of Monte Carlo estimates of the original observable to the new observables defined by contour deformation. All estimates are measured using 10 000 Monte Carlo samples and are plotted relative to the exact value $\langle e^{3ix} \rangle = e^{-9/2}$. Constant-shift manifolds give significant improvements in precision, with the optimal choice of $k' = 3$ giving a zero-variance result. The observable associated with the Gaussian-shift manifold also improves upon the original observable. An incorrect version of the Gaussian-shift observable that excludes the Jacobian factor is also plotted for comparison; without the Jacobian factor, the observable no longer has an identical expectation value.

the original coordinates is given by

$$\frac{1}{Z} \int dx \, \frac{\partial z}{\partial x} e^{ikz(x)} e^{-\frac{z(x)^2}{2}}, \quad \text{where } \frac{\partial z}{\partial x} = 1 - ix\alpha e^{-\frac{x^2}{2}}. \tag{5.14}$$

No absolute value is taken over the Jacobian factor. Factoring out the original Monte Carlo weight from this path integrand, the Gaussian shift deformations define another family of observables with identical expectation value to the original observable

$$\langle e^{ikx} \rangle = \left\langle \left( 1 - ix\alpha e^{-\frac{x^2}{2}} \right) e^{ikz(x)} e^{-\frac{z(x)^2}{2} + \frac{x^2}{2}} \right\rangle. \tag{5.15}$$

Figure 5.3 compares Monte Carlo estimates of the observable $e^{ikx}$ with $k = 3$ to equivalent observables defined by constant-shift manifolds and a Gaussian-shift manifold. The figure compares three choices of the constant shift $k' \in \{1, 2, 3\}$, and values of $k'$ closer to $k$ can be seen to provide the most precise estimates of the true expectation value. The observable associated with the Gaussian-shift manifold with $\alpha = 1$ also improves over the original observable, but, as shown in the figure, it is absolutely crucial to include the Jacobian factor in the definition. Excluding the Jacobian factor results in an incorrect estimate that significantly differs from the exact result. The need to efficiently compute and include this Jacobian factor in high-dimensional cases motivates the structure of families of path integral contour deformations introduced later in this chapter.

In explorations of path integral deformations to reduce observable sign and signal-to-noise problems in the remainder of this chapter, we find that when it is possible to identify integrands with a structure analogous to $e^{ikx}$, imaginary shifts of the corresponding variables effectively reduce the average integrand magnitude. Monte Carlo estimates based on the new observables derived from these manifold deformations are significantly more precise.

### 5.1.3 Deforming path integrals

With some intuition in hand from a simple one-dimensional example, we review how contour deformations can be applied to the more complex case of path integrals over many variables.

Though multivariate complex analysis is a rich field of study, generalizing contour deformations to this setting only requires a small subset of the machinery available in the field. In particular, deforming an integral over an $\mathscr{N}$-dimensional manifold $\mathcal{M}_A$ to an integral over a manifold $\mathcal{M}_B$ will result in the same integral value if (1) $\mathcal{M}_A$ and $\mathcal{M}_B$ taken with appropriate orientations bound a closed region $\mathcal{D}$ and (2) the integrand is holomorphic in a region containing $\mathcal{D}$. A proof based on Stokes' theorem is presented in Ref. [209]. This allows the integration domain of high-dimensional path integrals in lattice field theory to be deformed when the path integral weights and observables are jointly holomorphic. Using notation tailored to our lattice gauge theory applications, this means that

$$\langle \mathcal{O}(U) \rangle = \frac{1}{Z} \int_{\mathcal{M}} \mathcal{D}U \, \mathcal{O}(U) \, e^{-S(U)} = \frac{1}{Z} \int_{\widetilde{\mathcal{M}}} \mathcal{D}\widetilde{U} \, \mathcal{O}(\widetilde{U}) \, e^{-S(\widetilde{U})} \tag{5.16}$$

when $\mathcal{O}(U)e^{-S(U)}$ is a holomorphic integrand and $\widetilde{\mathcal{M}}$ is a deformation of $\mathcal{M}$. In general, we write integrals over deformed contours using integration variables with tildes, in analogy to moving from integration over the real variable $x$ to the complex variable $z$ in one-dimensional complex analysis. It is worth highlighting that this is simply a notational choice, and the content of the contour deformation is entirely contained in the fact that we now integrate over $\widetilde{\mathcal{M}}$ rather than $\mathcal{M}$. In the following, we also leave $\mathcal{M}$ implicit when it is clear from context that an integral is evaluated on the original manifold.

Depending on the theory under study, the original $\mathscr{N}$-dimensional manifold of integration may be $\mathbb{R}^{\mathscr{N}}$ (e.g. in the case of scalar field theory) or a more complicated manifold. In the latter case, we restrict to considering manifolds that can be given global coordinates. This allows the original integral to be written over (a subset of) $\mathbb{R}^{\mathscr{N}}$ in all cases. Using the notation $\Omega$ for the set of coordinates to be integrated over, the path integral given in Eq. (5.16) can be written as an explicit integral over real degrees of freedom:

$$\frac{1}{Z} \int \mathcal{D}U \, \mathcal{O}(U) \, e^{-S(U)} = \frac{1}{Z} \int_{\mathcal{B} \subset \mathbb{R}^{\mathscr{N}}} \mathcal{D}\Omega \, J(\Omega) \mathcal{O}(U(\Omega)) \, e^{-S(U(\Omega))}, \tag{5.17}$$

where $J(\Omega) = \det \partial U / \partial \Omega$ is the Jacobian factor relating the measure on the coordinates

$\mathcal{D}\Omega$ to the measure on the manifold $\mathcal{D}U$, and $\mathcal{B}$ is the domain over which the coordinates range. In all applications considered in this chapter, including non-Abelian lattice gauge theory, it is possible to choose such global coordinates. We also choose to work with identical coordinates $\Omega$ for the deformed and undeformed manifold, although this is not required, allowing us to use the same underlying degrees of freedom to specify the deformed integral as

$$\frac{1}{Z}\int_{\widetilde{\mathcal{M}}}\mathcal{D}\widetilde{U}\,\mathcal{O}(\widetilde{U})\,e^{-S(\widetilde{U})} = \frac{1}{Z}\int_{\mathcal{B}\subset\mathbb{R}^{\mathscr{N}}}\mathcal{D}\Omega\,\widetilde{J}(\Omega)\mathcal{O}(\widetilde{U}(\Omega))e^{-S(\widetilde{U}(\Omega))}. \tag{5.18}$$

The only distinction is in the choice of map from coordinates to the manifold variables and the corresponding Jacobian factors $\widetilde{J}(\Omega) = \det \partial\widetilde{U}/\partial\Omega$. For gauge theories with gauge group $G$, the map defining the original manifold produces link variables $U(\Omega) \in G$, whereas the map defining the deformed manifold produces link variables $\widetilde{U}(\Omega)$ in the complexification of $G$. This is discussed further in the context of the groups U(1) and SU($N$) in Sec. 5.3.

Combining Eqs. (5.17) and (5.18), one can write the deformed path integral using the original measure,

$$\begin{aligned}\frac{1}{Z}\int_{\widetilde{\mathcal{M}}}\mathcal{D}\widetilde{U}\,\mathcal{O}(\widetilde{U})\,e^{-S(\widetilde{U})} &= \frac{1}{Z}\int\mathcal{D}U\,\left[\frac{\widetilde{J}(\Omega(U))}{J(\Omega(U))}\right]\mathcal{O}(\widetilde{U}(U))\,e^{-S(\widetilde{U}(U))}\\ &= \frac{1}{Z}\int\mathcal{D}U\,\widetilde{J}(U)\,\mathcal{O}(\widetilde{U}(U))\,e^{-S(\widetilde{U}(U))},\end{aligned} \tag{5.19}$$

where $\widetilde{U}(U)$ can be concretely computed by sending $U$ back through the coordinate map to get $\Omega(U)$ and then $\widetilde{U}(\Omega(U))$, and the Jacobian factors $\widetilde{J}(\Omega(U))$ and $J(\Omega(U))$ are computed similarly. The ratio of Jacobian factors in square brackets is denoted by $\widetilde{J}(U) = \widetilde{J}(\Omega(U))/J(\Omega(U))$, and can be interpreted as the total Jacobian of the map $\widetilde{U}(U)$. The rewriting in Eq. (5.19) allows the path integral contour deformation to be interpreted in terms of the original path integral structure without the need to make a cumbersome translation to an explicit set of coordinates except when initially defining $\widetilde{U}(U)$ and $\widetilde{J}(U)$.

When integrating according to the deformed path integral, one must be able to compute the Jacobian ratio $\widetilde{J}(U)$, or equivalently $\widetilde{J}(\Omega)$ and $J(\Omega)$ in the coordinate picture. Computing the Jacobian determinant requires $O(\mathscr{N}^3)$ operations in general for $\mathscr{N}$-dimensional integrals; in some cases, net gains have been made by accepting this cost while making exponential progress in reducing the extensive sign problem (see Sec. 5.1.4). Since $\mathscr{N}$ scales with the lattice volume, however, this is a prohibitive cost for state-of-the-art lattice calculations. Stochastic estimators have been proposed to circumvent this cost in some settings [428–431]. The cost of exactly computing the Jacobian determinant can also be reduced to $O(\mathscr{N})$ when the Jacobian is diagonal or triangular [432, 433]. Ref. [433] in particular proposes the use of coupling layers analogous to the approach applied for ensemble generation in the previous two chapters. In this chapter, we use the similar method of constructing manifolds based on an autoregressive architecture that limits the transformation of each degree of freedom to depend only on previous degrees of freedom in a canonical order. See Chapter 3 for a discussion of constructing invertible transformations with tractable Jacobian determinant factors using coupling layers or autoregressive transformations. In contrast to the work in Chapter 3, contour deformations applied in this chapter do not need to be constructed to exactly satisfy translational invariance because our approach is to target observables which are localized on the spacetime lattice. In particular, we find that it is important to deform degrees of freedom local to the region in which an observable is defined. This motivates the choice of an autoregressive structure rather than coupling layers in this chapter. Nonetheless, coupling layers could also be constructed without respecting translational symmetry, and it is an interesting future line of study to consider replacing the autoregressive structure used here with other choices of transformations including coupling layers.

### 5.1.4 Related work

The path integral contour deformations defined in the previous section have been applied in prior works to mitigating or solving extensive sign problems in several lattice field theories. Extensive sign problems broadly occur in two cases: either when working with a Euclidean path integral with a complex action, for example arising from a chemical potential coupled to a fluctuating conserved charge, or when working with a path integral involving regions with Minkowski signature (i.e. involving real-time evolution). In the first case, the path integral weights $e^{-S}$ have a complex phase set by $\mathrm{Im}\,S$, which scales extensively with the spacetime volume. In the second case, the path integral weights $e^{iS}$ have a complex phase set by $S \in \mathbb{R}$, which also has fluctuations generically scaling with the spacetime volume.

Both situations are problematic for Monte Carlo approaches to path integrals. A Monte Carlo calculation of expectation values of observables relies on interpreting the path integral weights as a positive-definite probability measure that can be sampled. When this is not the case, reweighting can potentially be applied to give estimates. For example, in the case of a Euclidean path integral with complex action, the expectation value of an observable $\mathcal{O}$ can be written using reweighting as

$$\langle \mathcal{O} \rangle_S = \frac{\int \mathcal{D}U \,\left[ \mathcal{O}(U) e^{-i\,\mathrm{Im}\,S} \right] e^{-\,\mathrm{Re}\,S}}{\int \mathcal{D}U \,\left[ e^{-i\,\mathrm{Im}\,S} \right] e^{-\,\mathrm{Re}\,S}} = \frac{\langle \mathcal{O}(U) e^{-i\,\mathrm{Im}\,S} \rangle_{\mathrm{Re}\,S}}{\langle e^{-i\,\mathrm{Im}\,S} \rangle_{\mathrm{Re}\,S}}, \qquad (5.20)$$

where $\langle \cdot \rangle_{\mathrm{Re}\,S}$ indicates taking the average with respect to the positive-definite weights $e^{-\,\mathrm{Re}\,S}$. Both the numerator and denominator of Eq. (5.20) can in principle be computed using Monte Carlo sampling. However, interpreting the path integral defined by the action $\mathrm{Re}\,S$ as a 'phase-quenched' theory, one can argue that the denominator is the ratio of partition functions $Z/Z_{pq}$, where $Z$ is the partition function of the original theory and $Z_{pq}$ is the phase-quenched partition function (see for example Ref. [434]). The ratio of two partition functions will scale exponentially in the volume. The Monte Carlo estimator of this quantity, however, is given by the sample mean of unit-norm complex numbers, and this results in an exponentially severe sign problem.

The past decade has seen the development of several methods that utilize complexi-fied field space and reduce extensive sign problems in some lattice field theories. Below, we discuss these methods and their relation to the present work.

**Lefschetz thimbles.**    Methods of complex contour deformations have close ties with computing properties of field theories on Lefschetz thimbles [435], which correspond to particular manifolds in complexified field space on which the phase of the integrand is constant. Several techniques have been developed to directly simulate on these thimbles with promising results demonstrated in theories with extensive sign problems arising from non-zero chemical potential and real-time dynamics [429, 436–458]. In some cases these methods face a residual sign problem from combining the results of integrating on many thimbles or from the Jacobian factor that must be introduced to integrate on such surfaces.

**Generalized Lefschetz thimbles.**    To circumvent solving for the description of Lefschetz thimbles and simulating exactly on the resulting manifold, a method of generalized Lefschetz thimbles has also been developed, in which contours approaching the Lefschetz thimbles can be used to exactly reproduce results on the original manifold while mostly removing sign and complex phase fluctuations [459, 460]. This generalization makes it possible to apply Lefschetz thimble methods in a broader range of contexts, and extensive sign problems have been addressed in several systems at non-zero chemical potential [432, 433, 459–467], with complex couplings [468–470], and in real time [431, 471].

Within the generalized Lefschetz thimbles methods, some works apply 'sign optimization' or 'path optimization' to search for good choices of integration manifolds numerically [432, 463, 466–468, 471–475]. This approach is based on constructing a family of manifolds written in terms of a set of free parameters that can be optimized to improve a measure of quality of the resulting manifold. In the context of extensive sign problems, the measure of quality is given by the 'average sign', i.e. the expectation value over the complex phase factors that must be included by reweighting. Increasing this average sign factor reduces the severity of the sign problem.

The work presented in this chapter uses path integral contour deformations with exactness guarantees in much the same way as the method of generalized Lefschetz thimbles. However, the focus of our work is on finding choices of manifolds that reduce sign problems in observables under real distributions rather than ones that converge to Lefschetz thimbles reducing sign problems for complex distributions. To this end, we follow a similar approach of parameterizing and numerically optimizing the choice of manifold as in the case of the path- and sign-optimized approaches, but construct distinct measures of quality aimed at quantifying the effect of manifolds on signal-to-noise problems in observables.

**Complex Langevin.**    Rather than simulating on a manifold with the same real dimension as the original manifold of integration, it is possible to relate complex-weighted path integrals to expectation values under field configurations traversed by complex Langevin dynamics ranging over the whole of complexified field space. Refs. [199, 476, 477] provide several comprehensive reviews on the status of this approach. Ref. [478] contrasts this approach to the method of simulating on Lefschetz thimbles in the context of a one-variable model, highlighting the relation between the thimble structure and the distribution sampled by Langevin dynamics in complexified field space.

## 5.2 Manifold deformations for observables

Our focus in this chapter is on addressing observables with difficult signal-to-noise problems in the context of path integrals with positive-definite weights. This is a significant obstacle to state-of-the-art Euclidean path integral estimates of observables in theories with zero chemical potential. By assumption, there is no extensive sign problem associated with fluctuations in $\mathrm{Im}\, S$ in this setting, and as a result we use path integral contour deformations to define new observables under a fixed Monte Carlo sampling scheme as detailed below.

### 5.2.1 Deformed observables

The key insight in this approach is that one can choose to deform the numerator alone in the path integral definition of the expectation value of an observable $\mathcal{O}$,

$$
\begin{aligned}
\langle \mathcal{O} \rangle &= \frac{\int \mathcal{D}U \; \mathcal{O}(U) \, e^{-S(U)}}{\int \mathcal{D}U \; e^{-S(U)}} = \frac{\int_{\widetilde{\mathcal{M}}} \mathcal{D}\widetilde{U} \; \mathcal{O}(\widetilde{U}) \, e^{-S(\widetilde{U})}}{\int \mathcal{D}U \; e^{-S(U)}} \\
&= \frac{\int \mathcal{D}U \; \widetilde{J}(U) \mathcal{O}(\widetilde{U}(U)) \, e^{-S(\widetilde{U}(U))}}{\int \mathcal{D}U \; e^{-S(U)}}.
\end{aligned}
\tag{5.21}
$$

Unlike in the setting of extensive sign problems, the denominator $Z = \int \mathcal{D}U \; e^{-S(U)}$ is an integral over positive-definite weights that is already factored into the Monte Carlo weights. Comparing this to the reweighting approach for theories with an extensive sign problem in Eq. (5.20), this numerator-only deformation would not be suitable in that setting due to the extensive sign problem also arising in the estimation of the denominator $\left\langle e^{-i\,\mathrm{Im}\, S} \right\rangle_{\mathrm{Re}\, S}$.

Further manipulating Eq. (5.21) to factor out the original Monte Carlo weights, we can identify the deformed path integral as the expectation value of a new observable under the same physical theory,

$$
\langle \mathcal{O} \rangle = \frac{1}{Z} \int \mathcal{D}U \; \left\{ \widetilde{J}(U) \mathcal{O}(\widetilde{U}(U)) e^{-S(\widetilde{U}(U))+S(U)} \right\} e^{-S(U)} \equiv \langle \mathcal{Q} \rangle,
\tag{5.22}
$$

where we define $\mathcal{Q}$ to be the *deformed observable* given by

$$
\mathcal{Q}(U) = \widetilde{J}(U) \mathcal{O}(\widetilde{U}(U)) e^{-S(\widetilde{U}(U))+S(U)}.
\tag{5.23}
$$

We use the terms *base observable* or *original observable* for the original choice of $\mathcal{O}$ associated with a deformed observable $\mathcal{Q}$. For any base observable $\mathcal{O}$, each choice of manifold deformation gives rise to a deformed observable $\mathcal{Q}$. Any such deformed observable will in general look nothing like the base observable under study, but by the properties of holomorphic complex integrals it is guaranteed to have an identical expectation value.

On the other hand, the variance of the Monte Carlo estimator associated with a deformed observable may differ significantly from the variance of estimates of the original observable. The variance of the real and imaginary components of $\mathcal{O}$ are defined by

$$
\begin{aligned}
\text{Var}[\text{Re}\,\mathcal{O}] &= \frac{1}{2}\left\langle|\mathcal{O}^2|\right\rangle + \frac{1}{2}\text{Re}\left\langle\mathcal{O}^2\right\rangle - [\text{Re}\left\langle\mathcal{O}\right\rangle]^2, \\
\text{Var}[\text{Im}\,\mathcal{O}] &= \frac{1}{2}\left\langle|\mathcal{O}^2|\right\rangle - \frac{1}{2}\text{Re}\left\langle\mathcal{O}^2\right\rangle - [\text{Im}\left\langle\mathcal{O}\right\rangle]^2.
\end{aligned}
\tag{5.24}
$$

In comparison, the variance of the real and imaginary components of $\mathcal{Q}$ are

$$
\begin{aligned}
\text{Var}[\text{Re}\,\mathcal{Q}] &= \frac{1}{2}\left\langle|\mathcal{Q}^2|\right\rangle + \frac{1}{2}\text{Re}\left\langle\mathcal{Q}^2\right\rangle - [\text{Re}\left\langle\mathcal{Q}\right\rangle]^2, \\
\text{Var}[\text{Im}\,\mathcal{Q}] &= \frac{1}{2}\left\langle|\mathcal{Q}^2|\right\rangle - \frac{1}{2}\text{Re}\left\langle\mathcal{Q}^2\right\rangle - [\text{Im}\left\langle\mathcal{Q}\right\rangle]^2.
\end{aligned}
\tag{5.25}
$$

Though we have analytically shown that $\langle\mathcal{O}\rangle = \langle\mathcal{Q}\rangle$, the other terms in Eq. (5.25) cannot be simply related to the terms in Eq. (5.24). Writing out the terms $\left\langle|\mathcal{Q}^2|\right\rangle$ and $\left\langle\mathcal{Q}^2\right\rangle$ explicitly,

$$
\begin{aligned}
\left\langle|\mathcal{Q}^2|\right\rangle &= \frac{1}{Z}\int\mathcal{D}U\left|\widetilde{J}(U)\mathcal{O}(\widetilde{U}(U))\right|^2 e^{-2\,\text{Re}\,S(\widetilde{U}(U))+S(U)} \\
\left\langle\mathcal{Q}^2\right\rangle &= \frac{1}{Z}\int\mathcal{D}U\widetilde{J}(U)^2\mathcal{O}(\widetilde{U}(U))^2 e^{-2S(\widetilde{U}(U))+S(U)},
\end{aligned}
\tag{5.26}
$$

one can see that the factor of $\widetilde{J}(U)\exp[-S(\widetilde{U}(U))+S(U)]$ in the definition of $\mathcal{Q}$ is squared in these terms and prevents identifying either of the expressions in Eq. (5.26) as deformations of the corresponding integrals for the base observable. It is also these terms that result in exponential degradation of the signal-to-noise ratio for observables with a StN problem, arising from the fact that they fall off exponentially more slowly than the expectation value itself. For each observable under study, our aim is thus to find a deformed observable for which the terms in Eq. (5.26) are significantly smaller than the corresponding terms for $\mathcal{O}$, and therefore the variance of the real/imaginary estimators of $\mathcal{Q}$ are minimized.

Using positivity of the variance, the two terms in Eq. (5.26) can be related. Considering the variance of the imaginary component of $\mathcal{Q}$, we can upper bound $\left\langle\mathcal{Q}^2\right\rangle$ by

$$
\begin{aligned}
0 \leq \text{Var}[\text{Im}\,\mathcal{Q}] &= \frac{1}{2}\left\langle|\mathcal{Q}^2|\right\rangle - \frac{1}{2}\text{Re}\left\langle\mathcal{Q}^2\right\rangle - [\text{Im}\left\langle\mathcal{Q}\right\rangle]^2 \\
0 &\leq \frac{1}{2}\left\langle|\mathcal{Q}^2|\right\rangle - \frac{1}{2}\text{Re}\left\langle\mathcal{Q}^2\right\rangle \\
\text{Re}\left\langle\mathcal{Q}^2\right\rangle &\leq \left\langle|\mathcal{Q}^2|\right\rangle.
\end{aligned}
\tag{5.27}
$$

Considering the variance of the real component of $\mathcal{Q}$, we can then lower bound $\left\langle\mathcal{Q}^2\right\rangle$

by

$$0 \leq \text{Var}[\text{Re } \mathcal{Q}] = \frac{1}{2} \left\langle |\mathcal{Q}^2| \right\rangle + \frac{1}{2} \text{Re} \left\langle \mathcal{Q}^2 \right\rangle - [\text{Re} \left\langle \mathcal{Q} \right\rangle]^2$$

$$0 \leq \frac{1}{2} \left\langle |\mathcal{Q}^2| \right\rangle + \frac{1}{2} \text{Re} \left\langle \mathcal{Q}^2 \right\rangle \qquad (5.28)$$

$$-\left\langle |\mathcal{Q}^2| \right\rangle \leq \text{Re} \left\langle \mathcal{Q}^2 \right\rangle.$$

This implies that minimizing the positive-definite quantity $\left\langle |\mathcal{Q}^2| \right\rangle$ is sufficient to also reduce the magnitude of $\text{Re} \left\langle \mathcal{Q}^2 \right\rangle$ and thus both the variance of the real and imaginary components of $\mathcal{Q}$. Intuitively, the term $\left\langle |\mathcal{Q}^2| \right\rangle$ represents the average radial extent of the distribution of $\mathcal{Q}$ in the complex plane, while $\text{Re} \left\langle \mathcal{Q}^2 \right\rangle$ gives a measure of the extent of the distribution along the real axis vs. the imaginary axis; reducing the radial extent is clearly sufficient to reduce the variance in all directions.

These bounds also match the intuition given in Sec. 5.1.2. Deformations that reduce the average magnitude of samples of $\mathcal{Q}$ reduce the severity of phase cancellations required to produce an exponentially small expectation value, and the bounds above give a quantitative basis for this statement. These bounds and the intuition from the one-dimensional example thus motivate the contour deformations considered in the applications presented in this chapter.

### 5.2.2 Optimizing observifolds

Having identified the properties of $\mathcal{Q}$ that result in a reduced StN problem and better statistical precision, our aim is to discover manifold deformations for which (1) the corresponding deformed observable $\mathcal{Q}$ is tractable to compute (in particular the Jacobian factor must be tractable) and (2) the variances of the components of $\mathcal{Q}$ that we aim to measure (real, imaginary, or both) are significantly smaller than the corresponding variances of components of $\mathcal{O}$. To do so, we follow the general approach of the method of sign- or path-optimized manifolds [432, 463, 473, 474]:

1. **Construct a family of manifolds** parameterized by a collection of real numbers. We denote by $\widetilde{M}_\omega$ the member of this family of manifolds identified by the parameters $\omega$. In practice such manifolds are defined by the parameterized map $\widetilde{U}(U; \omega)$ and associated Jacobian factor $\widetilde{J}(U; \omega)$.

2. **Iteratively optimize the manifold parameters** to minimize a chosen loss function providing a measure of the significance of the sign / StN problem on the current manifold.

This chapter details an approach based on independently optimizing an observifold, and the resulting deformed observable, per observable under study, where the term 'observifold' indicates a manifold targeting a specific observable or set of observables. Though this potentially requires optimizing many distinct contour deformations, the use of fixed Monte Carlo weights for all choices of integration manifold means that this optimization step can be performed as part of measuring observables, after ensembles have been generated (at significant cost) once and for all. This is advantageous for observables where the measurement cost is smaller than or comparable to the generation cost. Though one does not expect a common deformed manifold to result in optimal deformed observables associated with all observables of interest, phase fluctuations of similar observables are highly correlated [6] and similar observables are expected to receive similar reductions in their variance from a particular deformation. This can be exploited to reduce the practical cost of optimization. For example, it is possible to choose a subset of representative observables and identify optimized manifolds for this subset, giving suboptimal but significant improvements for similar observables not included in the set. Alternatively, one can use 'transfer learning' to take optimized manifolds for one observable as the starting point for optimizing manifolds for similar observables. In our application to $SU(N)$ lattice gauge theory in Sec. 5.5, the transfer learning approach was found to significantly reduce the total number of iterations required to optimize observifolds for Wilson loops.

Optimizing distinct observifolds for each observable requires the introduction of one loss function per observable studied. Manifold deformations aiming to solve extensive sign problems have previously used a single loss function written in terms of the average sign $\left\langle e^{-i\,\mathrm{Im}\,S} \right\rangle_{\mathrm{Re}\,S}$, reflecting the fact that there is a single source of the sign problem in that setting. Here, we instead directly use the first two terms in the variance (i.e. the terms that can be modified by deformation) of each observable of interest as the loss function for the observifold targeting that observable. Because the variance of the real and imaginary components of $\mathcal{O}$ differ, we define a distinct loss function for these two cases,

$$
\begin{aligned}
\mathcal{L}_{\mathrm{Re}\,\mathcal{O}} &\equiv \left\langle \mathrm{Re}[\mathcal{Q}_\omega]^2 \right\rangle = \frac{1}{2}\left\langle |\mathcal{Q}_\omega^2| \right\rangle + \frac{1}{2}\mathrm{Re}\left\langle \mathcal{Q}_\omega^2 \right\rangle, \\
\mathcal{L}_{\mathrm{Im}\,\mathcal{O}} &\equiv \left\langle \mathrm{Im}[\mathcal{Q}_\omega]^2 \right\rangle = \frac{1}{2}\left\langle |\mathcal{Q}_\omega^2| \right\rangle - \frac{1}{2}\mathrm{Re}\left\langle \mathcal{Q}_\omega^2 \right\rangle,
\end{aligned}
\tag{5.29}
$$

where $\mathcal{Q}_\omega$ is the deformed observable corresponding to $\mathcal{O}$ on the deformed manifold $\widetilde{\mathcal{M}}_\omega$. When we are only interested in the real or imaginary component, the corresponding loss function can be used to construct an observifold for that component of the observable alone. In cases where both components are under study, they can either be measured independently using two independently optimized deformed observables, or a single deformed observable can be chosen by minimizing $\mathcal{L}_{\mathrm{Re}\,\mathcal{O}} + \mathcal{L}_{\mathrm{Im}\,\mathcal{O}} = \left\langle |\mathcal{Q}_\omega|^2 \right\rangle$. The latter choice may be more efficient in cases where the most significant improvements are expected from the term $\left\langle |\mathcal{Q}_\omega|^2 \right\rangle$ in the variance, for which one would expect a similar optimal choice of deformed observable for the estimate of the real and imaginary components of $\mathcal{O}$.

Gradient-based techniques can be applied to iteratively optimize the parameters $\omega$ defining the manifold $\widetilde{\mathcal{M}}_\omega$ and corresponding deformed observable $\mathcal{Q}_\omega$. To apply such methods, the gradients need to be computed for arbitrary deformed observables $\mathcal{Q}_\omega$. Keeping in mind that the expectation value $\langle \cdot \rangle$ in Eq. (5.29) is always taken with respect to the original Monte Carlo weights, gradients can be taken inside these expectation values by linearity,

$$\nabla_\omega \mathcal{L}_{\text{Re}\,\mathcal{O}} = \frac{1}{2} \left\langle \nabla_\omega |\mathcal{Q}_\omega|^2 \right\rangle + \frac{1}{2} \text{Re} \left\langle \nabla_\omega \mathcal{Q}_\omega^2 \right\rangle,$$
$$\nabla_\omega \mathcal{L}_{\text{Im}\,\mathcal{O}} = \frac{1}{2} \left\langle \nabla_\omega |\mathcal{Q}_\omega|^2 \right\rangle - \frac{1}{2} \text{Re} \left\langle \nabla_\omega \mathcal{Q}_\omega^2 \right\rangle. \tag{5.30}$$

The deformed observable itself is defined by the parameterized deformation to be

$$\mathcal{Q}_\omega(U) = \widetilde{J}(U; \omega) \mathcal{O}(\widetilde{U}(U; \omega)) e^{-S(\widetilde{U}(U;\omega)) + S(U)}. \tag{5.31}$$

Combining Eqs. (5.30) and (5.31), the gradients of loss functions can be straightforwardly computed by application of the chain rule when the explicit manifold parameterziation is known. Because gradients only act within the expectation values in Eq. (5.30), stochastic estimators based on Monte Carlo samples can be used for both the loss function itself and the gradients. For a set of samples $\{U^{(i)}\}_{i=1}^n$ of size $n$, unbiased loss estimates are explicitly given by

$$\widehat{\mathcal{L}_{\text{Re}\,\mathcal{O}}} = \frac{1}{2n} \sum_{i=1}^n |\mathcal{Q}_\omega(U^{(i)})|^2 + \frac{1}{2n} \text{Re} \sum_{i=1}^n \mathcal{Q}_\omega^2(U^{(i)}),$$
$$\widehat{\mathcal{L}_{\text{Im}\,\mathcal{O}}} = \frac{1}{2n} \sum_{i=1}^n |\mathcal{Q}_\omega(U^{(i)})|^2 - \frac{1}{2n} \text{Re} \sum_{i=1}^n \mathcal{Q}_\omega^2(U^{(i)}). \tag{5.32}$$

For each stochastic estimate, corresponding unbiased gradient estimates can be computed in practice using automatic differentiation in any machine learning framework; in all applications presented in this chapter, the JAX [354] framework was applied to compute gradients and to optimize manifold parameters. While we use the loss function in Eq. (5.32) for optimization in our main applications, we note that taking the log of analogous loss functions has been suggested for numerical optimization of manifold parameters for extensive sign problems [472]. This rescales gradient estimates to have a more uniform magnitude (without changing their direction) as the loss function varies over many orders of magnitude. This may be a useful practical modification for future applications.

Though the original observables may be affected by severe signal-to-noise problems, the loss functions defined in Eq. (5.29) are the expectation values of positive-definite quantities which can be precisely estimated. Gradients of the loss function are not necessarily positive definite quantities, but in cases in which the manifold parameterization allows progress to be made, average gradients should not be expected to be exponentially small relative to the magnitude of each sample. Though it is not possible *a priori* to determine that such progress is possible, empirical results for the applications considered in Sections 5.4 and 5.5 demonstrate that good choices of manifolds can already be found using simple parameterizations in specific lattice gauge theories using gradient-based optimization.

When applying contour deformations in the context of fixed Monte Carlo sampling, it is also possible that one encounters an *overlap problem* which gives high variance in cases in which the term $\widetilde{J}(U)e^{-S(\widetilde{U}(U))+S(U)}$ has large magnitude fluctuations. Such magnitude fluctuations are penalized by the loss functions defined in Eq. (5.29), but when significant magnitude fluctuations occur it is also difficult to measure the loss function accurately. In this case, the stochastic estimates of the loss function given in Eq. (5.32) may fail to sufficiently penalize these fluctuations on sampled batches of configurations. To monitor for an overlap problem, we measure the effective sample size (ESS)[1] defined in Chapter 3 as

$$\text{ESS} \equiv \frac{\left(\frac{1}{n}\sum_i w_i\right)^2}{\frac{1}{n}\sum_i w_i^2}, \tag{3.34}$$

where $w_i = |\widetilde{J}(U^{(i)})\exp[-S(\widetilde{U}(U^{(i)})) + S(U^{(i)})]|$ for an ensemble of $n$ configurations $\{U^{(i)}\}_{i=1}^n$. Values of the ESS close to 1 indicate a good overlap and correspondingly small magnitude fluctuations. We do not in practice find a significant overlap problem in the applications considered below, even when freely optimizing the manifold parameterizations considered.

Stochastic estimates of the variance on a particular set of samples in general differs from the actual variance of the deformed observable. When many optimization steps are applied for a fixed ensemble of samples used for estimates of the loss function, the choice of manifold may become *overtrained* [479–481] to minimize the sample variance rather than true variance of the observable. A potential overlap problem is simply one particular instantiation of this issue. Though we did not encounter overlap problems in the applications considered in this chapter, optimizing manifolds using a fixed batch of samples for all optimization steps did result in overtraining. To address this overtraining, two approaches can be applied:

- The effect of overtraining can be monitored by reserving a 'test set' of samples that is used only to measure the loss function but never to compute gradients. Overtraining can be detected as a divergence of the loss function measured on this test set compared to the loss function measured on the 'training set' of samples used to measure gradients and optimize the manifold parameters. Training should be stopped when the loss function measured on the test set is no longer decreasing.

- To reduce the rate at which overfitting occurs, a mini-batching approach can be applied in which subsets of size $m \leq n$ are resampled from the total training set to estimate the loss and its gradient on each iteration of optimization. This has the advantage of introducing fluctuations in the estimate of the sample variance proportional to the statistical uncertainty on the measurement of the variance. This applies even in the case of $m = n$, in which case the resampling is exactly equivalent to bootstrap resampling for uncertainty estimation.

Though overlap and overtraining problems should be monitored and counteracted to improve the quality of deformed observables produced by the method of observifolds, it must be stressed that none of these difficulties can compromise the exactness of the method. Obstacles encountered during training modify the variance of the deformed observable, and thus the efficiency of the method, but any choice of manifold gives a deformed observable with identical expectation value to the original observable of interest for a holomorphic path integral.

---

[1]Related works also use the term 'statistical power' for this quantity.

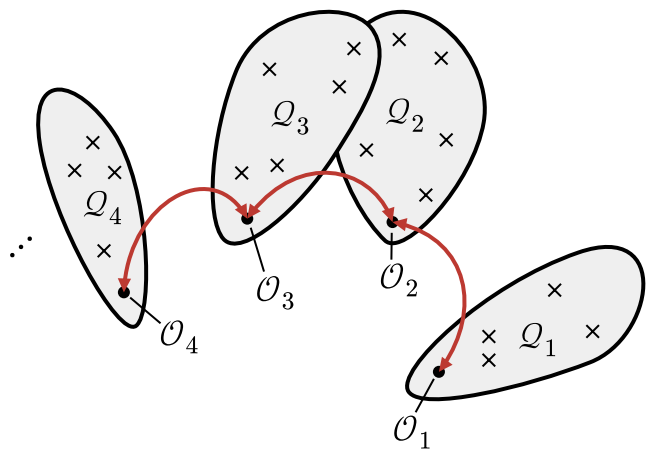

**Figure 5.4:** Schematic depiction of observables that can be related by contour deformations or symmetry considerations to have identical expectation values. Base observables $\mathcal{O}_i$ have identical expectation values $\langle \mathcal{O}_1 \rangle = \langle \mathcal{O}_2 \rangle = \dots$ based on symmetries (indicated by red arrows). Shaded regions indicate continuous sets of observables $\mathcal{Q}_i$ related by deformations to the base observable $\mathcal{O}_i$, and particular observables that may be selected are marked for example by crosses. Rewriting observables using symmetry (i.e. following red arrows in the diagram) may in general give rise to base observables that are associated with different families of deformed observables. Rewriting may thus give choices of base observables that are better or worse in terms of finding deformed observables with improved signal-to-noise properties.

### 5.2.3 Rewriting base observables before deformation

Contour deformations give an approach to discovering equivalent path integral representations for any holomorphic observable, but other techniques can also be applied to discover observables with equivalent expectation values. When a theory possesses a continuous or discrete symmetry, considering observables related by the symmetry provides another way to discover equivalent path integral representations. Using these distinct observables as base observables for deformations may in general result in different families of deformed observables, as shown schematically in Figure 5.4.

In some cases, base observables related by symmetries can give rise to deformed observables with very different signal-to-noise properties. This can be demonstrated in the simple example of a single-plaquette U(1) Euclidean lattice gauge theory in $(1+1)$ dimensions using a Wilson gauge action. The path integral representation of an observable $\mathcal{O}$ in this theory is given by integrating over the single plaquette variable,

$P = e^{i\phi}$, as

$$\langle \mathcal{O} \rangle = \frac{1}{Z} \int dP \, \mathcal{O}(P) \, e^{\beta \operatorname{Re} P} = \frac{1}{Z} \int_0^{2\pi} \frac{d\phi}{2\pi} \, \mathcal{O}(\phi) \, e^{\beta \cos \phi}. \tag{5.33}$$

One path integral representation of the expectation value of the $1 \times 1$ Wilson loop consisting of this single plaquette is given by

$$\left\langle e^{i\phi} \right\rangle = \frac{\int_0^{2\pi} \frac{d\phi}{2\pi} e^{i\phi} e^{\beta \cos \phi}}{\int_0^{2\pi} \frac{d\phi}{2\pi} e^{\beta \cos \phi}} = \frac{I_1(\beta)}{I_0(\beta)}, \tag{5.34}$$

where $I_n(\cdot)$ is the modified Bessel function of the first kind with rank $n$. We use the integral expression $I_n(\beta) = \int_0^{2\pi} \frac{d\phi}{2\pi} e^{in\phi} e^{\beta \cos \phi}$ for these Bessel functions here and in derivations below. This quantity is real in expectation, and an equivalent path integral representation is given by

$$\operatorname{Re} \left\langle e^{i\phi} \right\rangle = \langle \cos(\phi) \rangle = \frac{1}{Z} \int_0^{2\pi} \frac{d\phi}{2\pi} \, \cos(\phi) \, e^{\beta \cos \phi}. \tag{5.35}$$

We can also arrive at this alternative path integral representation by symmetry considerations: the theory is invariant under $\phi \to -\phi$ which physically corresponds to a charge conjugation symmetry. The observable $\cos(\phi)$ is given by taking the symmetry-averaged version of the observable $e^{i\phi}$.

The difference between these two path integral representations has no effect on Monte Carlo estimates of the real part of the unit-area Wilson loop because the sample mean of $\operatorname{Re}[e^{i\phi}] = \cos(\phi)$ is equivalent to the sample mean of $\operatorname{Re}[\cos(\phi)] = \cos(\phi)$. However, deforming the path integral in Eq. (5.34) has a vastly different effect than deforming the path integral in Eq. (5.35). Using a constant shift deformation allows significant improvements to be made in the signal-to-noise of estimates of $\langle e^{i\phi} \rangle$ while no improvements can be made for estimates of $\langle \cos(\phi) \rangle$, as shown in Figure 5.5 and derived below.

First, we can compute the variance $\operatorname{Var}_\beta[\operatorname{Re} e^{i\phi}] = \operatorname{Var}_\beta[\cos \phi]$ of the original Monte Carlo estimator given by the sample mean of $\operatorname{Re} e^{i\phi} = \cos \phi$. We use the subscript $\beta$ to indicate the choice of coupling parameter for variance and expectation value estimates. The variance can be derived in terms of Bessel functions as

$$\begin{aligned}
\operatorname{Var}_\beta[\operatorname{Re} e^{i\phi}] = \operatorname{Var}_\beta[\cos(\phi)] &= \left\langle \cos(\phi)^2 \right\rangle_\beta - \langle \cos(\phi) \rangle_\beta^2 \\
&= \frac{1}{2} \left[ 1 + \frac{I_2(\beta)}{I_0(\beta)} \right] - \left[ \frac{I_1(\beta)}{I_0(\beta)} \right]^2.
\end{aligned} \tag{5.36}$$

Applying the constant imaginary-shift deformation $\phi \to \widetilde{\phi} = \phi + if$ to the path integral representations in terms of $e^{i\phi}$ and $\cos \phi$ results in two distinct families of deformed observables. The deformed observables $\mathcal{Q}_e(\phi)$ and $\mathcal{Q}_c(\phi)$ associated with $e^{i\phi}$ and $\cos \phi$,

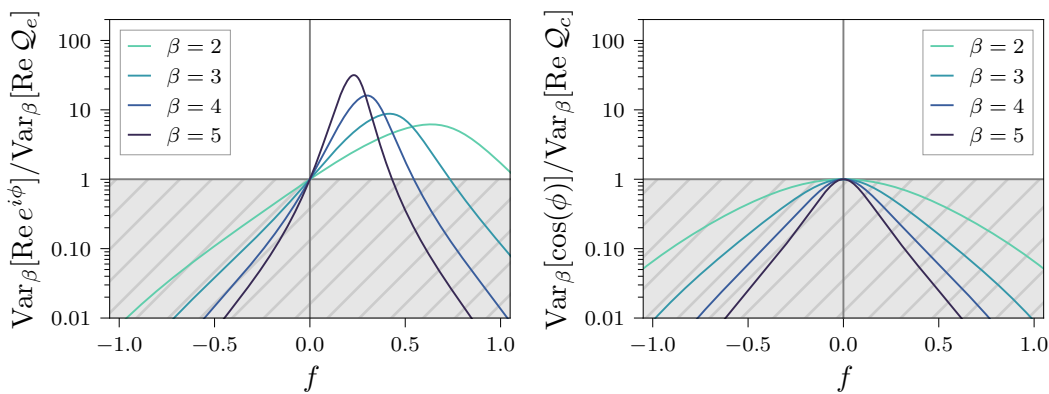

**Figure 5.5:** A comparison of the variance reduction achieved by applying the deformation $\phi \to \phi + if$ to the path integral definition of $\left\langle e^{i\phi} \right\rangle$ (left) and to the path integral definition of $\langle \cos(\phi) \rangle$ (right) in the one-plaquette U(1) model considered in the main text. The plots depict ratios of the original variance $\mathrm{Var}_\beta[\mathrm{Re}[e^{i\phi}]] = \mathrm{Var}_\beta[\cos(\phi)]$ to the variance of the deformed observable for a range of choices of coupling parameter $\beta$ and deformation parameter $f$. Significant improvements are possible for optimal choices of $f$ when deforming $e^{i\phi}$, whereas no improvements are possible when deforming $\cos(\phi)$ by this constant-shift deformation. The region in which the variance of the deformed observable is worse than the original variance is indicated by the gray hatching in both panels. Figure adapted from Fig. 2 of Pub. [1].

respectively, are given by

$$
\begin{aligned}
\mathcal{Q}_e(\phi) &= e^{i\widetilde{\phi}(\phi)}\, e^{\beta \cos \widetilde{\phi}(\phi) - \beta \cos \phi} = e^{-f} e^{i\phi}\, e^{\beta \cos(\phi + if) - \beta \cos \phi}, \\
\mathcal{Q}_c(\phi) &= \cos \widetilde{\phi}(\phi)\, e^{\beta \cos \widetilde{\phi}(\phi) - \beta \cos \phi} = \cos(\phi + if)\, e^{\beta \cos(\phi + if) - \beta \cos \phi}.
\end{aligned}
\tag{5.37}
$$

The expectation value of both deformed observables is equivalent to the original observables under consideration,

$$
\langle \mathcal{Q}_e(\phi) \rangle = \langle \mathcal{Q}_c(\phi) \rangle = \left\langle e^{i\phi} \right\rangle = \langle \cos(\phi) \rangle .
\tag{5.38}
$$

However, using the real component of either $\mathcal{Q}_e(\phi)$ or $\mathcal{Q}_c(\phi)$ as a Monte Carlo estimator of this expectation value results in estimates with significantly different variance. We can derive the variance in both cases in terms of Bessel functions as well, giving

$$
\begin{aligned}
\mathrm{Var}_\beta[\mathrm{Re}\,\mathcal{Q}_e] &= e^{-2f}\left[\frac{R(\beta, f)}{2} + V(\beta, f)\right] - \left[\frac{I_1(\beta)}{I_0(\beta)}\right]^2 \\
\mathrm{Var}_\beta[\mathrm{Re}\,\mathcal{Q}_c] &= \frac{1}{2}e^{-2f}V(\beta, f) + \frac{1}{2}e^{2f}V(\beta, -f) + \frac{1}{2}R(\beta, f) - \left[\frac{I_1(\beta)}{I_0(\beta)}\right]^2 ,
\end{aligned}
\tag{5.39}
$$

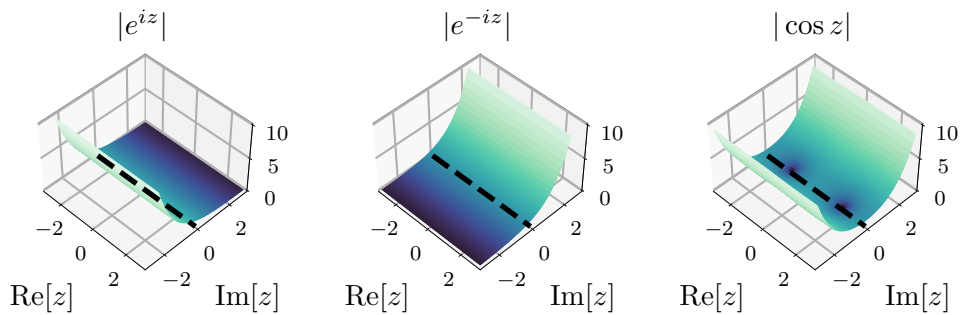

**Figure 5.6:** Comparison of the magnitude (indicated by both height and shading in the plot) of $e^{iz}$, $e^{-iz}$, and $\cos z$ for complex $z$. Both $e^{iz}$ and $e^{-iz}$ can be deformed from the real line (dashed black) into the complex plane to significantly reduce their magnitude. For $e^{iz}$ this occurs when shifting $z$ in the positive imaginary direction, while for $e^{-iz}$ this occurs when shifting $z$ in the negative imaginary direction. On the other hand, no significant magnitude reduction is possible by evaluating $\cos z$ for complex arguments because the magnitude of $\cos z = \frac{1}{2}e^{iz} + \frac{1}{2}e^{-iz}$ is dominated by the larger of the two terms when evaluating at either positive or negative offsets in the imaginary direction.

where

$$V(\beta, f) = \left(\frac{e^f - \frac{1}{2}}{e^{-f} - \frac{1}{2}}\right)\frac{I_2(\beta\sqrt{5 - 4\cosh(f)})}{2I_0(\beta)},$$
$$R(\beta, f) = \frac{I_0(\beta(2\cosh(f) - 1))}{I_0(\beta)}. \tag{5.40}$$

This variance is compared to the variance of the undeformed Monte Carlo estimate in Fig. 5.5 for a range of choices of $f$ and $\beta$. Significant reduction in the variance is possible for optimal choices of $f$ in the deformed observable $\mathcal{Q}_e$, while no reduction in variance is possible for $\mathcal{Q}_c$, and the optimal choice of manifold in this family is the original manifold corresponding to $f = 0$. This can be qualitatively explained by the fact that deforming the exponential $e^{i\phi}$ by a constant imaginary shift can exponentially reduce the average magnitude while no such reduction in magnitude is possible for $\cos(\phi)$. In Eq. (5.39), this difference in behavior can be seen by the appearance of a factor of $e^{-2f}$ alone for the variance of $\mathcal{Q}_e$ whereas competing factors of $e^{-2f}$ and $e^{2f}$ appear in the variance of $\mathcal{Q}_c$. In fact, for a fixed real part of $\phi$ the smallest possible magnitude for $\cos(\phi)$ occurs exactly on the real line. Figure 5.6 depicts the magnitudes of $e^{iz}$, $e^{-iz}$, and $\cos z$ over a range of choices of $z \in \mathbb{C}$. As one can see from the figure, integrands of the form $e^{iz}$ or $e^{-iz}$ both have a direction in the complex in which the average magnitude is significantly reduced, whereas the magnitude of $\cos(\phi)$ grows when moving off the real line. This suggests that this issue extends beyond merely failing to find a useful constant-shift deformation — even non-constant deformations into the complex plane encounter regions of exponentially larger magnitudes which can be expected to introduce additional noise.

Though this analysis can only be carried out analytically for simple theories, it highlights the importance of considering *rewritings* of the base observable before deforming.

In this one-plaquette U(1) example, choosing the form of the base observable that is not invariant under the symmetry actually results in the most variance reduction from contour deformation. For the application to U(1) lattice gauge theory with non-trivial spacetime volumes in Sec. 5.4, rewriting observables to take a form analogous to the choice $e^{i\phi}$ allows significant progress to be made by constant-shift deformations. A careful choice of the form of the observable is similarly necessarily in the application to SU($N$) lattice gauge theory presented in Sec. 5.5.

## 5.3 Families of manifolds for lattice gauge theory

We next define a particular family of observables suitable for deformations of variables in U(1) and SU($N$) lattice gauge theory. In both cases, the underlying manifold is compact and can be parameterized using angular coordinates that form a subset of $\mathbb{R}^{\mathcal{N}}$. We distinguish between two types of angular coordinates for the purpose of contour deformations:

1. **Zenith angles** are defined by convention in the interval $[0, \pi/2]$ with distinguished endpoints.[2] These endpoints must remain fixed when deforming the contour of integration for any zenith angle. We conventionally use the notation $\theta$ for zenith angles.

2. **Azimuthal angles** are defined in the interval $[0, 2\pi]$ with the endpoints 0 and $2\pi$ identified. This implies that the action and all possible observables take identical values at 0 and $2\pi$ and that under deformations these endpoints only need to remain identified, rather than fixed. We conventionally use the notation $\phi$ for azimuthal angles.

Figure 5.7 depicts the difference in allowed contour deformations for these two types of angular variables. The deformed contours must also be smoothly connected to the original contour of integration. This property is most easily enforced when constructing explicit families of manifolds by ensuring that the parameters define a continuous space of manifolds that includes the original manifold.

In the remainder of this section, we define families of manifolds describing deformations of single U(1) and SU($N$) variables, then connect these choices to families of manifolds applicable to lattice gauge theories.

### 5.3.1 Deformations of U(1) variables

The U(1) group manifold can be parameterized by a single azimuthal angle $\phi$ related to group elements by $U = e^{i\phi}$. The original domain of integration $\phi \in [0, 2\pi]$ corresponds to integrating $U$ over the unit circle in the complex plane. Any functions that can be written as holomorphic functions of $U$ and $U^{-1}$ will be holomorphic functions of $\phi$, because both $U = e^{i\phi}$ and $U^{-1} = e^{-i\phi}$ are themselves holomorphic functions of $\phi$.

---

[2]Any interval would be suitable because the endpoints are distinct. This choice follows the conventions of Ref. [482].

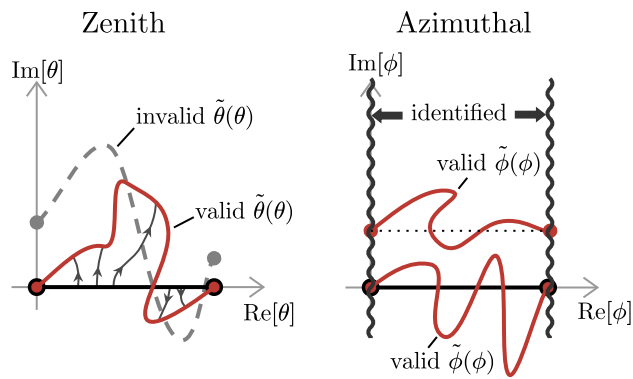

**Figure 5.7:** A comparison of the allowed contour deformations for zenith angles $\theta$ (left) and azimuthal angles $\phi$ (right). The original contour of integration is depicted by the black horizontal contour in both cases. For zenith angles, the endpoints must remain fixed, and the gray dashed contour is therefore invalid. For azimuthal angles, we only require that the endpoints remain identified, so the upper red contour is valid even though the endpoints are deformed relative to the original contour.

This includes the definitions of common actions and observables for U(1) lattice gauge theory.

In terms of the coordinate $\phi$, we define possible deformed contours of integration by *vertical shift deformations* that modify only the imaginary components of each coordinate to give a deformed coordinate $\widetilde{\phi}$ that can generically be written as

$$\widetilde{\phi} = \phi + if(\phi), \quad f(\phi) \in \mathbb{R}. \tag{5.41}$$

To satisfy the boundary conditions of the azimuthal angle $\phi$, we further require that $f(0) = f(2\pi)$. Vertical shift deformations have been applied in a number of previous applications of contour deformations to extensive sign problems [432, 433, 463, 466–468, 472–474], and complexified U(1) variables written in terms of angular parameters have been used in the context of one-plaquette and one-link models as well as the Thirring model [432, 437, 438, 441, 449, 459–461, 463, 465, 468–470]. Vertical shift deformations are an appealing choice because they describe a wide variety of possible manifolds with a relatively stable Jacobian factor, though we note that they are not fully general even for arbitrary functions $f$, because one cannot describe contours for which multiple points share the same real component. This distinction is schematically depicted in Figure 5.8.

For U(1) variables, a shift in the imaginary direction corresponds to rescaling the magnitude. When observables can be written analogously to $e^{i\phi}$, we expect constant shifts in the imaginary direction to be a useful choice of deformation, resulting in a decrease in the average magnitude and a reduction in phase fluctuations of the deformed observable relative to the base observable. To systematically access more general vertical deformations, we choose to expand the function $f(\phi)$ as a Fourier series, keeping

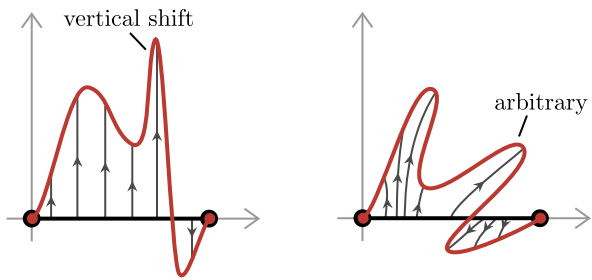

**Figure 5.8:** Vertical shift deformations vs. general contour deformations. For vertical shift contour deformations, each point on the original contour is mapped to a point with identical real component, as shown on the left. This excludes some choices of valid contours, including for example the contour shown on the right.

only terms that are periodic in the endpoints,

$$f(\phi; \lambda, \chi) = \sum_{n=0}^{\Lambda} \lambda_n \sin(n\phi + \chi_n).$$ (5.42)

The cutoff $\Lambda$ determines the highest mode included in the expansion and the real numbers $\lambda = (\lambda_1, \ldots, \lambda_\Lambda)$ and $\chi = (\chi_1, \ldots, \chi_\Lambda)$ parameterize the scaling and phase offset of the Fourier modes included. As $\Lambda$ is increased, the expressivity of the function $f$ is systematically improved at the cost of the inclusion of additional terms involving additional parameters that must be optimized.

The Jacobian for vertical shift deformations using a Fourier expansion can be straightforwardly calculated for one U(1) variable, giving

$$\widetilde{j}(U = e^{i\phi}) = 1 + if'(\phi) = 1 + i \sum_{n=0}^{\Lambda} n\lambda_n \cos(n\phi + \chi_n),$$ (5.43)

where we use lowercase notation to distinguish single-variable Jacobians from the full Jacobian for transformations in lattice gauge theory below. For lattice gauge theory, the function $f$ deforming each component of the gauge field can generally depend on other components, and a description of this dependence and the computation of the Jacobian are described in Sec. 5.3.3 below.

### 5.3.2 Deformations of SU($N$) variables

Previous works have performed complex contour deformations and calculations of Lefschetz thimbles for SU($N$) variables in $(0 + 1)$D QCD and QCD in the heavy-dense limit [456, 458, 475, 483, 484]. Parameterizations of manifolds used in these works either rely on working in a diagonal gauge or left-multiplying by a complexified group element that is a constant with respect to the gauge variables. We here introduce an approach to deforming SU($N$) variables based instead on an angular parameterization of the group manifold. This angular parameterization allows a simple definition of a wide variety of manifold deformations despite the non-trivial nature of the underlying SU($N$) manifold. We demonstrate in the application to SU($N$) lattice gauge theory in Sec. 5.5 that this angular parameterization also allows Wilson loop observables to be written in terms of $e^{i\phi}$ factors where $\phi$ are combinations of angular parameters that can be shifted to reduce the average magnitude; this results in significant signal-to-noise improvements, as discussed in that section.

For any choice of $N$, an explicit coordinatization of the group manifold of SU($N$) is possible in terms of $N^2 - 1$ real angular variables [482]. This allows the domain of integration to be identified with a subset of $\mathbb{R}^{N^2-1}$. This domain of integration can then be complexified and deformed into a submanifold of $\mathbb{C}^{N^2-1}$ by independently considering deformations of each angular coordinate, while allowing interdependence between each coordinate. Using the explicit parameterization provided in Ref. [482], the SU($N$) group manifold is given in terms of $J = (N^2 + N - 2)/2$ azimuthal angles $\phi_1, \ldots, \phi_J \in [0, 2\pi]$ and $K = (N^2 - N)/2$ zenith angles $\theta_1, \ldots, \theta_K$. We use the notation $\Omega = (\phi_1, \ldots, \phi_J, \theta_1, \ldots, \theta_K)$ to collectively label the $(N^2 - 1)$-dimensional coordinate uniquely identifying a group element by the map $U(\Omega)$.

Writing the integral over an arbitrary function $g(U)$ of an SU($N$) variable $U$ as $\mathcal{I} = \int dU\, g(U)$, we can give an explicit coordinatized version of this integral using the azimuthal and zenith angles associated with the group as

$$\mathcal{I} = \prod_{j=1}^{J} \left[ \int_0^{2\pi} d\phi_j \right] \prod_{k=1}^{K} \left[ \int_0^{\pi/2} d\theta_k \right] h(\Omega) g(U(\Omega)), \tag{5.44}$$

where $h(\Omega)$ relates the Haar measure $dU$ on the group SU($N$) to the measure $\prod_j d\phi_j$ $\prod_k d\theta_k$ over angular coordinates. The deformed integral is then given by

$$\begin{aligned}
\mathcal{I} &= \prod_{j=1}^{J} \left[ \int_0^{2\pi} d\phi_j \right] \prod_{k=1}^{K} \left[ \int_0^{\pi/2} d\theta_k \right] h(\widetilde{\Omega}) \widetilde{j}(\Omega) g(U(\widetilde{\Omega})) \\
&= \prod_{j=1}^{J} \left[ \int_0^{2\pi} d\phi_j \right] \prod_{k=1}^{K} \left[ \int_0^{\pi/2} d\theta_k \right] h(\Omega) \left\{ \frac{h(\widetilde{\Omega})}{h(\Omega)} \widetilde{j}(\Omega) g(U(\widetilde{\Omega})) \right\} \\
&= \int dU\, \widetilde{j}(U)\, g(\widetilde{U}(U)),
\end{aligned} \tag{5.45}$$

where $\widetilde{j}(\Omega) \equiv \det \frac{\partial \widetilde{\Omega}_\alpha}{\partial \Omega_\beta}$ is the coordinate-representation Jacobian determinant associated with the deformation, $h(\widetilde{\Omega})$ is the Haar measure factor evaluated on the complexified coordinate, and by combining the Jacobian and Haar factors we define the Jacobian with respect to the original integral measure as

$$\widetilde{j}(U) = \frac{\widetilde{j}(\Omega(U)) h(\widetilde{\Omega}(U))}{h(\Omega(U))}. \tag{5.46}$$

The final line of Eq. (5.45) is a deformed integral defined in terms of the original measure and manifold by $\widetilde{U}(U)$ and $\widetilde{j}(U)$ as described for the general case in Eq. (5.19)

above. In practice, the coordinate description can thus be used to specify the deformation, determining $\widetilde{U}(U)$ and $\widetilde{j}(U)$, and can then be ignored for the remainder of the calculation.

To give a concrete handle on the angular coordinates for SU($N$), we present the explicit coordinates for the groups SU(2) and SU(3); see Ref. [482] for a derivation and the general case. These coordinates are also applied in Sec. 5.5 to SU(2) and SU(3) lattice gauge theory. The group manifold of SU(2) is parameterized in terms of two azimuthal angles $\phi_1$ and $\phi_2$ and a single zenith angle $\theta$. Group elements are associated with coordinates $\Omega = (\phi_1, \phi_2, \theta)$ by the coordinate map

$$U(\Omega) = \begin{pmatrix} \sin\theta e^{i\phi_1} & \cos\theta e^{i\phi_2} \\ -\cos\theta e^{-i\phi_2} & \sin\theta e^{-i\phi_1} \end{pmatrix}, \tag{5.47}$$

where the group elements are written in the fundamental $2 \times 2$ matrix representation. This representation is faithful and the coordinates $\Omega$ are thus sufficient to express $U$ in any representation. The Haar measure of SU(2) is related to the measure on these coordinates by

$$dU = h(\Omega)\,d\Omega \equiv \frac{1}{4\pi^2}\sin(2\theta)\,d\theta\,d\phi_1\,d\phi_2. \tag{5.48}$$

The analogous parameterization of SU(3) is given in terms of five azimuthal angles and three zenith angles, $\Omega = (\phi_1, \ldots, \phi_5, \theta_1, \theta_2, \theta_3)$. Group elements in the fundamental representation are given by the coordinate map

$$U(\Omega) = \begin{pmatrix} c_1 c_2 e^{i\phi_1} & s_1 e^{i\phi_3} & c_1 s_2 e^{i\phi_4} \\[4pt] \begin{matrix} s_2 s_3 e^{-i(\phi_4+\phi_5)} - \\ s_1 c_2 c_3 e^{i(\phi_1+\phi_2-\phi_3)} \end{matrix} & c_1 c_3 e^{i\phi_2} & \begin{matrix} -c_2 s_3 e^{-i(\phi_1+\phi_5)} - \\ s_1 s_2 c_3 e^{i(\phi_2-\phi_3+\phi_4)} \end{matrix} \\[4pt] \begin{matrix} -s_1 c_2 s_3 e^{i(\phi_1-\phi_3+\phi_5)} - \\ s_2 c_3 e^{-i(\phi_2+\phi_4)} \end{matrix} & c_1 s_3 e^{i\phi_5} & \begin{matrix} c_2 c_3 e^{-i(\phi_1+\phi_2)} - \\ s_1 s_2 s_3 e^{-i(\phi_3-\phi_4-\phi_5)} \end{matrix} \end{pmatrix}, \tag{5.49}$$

where $c_i \equiv \cos\theta_i$ and $s_i \equiv \sin\theta_i$. The Haar measure of SU(3) is related to the measure on these coordinates by

$$dU = h(\Omega)\,d\Omega = \frac{1}{2\pi^5}s_1(c_1)^3 s_2\,c_2\,s_3\,c_3\,d\theta_1\,d\theta_2\,d\theta_3\,d\phi_1\ldots d\phi_5. \tag{5.50}$$

For any SU($N$), we deform each azimuthal and zenith angle using vertical deformations as described for the single azimuthal angle of U(1) in the previous section. In contrast to the group U(1), however, we allow the vertical deformations of each angle to depend on the values of other angles. A generic vertical deformation of an angular parameterization of SU($N$) thus takes the form

$$\widetilde{\Omega} = \Omega + if(\Omega). \tag{5.51}$$

where $f(\Omega) \in \mathbb{R}^{N^2-1}$ and the deformation must satisfy periodicity for azimuthal variables and must keep the endpoints fixed for zenith variables. This can be achieved using a Fourier deformation basis, analogously to the case of U(1), up to minor differences in the treatment of azimuthal and zenith variables. We give a Fourier-basis definition of $f$ truncated at a cutoff mode $\Lambda$ by

$$f(\Omega; \lambda, \chi) = \sum_{\{n_i\}=0}^{\Lambda} \sum_{\{m_j\}=1}^{\Lambda} \lambda_I \, T_I(\Omega; \chi_I^1, \ldots, \chi_I^J), \qquad (5.52)$$

where $I \equiv (n_1 \ldots n_J, m_1 \ldots m_K)$, and

$$T_I(\Omega; \chi^1 \ldots \chi^J) \equiv \prod_{i=1}^{J} \sin(\phi_i n_i + \chi^i) \prod_{j=1}^{K} \sin(2\theta_j m_j). \qquad (5.53)$$

This form of the Fourier decomposition gives a complete basis for functions that are periodic in the $\phi_i$ angles and fixed to zero at the endpoints of the $\theta_i$ angles. In practice, this basis already involves a large number of parameters for small cutoffs, and for simplicity in our proof-of-principle application to SU($N$) lattice gauge theory only a subset of the possible modes are included. In future work it may be interesting to consider other possible parameterizations of the function $f(\Omega)$ which are more parameter efficient; for example, we used invertible splines [8, 402–404] to efficiently parameterize invertible maps in Chapter 4 and a modification of that approach could be applied to the deformations of azimuthal and zenith angles described here. Given an explicit parameterization of the vertical shift deformation $\widetilde{\Omega}(\Omega)$, the coordinate-representation Jacobian determinant $\widetilde{j}(\Omega)$ of this transformation can be computed by explicitly constructing the Jacobian matrix and evaluating the determinant as

$$\widetilde{j}(\Omega) = \det \frac{\partial \widetilde{\Omega}_\alpha}{\partial \Omega_\beta} = \det \left( \delta_{\alpha\beta} + i \frac{\partial f(\Omega)_\alpha}{\partial \Omega_\beta} \right). \qquad (5.54)$$

For the small SU($N$) groups involved in phenomenologically relevant lattice gauge theories, this is a completely tractable calculation.

The complexification of real Lie groups is well understood, and in particular the complexification of SU($N$) is the group $SL(N, \mathbb{C})$ consisting of all $N \times N$ complex matrices with unit determinant [485]. It is thus straightforward to analytically continue the action and any observables as long as they are defined in terms of the fundamental matrix representation of gauge links.[3] Any instance of a gauge link variable $U$ can simply be replaced by $\widetilde{U}$; the adjoint gauge link $U^\dagger$ must first be rewritten as $U^{-1}$, and then can also be evaluated on the deformed contour by replacing with $\widetilde{U}^{-1}$ [477, 486, 487]. All matrix components of the fundamental representation of SU($N$) are simply given by holomorphic functions of the coordinates $\Omega$. The matrix elements of the inverse $U^{-1}$ can also be straightforwardly shown to be holomorphic functions of $\Omega$.[4] Therefore actions and observables involving $U^\dagger$ and its deformed complexification

---

[3] Analytical continuation of functions involving other representations is also possible using the full theory of complexified Lie groups [485].

[4] By holomorphy of the matrix elements of $U$, the Wirtinger derivatives with respect to the conjugate coordinates are zero, $\partial U_{ij}(\Omega)/\partial \overline{\Omega}_\alpha = 0$ [488]. Differentiating the equation $U_{ij}U_{jk}^{-1} = \delta_{ik}$ implies that

$\widetilde{U}^{-1}$ are holomorphic despite the matrix inversion. Standard gauge actions used for lattice gauge theory satisfy this form. Fermionic actions result in the presence of a determinant over the lattice-discretized Dirac operator. Unrooted determinants are simply polynomial functions of $U$ and $U^{-1}$, as are the integrands associated with many fermionic observables [465], and as such path integrals defining these theories may be deformed.

### 5.3.3 Deformations for lattice gauge theory

Deformations for single variables in the gauge group can be extended to lattice gauge theory by applying deformations to all group-valued integration variables in the path integral. These variables correspond to gauge links $U_\mu(x)$ in the standard representation of lattice gauge theory. In our applications to two-dimensional lattice gauge theory in the remainder of the chapter, we choose to instead work with matrix-valued plaquettes $P_{01}(x) = U_0(x)U_1(x+\hat{0})U_0^{-1}(x+\hat{1})U_1^{-1}(x)$. Working with deformations of plaquettes results in definitions of Wilson loop observables involving factors analogous to $e^{i\phi}$ in terms of the coordinate description of each plaquette. This allows significant signal-to-noise improvements to be made with simple contour deformations. An exact change of variables to plaquettes in the path integral is only possible in these theories because of the two-dimensional geometry and because we choose to use open boundary conditions. However, this approach can be applied in general by working with lattice degrees of freedom in a maximal-tree gauge fixing scheme. Benefits of working with open loops corresponding to gauge-fixed variables, rather than gauge-variant links, have also been seen in previous works addressing few-variable gauge theories [468–471]. A downside of working with maximal-tree gauge fixing schemes is that primary variables in such schemes are equivalent to non-local loops, as discussed in Chapter 4. It is alternatively possible to apply the constructions utilized to transform open loops of links in arbitrary spacetime dimensions in Chapter 4 to capture the benefits of working with local variables. Extensions to using gauge-fixed representations or transformations of subsets of open loops are interesting directions for future work.

If single-variable deformations are applied independently for each group-valued variable, deformed configurations can be straightforwardly computed on a per-link basis, the overall Jacobian determinant can be computed as the product over each Jacobian factor per link, and the deformed observable associated with each base observable can be measured. However, deformations with greater expressivity can be constructed by allowing interdependence between the vertical shifts defining the deformations of each gauge variable. There is no theoretical obstacle to simply extending the vertical shift functions $f(\cdot)$ for each gauge variable to depend on other variables, but the Jacobian determinant of general transformations for lattice gauge theory involve too many degrees of freedom to be tractable to compute.

In this chapter, we apply an autoregressive approach to guarantee that the Jacobian determinant of our deformations remains tractable to compute. Specifically, plaquette variables are arranged in a canonical order, $P_1, \dots, P_{\mathcal{N}}$, and the deformation of each

---

the Wirtinger derivatives of the inverse matrix components with respect to the conjugate coordinates are also zero, $\partial U_{ij}^{-1}(\Omega)/\partial \bar{\Omega}_\alpha = 0$.

variable $P_i$ is restricted to depend only on previous variables in the ordering. Labeling the coordinate representation for the $i$th plaquette as $\Omega_i$, the transformation of this set of coordinates is given in our autoregressive approach by

$$\widetilde{\Omega}_i = \Omega_i + if(\Omega_i; \{\Omega_j : j < i\}). \tag{5.55}$$

This restricts the Jacobian to be triangular, and the Jacobian determinant is given in terms of the diagonal components by

$$\widetilde{J} = \prod_i \widetilde{j}_i(\Omega_i; \{\Omega_j : j < i\}), \tag{5.56}$$

where $\widetilde{j}_i(\Omega_i)$ is the Jacobian determinant of the deformation for the $i$th variable alone.

There are many ways that variables earlier in the ordering can be factored into the deformation of the $i$th variable. To extend the Fourier deformation basis to this setting, a generic approach would be to include Fourier modes associated with all components of $\Omega_j$ for $j < i$ in the expansion given in Eq. (5.53). However, this quickly results in a very large number of parameters describing the deformed contour as the cutoff parameter $\Lambda$ is taken to infinity. As such, we restrict to a significantly smaller subset of the terms in Eq. (5.53) in the applications presented below to reduce the number of parameters involved.

## 5.4 U(1) gauge theory in (1+1)D

In this and the following section, we construct and optimize manifolds for lattice gauge theory and study their ability to alleviate signal-to-noise problems afflicting observables of interest. This section describes the application to U(1) lattice gauge theory.

### 5.4.1 Action, observables, and parameters

We consider U(1) lattice gauge theory in $(1 + 1)$ dimensions with open boundary conditions for a proof-of-concept demonstration in a context with analytically solvable results. This theory is nearly identical to the one used in the study of U(1) flow-based ensemble generation (see Sec. 4.6), up to the physically inconsequential choice of boundary conditions. The choice of dimensionality and boundary conditions here allows a change of variables from links to plaquettes (see Appendix A), which simplifies the definition of useful contour deformations for this theory.

**Action and parameters.** The action for U(1) lattice gauge theory in $(1+1)$ dimensions is identical to the action given previously in Eq. (4.45). We rewrite Eq. (4.45) to be better suited for contour deformations by replacing $P^\dagger \to P^{-1}$ and define the action using plaquettes as the primary variables, giving

$$S(P; \beta) = -\beta \sum_x \frac{1}{2} \left[ P_{01}(x) + P_{01}^{-1}(x) \right]. \tag{5.57}$$

Rewriting in this way allows the action to be trivially analytically continued to complexified field configurations. The path integral for an observable $\mathcal{O}$ represented in terms of plaquette variables is given by

$$\langle \mathcal{O} \rangle = \frac{1}{Z} \int \mathcal{D}P \; \mathcal{O}(P) \, e^{-S(P;\beta)}, \qquad (5.58)$$

where $Z = \int \mathcal{D}P \; e^{-S(P;\beta)}$ and $\mathcal{D}P = \prod_x dP_{01}(x)$ in terms of the Haar measure $dP_{01}(x)$ of each U(1)-valued plaquette. In the plaquette representation, it is clear that the path integral for this theory actually factors completely over the individual plaquettes. For an observable that can be composed as products of functions of each plaquette, the path integral average can therefore be divided into a product over single-variable averages as

$$\langle \mathcal{O}_1(P_1)\mathcal{O}_2(P_2)\ldots \rangle = \langle \mathcal{O}_1 \rangle_{1\mathrm{var}} \langle \mathcal{O}_2 \rangle_{1\mathrm{var}} \ldots, \qquad (5.59)$$

where $P_1, P_2, \ldots$ label the plaquettes enumerated in an arbitrary order across the lattice, and $\langle \cdot \rangle_{1\mathrm{var}}$ denotes the average with respect to the single-plaquette distribution,

$$\langle \mathcal{O} \rangle_{1\mathrm{var}} = \frac{\int dP \, \mathcal{O}(P) \, e^{\frac{\beta}{2}[P+P^{-1}]}}{\int dP \, e^{\frac{\beta}{2}[P+P^{-1}]}}. \qquad (5.60)$$

This factorization is used to analytically determine properties of some deformations below.

Each plaquette can be given an explicit coordinate representation $P_{01}(x) = e^{i\phi(x)}$ in terms of the angular variable $\phi(x)$. The Haar measure is related to the measure over these coordinates by $dP_{01}(x) = \frac{d\phi(x)}{2\pi}$. The action takes a manifestly holomorphic form when written in terms of these coordinates:

$$S(\phi; \beta) = -\beta \sum_x \frac{1}{2} \left[ e^{i\phi(x)} + e^{-i\phi(x)} \right] = -\beta \sum_x \cos\phi(x). \qquad (5.61)$$

We are thus free to deform the azimuthal angles $\phi(x)$ without modifying observable expectation values.

To study the effect of deformations over a range of parameters, we worked with the four choices of $\beta$ given in Table 5.1, corresponding to string tensions $\sigma$ (see the discussion of Wilson loops below) ranging over the values 0.1–0.4. The four values of $\beta$ were analytically determined to reproduce these target values of $\sigma$. In all cases, we used lattices with an extent of $64 \times 64$ lattice units.

**Wilson loops.** The Wilson loop $W_{\mathcal{A}}$ over a rectangular region $\mathcal{A}$ with area $A$ can be interpreted as the correlation function of a static quark-antiquark pair. Wilson loops in this theory scale with an exact area law, $W_{\mathcal{A}} = e^{-\sigma A}$, where the exponent is set by the *string tension* $\sigma$. Area law scaling indicates that static charges are confined in this theory and the string tension determines the strength of the confining force. This scaling can be seen directly in the plaquette representation, in which Wilson loops can be defined by a product over all plaquette variables contained within the region,

$$\langle W_{\mathcal{A}} \rangle = \left\langle \prod_{x \in \mathcal{A}} e^{i\phi(x)} \right\rangle . \tag{5.62}$$

| $\beta$ | 1.843 | 2.296 | 3.124 | 5.555 |
|---------|-------|-------|-------|-------|
| $\sigma$ | 0.4 | 0.3 | 0.2 | 0.1 |
| $\delta$ | 0.65 | 0.52 | 0.37 | 0.20 |

**Table 5.1:** Choices of the coupling parameter $\beta$ and corresponding string tension $\sigma$ studied for the application of observifolds methods to U(1) lattice gauge theory in (1 + 1)D. Exact results for the string tension were used to choose the values of $\beta$. In addition, we report the optimal contour deformation parameter $\delta$ for the construction of deformed observables for each $\beta$, as described in the main text.

By the factorization of path integral estimates over plaquettes, this expectation value is equivalent to the $A$th power of a one-plaquette expectation value,

$$\langle W_{\mathcal{A}} \rangle = \left[ \left\langle e^{i\phi} \right\rangle_{1\mathrm{var}} \right]^A = e^{-\sigma A} . \tag{5.63}$$

The string tension can be exactly computed by evaluating the one-variable integral, giving

$$\sigma = \left[ \frac{I_0(\beta)}{I_1(\beta)} \right] , \tag{5.64}$$

in terms of the modified Bessel functions $I_n(\beta)$. The derivations of this and other exact results used below are provided in Appendix A.

Though the theory is solvable, Monte Carlo estimates of Wilson loops are still affected by an exponentially severe signal-to-noise problem, providing a useful testing ground for novel algorithms. The action in Eq. (5.61) is invariant under a local charge conjugation symmetry that transforms $\phi(x) \to -\phi(x)$. This symmetry consideration has analogues in general lattice gauge theories that allow us to determine that Wilson loop expectation values are real, thus we focus on Monte Carlo estimates of $\langle \mathrm{Re}[W_{\mathcal{A}}] \rangle$. We have just shown that the (real) expectation values of Wilson loops decrease exponentially with area. The variance of the real component of the Wilson loop, on the other hand, can be computed using the variance correlator to be $O(1)$ at all choices of area,

$$\begin{aligned}
\mathrm{Var}[\mathrm{Re}\, W_{\mathcal{A}}] &= \frac{1}{2} \left\langle |W_{\mathcal{A}}|^2 \right\rangle + \frac{1}{2} \left\langle W_{\mathcal{A}}^2 \right\rangle - \langle \mathrm{Re}\, W_{\mathcal{A}} \rangle^2 \\
&= \frac{1}{2} + \frac{1}{2} e^{-\sigma' A} - e^{-2\sigma A} ,
\end{aligned} \tag{5.65}$$

where $\sigma' = \ln \left[ I_0(\beta)/I_2(\beta) \right]$. Combining the variance and signal results, the signal-to-noise ratio is exactly given by

$$\mathrm{StN}[W_{\mathcal{A}}] = \frac{e^{-\sigma A}}{\sqrt{\frac{1}{2} + \frac{1}{2}e^{-\sigma' A} - e^{-2\sigma A}}}. \tag{5.66}$$

This asymptotically falls off as $e^{-\sigma A}$, and the Wilson loop estimator therefore has a signal-to-noise problem for loops with large area.

### 5.4.2 Contour deformations

We apply contour deformations of the path integral to address the exponential signal-to-noise problem of Wilson loops in this theory.

**Family of manifolds.** The observable written in terms of plaquette coordinates in Eq. (5.62) already takes the form of an exponential analogous to $e^{i\phi}$ considered in toy theories earlier in this chapter. This suggests that constant imaginary shifts in the positive imaginary direction will be effective at reducing the average magnitude. For this proof-of-principle study, we therefore consider families of constant-shift manifolds defined by

$$\widetilde{\phi}(x) = \phi(x) + i\delta(x), \tag{5.67}$$

where $\delta(x) \in \mathbb{R}$ defines the shift in the imaginary direction for the angular variable associated with each plaquette. The Jacobian determinant of the transformation applied to each variable $\phi(x)$ is trivial, i.e. $\widetilde{j}(\phi(x)) = 1$, and therefore $\widetilde{J}(\phi) = 1$ for the total Jacobian determinant of the transformation across all sites. The parameterization used in Eq. (5.67) corresponds to a Fourier-series manifold with the cutoff fixed to $\Lambda = 0$, giving only contributions from the constant Fourier mode for the deformation of each variable. More general deformations are possible for this theory, but as demonstrated below the choice $\Lambda = 0$ is already sufficient to almost entirely remove the signal-to-noise problem in this simple theory.

The manifold deformation given in Eq. (5.67) defines a deformed observable $\mathcal{Q}_{\mathcal{A}}$ corresponding to each base observable $W_{\mathcal{A}}$. Explicitly evaluating the deformed observable for an arbitrary choice of $\delta(x)$ gives

$$\begin{aligned}
\mathcal{Q}_{\mathcal{A}} &= W_{\mathcal{A}}(\widetilde{\phi}) \, \widetilde{J}(\phi) \, e^{-S(\widetilde{\phi}(\phi)) + S(\phi)} \\
&= \left[ \prod_{x \in \mathcal{A}} e^{i\phi(x)} e^{-\delta(x)} \right] \exp\left( \beta \sum_x \cos(\phi(x) + i\delta(x)) - \cos\phi(x) \right).
\end{aligned} \tag{5.68}$$

The deformed observable has identical expectation value to the base Wilson loop observable for all choices of area, i.e. $\langle \mathcal{Q}_{\mathcal{A}} \rangle = \langle W_{\mathcal{A}} \rangle$, and $\mathcal{Q}_{\mathcal{A}}$ can be used as an equally valid Monte Carlo estimator of the Wilson loop.

Given that the path integral for this theory factorizes individually over plaquettes, we can make several further analytical simplifications to determine a good choice of contour deformation. First, we note that the signal-to-noise problem of the original observable $W_{\mathcal{A}}$ can be attributed to the product of the independently fluctuating variables $e^{i\phi(x)}$ for $x \in \mathcal{A}$. Because these phase fluctuations only arise from plaquettes within $\mathcal{A}$, there is no advantage to deforming plaquettes outside this region when constructing the deformed observable $\mathcal{Q}_{\mathcal{A}}$. We thus fix $\delta(y) = 0$ for all $y \notin \mathcal{A}$. Secondly, since the variance arises independently from each plaquette, we can optimize the deformation for the single-site integral once and for all, then apply that deformation independently to each plaquette contained in $\mathcal{A}$. Our goal is thus to optimize the choice of a constant $\delta$ which will then define the parameters $\delta(x) = \delta$ for all $x \in \mathcal{A}$.

**Optimization.** The reduced family of possible one-parameter deformations defined by $\delta$ could be optimized numerically by applying the iterative optimization procedure outlined in Sec. 5.2.2. The simplicity of this deformation allows us to instead analytically determine the signal-to-noise ratio of the deformed observable and analytically identify the optimal choice of $\delta$ for all Wilson loop observables. We consider both the expectation value and variance of the Monte Carlo estimator $\mathrm{Re}\,\mathcal{Q}_{\mathcal{A}}$ for the real component of the Wilson loop. The expectation value is identical to the undeformed observable,

$$\langle \mathrm{Re}\,\mathcal{Q}_{\mathcal{A}} \rangle = e^{-\sigma A}. \tag{5.69}$$

The variance can be computed analytically by reducing to tractable one-plaquette integrals, giving

$$\begin{aligned}
\mathrm{Var}[\mathrm{Re}\,\mathcal{Q}_{\mathcal{A}}] &= \frac{1}{2}\left\langle |\mathcal{Q}_{\mathcal{A}}|^2 \right\rangle + \frac{1}{2}\,\mathrm{Re}\left\langle \mathcal{Q}_{\mathcal{A}}^2 \right\rangle - \langle \mathrm{Re}\,\mathcal{Q}_{\mathcal{A}} \rangle^2 \\
&= \frac{1}{2}\left\langle e^{-2\delta} e^{2\beta\,\mathrm{Re}\cos(\phi+i\delta) - 2\beta\cos\phi} \right\rangle_{1\mathrm{var}}^A \\
&\quad + \frac{1}{2}\left\langle e^{-2\delta} e^{2i\phi} e^{2\beta\cos(\phi+i\delta) - 2\beta\cos\phi} \right\rangle_{1\mathrm{var}}^A - e^{-2\sigma A} \\
&= \frac{1}{2} e^{-(2\delta+\sigma_\delta)A} + \frac{1}{2} e^{-(2\delta+\sigma'_\delta)A} - e^{-2\sigma A},
\end{aligned} \tag{5.70}$$

where

$$\begin{aligned}
\sigma_\delta &= -\ln\left[\frac{I_0(\beta(2\cosh\delta - 1))}{I_0(\beta)}\right], \\
\sigma'_\delta &= -\ln\left[\frac{I_2(\beta\sqrt{5 - 4\cosh\delta})}{I_0(\beta)}\left(\frac{e^\delta - \frac{1}{2}}{e^{-\delta} - \frac{1}{2}}\right)\right].
\end{aligned} \tag{5.71}$$

These expressions are complicated, but a key feature of the deformed variance is that all terms are exponentially suppressed at large area: the first two terms in Eq. (5.70) scale proportionally to $e^{-2\delta A}$ while the last term is the expectation value of the observable squared which scales as $e^{-2\sigma A}$.

An optimal choice of $\delta$ maximizes the signal-to-noise ratio for $\mathrm{Re}\,\mathcal{Q}_{\mathcal{A}}$, which is given in terms of Eqs. (5.69) and (5.70) as

$$\mathrm{StN}[\mathrm{Re}\,\mathcal{Q}_{\mathcal{A}}] = \frac{e^{-\sigma A}}{\sqrt{\frac{1}{2} e^{-(2\delta+\sigma_\delta)A} + \frac{1}{2} e^{-(2\delta+\sigma'_\delta)A} - e^{-2\sigma A}}}. \tag{5.72}$$

Because the numerator is independent of $\delta$ (the expectation value is unchanged by deformation), maximizing the signal-to-noise ratio can be achieved by minimizing the variance given in Eq. (5.70). This can be performed numerically, and the result demonstrates surprisingly little dependence on the area $A$. For example, the optimal $\delta$ for the choice $\beta = 5.555$ (corresponding to $\sigma = 0.1$) varies between $\delta \approx 0.204$ for $A = 1$ and $\delta \approx 0.197$ for $A = 1000$, and empirically the optimal choices of $\delta$ converge to a fixed value as $A$ is taken large. One expects from the qualitative arguments given earlier that the average magnitude should be reduced to approximately $e^{-\sigma A}$ in order for the deformed observable to have significantly reduced phase fluctuations and an improved signal-to-noise ratio. This is approximately the case for the optimal choice of $\delta$: the deformed observable includes a factor of $e^{-\delta A}$ and the optimal choice of $\delta$ is roughly equivalent to $\sigma$ for this choice of parameters. To more fully understand this result, we must also account for contributions to the magnitude from the term $e^{-S(\tilde{\phi}(\phi))+S(\phi)}$ in the deformed observable. Expanding the real part of the term in the exponent explicitly gives

$$\beta \sum_x \text{Re} \cos(\phi(x) + i\delta(x)) - \cos\phi(x) = \beta \sum_x \cos\phi(x)[\cosh(\delta(x)) - 1]. \quad (5.73)$$

This contributes a factor on the order of $\beta[\cosh(\delta) - 1]A$ to the log-magnitude of the deformed Wilson loop observable. Altogether the average magnitude of the deformed observable scales approximately as $e^{-\delta A + \beta[\cosh(\delta)-1]A}$ when $\cos\phi \approx 1$. For the optimal choices of $\delta$ found for $\beta = 5.555$ this equals $e^{-0.089A}$, very nearly matching the scaling of the expectation value $e^{-\sigma A} = e^{-0.1A}$. To refine this analysis further, one would need to account for the average fluctuations in $\cos\phi$, but not much more insight can be gained from more complete calculations of the average magnitude.

### 5.4.3 Monte Carlo studies

To confirm exactness and investigate the scale of improvements for practical calculations, we performed Monte Carlo calculations of the base and deformed observables for Wilson loops considered for this theory. For each of the four choices of $\beta$ given in Table 5.1, we produced ensembles of $n = 10000$ samples of $64 \times 64$ lattices. In particular, we independently generated $64 \times 64 \times 10000$ Monte Carlo samples of individual plaquette variables according to the von Mises distribution

$$p(\phi; \beta) = \frac{1}{2\pi I_0(\beta)} e^{\beta \cos(\phi)}, \quad (5.74)$$

then composed these into $10\,000$ lattices with $64 \times 64$ geometry. Utilizing the factorized nature of the path integral in this way resulted in independent Monte Carlo samples of lattice configurations, eliminating the need to measure and quantify autocorrelations for this proof-of-principle study.

For all four ensembles, we measured $W_\mathcal{A}$ and $\mathcal{Q}_\mathcal{A}$ for regions $\mathcal{A}$ consisting of between 1 and 4096 plaquettes. Figures 5.9 and 5.10 compare the results for the base observable $W_\mathcal{A}$ versus the deformed observable $\mathcal{Q}_\mathcal{A}$ for a range of Wilson loop areas $A$ for which the signal remains reliable. As shown in the figures, the estimates of $\mathcal{Q}_{\mathcal{A}}$ are far more precise than the corresponding estimates of $W_{\mathcal{A}}$, and the results are consistent within statistical uncertainties in all cases. The theoretical signal-to-noise ratios for both the base (Eq. (5.66)) and deformed (Eq. (5.72)) observables are shown in the right panels of both figures. The theoretical signal-to-noise ratio was found to agree with the Monte Carlo results where those results could be precisely measured. Neither the signal-to-noise ratio nor the Wilson loop estimates can be precisely estimated when the signal-to-noise ratio is smaller than approximately $1/\sqrt{n} = 10^{-2}$, but the theoretical results allow extrapolating the base observables results to larger values of the area for comparison to the much more precise deformed observables. The improved signal-to-noise ratio of the deformed observable enables controlled estimates of Wilson loops with areas that are orders of magnitude larger than the maximum Wilson loop area accessible using the base observables given the same statistics and computational cost.

An estimate $\sigma^{\text{eff}}$ of the string tension can be defined based on Wilson loop estimates as

$$\sigma^{\text{eff}}(A) = \ln \langle W_{\mathcal{A}} \rangle - \ln \langle W'_{\mathcal{A}} \rangle, \tag{5.75}$$

where $\mathcal{A}$ and $\mathcal{A}'$ are regions with areas related by $A' = A + 1$. Using $\langle \mathcal{Q}_{\mathcal{A}} \rangle$ in place of $\langle W_{\mathcal{A}} \rangle$ results in a different estimator with distinct signal-to-noise properties. Analytical results for the signal-to-noise ratio are difficult to obtain directly for $\sigma^{\text{eff}}$, however Monte Carlo estimates can be used to confirm that the deformed observable also results in significant improvements in the statistical precision of these quantities. Figure 5.11 shows the measured $\sigma^{\text{eff}}(A)$ using both the base and deformed observable for the ensemble with $\beta = 5.555$. Both estimates are consistent with the exact value of $\sigma$, but the deformed observable provides reliable estimates up to choices of $A$ that are orders of magnitude larger than those measurable using the base observable using the same statistics. The significantly increased precision and quantity of $\sigma^{\text{eff}}$ measurements could in principle be fit to a 'plateau' to provide very precise estimates of $\sigma$. In this theory $\sigma$ is exactly known, but it is a physically relevant quantity that must be measured in more general theories using such fits.

The analytical and numerical results both confirm that simple contour deformations based on shifts in the imaginary direction in complexified field space are sufficient to significantly increase the signal-to-noise ratio of Wilson loops estimates in this $(1+1)$D lattice gauge theory with open boundary conditions. The statistics of two-dimensional lattice gauge theory with open boundary conditions are equivalent to those in the case of periodic boundary conditions, up to exponentially small finite volume effects. Similar deformations are therefore expected to improve the signal-to-noise ratio of these observables in the case of periodic boundary conditions as well. However, when periodic boundary conditions are applied it is no longer possible to exactly change variables from gauge links to plaquettes. In this case one could instead choose to work with links in the original configuration, but it has been demonstrated that it is difficult to identify useful transformations of links in non-gauge-fixed representations [468]. Though deformations with arbitrary expressivity must in principle be able to capture similar transformations to those applied to plaquettes, this suggests that deformations directly acting on this basis may not be a useful practical starting point for contour deformations for gauge

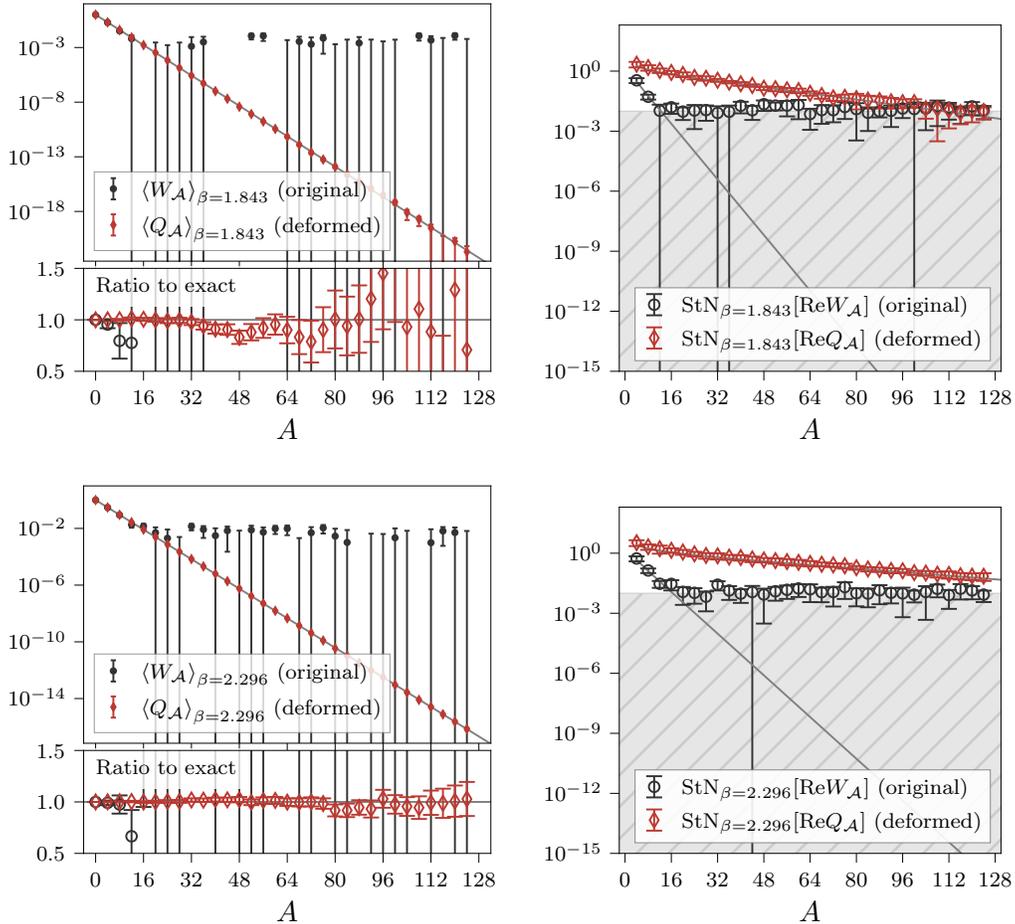

**Figure 5.9:** Measurements of the Wilson loop Monte Carlo expectation values (left) and signal-to-noise ratios (right) using both the base observable $W_{\mathcal{A}}$ and deformed observable $\mathcal{Q}_{\mathcal{A}}$ on ensembles with $\beta \in \{1.843, 2.296\}$. In all cases, the analytically determined values are indicated by the solid gray lines. The lower panel of each figure on the left indicates the ratio of measured expectation values to the analytically determined exact results, demonstrating agreement within statistical errors for all measurements. In each of the figures on the right, the gray hatched regions indicate a signal-to-noise ratio below $1/\sqrt{n} = 10^{-2}$ in terms of the ensemble size $n = 10000$ used for this study. Below this threshold, neither the signal-to-noise ratio nor the Wilson loop expectation values could be reliably estimated by the Monte Carlo values, but theoretical curves allow a comparison to the base observable at larger areas than are accessible via Monte Carlo data alone. Figure adapted from Fig. 1 of Pub. [4].

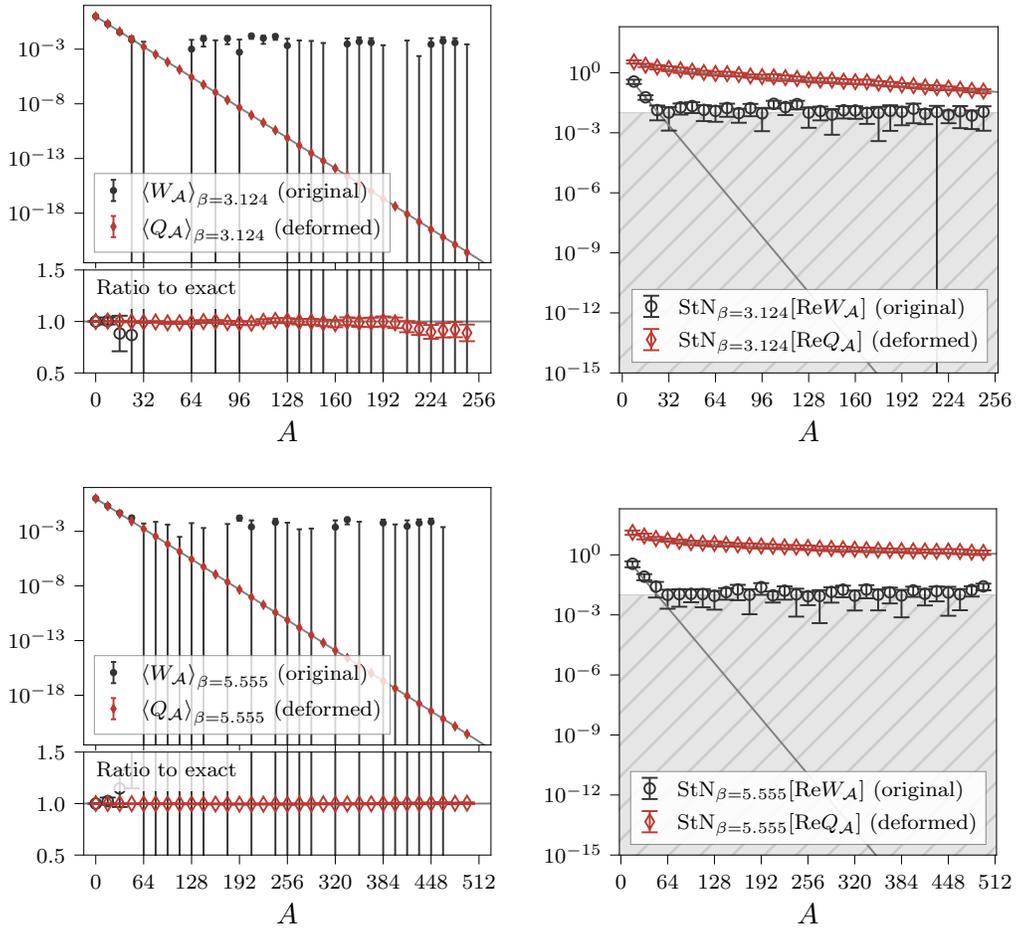

**Figure 5.10:** Measurements of the Wilson loop Monte Carlo expectation values (left) and signal-to-noise ratios (right) using both the base observable $W_{\mathcal{A}}$ and deformed observable $\mathcal{Q}_{\mathcal{A}}$ on ensembles with $\beta \in \{3.124, 5.555\}$. See Figure 5.9 for details.

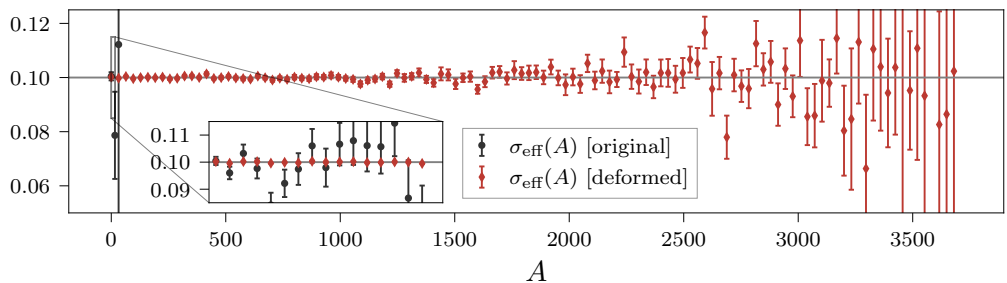

**Figure 5.11:** A comparison of $\sigma^{\mathrm{eff}}$ measured using Monte Carlo estimates of the base observable $W_{\mathcal{A}}$ (gray circle markers) versus $\sigma^{\mathrm{eff}}$ measured using Monte Carlo estimates of the deformed observable $\mathcal{Q}_{\mathcal{A}}$ (red diamond markers). In the outer plot, measurements of $\sigma^{\mathrm{eff}}$ are plotted for values of $A$ that are multiples of 16 and 32 for the base and deformed observables, respectively. Reliable estimates of $\sigma^{\mathrm{eff}}(A)$ are possible for much larger values of $A$ using the deformed observable, and estimates derived from the deformed observable have much higher precision throughout, as shown for $0 \leq A < 16$ in the inset. Figure adapted from Fig. 1 of Pub. [4].

theory observables.

To circumvent this problem in future studies, it is an intriguing possibility to apply the coupling layer architectures described in Sec. 5.3.3 to describe deformations in terms of loops. Though the coupling layers defined in that section are constructed to provide invertible transformations within the real domain of integration of the path integral, rather than complexified deformations, the aims of these transformations are quite similar. It has been shown that normalizing flows and contour deformations may be defined in a unified framework [489]. Modifying gauge-equivariant coupling layers to extend these methods to observifold deformations may provide a useful path forward when periodic boundary conditions are applied or when working in higher spacetime dimensions, where similar obstacles arise.

Though it is not yet clear which families of manifolds will allow deformations when working in a basis other than the plaquette basis for gauge theories, this is a practical obstacle rather than a theoretical one. To demonstrate that periodic boundary conditions do not present an obstacle to the successful applications of observifolds and to consider a more complicated extension of the deformations described in this section, we next investigate contour deformations for correlation functions in the $(0 + 1)$D theory of a complex scalar.

### 5.4.4 Extension to a complex scalar in (0+1)D

Though the action in Eq. (5.61) was derived for $(1 + 1)$D U(1) lattice gauge theory, there is no dependence on the two-dimensional geometry of the lattice in the action. This is the case whether or not the path integral is explicitly transformed to integrate over plaquettes, which is possible only for open boundary conditions. An interesting connection can thus be made to the theory of a complex scalar $\Phi(t) \in \mathbb{C}$ in $(0 + 1)$ dimensions, where $t$ labels the single temporal coordinate in this spacetime geome-try. Though this is superficially a significantly different theory, taking the limit of an infinitely deep quartic potential yields a non-linear sigma model with identical path integral representation as the U(1) gauge theory. In particular, the polar decomposition $\Phi(t) = R(t)e^{i\phi(t)}$ allows one to write a lattice description of the theory with an arbitrary radial potential using the discretized action

$$
\begin{aligned}
S(\Phi) &= \sum_t \Phi^*(t)[2\Phi(t) - \Phi(t+1) - \Phi(t-1)] + V(|\Phi(t)|) \\
&= \sum_t -2R(t)R(t+1)\cos[\phi(t+1) - \phi(t)] + V(R(t)) + 2R^2(t).
\end{aligned}
\tag{5.76}
$$

Applying a quartic potential $V(R(t)) = m^2 R^2(t) + \lambda R^4(t)$ and taking the non-linear sigma limit given by $m^2 \to -\infty$ and $\lambda > 0$ restricts the degrees of freedom to the angular variable $\phi(t)$ alone. This yields the same action as Eq. (5.61) if the differences $\phi(t+1) - \phi(t)$ are identified with U(1) plaquette degrees of freedom and $\beta$ is identified with $2\langle R\rangle^2$ in terms of the vacuum expectation value $\langle R\rangle$ of the frozen radial coordinate. Outside of the non-linear sigma limit, both $\phi(t)$ and $R(t)$ are dynamical degrees of freedom. We consider signal-to-noise problems in this general case to extend the results derived earlier in this section. Path integral expectation values will be denoted here by $\langle\cdot\rangle_\Phi$ to distinguish this theory.

Two-point correlation functions of the complex scalar in this general case provide an analogous observable to Wilson loops in U(1) lattice gauge theory. We define the two-point correlation function between time 0 and $t$ by

$$
C(t) = \langle\Phi(t)\Phi^*(0)\rangle_\Phi, \quad \text{where} \quad \Phi(t)\Phi^*(0) = R(t)R(0)e^{i\phi(t)-i\phi(0)}.
\tag{5.77}
$$

A charge conjugation symmetry implies that this quantity is real, and we restrict focus to the Monte Carlo estimator $\mathrm{Re}[\Phi(t)\Phi^*(0)]$. Though an analytical calculation of the signal-to-noise ratio of this estimator is not possible in this theory, a Parisi-Lepage argument based on the physical content of the variance correlator (see Sec. 2.5.2) can be applied instead to determine the asymptotic behavior. The scaling of the variance is set by

$$
\mathrm{Var}[\mathrm{Re}[\Phi(t)\Phi^*(0)]] \sim \langle[\Phi(t)\Phi^*(0)][\Phi^*(t)\Phi(0)]\rangle_\Phi = \langle|\Phi(t)|^2|\Phi(0)|^2\rangle_\Phi.
\tag{5.78}
$$

The operators $|\Phi(0)|^2$ and $|\Phi(t)|^2$ overlap onto the vacuum state of this theory and this implies that the variance asymptotically scales as a constant in the temporal separation $t$. On the other hand, the two-point correlation function does not overlap the vacuum state, and the signal-to-noise will therefore decay exponentially with a scale set by the energy gap of this state accessed by the two-point correlation function.

The exponential decay of the two-point correlation with temporal separation $t$ gives information about the spectrum of this theory, and the large-$t$ behavior in particular scales as $C(t) \sim e^{-m_\Phi t}$ in terms of the mass $m_\Phi$ of the scalar particle. An effective mass estimator can be defined analogously to the effective string tension $\sigma^{\mathrm{eff}}$ in the U(1) lattice gauge theory study. We use an estimator based on an arccosh form accounting for thermal effects from the periodic boundary conditions in this case,

$$m^{\text{eff}}(t) = \text{arccosh}\left(\frac{C(t-1) + C(t+1)}{2C(t)}\right). \tag{5.79}$$

This estimator will also be afflicted by severe noise if estimates of $C(t)$ are noisy.

The similarities between Eq. (5.76) and the U(1) lattice gauge theory action and between Eq. (5.77) and the U(1) Wilson loop observable suggest that analogous contour deformations of the $\phi(t)$ degrees of freedom will achieve similar results in this case. We therefore define a family of manifolds by

$$\widetilde{\phi}(t) = \phi(t) + i\delta(t). \tag{5.80}$$

In this parameterization, the differences $\delta(t+1) - \delta(t)$ correspond to the imaginary shifts of plaquette variables in the case of U(1) lattice gauge theory. These differences are automatically constrained to sum to zero when summed over all $t$ in the periodic boundary conditions employed here.

The action in Eq. (5.76) and correlation function in Eq. (5.77) are both holomorphic functions of $\phi(t)$. The path integral measure must also be rewritten in terms of polar coordinates, giving

$$\mathcal{D}\Phi = \prod_t d\,\text{Re}\,\Phi(t)\,d\,\text{Im}\,\Phi(t) = \prod_t R(t)\,dR(t)\,d\phi(t). \tag{5.81}$$

The extra factor of $R(t)$ is also a holomorphic function of $R(t)$ and $\phi(t)$ and therefore the path integral over the $R(t)$ and $\phi(t)$ degrees of freedom can be safely deformed by Eq. (5.80). Deformations could additionally be applied to $R(t)$ if the limits $R(t) \to 0$ and $R(t) \to \infty$ were appropriately held fixed, however based on intuition from U(1) lattice gauge theory and for simplicity we focus here on deformations of $\phi(t)$ alone.

The deformation in Eq. (5.80) can be used to define a deformed observable

$$\mathcal{Q}(t) = R(t)R(0)e^{i\widetilde{\phi}(t) - i\widetilde{\phi}(0)}\frac{e^{-S(\widetilde{\phi},R)}}{e^{-S(\phi,R)}}, \tag{5.82}$$

where the Jacobian determinant is included but is trivial for this deformation. The expectation value of this deformed observable equals $C(t)$ for any deformation by holomorphy. Estimates of $C(t)$ using deformed observables can thus also be used to compute $m^{\text{eff}}(t)$.

To study the improvement made possible by this family of contour deformations, we performed a Monte Carlo study over the four choices of $\lambda \in \{0, 1, 2, 3\} \times 10^{-3}$ and the fixed choice of $m^2 = (0.15)^2$ for $(0+1)$D lattices with temporal length $L_t = 64$ lattice sites and periodic temporal boundaries. At the largest value of the quartic coupling, $\lambda = 3 \times 10^{-3}$, the theory is well outside the perturbative regime based on observations of $O(1)$ corrections to the scalar particle mass $m_\Phi \not\approx m$. We generated ensembles of $10\,000$ samples using HMC at all four choices of parameters, with HMC parameters tuned to keep autocorrelation times shorter than 1 for measured correlation functions. We also confirmed that the topological charge corresponding to the winding of the phase $\phi(t)$ around the periodic boundaries of the lattice was well sampled in the generated ensembles. Each ensemble of 10 000 samples was divided into a training set of 1000 configurations, a test set of 1000 configurations, and a measurement set consisting of the remaining 8000 configurations. The training set was used exclusively for optimization, the test set was used to measure the loss function and determine an early stopping time to avoid overtraining, and the measurement set was used to measure observables for our comparisons below.

For each choice of parameters and each value of the temporal separation $t$, we used the Adam optimizer [352] to numerically optimize the parameters $\delta(t')$ defining a deformed manifold such that the loss function associated with $\mathrm{Re}[\mathcal{Q}(t)]$ given in Eq. (5.29) was minimized. In each step of optimization, we computed a stochastic estimate of the loss function given in Eq. (5.32) using a batch of configurations sampled from the training set, used automatic differentiation in a JAX [354] implementation to compute its gradients, then applied one step of the Adam optimizer to update manifold parameters. Manifold parameters were iteratively updated until a plateau of the test loss function was reached,[5] at which point we halted the optimization to avoid overtraining.

The resulting optimized choices of $\delta(t')$ associated with each possible base observable $\Phi(t)\Phi^*(0)$ were then used to measure $\langle \mathcal{Q}(t) \rangle_\Phi = \langle \Phi(t)\Phi^*(0) \rangle_\Phi = C(t)$. Figure 5.12 compares the variance of the base observable to that of the deformed observable across the four choices of physical parameters. In all cases the variance of the base observable is asymptotically constant for large separations $0 \ll t \ll L_t$, where the latter inequality is due to the periodic boundary conditions. This matches our expectations based on the Parisi-Lepage scaling arguments above. On the other hand, the variance of the deformed observable decreases consistently until the maximum separation $t = L_t/2$ is reached. At this maximum separation, the variance is reduced by an order of magnitude or more compared to the base observable. In addition, we find that the variance reduction is more significant for the ensembles with larger quartic couplings $\lambda$. These are the ensembles with larger renormalized masses $m_\Phi$ and thus a more severe signal-to-noise problem for the base observable. We did not examine this effect with further parameter scans, but this is suggestive of a scale of variance reduction that grows with the renormalized mass itself.

We further used Monte Carlo estimates of $C(t)$ derived from both the base and deformed observables to compute the effective mass $m^{\mathrm{eff}}(t)$, as shown for the ensemble with the largest quartic coupling in Figure 5.13. Each trace is truncated where the uncertainties became larger than 75 % of the central value. The variance is significantly reduced for estimates of $C(t)$ derived from the deformed observables and as a result estimates of $m^{\mathrm{eff}}(t)$ at larger values of $t$ have significantly reduced uncertainties with respect to $m^{\mathrm{eff}}(t)$ estimated using the base observable. In particular, the uncertainties were reduced by a factor of 1.5–2 in the range $4 \leq t \leq 9$, and measurements below the 75 % noise threshold were possible for several additional values of $t$ using the deformed observable estimates. In many lattice field theory applications, excited state contami-nation prevents fits to effective masses at small temporal separation $t$. Additional and more precise estimates of the effective mass at larger values of the temporal separation are thus a significant improvement that can result in more precisely constrained fits.

---

[5]The precise halting procedure is described more fully in the application to $\mathrm{SU}(N)$ gauge theory in Sec. 5.5.

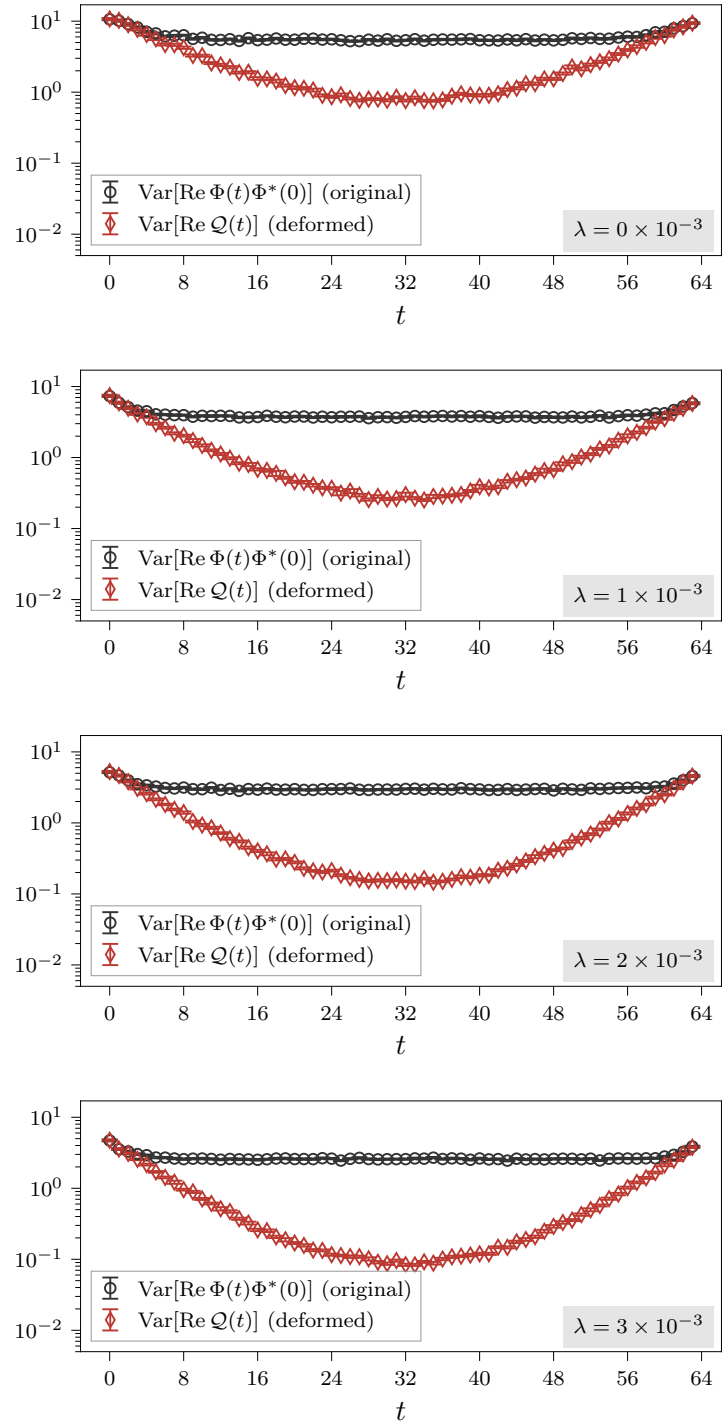

**Figure 5.12:** Variance of the optimized deformed observable $\mathrm{Re}[\mathcal{Q}(t)]$ vs. the base observable $\mathrm{Re}[\Phi(t)\Phi^*(0)]$ measured on ensembles generated using each of the four choices of quartic coupling $\lambda \in \{0, 1, 2, 3\} \times 10^{-3}$.

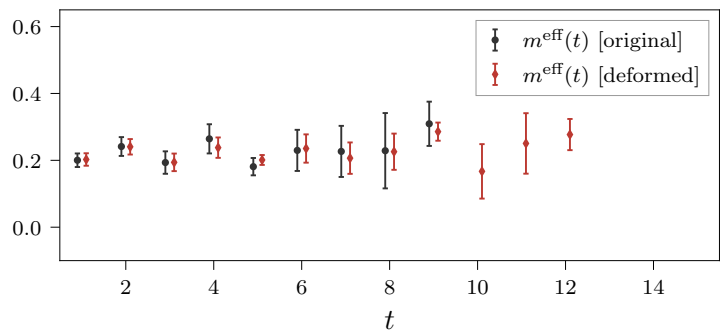

**Figure 5.13:** Effective mass $m^{\text{eff}}(t)$ computed from Monte Carlo estimates of $C(t)$ given by the original and deformed observables applied to the same ensemble with coupling $\lambda = 3 \times 10^{-3}$.

Finally, for this simple $(0+1)$D theory we can analyze the learned manifold parameters explicitly. Based on intuition from the case of U(1) lattice gauge theory, we expect constant imaginary shifts of differences $\phi(t' + 1) - \phi(t')$ in the interval $[0, t]$ to result in significant reduction in the average magnitude and correspondingly in the variance of the deformed observable $\mathcal{Q}(t)$. This would correspond to a linearly increasing set of $\delta(t')$ within the interval $[0, t]$. Unlike our study of U(1) lattice gauge theory, periodic boundary conditions here prevent a constant value of $\delta(t')$ for $t' > t$ without resulting in a sharp jump at $t' = L-1$. Instead, one might expect that the value of $\delta(t)$ and $\delta(0)$ smoothly connect over the remaining interval and through the periodic boundaries. This is indeed the case for the learned manifold parameters, as shown in Figure 5.14.

These results demonstrate that periodic boundary conditions and the presence of additional fields — in this case the fluctuating field magnitude $R(t)$ — do not obstruct the application of the method of observifolds. Significant improvements in the variance of Monte Carlo estimates of $C(t)$ are possible without introducing bias, and the resulting correlation function also can be used to derive precise estimates of quantities such as the effective mass. We do not see an improvement that is as striking as the case of U(1) lattice gauge theory, but this is not entirely surprising: in this study we extended the simple deformations of angular coordinates $\phi(t)$ to the more complex case of the scalar field and simply ignored the radial coordinates $R(t)$. In the non-linear sigma limit of the model, we argued that the theory reduces to the U(1) lattice gauge theory with periodic boundary conditions and the effective coupling parameter $\beta = 2 \langle R \rangle^2$. Away from this limit, however, fluctuations in $R(t)$ prevent the identification of a fixed coupling parameter and an optimal constant imaginary shift of each $\phi(t)$. Instead, one expects an optimal deformation to depend on $R(t)$ as well as $\phi(t)$. Pub. [4] discusses additional deformations that do depend on $R(t)$ which make further exponential improvements in the signal-to-noise ratio of correlation functions and the precision of the effective mass. Because our main interest is in applications to lattice gauge theories, we do not detail these more complicated deformations here, but pursuing improved deformations along those lines may be fruitful for extensions of this work to scalar field theory in higher dimensions.

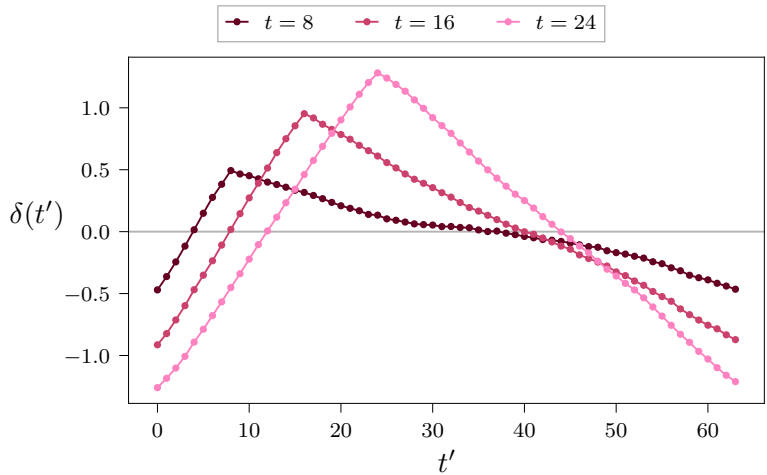

**Figure 5.14:** Learned manifold parameters $\delta(t')$ for three choices of correlator separation $t$ corresponding to three deformed observables $\mathcal{Q}(t)$. A linearly increasing value of $\delta(t')$ can be seen in the respective intervals $[0, t]$, while the remaining values of $\delta(t')$ smoothly connect the endpoints through the periodic boundaries. This structure matches the intuition from the U(1) lattice gauge theory results extended to the case of periodic boundary conditions, as discussed in the main text.

## 5.5 SU(N) gauge theory in (1+1)D

The contour deformation approaches applied to U(1) lattice gauge theory in the previous section relied on a constant imaginary shift that was already sufficient to make progress given the simple structure of the theory and observables. We next apply observifolds to SU(2) and SU(3) lattice gauge theory and investigate the improvements in observable signal-to-noise attained by applying contour deformations suitable for the more complicated SU(N) group manifolds.

### 5.5.1 Action, observables, and parameters

We consider SU(N) lattice gauge theory in (1 + 1) dimensions with open boundary conditions. In our construction of contour deformations and Monte Carlo studies, we specialize to the cases of $N = 2$ and $N = 3$ which have the most relevance for the application of lattice gauge theory to phenomenology. As for the case of the U(1) gauge group, this theory is analytically solvable [38, 39, 392]. We utilize analytically derived results to confirm the accuracy of Monte Carlo quantities. It is also similarly possible to exactly change variables from gauge links to (untraced) plaquettes (see Appendix A). We exploit this property to simplify exact calculations and develop an understanding of observable signal-to-noise problems below.

**Action and parameters.** The action for SU($N$) lattice gauge theory in $(1 + 1)$ dimensions is identical to the action given previously in Eq. (4.48). We rewrite Eq. (4.48) to be better suited for contour deformations by replacing $P^\dagger \to P^{-1}$ as in the case of the U(1) action, giving

$$S(P; \beta) = -\frac{\beta}{2N} \sum_x \text{tr} \left[ P_{01}(x) + P_{01}^{-1}(x) \right].$$

(5.83)

The path integral over plaquette variables is written identically to the case of U(1) lattice gauge theory given in Eq. (5.58), though the path integral in this case is over the SU($N$) Haar measure $dP_{01}(x)$ for each plaquette variable $P_{01}(x) \in$ SU($N$). A similar factorization can also be applied to the path integral, in this case written in terms of the SU($N$) one-plaquette averages $\langle \cdot \rangle_{\text{1var}}$ given by

$$\langle \mathcal{O} \rangle_{\text{1var}} = \frac{\int dP \, \mathcal{O}(P) \, e^{\frac{\beta}{2N} \, \text{tr}[P + P^{-1}]}}{\int dP \, e^{\frac{\beta}{2N} \, \text{tr}[P + P^{-1}]}}.$$

(5.84)

For our numerical studies, we selected three choices of the parameter $\beta$ corresponding to a string tension $\sigma$ ranging over the values 0.1–0.4 for each of the gauge groups SU(2) and SU(3). The total lattice volume $V$ given in lattice units was chosen to fix $\sigma V = 6.4$ across all choices, corresponding to a fixed physical volume. The factorization of the path integral over plaquettes implies that the lattice geometry and arrangement of these plaquettes is irrelevant for computing expectation values in this theory. In the Monte Carlo studies described below, each sampled lattice configuration was described by a linear list of $V$ plaquette values. Furthermore, there are no finite volume effects for this theory, and the total lattice volume does not affect the expectation values of Wilson loops. The choices of $V$ defined here, however, do set the maximum area of the Wilson loops measured in each case. Fixing $\sigma V$ allows a comparison of Wilson loops with identical choices of physical area across all ensembles. The values of $\beta$, the corresponding bare gauge coupling $g = \sqrt{2N/\beta}$, the string tension $\sigma$, and the lattice volume $V$ are reported in Table 5.2.

**Wilson loops.** The traced Wilson loop $\frac{1}{N} \text{tr} \, W_{\mathcal{A}}$ over a rectangular region $\mathcal{A}$ with area $A$ has an interpretation as a correlation function of a static quark-antiquark pair, as in the case of U(1) lattice gauge theory. This Wilson loop is written in terms of the observable $W_{\mathcal{A}} \in$ SU($N$) that we define as the untraced loop opening conventionally in the lower-left corner of the rectangle. It is useful to work with the untraced Wilson loops at various points when studying the signal-to-noise and deformations of Wilson loops. This group-valued Wilson loop observable is written in terms of the plaquette variables as in Appendix A as

$$W_{\mathcal{A}} = \prod_{x \in \mathcal{A}} P_{01}(x).$$

(5.85)

| | | SU(2) | | SU(3) | |
|---|---|---|---|---|---|
| $\sigma$ | $V$ | $g$ | $\beta$ | $g$ | $\beta$ |
| 0.4 | 16 | 0.98 | 4.2 | 0.72 | 11.7 |
| 0.2 | 32 | 0.71 | 8.0 | 0.53 | 21.7 |
| 0.1 | 64 | 0.51 | 15.5 | 0.38 | 41.8 |

**Table 5.2:** The choices of coupling parameter $\beta$ and corresponding bare gauge coupling $g = \sqrt{2N/\beta}$ used in our study of contour deformations for observables in SU($N$) lattice gauge theory. Coupling parameters for both $N = 2$ and $N = 3$ were computed using exact results to fix $\sigma \in \{0.1, 0.2, 0.4\}$ and the total lattice volume was correspondingly chosen to give a fixed total physical lattice volume $\sigma V = 6.4$. Table adapted from Table I of Pub. [1].

Unlike for U(1) Wilson loops, the product must be ordered because the plaquette variables are non-commuting. The precise ordering is fixed by the choices made in factorizing the path integral, but for the purposes of studying the signal-to-noise and deformation properties of Wilson loops the order is irrelevant.

The expectation value of matrix elements $W_{\mathcal{A}}^{ab}$ can be exactly evaluated using factorization and known single-plaquette integrals, as detailed in Appendix A, giving

$$
\begin{aligned}
\left\langle W_{\mathcal{A}}^{ab} \right\rangle &= \left\langle P_{01}^{az_1}(x_1) P_{01}^{z_1 z_2}(x_2) \dots P_{01}^{z_{A-1}b}(x_A) \right\rangle \\
&= \left\langle P^{az_1} \right\rangle_{1\mathrm{var}} \left\langle P^{z_1 z_2} \right\rangle_{1\mathrm{var}} \dots \left\langle P^{z_{A-1}b} \right\rangle_{1\mathrm{var}} \\
&= \delta^{ab}[f_1(\beta)]^A,
\end{aligned}
\tag{5.86}
$$

where summation over repeated indices is implied, $x_1, \dots, x_A$ are the sites contained in the product in Eq. (5.85), and in the last step we applied $\left\langle P^{ab} \right\rangle_{1\mathrm{var}} = \delta^{ab} f_1(\beta)$ in terms of the fundamental character expansion coefficient $f_1(\beta)$ given in Appendix A. Applying $\frac{1}{N} \mathrm{tr}(\cdot)$ to Eq. (5.86) shows that traced Wilson loop expectation values follow area law scaling as $e^{-\sigma A}$ with the string tension given by

$$
\sigma = -\ln f_1(\beta).
\tag{5.87}
$$

This implies that static charges are also confined in SU($N$) lattice gauge theory in $(1+1)$D with open boundary conditions. Though the exact form of $f_1(\beta)$ is not relevant for this study, we note that $f_1(\beta) < 1$ for all choices of $\beta$, giving a positive string tension, and that $\lim_{\beta \to \infty} f_1(\beta) = 1$ corresponding to $\sigma \to 0$ in lattice units. We thus interpret larger values of $\beta$ as corresponding to smaller lattice spacings (as set by the string tension) with a continuum limit associated with $\beta \to \infty$.

Exponential scaling of the Wilson loop expectation value also results in an exponentially severe signal-to-noise problem for these quantities. The variance of Wilson loops can be exactly computed following a similar derivation based on factorization. We begin by computing the variance of the trace $\frac{1}{N} \mathrm{tr}\, W_{\mathcal{A}}$. The averages $\left\langle |\mathrm{tr}\, W_{\mathcal{A}}|^2 \right\rangle$ and $\left\langle (\mathrm{tr}\, W_{\mathcal{A}})^2 \right\rangle$ are necessary for the variance and can be computed using the single-site integrals given in Appendix [A] as

$$\begin{aligned}
\langle |\operatorname{tr} W_{\mathcal{A}}|^2 \rangle &= \Big\langle P^{az_1}(x_1) P^{z_1 z_2}(x_2) \ldots P^{z_{A-1} a}(x_A) \\
&\qquad \times [P^{z'_{A-1} a'}(x_A)]^* \ldots [P^{z'_1 z'_2}(x_2)]^* [P^{a' z'_1}(x_1)]^* \Big\rangle \\[4pt]
&= [\hat{P}_0^{aa'; z_1 z'_1} f_0(\beta) + \hat{P}_{2,\bar{1}}^{aa'; z_1 z'_1} f_{2,\bar{1}}(\beta)] \\
&\qquad \times [\hat{P}_0^{z_1 z'_1; z_2 z'_2} f_0(\beta) + \hat{P}_{2,\bar{1}}^{z_1 z'_1; z_2 z'_2} f_{2,\bar{1}}(\beta)] \times \ldots \\
&\qquad \times [\hat{P}_0^{z_{A-1} z'_{A-1}; aa'} f_0(\beta) + \hat{P}_{2,\bar{1}}^{z_{A-1} z'_{A-1}; aa'} f_{2,\bar{1}}(\beta)] \\[4pt]
&= [f_0(\beta)]^A \operatorname{tr} \hat{P}_0 + [f_{2,\bar{1}}(\beta)]^A \operatorname{tr} \hat{P}_{2,\bar{1}} \\[4pt]
&= 1 + (N^2 - 1)[f_{2,\bar{1}}(\beta)]^A
\end{aligned} \tag{5.88}$$

and

$$\begin{aligned}
\langle (\operatorname{tr} W_{\mathcal{A}})^2 \rangle &= \Big\langle P^{az_1}(x_1) P^{z_1 z_2}(x_2) \ldots P^{z_{A-1} a}(x_A) \\
&\qquad \times P^{z'_{A-1} a'}(x_A) \ldots P^{z'_1 z'_2}(x_2) P^{a' z'_1}(x_1) \Big\rangle \\[4pt]
&= [\hat{P}_2^{aa'; z_1 z'_1} f_2(\beta) + \hat{P}_{1,1}^{aa'; z_1 z'_1} f_{1,1}(\beta)] \\
&\qquad \times [\hat{P}_2^{z_1 z'_1; z_2 z'_2} f_2(\beta) + \hat{P}_{1,1}^{z_1 z'_1; z_2 z'_2} f_{1,1}(\beta)] \times \ldots \\
&\qquad \times [\hat{P}_2^{z_{A-1} z'_{A-1}; aa'} f_2(\beta) + \hat{P}_{1,1}^{z_{A-1} z'_{A-1}; aa'} f_{1,1}(\beta)] \\[4pt]
&= [f_2(\beta)]^A \operatorname{tr} \hat{P}_2 + [f_{1,1}(\beta)]^A \operatorname{tr} \hat{P}_{1,1} \\[4pt]
&= \frac{1}{2}(N^2 + N)[f_2(\beta)]^A + \frac{1}{2}(N^2 - N)[f_{1,1}(\beta)]^A.
\end{aligned} \tag{5.89}$$

Above, the notation $\hat{P}_r$ indicates a projector from the generic two-variable tensor product space into the irreducible representation $r$, as described in the appendix. The defining property of the projectors, $\hat{P}_r \hat{P}_r = \hat{P}_r$, allows the simplification to final results in both Eq. (5.88) and Eq. (5.89). Traces of these projectors give the dimensions of the respective representations, giving the $N$-dependence of the result, while the prefactors $[f_r(\beta)]^A$ arise from the single-site integrals and carry the dependence on $\beta$ and $A$. The variance can be written in terms of Eqs. (5.88) and (5.89) as

$$\begin{aligned}
\operatorname{Var}\left[ \frac{1}{N} \operatorname{Re} \operatorname{tr} W_{\mathcal{A}} \right] &= \frac{1}{2N^2} \langle |\operatorname{tr} W_{\mathcal{A}}|^2 \rangle + \frac{1}{2N^2} \langle (\operatorname{tr} W_{\mathcal{A}})^2 \rangle - \left\langle \frac{1}{N} \operatorname{tr} W_{\mathcal{A}} \right\rangle^2 \\
&= \frac{1}{2N^2} \left( 1 + (N^2 - 1)[f_{2,\bar{1}}(\beta)]^A \right) \\
&\quad + \frac{1}{4N^2} \left( (N^2 + N)[f_2(\beta)]^A + (N^2 - N)[f_{1,1}(\beta)]^A \right) - e^{-2\sigma A}.
\end{aligned} \tag{5.90}$$

All $f_r(\beta)$ are strictly less than 1 for finite $\beta$ and any non-trivial irrep $r$. The terms involving these factors in the variance thus fall off exponentially with $A$, and the leading factor of 1 in Eq. (5.88) dominates the variance at large $A$, giving the asymptotic scaling

$$\text{Var}\left[\frac{1}{N}\operatorname{Re}\operatorname{tr}W_{\mathcal{A}}\right] \sim \frac{1}{2N^2}. \tag{5.91}$$

This means that the signal-to-noise ratio falls exponentially as $e^{-\sigma A}$, exactly matching the scaling of the U(1) theory.

**Rewriting the Wilson loop observable.** Section 5.2.3 discussed the idea that contour deformations can make wildly varying amounts of progress in reducing the variance of observables depending on which of a class of symmetry-related observables one chooses as a base observable to deform. The traced Wilson loop can be related to a number of equivalent observables by gauge symmetry considerations. Gauge transformations act on the matrix-valued Wilson loop $W_{\mathcal{A}}$ by matrix conjugation. All $N \times N$ permutation matrices are included in SU($N$), up to a sign fixed by the unit-determinant condition. Thus gauge transformations include permutations of the matrix of elements of $W_{\mathcal{A}}$ as a subgroup, generically transforming each matrix element as

$$W_{\mathcal{A}}^{ab} \to W_{\mathcal{A}}^{\alpha(a)\alpha(b)}, \tag{5.92}$$

where $\alpha$ is any possible permutation on $N$ indices. This in particular implies that the expectation values of all diagonal elements are equal to each other and are equal to the normalized expectation value of the trace,

$$\langle W_{\mathcal{A}}^{11} \rangle = \langle W_{\mathcal{A}}^{22} \rangle = \cdots = \langle W_{\mathcal{A}}^{NN} \rangle = \left\langle \frac{1}{N}\operatorname{tr}W_{\mathcal{A}} \right\rangle. \tag{5.93}$$

Any of these observables could be used as base observables for deformation and do not result in the same families of deformed observables, for the reasons discussed in Sec. 5.2.3.

The distinction can be most clearly seen by considering the group SU(2). Using the angular parameterization given in Sec. 5.3.2, the trace of a generic group element $U$ is given by

$$\operatorname{tr}U(\theta, \phi_1, \phi_2) = 2\sin\theta\cos\phi_1. \tag{5.94}$$

This does not take the form $e^{i\phi}$ and as discussed in Sec. 5.2.3, the cosine function — and equivalently the sine function — has an increased magnitude when deforming its argument off of the real line. Increased average magnitudes are expected to lead to increased phase fluctuations and decreased signal-to-noise ratios. This is a qualitative argument, and in practice the underlying coordinates being deformed are those of the plaquettes rather than the Wilson loops, but this motivates considering the base observable $W_{\mathcal{A}}^{11}$ which has no such problem. For SU(2) variables, the $(1,1)$ component of a generic group element $U$ is given in coordinate form by

$$U^{11}(\theta, \phi_1, \phi_2) = \sin\theta e^{i\phi_1}. \tag{5.95}$$

Deformations corresponding to an imaginary shift of $\phi_1$ can be expected to give signif-icantly better signal-to-noise results based on the discussion in Sec. 5.2.3.

A similar qualitative argument can be made for general SU($N$) variables by considering the definition of the trace in terms of eigenvalues. These eigenvalues must collectively multiply to 1 by the unit-determinant condition, even after complexifying the group to $SL(N, \mathbb{C})$, and it is impossible to exponentially reduce the magnitude of all $N$ eigenvalues simultaneously. The magnitude of the trace can thus only be made exponentially small for particular choices of the eigenvalues by a delicate cancellation of the complex phases of the eigenvalues. For general group elements, the magnitude will be $O(1)$ or greater. Finding contour deformations that significantly reduce the average magnitude is therefore expected to either be difficult (requiring careful parameterization in terms of the underlying group element) or impossible. No such difficulties are encountered for the $(1, 1)$ component of $SL(N, \mathbb{C})$ matrices, which are generally unconstrained.

In our study of contour deformations for SU($N$) lattice gauge theory, we therefore choose to work with $W_{\mathcal{A}}^{11}$ as the base observable based on these arguments. The variance of the Monte Carlo estimator $\text{Re}[W_{\mathcal{A}}^{11}]$ can be derived by nearly identical calculations to those given in Eqs. (5.88)–(5.90) for $\frac{1}{N} \text{Re}[\text{tr}\, W_{\mathcal{A}}]$. The results are

$$
\begin{aligned}
\langle |W_{\mathcal{A}}^{11}|^2 \rangle &= [f_0(\beta)]^A \hat{P}_0^{11;11} + [f_{2,\bar{1}}(\beta)]^A \hat{P}_{2,\bar{1}}^{11;11} \\
&= \frac{1}{N} + \left(1 - \frac{1}{N}\right) [f_{2,\bar{1}}(\beta)]^A \\
\langle (W_{\mathcal{A}}^{11})^2 \rangle &= [f_2(\beta)]^A \hat{P}_2^{11;11} + [f_{1,1}(\beta)]^A \hat{P}_{1,1}^{11;11} \\
&= [f_2(\beta)]^A \\
\text{Var}[\text{Re}\, W_{\mathcal{A}}^{11}] &= \frac{1}{2N} + \left(\frac{1}{2} - \frac{1}{2N}\right) [f_{2,\bar{1}}(\beta)]^A + \frac{1}{2}[f_2(\beta)]^A - e^{-2\sigma A}.
\end{aligned}
\tag{5.96}
$$

For large $A$ the variance of $W_{\mathcal{A}}^{11}$ also scales as a constant, $\text{Var}[\text{Re}\, W_{\mathcal{A}}^{11}] \sim \frac{1}{2N}$.[6] This estimator then has the same asymptotic signal-to-noise ratio as the traced Wilson loop.

### 5.5.2 Contour deformations

We apply contour deformations using the coordinate description given in Sec. 5.3.2 to tackle the signal-to-noise problem of the Wilson loops in this SU($N$) theory. A general Fourier series manifold could in principle be applied in both the case of $N = 2$ and $N = 3$. We instead choose to restrict the Fourier modes included in the definition of contour deformations in both cases to reduce the total number of parameters and make it feasible to explore cutoffs up to $\Lambda = 2$ numerically, enabling a study of whether Fourier mode dependence affects the variance reduction achievable using deformed observables.

---

[6]Interestingly, the asymptotic value differs by a factor of $N$, which is the expected factor if all diagonal entries are treated as independent in the trace. We can interpret this relation as an indication that correlations between diagonal elements of $W_{\mathcal{A}}$ are far smaller than the variance of the elements themselves, and that the average over diagonal elements provided by the trace is a minor form of variance reduction.

**Family of manifolds for** SU(2). Each SU(2) group element is parameterized by one zenith and two azimuthal angles. We apply a subset of the Fourier series terms defining the general vertical deformation for this parameterization, specifically restricting to terms that include at most two angular coordinates simultaneously. We also apply an autoregressive approach by deforming each plaquette variable $P_x$ as a function of the $P_y$ with $y \leq x$. Altogether, the deformation is parameterized as

$$
\begin{aligned}
\tilde{\theta}_x &\equiv \theta_x + i \sum_{y \leq x} f_\theta(\theta_y, \phi_y^1, \phi_y^2; \kappa^{xy}, \lambda^{xy}, \eta^{xy}, \chi^{xy}, \zeta^{xy}) \\
\tilde{\phi}_x^1 &\equiv \phi_x^1 + i \kappa_0^{x;\phi^1} + i \sum_{y \leq x} f_{\phi^1}(\theta_y, \phi_y^1, \phi_y^2; \kappa^{xy}, \lambda^{xy}, \eta^{xy}, \chi^{xy}, \zeta^{xy}) \\
\tilde{\phi}_x^2 &\equiv \phi_x^2 + i \kappa_0^{x;\phi^2} + i \sum_{y \leq x} f_{\phi^2}(\theta_y, \phi_y^1, \phi_y^2; \kappa^{xy}, \lambda^{xy}, \eta^{xy}, \chi^{xy}, \zeta^{xy})
\end{aligned}
\tag{5.97}
$$

in terms of the global shifts $\kappa_0^x$ and the five parameter tensors $\kappa^{xy}$, $\lambda^{xy}$, $\eta^{xy}$, $\chi^{xy}$, and $\zeta^{xy}$. The functions $f_\theta$, $f_{\phi^1}$, and $f_{\phi^2}$ are defined by a Fourier series expansion, where the amplitudes and phases of the Fourier series are parameterized by the components of the five parameter tensors as

$$
\begin{aligned}
f_\theta(\dots) &= \sum_{m=1}^{\Lambda} \kappa_m^{xy;\theta} \sin(2m\theta_y) \Big\{ 1 + \\
&\qquad \sum_{n=1}^{\Lambda} \Big[ \eta_{mn}^{xy;\theta,\phi^1} \sin(n\phi_y^1 + \chi_{mn}^{xy;\theta,\phi^1}) + \eta_{mn}^{xy;\theta,\phi^2} \sin(n\phi_y^2 + \chi_{mn}^{xy;\theta,\phi^2}) \Big] \Big\} \\
f_{\phi^1}(\dots) &= \sum_{m=1}^{\Lambda} \kappa_m^{xy;\phi^1} \sin(m\phi_y^1 + \zeta_m^{xy;\phi^1}) \Big\{ 1 + \\
&\qquad \sum_{n=1}^{\Lambda} \Big[ \lambda_{mn}^{xy;\phi^1,\theta} \sin(2n\theta_y) + \eta_{mn}^{xy;\phi^1,\phi^2} \sin(n\phi_y^2 + \chi_{mn}^{xy;\phi^1,\phi^2}) \Big] \Big\} \\
f_{\phi^2}(\dots) &= \sum_{m=1}^{\Lambda} \kappa_m^{xy;\phi^2} \sin(m\phi_y^2 + \zeta_m^{xy;\phi^2}) \Big\{ 1 + \\
&\qquad \sum_{n=1}^{\Lambda} \Big[ \lambda_{mn}^{xy;\phi^2,\theta} \sin(2n\theta_y) + \eta_{mn}^{xy;\phi^2,\phi^1} \sin(n\phi_y^1 + \chi_{mn}^{xy;\phi^2,\phi^1}) \Big] \Big\}.
\end{aligned}
\tag{5.98}
$$

Above, the zero modes of the Fourier expansion have no dependence on $y$ and have been collected into the $\kappa_0$ parameters defining imaginary shifts of the azimuthal angles in Eq. (5.97). In our numerical study, we consider the cases of $\Lambda \in \{0, 1, 2\}$ to investigate the effect of higher Fourier modes on the performance of these contour deformations.

The use of an autoregressive structure allows us to compute the Jacobian determi-nant of the whole deformation as a product over determinants of each diagonal block,

$$\widetilde{J} = \prod_x \widetilde{j}_x, \quad \text{where} \quad \widetilde{j}_x = \det \begin{pmatrix} \frac{\partial f_\theta}{\partial \theta_x} & \frac{\partial f_\theta}{\partial \phi_x^1} & \frac{\partial f_\theta}{\partial \phi_x^2} \\ \frac{\partial f_{\phi^1}}{\partial \theta_x} & \frac{\partial f_{\phi^1}}{\partial \phi_x^1} & \frac{\partial f_{\phi^1}}{\partial \phi_x^2} \\ \frac{\partial f_{\phi^2}}{\partial \theta_x} & \frac{\partial f_{\phi^2}}{\partial \phi_x^1} & \frac{\partial f_{\phi^2}}{\partial \phi_x^2} \end{pmatrix}. \tag{5.99}$$

The individual derivatives in Eq. (5.99) can be straightforwardly derived from Eq. (5.98). In addition to the Jacobian determinant of the deformation applied to coordinates, the ratio of Haar measure factors evaluated on the deformed and undeformed coordinates must be included to get the full Jacobian with respect to the Haar measure of the path integral. For SU(2) group elements, this ratio is given by

$$\prod_x \frac{h(\widetilde{\Omega}(x))}{h(\Omega(x))} = \prod_x \left[ \frac{\sin(2\widetilde{\theta}_x)}{\sin(2\theta_x)} \right]. \tag{5.100}$$

In total, the deformed observable associated with $W_{\mathcal{A}}^{11}$ is then given by

$$\mathcal{Q}_{\mathcal{A}}^{\text{SU(2)}} = W_{\mathcal{A}}^{11}(\widetilde{P}) \frac{e^{-S(\widetilde{P})}}{e^{-S(P)}} \prod_x \widetilde{j}_x \left[ \frac{\sin(2\widetilde{\theta}_x)}{\sin(2\theta_x)} \right], \tag{5.101}$$

where $\widetilde{P}$ collectively indicates the set of all deformed plaquettes computed by mapping through the coordinate map and the deformation map as $P(x) \to \Omega(x) \to \widetilde{\Omega}(x) \to \widetilde{P}(x)$.[7] By holomorphy of the path integrand, $\left\langle \mathcal{Q}_{\mathcal{A}}^{\text{SU(2)}} \right\rangle = \left\langle W_{\mathcal{A}}^{11} \right\rangle$.

**Family of manifolds for** SU(3). Each SU(3) group element is parameterized by three zenith angles $\theta^1, \theta^2, \theta^3$ and five azimuthal angles $\phi^1, \ldots, \phi^5$. We use the same restrictions as for SU(2) to define a family of possible vertical deformations for these angular variables using a subset of the general Fourier series expansion. The vertical shift deformations are defined as

$$\begin{aligned} \widetilde{\theta}_x^a &= \theta_x^a + i \sum_{y \le x} f_{\theta^a}(\Omega_y; \kappa^{xy}, \lambda^{xy}, \chi^{xy}), \\ \widetilde{\phi}_x^b &= \phi_x^b + i\kappa_0^{x;\phi^b} + i \sum_{y \le x} f_{\phi^b}(\Omega_y; \kappa^{xy}, \lambda^{xy}, \chi^{xy}, \zeta^{xy}) \end{aligned} \tag{5.102}$$

---

[7] Mapping from the plaquette $P(x)$ to the angular coordinates $\Omega(x)$ involves non-holomorphic functions of the matrix elements of $P(x)$. This is not an issue for deformations of the coordinates, which are valid as long as the path integrand is a holomorphic function of the coordinates $\Omega(x)$.

where $a \in \{1, 2, 3\}$, $b \in \{1, \ldots, 5\}$, and the Fourier mode expansions of $f_{\theta^a}$ and $f_{\phi^b}$ are chosen to be

$$
f_{\theta^a} = \sum_{m=1}^{\Lambda} \kappa_m^{xy;\theta^a} \sin(2m\theta_y^a) \Big\{ 1 +
$$
$$
\sum_{n=1}^{\Lambda} \Big[ \sum_{\substack{r \neq a \\ r=1}}^{3} \lambda_{mn}^{xy;\theta^a \theta^r} \sin(2n\theta_y^r) + \sum_{s=1}^{5} \eta_{mn}^{xy;\theta^a \phi^s} \sin(n\phi_y^s + \chi_{mn}^{xy;\theta^a \phi^s}) \Big] \Big\},
$$

$$
f_{\phi^b} = \sum_{m=1}^{\Lambda} \kappa_m^{xy;\phi^b} \sin(m\phi_y^b + \zeta_m^{xy;\phi^b}) \Big\{ 1 +
$$
$$
\sum_{n=1}^{\Lambda} \Big[ \sum_{r=1}^{3} \lambda_{mn}^{xy;\phi^b \theta^r} \sin(2n\theta_y^r) + \sum_{\substack{s \neq b \\ s=1}}^{5} \eta_{mn}^{xy;\phi^b \phi^s} \sin(n\phi_y^s + \chi_{mn}^{xy;\phi^b \phi^s}) \Big] \Big\}.
$$

(5.103)

The diagonal-block Jacobian factors can be straightforwardly computed as in the case of SU(2) deformations by assembling the $8 \times 8$ Jacobian matrix $\partial \widetilde{\Omega}_i / \partial \Omega_j$ and directly computing the determinant of this small matrix, $\widetilde{j}_x = \det(\partial \widetilde{\Omega}_i / \partial \Omega_j)$. The autoregressive structure then allows the full Jacobian determinant to be computed as $\widetilde{J} = \prod_x \widetilde{j}_x$. Finally, we extend the Jacobian determinant of the coordinate deformations to the Jacobian factor $\widetilde{J}(U)$ with respect to the original path integral measure by including the ratio of Haar measure factors

$$
\prod_x \frac{h(\widetilde{\Omega}_x)}{h(\Omega_x)} = \prod_x \left[ \frac{\sin \widetilde{\theta}_x^1 (\cos \widetilde{\theta}_x^1)^3 \sin \widetilde{\theta}_x^2 \cos \widetilde{\theta}_x^2 \sin \widetilde{\theta}_x^3 \cos \widetilde{\theta}_x^3}{\sin \theta_x^1 (\cos \theta_x^1)^3 \sin \theta_x^2 \cos \theta_x^2 \sin \theta_x^3 \cos \theta_x^3} \right].
$$

(5.104)

Combining these Jacobian factors with the action and observable evaluated on deformed variables gives the deformed observable

$$
\mathcal{Q}_{\mathcal{A}}^{\mathrm{SU}(3)} = W_{\mathcal{A}}^{11}(\widetilde{P}) \frac{e^{-S(\widetilde{P})}}{e^{-S(P)}} \prod_x \widetilde{j}_x \frac{h(\widetilde{\Omega}_x)}{h(\Omega_x)}.
$$

(5.105)

This observable has identical expectation value to the base observable.

### 5.5.3 Monte Carlo studies

The variance of $\mathcal{Q}_{\mathcal{A}}^{\mathrm{SU}(2)}$ and $\mathcal{Q}_{\mathcal{A}}^{\mathrm{SU}(3)}$ will in general differ from the respective base Wilson loop observables in SU(2) and SU(3) lattice gauge theory. We studied the ability of these families of manifolds and corresponding deformed observables to reduce variance versus the base observables using Monte Carlo studies for both gauge groups involving generating a set of ensembles at the three sets of parameters given in Table 5.2 and considering manifold deformations for each ensemble and a range of Wilson loop observables. All calculations of manifold deformations and optimization runs were performed using a JAX [354] implementation.

**Ensembles.** For each choice of parameters, we generated ensembles of 32 000 configurations each. Factorization of the path integral over plaquettes allowed a simplification of this ensemble generation step: rather than sample lattices with area $V$ in lattice units, we instead sampled $32000V$ individual plaquettes according to the one-plaquette distribution then composed these into $V$-plaquette lattices. This reduced the computational cost of this study and allowed us to nearly eliminate autocorrelations. Generating individual plaquettes was done using HMC with parameters tuned to limit the integrated autocorrelation time to 1–2 for all three ensembles. Before arranging the sampled plaquettes into full lattices, a random shuffle was applied to remove any spatial correlations that may have otherwise arisen from correlations in the Markov chain used to sample plaquettes. This ensured that ensembles had asymptotically correct statistics. The generated ensembles were divided into three subsets: a training set of 320 configurations reserved for optimizing manifold parameters, a test set of 320 configurations reserved for monitoring the loss function to determine the optimizer step size and early stopping time, and the remaining set of 31 360 configurations used to measure final results.

**Optimization.** At all three choices of couplings in both SU(2) and SU(3) gauge theory, we optimized manifold parameters for a manifold with the lowest possible Fourier cutoff, $\Lambda = 0$, for each possible Wilson loop area $A \leq V$, producing a distinct optimized manifold per choice of Wilson loop shape. These $\Lambda = 0$ families of manifolds are described entirely by constant shifts of the azimuthal parameters of each plaquette, generalizing the notion of constant imaginary shift deformations explored for U(1) lattice gauge theory. To further explore the effects of additional Fourier modes on expressivity, we compared these optimized manifolds to manifolds with higher cutoffs $\Lambda = 1$ and $\Lambda = 2$ for the ensembles with intermediate string tension $\sigma = 0.2$. We use the term 'training run' to describe the total optimization procedure applied for a given choice of manifold parameterization and Wilson loop area $A$.

In each training run, we optimized the manifold parameters iteratively using the Adam optimizer [352] until the loss function converged. At each step of optimization, the loss function in Eq. (5.32) and its gradients with respect to manifold parameters were estimated using a resampled batch of configurations from the training set. All Adam hyperparameters were fixed to the default provided in the JAX implementation except for the learning rate, which determined the step size on each iteration. Dynamically altering the step size in stochastic gradient-based optimization can result in more efficient training [377], and in this study we used a dynamic schedule for the Adam learning rate based on measurements of the loss function on the test set of configurations. The learning rate schedule was defined by (1) initializing the learning rate to a value of $s_0$, (2) permanently reducing by a factor of $F$ each time the loss function averaged over a window of $W$ steps failed to decrease for $P$ consecutive steps, and (3) halting training after the learning rate was reduced $N_r$ times. We fixed $F = 10$, $W = 50$, $P = 250$, $N_r = 2$, and $s_0 = 10^{-2}$ for the results presented here. Scheduling the learning rate in this way additionally served the purpose of avoiding overtraining by quickly halting training when the test set loss function was no longer improving.

Optimizing deformed observables requires separate manifold parameters to be de-termined for each choice of observable. For Wilson loops, this includes loops of each possible area, requiring a large number of independent training runs to be executed. As discussed in Sec. 5.2.2, correlations between similar observables suggests that the final manifold parameters determined for each observable may share significant similarities, especially for observables with nearly identical structure such as Wilson loops covering nearly the same region. To exploit this feature, we chose to initialize the optimization of manifold parameters for each Wilson loop of area $A$ from the final optimized parameters determined for the Wilson loop of area $A - 1$ for manifolds with Fourier cutoff $\Lambda = 0$. For manifolds with Fourier cutoff $\Lambda = 1$ and $\Lambda = 2$, we initialized manifold parameters for each Wilson loop of area $A$ using the optimized manifold associated with the Wilson loop of area $A$ and cutoff $\Lambda - 1$. Figure 5.15 compares this transfer learning approach to initializing from the undeformed manifold. The loss function converges much more quickly for the manifolds initialized by transferring from a previously optimized manifold, significantly reducing the number of training iterations required to converge to a minimum of the loss function, especially for deformed observables associated with Wilson loops of large area.

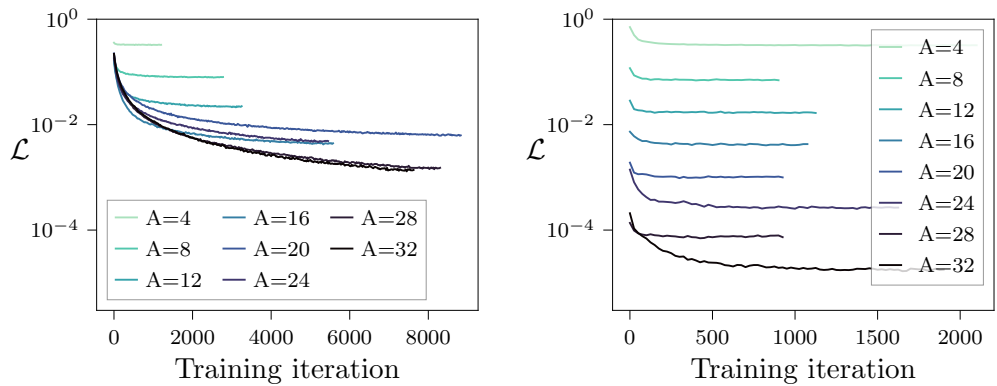

**Figure 5.15:** The loss function $\mathcal{L}$ measured during optimization of $\Lambda = 0$ manifolds for Wilson loops over various choices of area $A$ on the SU(2) ensemble with intermediate string tension $\sigma = 0.2$. In the left panel, manifold parameters were initialized from the original manifold. In the right panel, manifold parameters were instead initialized from the optimized manifold parameters for the Wilson loop of area $A - 1$. The loss function for all choices of Wilson loop begins from an $O(1)$ value on the left, reflecting the constant variance with respect to choices of area $A$ for the undeformed observable. For Wilson loops with area larger than approximately 16, the loss function did not converge within the 8000 iterations shown when initializing from the undeformed manifold. On the right, the optimization converges to a minimum of the loss function within 2000 iterations for all values of $A$ shown. The variability in stopping times is due to the stochastic nature of the automatic learning rate schedule. Figure adapted from Fig. 3 of Pub. [1].

**Results.** Deforming observables using the $\Lambda = 0$ family of manifolds is already sufficient to significantly reduce the variance of Wilson loops for all choices of parameters studied. Figures 5.16 and 5.17 compare base Wilson loop observables to $\Lambda = 0$ deformed observables measured on all ensembles studied. Expectation values of the deformed observables agree with the analytical results for all choices of area and agree with Monte Carlo estimates of the base observables where these can be reliably determined. Though this is guaranteed analytically by holomorphy of the path integral, it is nevertheless a useful cross-check on the finite-sample-size statistics of deformed observables. The figures also compare the variance measured for both base and deformed observables. In all cases, the variance of the base observable $W_{\mathcal{A}}^{11}$ plateaus to a constant value as $A$ is increased, agreeing with the analytically determined result. On the other hand, the variance of the deformed observable $\mathcal{Q}_{\mathcal{A}}$ decreases by many orders of magnitude as $A$ is increased. Though no consistent trend in the variance is guaranteed *a priori* for these individually optimized deformed observables at each choice of $A$, one can see a trend consistent with exponentially decreasing variance. For the Wilson loops with the largest areas considered, the variance was reduced by three to four orders of magnitude. Achieving this variance reduction by using measurements of deformed observables rather than simply increasing the number of measurements of the base observable could thus save orders of magnitude in the cost of additional ensemble generation if such precision were required.

The $\Lambda = 0$ family of manifolds consists of few enough parameters that it is possible to study the optimized values of parameters to understand the source of the observed variance reduction. As discussed in Sec. 5.2.1, it is the average magnitude of observables that sets the scale of the variance of both the base and deformed observables. We thus expect the optimal parameters to result in a reduction of the average magnitude of $\mathcal{Q}_{\mathcal{A}}$ with respect to $W_{\mathcal{A}}^{11}$. In simple examples, we found that for observables taking the form $e^{i\phi}$ this could be achieved by imaginary shifts of the variable $\phi$. A similar structure occurs for the SU(2) and SU(3) Wilson loop observables and we discuss how optimized constant shift parameters result in constant shifts of variables appearing with such an exponential form in $W_{\mathcal{A}}^{11}$.

First, we consider the value of $W_{\mathcal{A}}^{11}$ for an SU(2) Wilson loop of area $A = 1$ consisting of a plaquette $P_x$ and for a Wilson loop of area $A = 2$ consisting of plaquettes $P_x$ and $P_{x'}$. These are respectively given by

$$P_x^{11} = \sin\theta_x e^{i\phi_x^1} \tag{5.106}$$

and

$$(P_x P_{x'})^{11} = \sin\theta_x \sin\theta_{x'} e^{i\phi_x^1 + i\phi_{x'}^1} - \cos\theta_x \cos\theta_{x'} e^{i\phi_x^2 - i\phi_{x'}^2}. \tag{5.107}$$

In the area $A = 1$ case given in Eq. (5.106), a constant positive imaginary shift in $\phi_x^1$ will reduce the average magnitude. This shift applied to both $\phi_x^1$ and $\phi_{x'}^1$ also reduces the average magnitude of the first term for the $A = 2$ Wilson loop in Eq. (5.107). The average magnitude of the second term is reduced by shifting the linear combination $\phi_x^2 - \phi_{x'}^2$ in the positive imaginary direction. This qualitative argument can be generalized to $A > 2$ suggesting that shifting all $\phi_x^1$ and $\phi_x^2 - \phi_{x+1}^2$ in the positive imaginary direction will result in reduced average magnitudes and reduced variance for deformed Wilson loop observables. These qualitative results agree with the optimized constant shifts $\kappa_0^{x;\phi^1}$ and $\kappa_0^{x;\phi^2}$ depicted for three choices of Wilson loop area in Fig. 5.18.

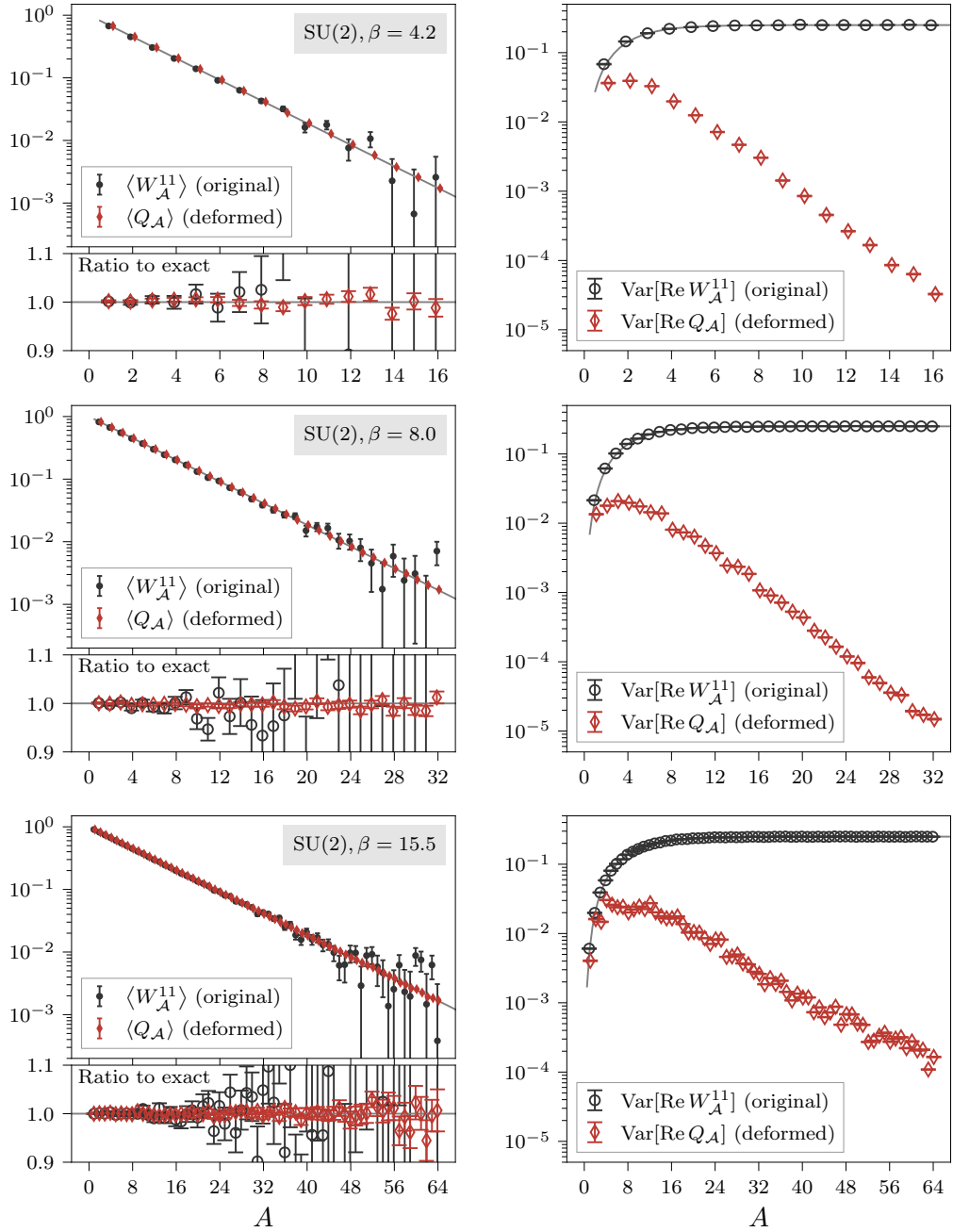

**Figure 5.16:** Left: A comparison of the expectation values of base Wilson loop observables $W_{\mathcal{A}}^{11}$ to $\mathcal{Q}_{\mathcal{A}}$ on the three sets of SU(2) parameters corresponding to string tension $\sigma = 0.4$ (top), $\sigma = 0.2$ (middle), and $\sigma = 0.1$ (bottom). The deformed observable expectation values agree with analytical results shown by the gray lines at all choices of Wilson loop area $A$. Right: A comparison of the variance of the Monte Carlo estimators Re $W_{\mathcal{A}}^{11}$ and Re $\mathcal{Q}_{\mathcal{A}}$ on the same three choices of parameters. The base observable variance is asymptotically constant with area, agreeing with the analytically derived results for the undeformed variance shown by the gray lines, while the deformed observable variance is orders of magnitude smaller at the largest Wilson loop areas and appears to be consistent with an asymptotically exponential decrease. Figure adapted from Fig. 4 of Pub. [1].

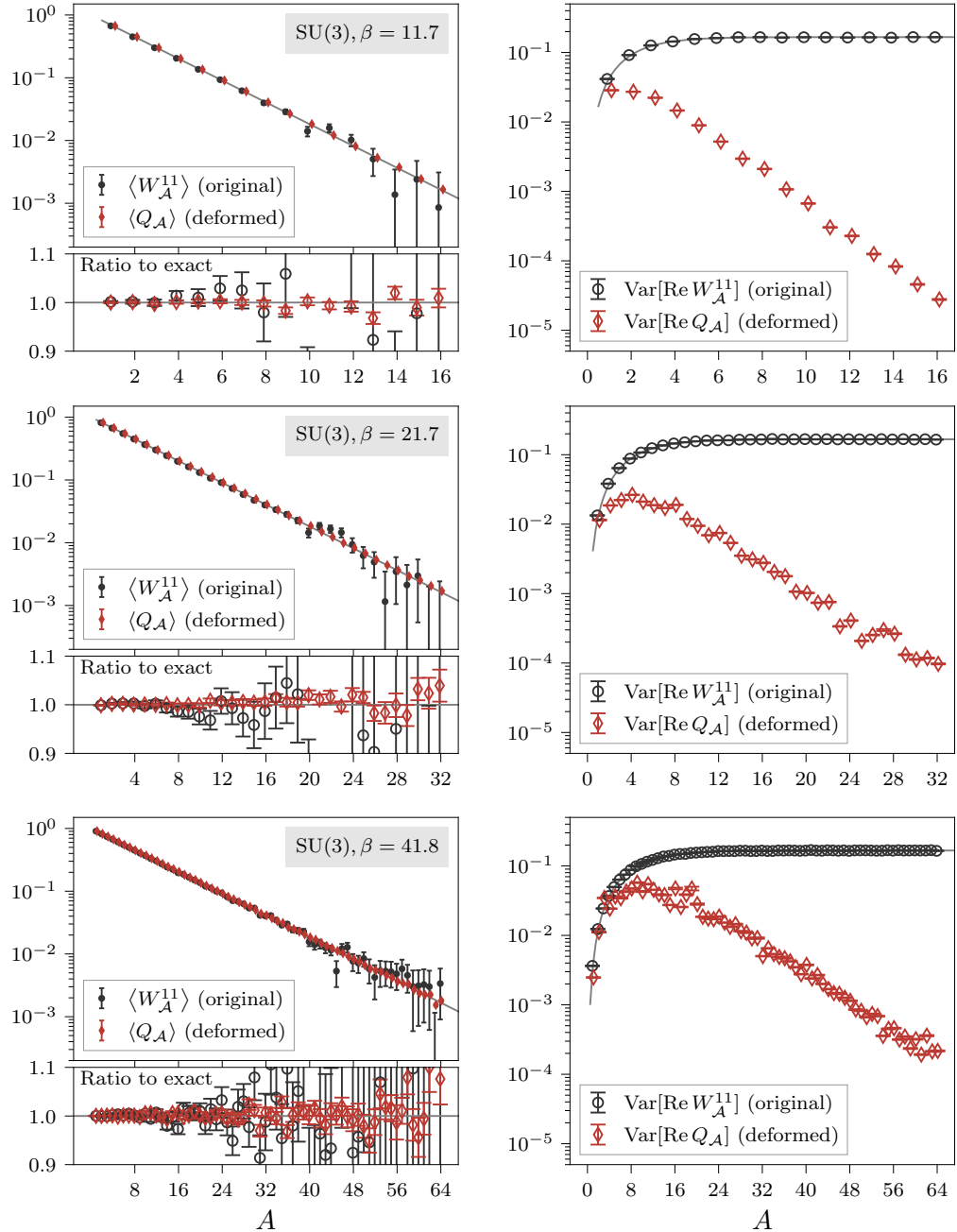

**Figure 5.17:** A comparison of the expectation values of base and deformed Wilson loop observables on the three sets of SU(3) parameters corresponding to string tensions $\sigma = 0.4$ (top), $\sigma = 0.2$ (middle), and $\sigma = 0.1$ (bottom). See Fig. 5.16 for additional details. Figure adapted from Fig. 8 of Pub. [1].

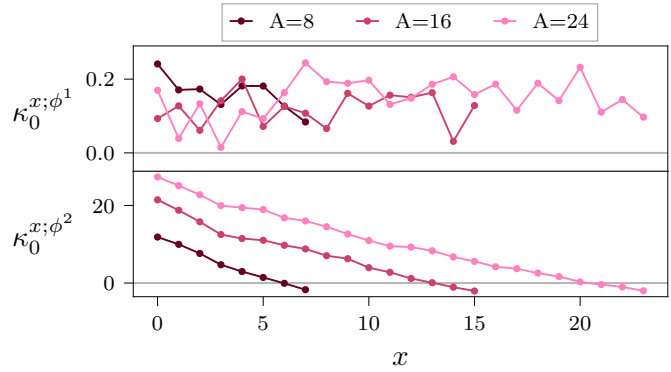

**Figure 5.18:** The optimized $\Lambda = 0$ manifold parameters for SU(2) Wilson loop observables with three choices of area $A$ on the ensemble with intermediate string tension $\sigma = 0.2$. The parameters $\kappa_0^{x;\phi^1}$ and $\kappa_0^{x;\phi^2}$ determine constant shifts of the azimuthal angles defining the plaquette $P_x$. Optimized values of these parameters across $x$ are consistent with the qualitative analysis based on average observable magnitude presented in the main text. Figure adapted from Fig. 6 of Pub. [1].

Secondly, we consider the value of $W_\mathcal{A}^{11}$ for an SU(3) Wilson loop of area $A = 1$ consisting of a plaquette $P_x$ and for a Wilson loop of area $A = 2$ consisting of plaquettes $P_x$ and $P_{x'}$. These are respectively given by

$$P_x^{11} = \cos\theta_x^1 \cos\theta_x^2 e^{i\phi_x^1} \tag{5.108}$$

and

$$
\begin{aligned}
(P_x P_{x'})^{11} = {}& e^{i\phi_x^1 + i\phi_{x'}^1} \cos\theta_x^1 \cos\theta_{x'}^1 \cos\theta_x^2 \cos\theta_{x'}^2 \\
& - e^{i\phi_{x'}^1 + i\phi_{x'}^2 + i(\phi_x^3 - \phi_{x'}^3)} \cos\theta_{x'}^2 \cos\theta_{x'}^3 \sin\theta_x^1 \sin\theta_{x'}^1 \\
& - e^{-i\phi_{x'}^2 + i(\phi_x^4 - \phi_{x'}^4)} \cos\theta_x^1 \cos\theta_{x'}^3 \sin\theta_x^2 \sin\theta_{x'}^2 \\
& - e^{i\phi_x^4 + i\phi_{x'}^1 - i\phi_x^3 + i\phi_{x'}^5} \cos\theta_x^1 \cos\theta_{x'}^2 \sin\theta_x^1 \sin\theta_x^2 \sin\theta_{x'}^3 \\
& + e^{i\phi_x^3 - i\phi_{x'}^4 - i\phi_{x'}^5} \sin\theta_x^1 \sin\theta_{x'}^2 \sin\theta_{x'}^3 .
\end{aligned}
\tag{5.109}
$$

From Eq. (5.108), it is clear that a positive imaginary shift of $\phi_x^1$ also achieves the desired reduction in average magnitude for the $A = 1$ Wilson loop in the SU(3) case. The form in Eq. (5.109) is more complicated, but a positive imaginary shift of $\phi_x^1$ and $\phi_{x'}^1$ will also reduce the magnitude of the first, second, and fourth terms for the $A = 2$ Wilson loop. Shifting $\phi_x^3 - \phi_{x'}^3$, $\phi_x^4 - \phi_{x'}^4$, $\phi_x^3 - \phi_{x'}^4$, and $\phi_x^4 - \phi_{x'}^3$ in the positive imaginary direction reduces the magnitude of the second, third, fourth, and fifth terms. Finally, $\phi_{x'}^2$ and $\phi_{x'}^5$ appear with opposing signs in different terms in Eq. (5.109), so there is no consistent choice of constant shift that globally reduces the magnitude in the $A = 2$ case. These structures can be generalized to the case of arbitrary $A$, and we expect simultaneous positive imaginary shifts in $\phi_x^1$, $\phi_x^3 - \phi_{x+1}^3$, $\phi_x^4 - \phi_{x+1}^4$ for all $x$ in the general case. These qualitative arguments agree with the optimized imaginary shift parameters $\kappa_0^{x;\phi^1}, \ldots, \kappa_0^{x;\phi^5}$ of the five azimuthal angles plotted in Figure 5.19.

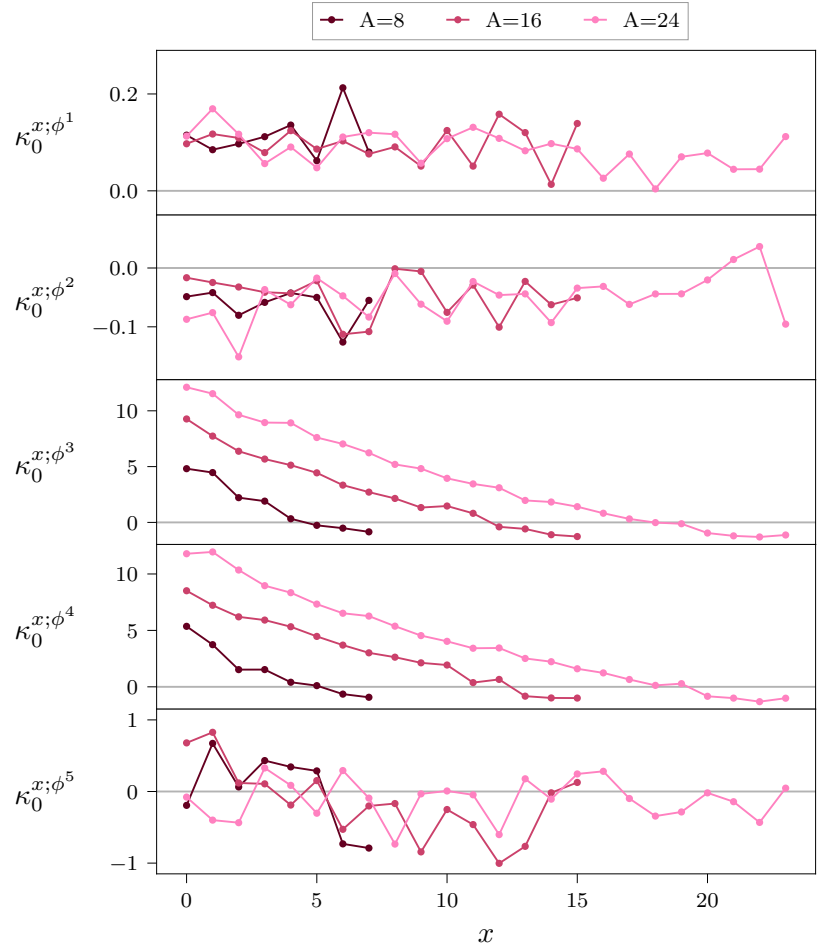

**Figure 5.19:** The optimized $\Lambda = 0$ manifold parameters for SU(3) Wilson loop observables with three choices of area $A$ on the ensemble with intermediate string tension $\sigma = 0.2$. The parameters $\kappa_0^{x;\phi^1}, \ldots, \kappa_0^{x;\phi^5}$ determine constant shifts of the azimuthal angles defining the plaquette $P_x$. Optimized values of these parameters across $x$ are consistent with the qualitative analysis based on average observable magnitude presented in the main text. The observed fluctuations in the particular optimized values on top of these trends arise from the stochastic nature of the optimization procedure. Figure adapted from Fig. 10 of Pub. [1].

These results give evidence in favor of choosing families of manifolds parameterized to allow imaginary shifts of linear combinations of variables that appear in complex exponential factors in observables, if possible. Though the change of variables to plaquettes employed here is specific to two dimensions and open boundary conditions, this suggests that deforming the corresponding degrees of freedom in higher spacetime dimensions will also reduce the average magnitude of Wilson loop observables. The additional complication arising in these theories is then to determine the corresponding deformations in terms of the underlying link variables. This can for example be achieved using gauge fixing or a scheme analogous to the invertible transformations of loops written in terms of link variables in Chapter 4.

Results at the three choices of string tension $\sigma \in \{0.1, 0.2, 0.4\}$, corresponding to three choices of the lattice spacing, allows a further investigation of the effect of the continuum limit on the performance of deformed observables. The axes of the plots shown in Figs. 5.16 and 5.17 are scaled to show a consistent range of physical Wilson loop areas in all cases. Qualitatively similar results for the variance can be seen for all three choices of parameters using these $\Lambda = 0$ deformed observables. Figures 5.20 and 5.21 show quantitative comparisons of the variance reduction achieved by $\Lambda = 0$ deformed observables as a function of physical Wilson loop area $\sigma A$. In these figures, the ratio $\mathrm{Var}[\mathrm{Re}\, W_{\mathcal{A}}^{11}] / \mathrm{Var}[\mathrm{Re}\, \mathcal{Q}_{\mathcal{A}}]$ measures the factor by which the deformed observable variance was reduced compared to the base observable variance. Broadly, we find that these variance ratios are consistent across the three choices of lattice spacings. At the larger choices of area, the SU(3) results do show variation by roughly an order of magnitude. Nonetheless, this variation is orders of magnitude smaller than the net variance reduction achieved at these Wilson loop areas. The effect of lattice spacing on the statistics of this two-dimensional lattice gauge theory may be quite different than the effects of lattice spacing in higher spacetime dimensions, and understanding both the scale and scaling of variance reduction possible in higher spacetime dimensions remains the subject of future work.

Finally, we discuss the effects of our choice of manifold parameterization on these results. As shown in the figures discussed above, the constant imaginary shift parameterization given by the $\Lambda = 0$ family of manifolds was already sufficient to produce deformed observables with significant reductions in variance compared to the base Wilson loop observables. Figures 5.22 and 5.23 compare these results to the optimized deformed observables given for higher cutoff manifolds with $\Lambda \in \{1, 2\}$. As shown in the figures, the optimized manifolds identified within these higher cutoff families do not meaningfully improve the variance reduction achieved by deformed observables. At some choices of Wilson loop area, slightly worse results can be seen for higher cutoff manifolds. The higher cutoff manifolds necessarily include lower cutoff manifolds as a subspace (one can project to a lower cutoff by simply setting the coefficients of higher cutoffs to zero), thus these differences are necessarily due to training effects. Higher cutoff manifolds include many additional parameters that can result in a higher degree of noise in the stochastic optimization of manifold parameters. Increasing the number of parameters can also cause overtraining effects to become more pronounced, and despite the early stopping employed in this study, these differences may be the result of a small amount of overtraining taking effect before training was halted. For this proof-of-principle study, we did not extensively explore possible training procedures and hyperparameters for these higher cutoff manifolds. Optimizing many-parameter models is a well-studied (but active) area of research in machine learning [490] and it may be possible to yield better results for these manifolds by implementing such methods.

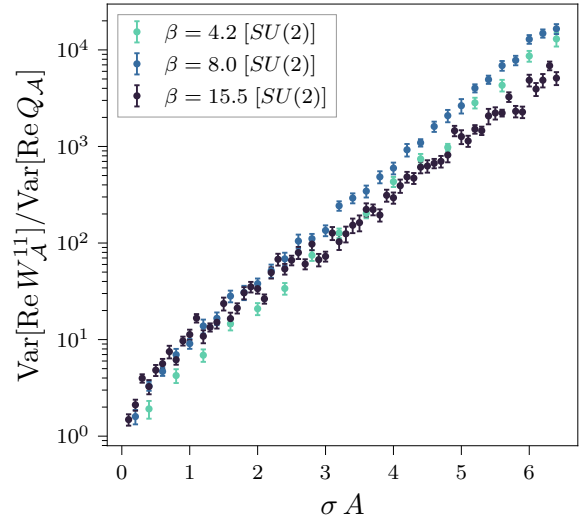

**Figure 5.20:** Ratios of the variance of base SU(2) Wilson loop observables to the variance of $\Lambda = 0$ deformed observables for ensembles at the three choices of parameter $\beta$ studied here. Figure adapted from Fig. 5 of Pub. [1].

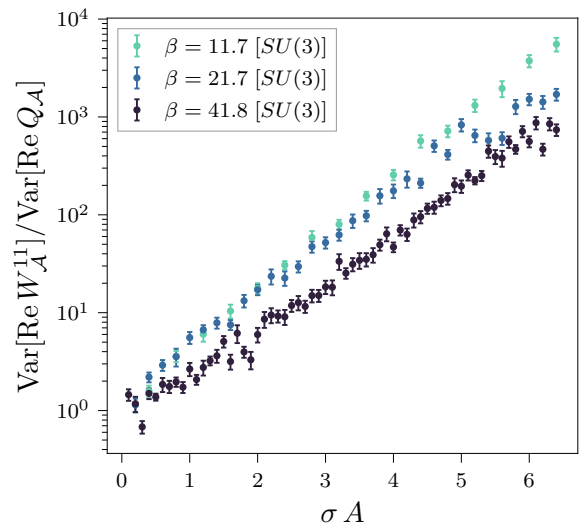

**Figure 5.21:** Ratios of the variance of base SU(3) Wilson loop observables to the variance of $\Lambda = 0$ deformed observables for ensembles at the three choices of parameter $\beta$ studied here. Figure adapted from Fig. 9 of Pub. [1].

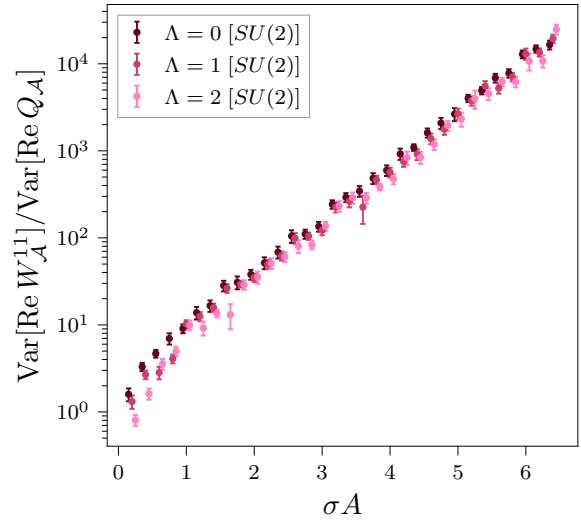

**Figure 5.22:** Ratios of the variance of base SU(2) Wilson loop observables to the variance of deformed observables determined from manifolds with three possible Fourier cutoffs $\Lambda \in \{0, 1, 2\}$. Results are shown for the ensemble with intermediate string tension $\sigma = 0.2$ corresponding to $\beta = 8.0$. For cutoffs $\Lambda \in \{1, 2\}$, a few outliers with larger uncertainties or downwards fluctuations relative to the $\Lambda = 0$ results arise from the stochastic nature of the optimization procedure. Figure adapted from Fig. 7 of Pub. [1].

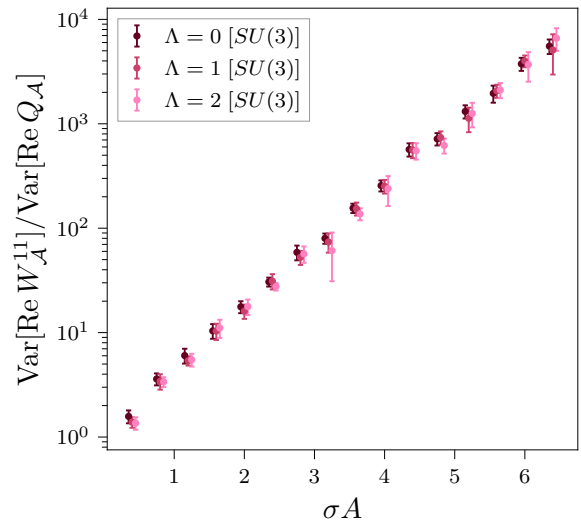

**Figure 5.23:** Ratios of the variance of base SU(3) Wilson loop observables to the variance of deformed observables determined from manifolds with three possible Fourier cutoffs $\Lambda \in \{0, 1, 2\}$. Results are shown for the ensemble with intermediate string tension $\sigma = 0.2$ corresponding to $\beta = 21.7$. A few outliers with larger uncertainties analogous to those in Fig. 5.22 can be seen for the $\Lambda \in \{1, 2\}$ results due to the stochastic nature of the optimization procedure. Figure adapted from Fig. 11 of Pub. [1].

On the other hand, the equivalence of results at all three Fourier cutoffs may also be due to a lack of expressivity in the choice of the Fourier modes included at higher cutoffs, the restriction to vertical deformations in the first place, or even due to theoretical limitations on how much variance reduction can be achieved by arbitrary deformations for these observables. It is not obvious that simplistic deformations analogous to the $\Lambda = 0$ manifolds will result in equivalent performance in applications in higher spacetime dimensions, and it would be quite surprising indeed if it were possible to take the zero-variance limit by contour deformations. Rather, one should consider this method a practical improvement upon the typical observable measurement procedure performed in lattice calculations, with choices of manifold deformations to be determined on an empirical basis. Ultimately, it is a significant practical gain to reduce the variances of observables by several orders of magnitude simply by the use of deformed observables which can be analytically shown to be equivalent to the observables under study. Determining how such results can be extended to the setting of state-of-the-art calculations is a natural and promising future direction of inquiry building upon this work.

# Chapter 6

# Conclusions and Outlook

Lattice quantum field theories provide the only general and non-perturbative approach to investigating properties of strongly coupled quantum fields. Lattice QCD in particular has played a major role in bridging the gap between the theoretical description of the QCD sector of the Standard Model and concrete experimental probes of its properties. As lattice QCD enters the era of precision calculations, control over every aspect of calculations is of utmost importance. This dissertation has addressed two main obstacles to precision lattice QFT calculations — critical slowing down and the signal-to-noise problem. The former causes the computational costs of ensemble generation to diverge as the lattice spacing is reduced, hindering controlled extrapolations to the continuum limit; the latter describes the exponentially vanishing statistical precision of estimates of most observables, including correlation functions used to extract many physical properties of interest.

To mitigate these issues in lattice QFT calculations, we introduced two new techniques in this dissertation: flow-based Markov chain Monte Carlo (Chapters 3 and 4) and observifold deformations of the path integral (Chapter 5). At their core, these are both parametric algorithms that use highly parameterized field transformations to respectively give a general approach to sampling field configurations and to computing Monte Carlo estimates of observables. Both flow-based MCMC and path integral deformations result in exact algorithms, in the sense that no systematic bias is introduced in either sampling or observable estimates. Moreover, this exactness is independent of the field transformations utilized. The choice of field transformation does, however, significantly effect the efficiency of the resulting algorithms, and optimizing the choice of field transformations is the main handle by which these parametric approaches can be tuned to give improvements over existing algorithms. In these methods, expressive families of field transformations with tractable Jacobians can be described using machine learning models, and these highly parameterized field transformations can be optimized (or 'trained') using gradient-based techniques also drawn from the machine learning domain.

In constructing flow-based models and observifold deformations, we have shown that encoding physical principles in the choice of field transformations is a key step towards identifying classes of transformations that result in efficient sampling procedures and observable estimates. For flow-based MCMC, translational symmetries and gauge symmetries correspond to large classes of degeneracies in the field theory distributions to be sampled. Exactly incorporating these symmetries into flow-based models is not necessary for correctness but improves the data efficiency of training. We also argued based on the path integral description of variance estimates that path integral deformations used to define deformed observables are more effective when observables and field variables are rewritten to give base observables the form $e^{i\phi}$, in terms of fundamental degrees of freedom $\phi$.

Physical insights also allow reductions in the overall cost of optimizing flow-based models and observifolds. We proposed several techniques to reduce the overall cost of training based on the notion of 'transfer learning', i.e. utilizing information from previously optimized models to reduce the cost of future training. Volume bootstrapping and (physical) parameter bootstrapping were introduced for flow-based models in Chapter 4, and respectively allow one to initialize model parameters from optimized models trained at smaller volumes or trained to target different choices of physical parameters (e.g. couplings or masses). For lattice field theories not strongly affected by finite-volume effects, volume bootstrapping is expected to be effective. On the other hand, parameter bootstrapping is expected to be effective when field theory distributions do not strongly depend on physical parameters or when targeting a range of related parameters, in which case training costs can be amortized. We applied a similar bootstrapping approach to optimize observifolds, in this case based on the fact that phase fluctuations of similar observables are highly correlated. Chapter 5 demonstrated that bootstrapping from manifold parameters optimized for one choice of observable to manifold parameters for other related observables could significantly reduce the number of optimization steps required for convergence of the latter.

There are several future directions in which physical principles may be utilized to improve the efficiency of the methods presented in this dissertation. For one, based on renormalization group principles, distributions over coarse-grained degrees of freedom are described by coarse-grained actions that may similarly be learned by flow-based models; using a hierarchical scheme may allow more efficient sampling by separating the task of learning correlations at widely varying length scales. Another potential direction is the use of physics-inspired prior distributions in such flow-based models; for example, gradient flow or simple free-theory distributions may provide better starting points for flows. The great deal of flexibility in the construction of field transformations for these parameterized algorithms, and the exactness guarantees independent of these choices, allows such physical insights to be freely leveraged in the flow-based and observifolds methods presented here.

Flow-based MCMC and observifolds were both demonstrated to provide promising gains in efficiency in a number of proof-of-principle applications to lattice field theories in $(1+1)$ dimensions. Chapter 3 demonstrated that flow-based MCMC could mitigate critical slowing down in a scalar $\phi^4$ theory and Chapter 4 demonstrated that topological freezing could be almost entirely resolved in a U(1) lattice gauge theory using gauge-equivariant flows. Gauge-equivariant flows were also applied to SU(2) and SU(3) lattice gauge theories in Chapter 4; though no equivalent topological freezing occurs for SU($N$) lattice gauge theories in $(1+1)$ dimensions, these flow-based models regardless achieved significant effective sample sizes corresponding to high quality approximations of the lattice gauge theory distributions. Chapter 5 then showed that observifolds could be applied to exponentially reduce the variance of Wilson loop observables in U(1), SU(2), and SU(3) lattice gauge theories.

The applications presented in this dissertation were restricted to lower dimensional spacetime lattices to investigate the benefits of these novel approaches at lower cost and in settings where many analytical results were available. However, there are no fundamental obstacles to extending the methods presented here to higher spacetime dimensions, and this is a natural continuation of this work. Along the way, there are several interesting practical questions to be studied. For flow-based models applied to lattice gauge theories in higher dimensions, it is not clear what masking patterns and choices of loops will result in the best efficiency. Extensions of the two-dimensional masking patterns applied in Chapter 4 are suitable for higher dimensions, but may not give the best coupling between degrees of freedom. The manifolds constructed for lattice gauge theory in Chapter 5 were also applied at the level of deformations to plaquette degrees of freedom. In higher spacetime dimensions, one cannot perform a one-to-one change of variables to plaquettes, and instead deformations likely must be applied at the level of link variables. Autoregressive deformations applied in this work for plaquette variables can be extended to gauge links, but it may be fruitful to instead apply techniques along the lines of the gauge-equivariant coupling layers introduced in Chapter 4 to define deformations in this case.

Applications presented here were also restricted to theories of only bosonic variables. The extension of flow-based and observifolds methods to theories with fermions is again a practical question rather than a theoretical one. In general, fermionic degrees of freedom in the path integral are analytically integrated out, resulting in purely bosonic path integrals over more complicated weights. This bosonic representation of the path integral can be sampled and deformed using identical approaches to those already applied to purely bosonic theories. However, the effects of the fermions are encoded in the determinant of a matrix with dimension scaling as the number of bosonic degrees of freedom. Computing this determinant directly to apply Metropolis corrective steps in flow-based MCMC or to evaluate deformed observables is impractical. Stochastic estimators may provide an efficient way to circumvent this issue, but studying the practical efficiency of these methods requires further investigation.

To access the continuum limit, one must also scale the number of degrees of freedom sampled by flow-based models. Section 3.7 argued that the expressivity of flow-based models must be increased in this limit, either by increasing the number of coupling layers or the expressivity of context functions. Though this is expected to increase the cost of evaluating such flow-based models, there are reasons for optimism. In particular, flow-based models based on convolutional neural networks can be described in terms of a local effective action, and increasing the expressivity of coupling layers and neural networks is expected to improve the agreement with the target action across all sites of the lattice simultaneously. In other words, the model expressivity is expected to scale extensively. How this gain in expressivity trades off against computational cost requires further empirical study in the context of particular theories of interest, but this feature suggests that an efficient scaling limit may be achieved.

If these scaling goals can be achieved — to higher spacetime dimensions, to theories with fermions, and towards the continuum limit — then we stand to make significant improvements in computational efficiency with respect to the state-of-the-art calculations of lattice field theories of phenomenological relevance. This opens the potential for finer and larger lattices to be accessed and more precise estimates of physical quantities to be made on ensembles generated at any such parameters, in all cases without the introduction of any additional systematic errors. If the improvement in statistics by several orders of magnitude achieved in the applications presented here are extended directly to such calculations, then the physics impact stated in Chapter 1 would easily be realized.

# Appendix A

# Group theory and exact integrals for U(1) and SU(N)

Due to their phenomenological relevance, the groups U(1) and SU(N) appear at various points throughout this dissertation. Properties of these groups are reviewed here to provide a single reference for the derivation of these related results. In the main text these results are used to give analytical results for comparison against the numerical approaches investigated in Chapters 3–5.

## A.1 Lie groups

Continuous groups with a manifold representation of the group space are known as *Lie groups*. The groups U(1) and SU(N) are both Lie groups. We review a few keys results in this section. This is far from a comprehensive study of the rich subject of Lie group theory; see for example Ref. [391] for one of the (many) excellent introductions to the subject.

### A.1.1 Highlights of Lie group theory

Any set $G$ is a group if it satisfies three key properties:

1. There is an associative group multiplication operation $U \cdot V = W$ under which $G$ is closed, i.e. $W \in G$ for all $U, V \in G$.[1]

2. There is an identity element 1 satisfying $1 \cdot U = U \cdot 1 = U$.

3. Every element $U$ has an inverse $U^{-1}$ defined by $U^{-1} \cdot U = U \cdot U^{-1} = 1$.

Lie groups are groups with a continuous manifold structure associated with their set of elements. Matrix groups like U(N) and SU(N) can be defined by interpreting matrices as group elements, matrix multiplication as the group multiplication operation, and matrix inversion as the group inverse operation. The group U(N) is defined by the set of $N \times N$ unitary matrices and SU($N$) is defined in the same way with the additional constraint that the determinant is 1. This gives a definition of U(1) in terms of $1 \times 1$ unitary matrices but one can equivalently define U(1) by the set of unit-norm complex scalars, as these possess an identical group structure.

---

[1] Our notation for group elements is selected to closely align with notation for group-valued variables in lattice gauge theory.

Group elements $U, V \in G$ *commute* if $U \cdot V = V \cdot U$. Groups in which all elements commute with each other are called Abelian, whereas other groups are called non-Abelian; for example, the group U(1) is an Abelian group whereas SU($N$) groups are non-Abelian. For non-Abelian groups, one can distinguish a non-trivial *center* of the group consisting of the set of elements that commute with every other group element. For $G = $ SU($N$), the center is given by a $\mathbb{Z}_N$ subgroup consisting of the elements $e^{i2\pi k/N}\mathbb{1}$ for $k \in \{1, \dots, N\}$.

For any group, *conjugacy classes* play an important role in describing the group structure. Conjugating $W \in G$ by $X \in G$ is defined by the operation $X \cdot W \cdot X^{-1}$. Conjugation is an equivalence relation, and conjugacy classes are the equivalence classes under conjugation. Any group element $W$ may be used to define a conjugacy class by the set of all elements accessible from it by conjugation,

$$C_W = \{X \cdot W \cdot X^{-1} : X \in G\}. \tag{A.1}$$

Any of the elements in $C_W$ could equivalently be used as the defining element for the same conjugacy class. In this dissertation, we use the terminology 'matrix conjugation' to mean group conjugation in a matrix group; in particular, the conjugation of the matrix $W$ by $X$ is defined to be $XWX^{-1}$, where matrix multiplication and the matrix inverse are used.

Functions that yield that same value across all elements within each conjugacy class are termed *class functions*. Common examples of class functions for matrix groups are the trace, determinant, and functions of these quantities. Characters of irreducible representations form a basis for class functions allowing a systematic decomposition, as discussed in Sec. A.1.3 below.

The group elements of a Lie group are also described by a differentiable manifold — a topological space that locally 'looks like' $\mathbb{R}^n$. The *dimension* of any Lie group is equal to the dimension $n$ of this manifold description. The group manifold can be given local coordinate charts, as with any manifold, and this plays an important role in our description of contour deformations for gauge theories in Chapter 5. Each Lie group is also associated with a Lie algebra that, loosely speaking, relates the tangent space at the identity to infinitesimal transformations about any element of the group, thus describing the local structure of the group.

Given the generally non-Euclidean structure of the group manifold, a natural question is how to assign a measure when integrating over the space. The *Haar measure*, which we simply write as $dU$, is the unique normalized measure that is invariant under left and right group multiplication,

$$dU = d(U \cdot V) = d(V \cdot U) \quad \text{and} \quad \int dU = 1, \quad \forall U, V \in G. \tag{A.2}$$

In this sense it gives the most natural uniform measure with respect to the group structure. Path integrals in lattice gauge theory are typically defined with respect to the Haar measure of gauge group elements appearing as the lattice-regularized degrees of freedom. Concrete realizations of the Haar measure are described for the groups U(1) and SU($N$) in Section A.2.

### A.1.2 Irreps, characters, and their properties

A representation $r$ of a group $G$ is defined by a homomorphic map that sends every group element $U$ to a corresponding matrix $\rho_r(U) \in \mathrm{GL}(d_r, \mathbb{C})$, where $\mathrm{GL}(d_r, \mathbb{C})$ is the group of all $d_r \times d_r$ invertible matrices and $d_r$ is called the dimension of the representation. The trivial representation is defined by the map that sends all groups elements to 1; we use the label $r = 0$ for this one-dimensional representation. Representations encountered in our application to lattice gauge theory are exclusively unitary representations, i.e. representations that maps all group elements specifically to unitary matrices $\rho_r(U) \in U(d_r) \subset \mathrm{GL}(d_r, \mathbb{C})$. Each representation $r$ can be associated with a dual representation $\bar{r}$, which for unitary representations is given by taking the complex conjugate (elementwise) of the matrices defining the representation,

$$\rho_{\bar{r}}(U) = \rho_r^*(U). \tag{A.3}$$

Irreducible representations (or 'irreps') are those representations for which the matrices in the image of the group cannot be simultaneously block-diagonalized by a change of basis, and the action of the group defined by an irrep thus cannot be factored into independent transformations of distinct subspaces. Irreps play an important role in Lie group theory as they form the building blocks of all representations. Categorizing irreps gives one a handle on arbitrary objects transforming under a Lie group and plays a particularly important role in decomposing the $(1 + 1)$D lattice gauge theory integrals discussed below.

The character of a group element $U$ in the representation $r$ is given by $\chi_r(U) \equiv \mathrm{tr}\, \rho_r(U)$. Characters are invariant under conjugation by arbitrary group elements and are thus class functions. Orthogonality properties of characters allow them to serve as 'fingerprints' of representations. Some key results are necessary for the following exact calculations:

1. The character of the dual representation $\bar{r}$ is given by complex conjugation of the character of $r$:

$$\chi_{\bar{r}}(U) = \mathrm{tr}\, \rho_{\bar{r}}(U) = \mathrm{tr}\, \rho_r^*(U) = \chi_r^*(U). \tag{A.4}$$

2. Characters of distinct irreps are orthogonal with respect to an inner product defined by integrating over the group, giving

$$\int dU \chi_r(U)\, \chi_{r'}^*(U) = \delta_{rr'} \tag{A.5}$$

for irreps $r$ and $r'$.

3. Schur orthogonality gives a more general orthogonality relation for matrix ele-

ments of irreps $r$ and $r'$:

$$\int dU \rho_r(U)_{ab} \, \rho_{r'}^*(U)_{a'b'} = \frac{1}{d_r} \delta_{aa'} \delta_{bb'} \delta_{rr'}. \tag{A.6}$$

### A.1.3  Character decompositions of class functions

The orthogonality of the characters of irreps allows any class function $f$ to be decomposed in terms of a linear combination of irrep characters as

$$f(U) = \sum_r f_r \chi_r(U). \tag{A.7}$$

The coefficients $f_r$ of the decomposition can be extracted in terms of group integrals by exploiting the orthogonality properties:

$$\int dU \chi_r^*(U) f(U) = \int dU \chi_r^*(U) \sum_{r'} f_{r'} \chi_{r'}(U) = f_r. \tag{A.8}$$

Inverting this relation, the character decomposition also allows the evaluation of integrals along the lines of the Eq. (A.8) if the coefficients $f_r$ can be extracted by some other means.

As a special case of this decomposition, an equivalent of the delta function on the space of group variables can be defined by evaluating its character coefficients according to its desired integration properties, giving

$$\delta_r = \int dU \chi_r^*(U) \delta(U) = \chi_r^*(1) = d_{\bar{r}} = d_r. \tag{A.9}$$

Thus the group delta function can be defined by its character decomposition as

$$\delta(U) \equiv \sum_r d_r \chi_r(U). \tag{A.10}$$

More details on this sort of Fourier analysis on groups can be found in Ref. [491], for example.

### A.1.4  Representations of U(1)

There is one representation of U(1) per integer $n$, given by

$$\rho_n(U) = U^n = e^{in\phi} \tag{A.11}$$

in terms of the defining unit-norm complex number representation $U = e^{i\phi}$. All functions of U(1) variables are class functions, because $XWX^{-1} = W$ for any $W, X \in$ U(1). The character decomposition in this case applies to any functions of group variables, and is just the Fourier decomposition of any function defined on the circle.

| $r$ | $d_r$ | Label |
|---|---|---|
| 0 | 1 | Trivial |
| $\{1\}$ | $N$ | Fundamental |
| $\{\bar{1}\}$ | $N$ | Anti-fundamental |
| $\{2,\bar{1}\}$ | $N^2 - 1$ | Adjoint |
| $\{2\}$ | $\frac{1}{2}(N^2 + N)$ | |
| $\{1,1\}$ | $\frac{1}{2}(N^2 - N)$ | |

**Table A.1:** Several particular 'low-lying' irreps $r$ of SU($N$) and their corresponding dimension $d_r$.

### A.1.5 Representations of SU($N$)

The representations and characters of the SU($N$) groups have been well studied in the context of strong coupling expansions; see for example Ref. [392] for a complete presentation. We here cover some of the key results required for the exact results employed in this dissertation.

Following the conventions of Ref. [392], we label irreducible representations of SU($N$) by a sequence of $N$ integers

$$r = \{r_1, r_2, \ldots, r_N\} \tag{A.12}$$

where $r_1 \geq r_2 \geq \ldots$ and $r_N = 0$.[2] For brevity, we elide any trailing zeros in the representation label, use a bar to indicate repeating labels (e.g. $\{2,\bar{1}\} \equiv \{2,1,1,\ldots,1,0\}$), and use the notation $r = 0$ for the trivial irrep $\{0,0,\ldots,0\}$. The dimension of each irrep is given in terms of this notation by

$$d_r = \prod_{i<j} \frac{r_i - r_j + j - i}{j - i}. \tag{A.13}$$

A few key irreps and their corresponding dimension are listed in Table A.1.

In terms of these labels, the character of an arbitrary group element $U \in \mathrm{SU}(N)$ with eigenvalues $\lambda_1, \ldots, \lambda_N$ is equal to

$$\chi_r(U) = \frac{\det_{ij}(\lambda_i^{r_j + N - j})}{\det_{ij}(\lambda_i^{N-j})}, \tag{A.14}$$

where $i, j \in \{1, \ldots, N\}$.

### A.2 One-dimensional integrals

We next introduce properties of the Haar measure and evaluate a few classes of one-variable group integrals with respect to the Haar measure for the groups U(1) and

---

[2]Taking $r_N$ to be arbitrary allows one to instead label the irreps of the group U($N$).

SU($N$).

### A.2.1   Integrals over U(1)

The Haar measure associated with the defining representation $U = e^{i\phi}$ is $\frac{d\phi}{2\pi}$, which satisfies the properties of the Haar measure by inspection (see Sec. 4.2 for an explicit confirmation). Integrals with respect to the Haar measure for U(1) thus reduce to integration over angular variables $\phi \in [0, 2\pi]$. We evaluate several one-dimensional integrals with respect to the Haar measure:

**Von Mises (VM) integrals.** Integrals over the von Mises distribution [492] arise commonly in U(1) lattice gauge theory in $(1+1)$D, for which the individual plaquettes follow this distribution. The distribution over $P \in$ U(1) is described by the single-variable action

$$S_{\text{VM}} = -\beta \operatorname{Re}[P] = -\frac{\beta}{2}[P + P^{-1}], \tag{A.15}$$

which gives rise to expectation values as

$$\langle \mathcal{O} \rangle_{\text{VM}} = \frac{\int dP \, \mathcal{O}(P) \, e^{\frac{\beta}{2}[P+P^{-1}]}}{\int dP \, e^{\frac{\beta}{2}[P+P^{-1}]}} = \frac{\int \frac{d\phi}{2\pi} \, \mathcal{O}(e^{i\phi}) \, e^{\beta \cos \phi}}{\int \frac{d\phi}{2\pi} \, e^{\beta \cos \phi}}. \tag{A.16}$$

The integral relation

$$\int \frac{d\phi}{2\pi} e^{in\phi} e^{\beta \cos \phi} = I_n(\beta) \tag{A.17}$$

gives the character coefficients of $e^{\beta \cos \phi}$ and allows many such expectation values to be evaluated in closed form, where $I_n(\beta)$ is the modified Bessel function of the first kind with rank $n$ [493]. We can immediately evaluate the denominator of Eq. (A.16) to get $I_0(\beta)$. The $n$th moments of the distribution can similarly be evaluated, giving

$$\left\langle e^{in\phi} \right\rangle_{\text{VM}} = \frac{I_n(\beta)}{I_0(\beta)}. \tag{A.18}$$

**Von Mises deformed observables.** Applying a vertical shift contour deformation $\phi \to \phi + i\delta$ to the action in Eq. (A.15) gives rise to deformed observables that involve factors of $e^{\beta \cos(\phi+i\delta) - \beta \cos \phi}$ (see Chapter 5). We evaluate a few common integrals required to determine exact results in this case:

1. Von Mises expectation values of the form

$$
\begin{aligned}
A_n &\equiv \left\langle e^{in\phi} | e^{\beta \cos(\phi+i\delta) - \beta \cos \phi} |^2 \right\rangle_{\text{VM}} \\
&= \frac{1}{I_0(\beta)} \int \frac{d\phi}{2\pi} \, e^{in\phi} e^{2\beta \operatorname{Re}[\cos(\phi+i\delta)] - \beta \cos \phi}
\end{aligned} \tag{A.19}
$$

can be determined by reducing to a regular Bessel function integral of the form given in Eq. (A.17). First, we manipulate the exponent to give

$$
\begin{aligned}
& 2\beta \, \mathrm{Re}[\cos(\phi + i\delta)] - \beta \cos \phi \\
& = 2\beta \, \mathrm{Re}[\cos \phi \cosh \delta - i \sin \phi \sinh \delta] - \beta \cos \phi \\
& = \beta(2 \cosh \delta - 1) \cos \phi.
\end{aligned}
\tag{A.20}
$$

Equation (A.19) can then be evaluated to give

$$
A_n = \frac{I_n(\beta(2\cosh\delta - 1))}{I_0(\beta)}.
\tag{A.21}
$$

2. Von Mises expectation values of the form

$$
\begin{aligned}
B_{2n} & \equiv \left\langle e^{2in\phi}(e^{\beta \cos(\phi + i\delta) - \beta \cos \phi})^2 \right\rangle_{\mathrm{VM}} \\
& = \frac{1}{I_0(\beta)} \int \frac{d\phi}{2\pi} \, e^{2in\phi} e^{2\beta \cos(\phi + i\delta) - \beta \cos \phi}
\end{aligned}
\tag{A.22}
$$

take a similar form to Eq. (A.19) but require somewhat more involved trigonometric manipulations. We restrict to evaluating only the even moments, as odd moments are not required for any results in this dissertation and this allows a simpler derivation. Expanding and collecting the terms in the second exponent gives

$$
\begin{aligned}
& 2\beta \cos(\phi + i\delta) - \beta \cos \phi \\
& = \beta \cos \phi(2 \cosh \delta - 1) - 2i\beta \sin \phi \sinh \delta.
\end{aligned}
\tag{A.23}
$$

Following a similar approach to the first class of integrals, we manipulate this to take the form $c_1 \cos(\phi + ic_2)$,

$$
\begin{aligned}
\beta \cos \phi(2 \cosh \delta - 1) - 2i\beta \sin \phi \sinh \delta & = c_1 \cos(\phi + ic_2) \\
& = c_1 \cos \phi \cosh c_2 - ic_1 \sin \phi \sinh c_2.
\end{aligned}
\tag{A.24}
$$

Identifying $\beta(2\cosh\delta - 1) = c_1 \cosh c_2$ and $2\beta \sinh \delta = c_1 \sinh c_2$, we can solve for both $c_1$ and $c_2$ as

$$
c_1^2 = \beta^2[(2\cosh\delta - 1)^2 - (2\sinh\delta)^2] \implies c_1 = \beta\sqrt{5 - 4\cosh\delta}
\tag{A.25}
$$

and

$$
\begin{aligned}
& c_1 e^{c_2} = \beta[(2\cosh\delta - 1) + 2\sinh\delta] \\
& \implies e^{c_2} = \frac{2e^\delta - 1}{\sqrt{5 - 4\cosh\delta}} = \sqrt{\frac{e^\delta - \frac{1}{2}}{e^{-\delta} - \frac{1}{2}}}.
\end{aligned}
\tag{A.26}
$$

The integral $B_{2n}$ can then be evaluated as

$$
\begin{aligned}
B_{2n} &= \frac{1}{I_0(\beta)} \int \frac{d\phi}{2\pi} e^{2in\phi} e^{c_1 \cos(\phi + ic_2)} = e^{2nc_2} \frac{I_{2n}(c_1)}{I_0(\beta)} \\
&= \left( \frac{e^\delta - \frac{1}{2}}{e^{-\delta} - \frac{1}{2}} \right)^n \frac{I_{2n}(\beta\sqrt{5 - 4\cosh\delta})}{I_0(\beta)}.
\end{aligned}
\tag{A.27}
$$

In this derivation we have ignored issues with taking roots of the potentially negative term $5 - 4\cosh\delta$ because the resulting ambiguities drop out of the definition of even moments $B_{2n}$, which is ultimately a real and smooth function of $\delta$.

### A.2.2 Integrals over $\mathrm{SU}(N)$

The Haar measure of an $\mathrm{SU}(N)$ variable $U$ can be marginalized to give a simple measure over the eigenvalues $\lambda_1 = e^{i\phi_1}, \ldots, \lambda_N = e^{i\phi_N}$ using the Weyl integration formula, which allows integration over class functions $f(U)$ to be written as [494]

$$
\int dU \, f(U) = \int \prod_{i=1}^N \frac{d\phi_i}{2\pi} \sum_k \delta\!\left(2\pi k - \sum_i \phi_i\right) \\
\times \mathrm{Haar}(\lambda_1, \ldots, \lambda_N) \, f(\mathrm{diag}(\lambda_1, \ldots, \lambda_N))
\tag{A.28}
$$

where

$$
\mathrm{Haar}(\lambda_1, \ldots, \lambda_N) \equiv \frac{1}{N!} \prod_{i<j} |\lambda_i - \lambda_j|^2.
\tag{A.29}
$$

This structure is used to evaluate the change in density when applying normalizing flows to the space of eigenvalues in Chapter 4. An alternative description of the Haar measure in terms of an angular parameterization of the group manifold is detailed for the groups $\mathrm{SU}(2)$ and $\mathrm{SU}(3)$ in Chapter 5.

We can sidestep the Weyl integration formula and integrate specific class functions using the character expansion and orthogonality relations:

**Matrix von Mises (MVM) integrals.** The analogue for $\mathrm{SU}(N)$ variables of the von Mises distribution is given by the single-variable action

$$
S_{\mathrm{MVM}} = -\frac{\beta}{N} \mathrm{tr}\,\mathrm{Re}[P] = -\frac{\beta}{2N} \mathrm{tr}[P + P^{-1}]
\tag{A.30}
$$

which gives rise to expectation values as

$$
\langle \mathcal{O} \rangle_{\mathrm{MVM}} = \frac{\int dP \, \mathcal{O}(P) e^{\frac{\beta}{2N} \mathrm{tr}[P + P^{-1}]}}{\int dP \, e^{\frac{\beta}{2N} \mathrm{tr}[P + P^{-1}]}}.
\tag{A.31}
$$

This can be considered an extension of the matrix von Mises distribution [495] to unitary matrices. The character coefficients of this distribution can be derived using character formulae and evaluating integrals over eigenvalues exactly to give [392]

$$f_r(\beta) \equiv \int dP \, \chi_r^*(P) \, e^{\frac{\beta}{2N} \, \text{tr}[P + P^{-1}]} = \sum_{k=-\infty}^{\infty} \det_{ij} \left( I_{r_j - j + i + k} \left( \frac{\beta}{N} \right) \right), \qquad \text{(A.32)}$$

where the indices $i, j$ iterate over $\{1, \ldots, N\}$, the components $r_j$ are elements of the list defining the irrep $r$ (see Eq. (A.12)), and as above the $I_n(\cdot)$ are modified Bessel functions of the first kind. Noting that the action in Eq. (A.30) is invariant under elementwise complex conjugation $P \to P^*$, we also have that the character coefficient of each representation is equal to the character coefficient of its dual, $f_r(\beta) = f_{\bar{r}}(\beta)$.

**Matrix von Mises non-class functions.** For explicit calculations of the SU($N$) Wilson loop signal-to-noise ratio in Chapter 5, we require integration over some non-class functions. First, expectation values of matrix-valued variables in an irrep $r$ can be computed using Schur orthogonality,

$$\begin{aligned}
\langle \rho_r(P) \rangle_{\text{MVM}}^{ab} &= \frac{1}{f_0(\beta)} \int dP \, \rho_r(P)^{ab} e^{\frac{\beta}{2N} \, \text{tr}[P + P^{-1}]} \\
&= \frac{1}{f_0(\beta)} \sum_{r'} f_{r'}(\beta) \int dP \, \rho_r(P)^{ab} \rho_{r'}(P)^{cc} \\
&= \frac{f_r(\beta)}{f_0(\beta)} \frac{1}{d_r} \delta^{ac} \delta^{bc} = \frac{f_r(\beta)}{f_0(\beta)} \frac{\delta^{ab}}{d_r}.
\end{aligned} \qquad \text{(A.33)}$$

Taking the trace over indices $ab$ recovers the character coefficient definition given above.

We also require the values of integrals over some two-variable quantities. To treat these integrals, we employ the method of birdtracks / projectors [496] to identify the index structure associated with each invariant subspace over the free indices, then utilize the Schur orthogonality property given in Eq. (A.6) to combine the results of the previously derived integrals with projectors into these subspaces. We address two such integrals:

1. We first evaluate matrix von Mises expectation values of the form

$$\begin{aligned}
C^{ab;a'b'} \equiv \left\langle P^{aa'} (P^{bb'})^* \right\rangle_{\text{MVM}} &= \frac{1}{f_0(\beta)} \int dP \, P^{aa'} (P^{bb'})^* e^{\frac{\beta}{2N} \, \text{tr}[P + P^{-1}]} \\
&= \frac{1}{f_0(\beta)} \int dP \, P^{aa'} (P^{bb'})^* \sum_r \chi_r(P) f_r(\beta).
\end{aligned} \qquad \text{(A.34)}$$

The product of matrix elements $P^{aa'} (P^{bb'})^*$ is equivalent to a matrix element of the tensor product representation $\{1\} \otimes \{\bar{1}\}$,

$$P^{aa'} (P^{bb'})^* = [\rho_1(P) \otimes \rho_{\bar{1}}(P)]^{ab;a'b'}. \qquad \text{(A.35)}$$

This tensor product representation decomposes into two invariant subspaces se-lected out by the projectors (using birdtrack notation [496])

$$\hat{P}_0^{ab;a'b'} = \frac{1}{N} \,\big)\!\big(\, = \frac{1}{N}\delta^{ab}\delta^{a'b'} \tag{A.36}$$

and

$$\hat{P}_{2,\bar{1}}^{ab;a'b'} = \;\overrightarrow{\phantom{xx}}\overleftarrow{\phantom{xx}}\; - \frac{1}{N}\,\big)\!\big(\, = \delta^{ab'}\delta^{ba'} - \frac{1}{N}\delta^{ab}\delta^{a'b'}. \tag{A.37}$$

By Schur orthogonality, these two invariant subspaces only give a non-zero integral when paired with the term with the corresponding irrep in the character decomposition in Eq. (A.34). In total, the integral can be evaluated in terms of these projectors as

$$C^{ab;a'b'} = \hat{P}_0^{ab;a'b'} + \hat{P}_{2,\bar{1}}^{ab;a'b'}\frac{f_{2,\bar{1}}(\beta)}{f_0(\beta)}, \tag{A.38}$$

where we used the fact that $f_r(\beta) = f_{\bar{r}}(\beta)$ for the matrix von Mises character coefficients.

2. We secondly evaluate matrix von Mises expectation values of the form

$$\begin{aligned} D^{ab;a'b'} \equiv \left\langle P^{aa'}P^{bb'}\right\rangle_{\mathrm{MVM}} &= \frac{1}{f_0(\beta)}\int dP\, P^{aa'}P^{bb'}e^{\frac{\beta}{2N}\mathrm{tr}[P+P^{-1}]}\\ &= \frac{1}{f_0(\beta)}\int dP\, P^{aa'}P^{bb'}\sum_r \chi_r(P)f_r(\beta). \end{aligned} \tag{A.39}$$

The tensor product of representations $\{1\} \otimes \{1\}$ similarly decomposes into two irreps selected out by the projectors

$$\begin{aligned} \hat{P}_2^{ab;a'b'} &= \frac{1}{2}\overrightarrow{\phantom{xx}}\overrightarrow{\phantom{xx}} + \frac{1}{2}\big)\!\!\times\!\!\big( = \frac{1}{2}\delta^{ab'}\delta^{ba'} + \frac{1}{2}\delta^{aa'}\delta^{bb'}\\ \hat{P}_{1,1}^{ab;a'b'} &= \frac{1}{2}\overrightarrow{\phantom{xx}}\overrightarrow{\phantom{xx}} - \frac{1}{2}\big)\!\!\times\!\!\big( = \frac{1}{2}\delta^{ab'}\delta^{ba'} - \frac{1}{2}\delta^{aa'}\delta^{bb'}. \end{aligned} \tag{A.40}$$

Applying Schur orthogonality, these two irreps extract two terms from the character expansion giving

$$D^{ab;a'b'} = \hat{P}_2^{ab;a'b'}\frac{f_2(\beta)}{f_0(\beta)} + \hat{P}_{1,1}^{ab;a'b'}\frac{f_{1,1}(\beta)}{f_0(\beta)}. \tag{A.41}$$

## A.3   Lattice gauge theory in (1+1)D

Finally, we use the properties covered in the previous sections to derive exact results for $(1+1)$D lattice gauge theory using the Wilson gauge action with the gauge group U(1) or SU($N$), following a similar approach to Refs. [38, 39, 392]. The Wilson gauge action specialized to two dimensions (see Sec. 2.2 for the general case) is defined for

either gauge group by

$$S(U; \beta) = -\frac{\beta}{2N} \sum_x \text{tr} \left[ P_{01}(x) + P_{01}^{-1}(x) \right], \tag{A.42}$$

where

$$P_{01}(x) = U_0(x) \, U_1(x + \hat{0}) \, U_0^\dagger(x + \hat{1}) \, U_1^\dagger(x). \tag{A.43}$$

For the $U(1)$ gauge group, the trace is assumed to act trivially on the scalar-valued plaquettes $P_{01}(x)$ and $N$ is set to 1. We derive results for two choices of the lattice geometry and boundary conditions:

1. We first consider $(L+1) \times (L+1)$ lattices with open boundary conditions (OBCs). The path integral degrees of freedom are the $L(L+1)$ temporal gauge links

$$\{U_0(x) : x_0 \in \{1, \ldots, L\}, x_1 \in \{1, \ldots, L+1\}\} \tag{A.44}$$

   and the $L(L+1)$ spatial gauge links

$$\{U_1(x) : x_0 \in \{1, \ldots, L+1\}, x_1 \in \{1, \ldots, L\}\}. \tag{A.45}$$

   In this case, the sum over $x$ in Eq. (A.42) runs over $x_0, x_1 \in \{1, \ldots, L\}$. This excludes any plaquettes that would extend beyond the last site of the lattice in each direction, resulting in the inclusion of $L^2$ plaquettes in the action.

2. We then consider $L \times L$ lattices with periodic boundary conditions (PBCs). The path integral degrees of freedom are the usual set of $2L^2$ gauge links $U_\mu(x)$ with $\mu \in \{0, 1\}$ and $x_0, x_1 \in \{1, \ldots, L\}$. In this case, the sum over $x$ in Eq. (A.42) runs over all $L^2$ sites of the lattice. The coordinates $x_0 = L+1$ and $x_1 = L+1$ are replaced by $x_0 = 1$ and $x_1 = 1$, respectively, in the definition of plaquettes at the edges of the lattice. This geometry also includes $L^2$ plaquettes in the action.

The lattice extents in both cases are chosen to result in an equivalent number of plaquettes for simplicity. Figure A.1 depicts the respective lattice geometries for the example of $L = 3$, for which the OBC and PBC lattices both involve 9 plaquettes. The number of sites differs in the two cases, however, with the PBC lattice containing 9 sites while the OBC lattice contains 16 sites.

**Gauge fixing.** Our strategy for demonstrating exact results in $(1+1)\text{D}$ is to gauge fix then change variables from the remaining links to plaquettes. In both cases, we apply a maximal-tree gauge fixing to fix a maximal set of temporal links and an additional set of spatial links at $x_0 = 0$ to unity. For OBCs, we can fix $U_0(x) = 1$ for all $x$ and additionally $U_1(x) = 1$ for sites with $x_0 = 0$. In the $(L+1) \times (L+1)$ geometry described above, a total of

$$N_{\text{fix}}^{\text{OBC}} = L(L+1) + L = L^2 + 2L \tag{A.46}$$

links are gauge fixed in this way. For PBCs, we can only fix $U_0(x) = 1$ for $x_0 < L$ (the additional link at $x_0 = L$ would form a closed loop around the periodic temporal boundary) and $U_1(x) = 1$ for $x_0 = 1$ and $x_1 < L$. In the $L \times L$ geometry described for PBCs, a total of

$$N_{\text{fix}}^{\text{PBC}} = L(L-1) + L - 1 = L^2 - 1 \tag{A.47}$$

links are gauge fixed in this way. Examples of these subsets of gauge-fixed links are shown for both OBCs and PBCs in Fig. A.1.

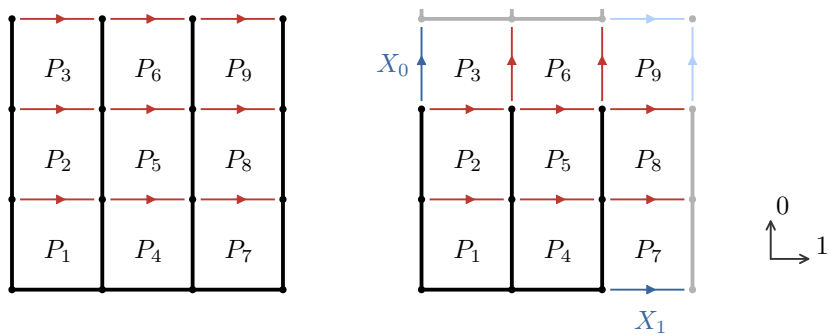

**Figure A.1:** Open boundary conditions (left) vs. periodic boundary conditions (right). Gauge-fixed links are indicated in bold in both cases. Remaining unfixed links corresponding to contractible Wilson loops are indicated in red, while unfixed links corresponding to non-contractible Polyakov loops are indicated in blue. In the right diagram, degrees of freedom related by the lattice periodicity to degrees of freedom already shown are indicated in lighter colors. In the case of periodic boundary conditions, the nine plaquette variables and two Polyakov loop variables are constrained by $P_3 P_2 P_1 P_6 P_5 P_4 P_9 P_8 P_7 X_1 X_0 X_1^\dagger X_0^\dagger = 1$, whereas the nine plaquette variables are independent in the case of open boundary conditions.

**Changing variables (OBCs).** For lattices with OBCs, there is a one-to-one correspondence between the $L^2$ plaquettes and the

$$2L(L+1) - N_{\text{fix}}^{\text{OBC}} = (2L^2 + 2L) - (L^2 + 2L) = L^2 \tag{A.48}$$

links remaining after gauge fixing. Gauge-fixed plaquettes are defined by evaluating the usual definition in Eq. (A.43) in terms of links $U_\mu(x)$ that have been gauge fixed, giving

$$P_{01}(x) = \begin{cases} U_1(x + \hat{0}) & x_0 = 1 \\ U_1(x + \hat{0}) U_1^\dagger(x) & x_0 \neq 1. \end{cases} \tag{A.49}$$

This relation can be inverted by sequentially solving for links $U_1(x)$ at sites with $x_0 = 2$, then at sites with $x_0 = 3$, etc. Explicitly, we use the recursion $U_1(x) = P_{01}(x-\hat{0})U_1(x-\hat{0})$ to find

$$U_1(x) = \prod_{i=1}^{x_0-1} P_{01}(\{x_0 - i, x_1\}), \tag{A.50}$$

where the order of the product indicates the order of matrix multiplications from left to right.

**Changing variables (PBCs).** There are a total of

$$2L^2 - N_{\text{fix}}^{\text{PBC}} = 2L^2 - (L^2 - 1) = L^2 + 1 \tag{A.51}$$

links remaining after applying gauge fixing in the PBC case. Two of these degrees of freedom must be identified with the non-contractible Polyakov loops extending through the periodic boundaries in the temporal and spatial directions. We choose

$$X_0 \equiv U_0(x)|_{x_0=L, x_1=1} \quad \text{and} \quad X_1 \equiv U_1(x)|_{x_0=1, x_1=L} \tag{A.52}$$

as the Polyakov loop variables in the $\hat{0}$ and $\hat{1}$ directions, respectively. These choices are shown in the examples in Fig. A.1.

Having subtracted two degrees of freedom, we are left with only $L^2 - 1$ remaining gauge links but have $L^2$ plaquette degrees of freedom. The difference is accounted for by one constraint between the set of all plaquettes and Polyakov variables. There are several ways to write this constraint; we choose a form that will result in the simplest manipulations for future calculations. First, we note that plaquettes $P_{01}(x)$ with $x_0, x_1 < L$ are related to links by the same change of variables presented in Eq. (A.50). A similar solution applies for $x_1 = L$, but requires the inclusion of $X_1$,

$$U_1(\{x_0, L\}) = \left[ \prod_{i=1}^{x_0-1} P_{01}(\{x_0 - i, L\}) \right] X_1. \tag{A.53}$$

The constraint can then be expressed as

$$1 = \left[ \prod_{x_1=1}^{L} P_{01}(\{L, x_1\}) \, U_1(\{L, x_1\}) \right] X_0 X_1^\dagger X_0^\dagger, \tag{A.54}$$

where the links $U_1(\cdot)$ can be substituted for plaquette and Polyakov variables using Eqs. (A.50) and (A.53). When the dust settles, we are left with

$$1 = \left[ P_{01}(\{L, 1\}) \, P_{01}(\{L-1, 1\}) \dots \right] X_1 X_0 X_1^\dagger X_0^\dagger, \tag{A.55}$$

where each plaquette appears exactly once in the term in square brackets. The order of these plaquette terms is ultimately irrelevant. To handle this constraint when changing variables to plaquettes in the path integral, we include the character sum $\sum_r d_r C_r$ as an additional factor in the integrand, where

$$C_r = \chi_r \big( P_{01}(\{L, 1\}) \, P_{01}(\{L-1, 1\}) \dots X_1 X_0 X_1^\dagger X_0^\dagger \big). \tag{A.56}$$

In this construction we have essentially related a definition of the $L \times L$ Wilson loop in terms of a product of plaquette variables to a definition in terms of the Polyakov loop variables $X_0$ and $X_1$. Physically, this constraint reflects Gauss' law acting on the single closed surface formed by the toroidal geometry of the PBC lattice. For the OBC lattice geometry, there are no closed surfaces and no constraints. This also clarifies why exact results cannot be straightforwardly extended to higher dimensions: in higher spacetime dimensions the number of Gauss' law constraints scales with the volume, and it would be exponentially difficult to evaluate analogous sums.

To complete the change of variables for PBCs, we write the path integral in terms of the $L^2$ plaquette variables and two Polyakov variables $X_0$ and $X_1$, giving

$$\langle \mathcal{O} \rangle = \frac{1}{Z} \sum_r d_r \int dX_0 dX_1 \prod_x dP_{01}(x) \, C_r \, \mathcal{O} \, e^{\frac{\beta}{2N} \operatorname{tr}[P + P^{-1}]} \tag{A.57}$$

and

$$Z = \sum_r d_r \int dX_0 dX_1 \prod_x dP_{01}(x) \, C_r \, e^{\frac{\beta}{2N} \operatorname{tr}[P + P^{-1}]}. \tag{A.58}$$

To extend OBC path integral results to the case of PBCs below, we include this character sum where relevant.

**Evaluating OBC observables.** For the OBC lattice geometry, the path integral can be written using integration over the set of plaquette variables $P_{01}(x)$ after gauge fixing. The Wilson gauge action also factorizes, and any gauge-invariant observables that can similarly be factored can be evaluated in terms of the one-dimensional integrals given in Sec. A.2. Observables must be gauge invariant to apply the change of variables without affecting the form of the observable itself, but more general observables can be computed if gauge fixing is combined with integration over the factored pure-gauge degrees of freedom. Such gauge-variant observables are not considered in this work.

This factorization immediately allows us to evaluate expectation values of rectangular Wilson loops, which are needed for some results in Chapter 5. These Wilson loops are defined by

$$W_{\mathcal{A}} = \prod_{x, \mu \in \partial \mathcal{A}} U_\mu(x), \tag{A.59}$$

where $\partial \mathcal{A}$ is the ordered set of links bounding a rectangular $l_0 \times l_1$ region $\mathcal{A}$. For simplicity, we exploit translational symmetry to shift the region $\mathcal{A}$ to be rooted at the origin. In gauge-fixed configurations, the only unfixed links in this loop are then given by $U_1(\{l_0, x_1\})$ for $x_1 \in \{1, \ldots, l_1\}$. Using Eq. (A.50) to replace $U_1(\{l_0, x_1\})$ with the associated plaquette degrees of freedom, the Wilson loop can be rewritten as a product over plaquette variables contained within the region $\mathcal{A}$ as

$$W_{\mathcal{A}} = \prod_{x \in \mathcal{A}} P_{01}(x), \tag{A.60}$$

where the product over $x$ is ordered based on Eq. (A.50) and Eq. (A.59). Ultimately, the order is irrelevant for the expectation value, which factorizes over each plaquette as

$$\begin{aligned} \left\langle W_{\mathcal{A}}^{ab} \right\rangle &= \left\langle P_{01}^{a z_1}(x_1) P_{01}^{z_1 z_2}(x_2) \ldots P_{01}^{z_{A-1} b}(x_A) \right\rangle \\ &= \langle P^{a z_1} \rangle_{1 \mathrm{var}} \left\langle P^{z_1 z_2} \right\rangle_{1 \mathrm{var}} \cdots \left\langle P^{z_{A-1} b} \right\rangle_{1 \mathrm{var}} \\ &= \delta^{ab} \left[ \frac{f_1(\beta)}{f_0(\beta)} \right]^A, \end{aligned} \tag{A.61}$$

where $A = l_0 l_1$ is the area of region $\mathcal{A}$ and $\langle \cdot \rangle_{1\text{var}}$ indicates an average with respect to the von Mises or matrix von Mises distributions for $U(1)$ and $SU(N)$ gauge groups, respectively. For the $U(1)$ gauge group, these results also apply to Wilson loops over more general regions $\mathcal{A}$ of the same area $A$ by the Abelian nature of the group.

**Evaluating PBC observables.** Extending exact results to the PBC lattice geometry requires handling the additional character sum constraining the plaquette and Polyakov degrees of freedom, as given in Eq. (A.57). The denominator $Z$ of this path integral can be evaluated once and for all. We first evaluate the integration over $X_0$ and $X_1$, using character orthogonality rules to arrive at

$$\int dX_0 dX_1 \, \rho_r(X_1)^{ab} \, \rho_r(X_0)^{bc} \, \rho_r(X_1^\dagger)^{cd} \, \rho_r(X_0^\dagger)^{de} = \frac{1}{d_r^2} \delta^{ae}. \quad (A.62)$$

The remainder of the path integral denominator can be evaluated by factorizing into one-variable integrals, giving

$$\begin{aligned}
Z &= \sum_r d_r \frac{1}{d_r^2} \operatorname{tr}\left[ \left( \int dP \, \rho_r(P) e^{\frac{\beta}{2N} \operatorname{tr}[P + P^{-1}]} \right)^{L^2} \right] \\
&= \sum_r \frac{1}{d_r} \operatorname{tr}\left[ \left( \mathbb{1} \frac{f_r(\beta)}{d_r} \right)^{L^2} \right] = \sum_r \left[ \frac{f_r(\beta)}{d_r} \right]^{L^2}.
\end{aligned} \quad (A.63)$$

This summation is exponentially dominated by the trivial irrep $r = 0$ because $f_r(\beta) < f_0(\beta)$ for all $r \neq 0$ [392]. For the choices of $\beta$ studied in the main text, $Z$ is approximated very precisely by the $r = 0$ term alone.

The numerator of the path integral can be computed similarly for particular choices of observables:

- Wilson loops $W_\mathcal{A}$: We make the same assumptions as in the case of OBCs to work with rectangular regions $\mathcal{A}$ rooted at the origin. The observable $W_\mathcal{A} = \prod_{x \in \mathcal{A}} P_{01}(x)$ thus introduces one extra fundamental-irrep factor $P_{01}(x)$ for each plaquette within $\mathcal{A}$ into the path integral numerator. Integration over all other plaquettes proceeds as in the case of the path integral denominator, giving factors of $\mathbb{1} f_r(\beta)/d_r$. The identity matrix factors drop out of the product in the constraint $C_r$. The terms involving $X_0$ and $X_1$ can also be resolved as in the case of the path integral denominator. Finally, the path integral reduces to integration over plaquettes restricted to the region $\mathcal{A}$,

$$\begin{aligned}
\langle W_\mathcal{A} \rangle &= \frac{1}{Z} \sum_r \frac{1}{d_r} \left[ \frac{f_r(\beta)}{d_r} \right]^{L^2 - A} \int \prod_{x \in \mathcal{A}} dP_{01}(x) \\
&\quad \left[ \prod_{x \in \mathcal{A}} P_{01}(x) \right] \chi_r \left( \prod_{x \in \mathcal{A}} P_{01}(x) \right) e^{\frac{\beta}{2N} \sum_{x \in \mathcal{A}} \operatorname{tr}[P_{01}(x) + P_{01}^{-1}(x)]}.
\end{aligned} \quad (A.64)$$

First, this expression can be evaluated straightforwardly for $G = U(1)$. The irreducible representations of $U(1)$ are one-dimensional and labeled by integers. Products in the expression for $W_{\mathcal{A}}$ can also be treated as unordered because $U(1)$ is Abelian. This allows a straightforward rewriting in terms of simple one-variable integrals, giving

$$
\begin{aligned}
\langle W_{\mathcal{A}} \rangle_{U(1)} &= \frac{1}{Z} \sum_r I_r(\beta)^{L^2-A} \left( \int dP \, P^{r+1} e^{\frac{\beta}{2}[P+P^{-1}]} \right) \\
&= \frac{\sum_r I_r(\beta)^{L^2-A} I_{r+1}(\beta)^A}{\sum_r I_r(\beta)^{L^2}}.
\end{aligned}
\tag{A.65}
$$

The summation over $r$ converges very quickly as irreps with larger $|r|$ are included in the sum, and can in practice be truncated based on the desired precision.

One can also evaluate for $G = SU(N)$ in terms of (somewhat tedious) one-variable integrals that can be resolved by the character decomposition, decomposition of tensor-product irreps, and Schur orthogonality. However, for the applications in the main text, we work with $A \ll L^2$ and values of $\beta$ small enough that this expression can be approximated to very high precision using just the trivial irrep $r = 0$. After truncating both the numerator and denominator to $r = 0$, the value of the PBC Wilson loop is the same as the value given for OBCs in Eq. (A.61).

- Powers of plaquettes $P^n$ [$G = U(1)$]: In the application to $U(1)$ lattice gauge theory in Chapter 4, we additionally measure $\langle P^n \rangle$ for several choices of the exponent $n$. Using similar simplifications to those applied in Eq. (A.65), we find

$$
\langle P^n \rangle_{U(1)} = \frac{\sum_r I_r(\beta)^{L^2-1} I_{r+n}(\beta)}{\sum_r I_r(\beta)^{L^2}}.
\tag{A.66}
$$

- Polyakov loops $\eta_0$ and $\eta_1$: We also investigate observables involving Polyakov loops winding around the temporal or spatial boundaries, given by

$$
\eta_0(x_1) = \text{tr} \left[ \prod_{x_0} U_0(\{x_0, x_1\}) \right] \quad \text{and} \quad \eta_1(x_0) = \text{tr} \left[ \prod_{x_1} U_1(\{x_0, x_1\}) \right].
\tag{A.67}
$$

After gauge fixing and translating to the origin, these Polyakov loops correspond to the traced Polyakov variables, $\eta_0(0) = \text{tr} \, X_0$ and $\eta_1(0) = \text{tr} \, X_1$. The Wilson gauge action satisfies center symmetry and the expectation values of the Polyakov loops must therefore vanish,

$$
\langle \eta_0 \rangle = \langle \eta_1 \rangle = 0.
\tag{A.68}
$$

We can also check this by direct integration. The numerator of the path integral for $\langle X_0^{ab} \rangle$ includes the integral

$$
\int dX_0 \, X_0^{ab} \, \rho_r(X_0)^{cd} \, \rho_r(X_0^{\dagger})^{ef}
\tag{A.69}
$$

which vanishes for any irrep $r$ for the groups $U(1)$ and $SU(N)$.

On the other hand, $\left\langle |\eta_0|^2 \right\rangle$ and $\left\langle |\eta_1|^2 \right\rangle$ do not vanish identically. Resolving all plaquette integrals in the path integral definition gives

$$
\begin{aligned}
\left\langle |\eta_0|^2 \right\rangle &= \frac{1}{Z} \sum_r d_r \left[ \frac{f_r(\beta)}{d_r} \right]^{L^2} \int dX_0 dX_1 \, | \operatorname{tr} X_0|^2 \, \chi_r \left( X_1 X_0 X_1^\dagger X_0^\dagger \right) \\
&= \frac{1}{Z} \sum_r \left[ \frac{f_r(\beta)}{d_r} \right]^{L^2} \int dX_0 \, | \operatorname{tr} X_0|^2 \, |\chi_r(X_0)|^2 .
\end{aligned}
\tag{A.70}
$$

A similar derivation gives $\left\langle |\eta_1|^2 \right\rangle = \left\langle |\eta_0|^2 \right\rangle$. The remaining one-dimensional integral is trivial for $G = U(1)$, in which case we find $\left\langle |\eta_0|^2 \right\rangle_{U(1)} = 1$. For $SU(N)$, we must decompose the tensor product irrep $\{1\} \otimes \{\bar{1}\} \otimes r \approx \bar{r}$ arising from the product in the integrand. This is straightforward for particular $r$, and the sum can be evaluated numerically by truncating to a highest weight irrep such that cutoff effects scaling as $[f_r(\beta)/(f_0(\beta) d_r)]^{L^2}$ are small. For the choices of $\beta$ and $L$ used in applications in the main text, $r = 0$ is sufficient to give a high-precision estimate. This leading term contributes $\int dX_0 | \operatorname{tr} X_0|^2 = 1$, giving $\left\langle |\eta_0|^2 \right\rangle_{SU(N)} \approx 1$.

- Topological charge $Q$ and susceptibility $\chi_Q$ [$G = U(1)$]: For the $U(1)$ gauge group, an integer topological charge can be defined by

$$
Q = \frac{1}{2\pi} \sum_x \arg\left( P_{01}(x) \right)
\tag{A.71}
$$

where $\arg(P) \in [-\pi, \pi]$. By charge conjugation symmetry, $\langle Q \rangle = 0$ for all values of the coupling parameter $\beta$. The topological susceptibility $\chi_Q = \langle Q^2/L^2 \rangle$ captures the width of this distribution about zero. We first define a procedure to numerically compute the marginal probability density $p(Q)$ more generally, which can then be integrated with respect to $Q^2/L^2$ to arrive at the topological susceptibility. This avenue allows other properties of the topological charge distribution to be calculated as well, if desired.

Defining $\phi_x \equiv \arg\left( P_{01}(x) \right)$, the probability density over $Q$ is given by the path integral

$$
p(Q) = \frac{1}{Z} \int \prod_x dP_{01}(x) \, \delta\left( \sum_x \phi_x - 2\pi Q \right) e^{\frac{\beta}{2} \sum_x [P_{01}(x) + P_{01}^{-1}(x)]} .
\tag{A.72}
$$

The Fourier representation of the Dirac delta allows the integrals over plaquette variables to be evaluated as

$$
\begin{aligned}
p(Q) &= \frac{1}{Z} \int \prod_x dP_{01}(x) \int_{-\infty}^{\infty} dk \, e^{i(\sum_x \phi_x - 2\pi Q)k} e^{\frac{\beta}{2} \sum_x [P_{01}(x) + P_{01}^{-1}(x)]} \\
&= \frac{1}{Z} \int_{-\infty}^{\infty} dk \, [I_k(\beta)]^{L^2} e^{2\pi i Q k} = \int_{-\infty}^{\infty} dk \left[ \frac{I_k(\beta)}{I_0(\beta)} \right]^{L^2} e^{2\pi i Q k},
\end{aligned}
\tag{A.73}
$$

where the Bessel functions $I_k(\cdot)$ with non-integer index $k$ are defined by analytic continuation [493]. The remaining one-dimensional integral can be evaluated numerically for any given $Q$. The density $p(Q)$ falls off rapidly with $|Q|$, and precise averages can be computed with respect to the distribution by truncating $|Q|$ to achieve the desired precision.

# Software References

The numerical work presented in this dissertation was made possible by several software packages and libraries. These are listed below and cited inline where appropriate.